\documentclass[aps,prd,secnumarabic,eqsecnum,titlepage,superscriptaddress,tightenlines,floatfix]{revtex4-2}
\usepackage{xcolor}
\usepackage[colorlinks=true,linkcolor=blue,filecolor=blue,urlcolor=blue,citecolor=blue,plainpages=false]{hyperref}
\usepackage{graphicx}
\usepackage{bm,amsmath,amssymb}
\usepackage[left]{lineno}
\newcommand{\ben}{\begin{eqnarray*}}
\newcommand{\een}{\end{eqnarray*}}

\newcommand{\dhd}{{\textstyle d}
\lower.03ex\hbox{\kern-0.38em$^{\scriptstyle-}$}\kern-0.05em{}}
\newcommand{\dbar}{{\textstyle \delta}
\lower.03ex\hbox{\kern-0.38em$^{\scriptstyle-}$}\kern-0.05em{}}

\begin{document}
\title{Snowmass 2021 White Paper: \\ Electron Ion Collider for High Energy Physics}
\hfill FERMILAB-PUB-22-125-QIS-SCD-T
\author{R.~Abdul~Khalek}
\affiliation{Thomas Jefferson National Accelerator Facility, Newport News, VA 23606, USA}
\author{U.~D'Alesio}
\affiliation{Dipartimento di Fisica, Universit\`a di Cagliari, Cittadella Universitaria, I-09042 Monserrato (CA), Italy}
\affiliation{INFN, Sezione di Cagliari, Cittadella Universitaria, I-09042 Monserrato (CA), Italy}
\author{Miguel Arratia}
\thanks{Editor}
\affiliation{Department of Physics and Astronomy, University of California, Riverside, California, 92521, USA}
\affiliation{Thomas Jefferson National Accelerator Facility, Newport News, Virginia 23606, USA}
\author{A.~Bacchetta}
\affiliation{Dipartimento di Fisica, Universi\`a di Pavia, INFN-Sezione Pavia, I-27100 Pavia, Italy}
\author{M.~Battaglieri}
\affiliation{INFN, Sezione di Genova, 16146 Genova, Italy}
\affiliation{Thomas Jefferson National Accelerator Facility, Newport News, VA 23606, USA}
\author{M.~Begel}
\affiliation{Physics Department, Brookhaven National Laboratory, Upton, New York 11973, USA}
\author{M.~Boglione}
\affiliation{Dipartimento di Fisica, Universit\`a di Torino, INFN-Sezione Torino, Italy}
\author{R.~Boughezal}
\affiliation{High Energy Physics Division, Argonne National Laboratory, Argonne, IL 60439, USA}
\author{R.~Boussarie}
\thanks{Editor}
\affiliation{CPHT, CNRS, Ecole Polytechnique, Institut Polytechnique de Paris, 91128 Palaiseau, France}
\author{G.~Bozzi}
\affiliation{Dipartimento di Fisica, Universit`a di Cagliari, Cittadella Universitaria, I-09042 Monserrato (CA), Italy}
\affiliation{INFN, Sezione di Cagliari, Cittadella Universitaria, I-09042 Monserrato (CA), Italy}
% ...
\date{\today}
\maketitle
%
%
%
%
\centerline{\textbf{ABSTRACT}}
\vspace{3ex}
Electron Ion Collider (EIC) is a particle accelerator facility planned for construction at Brookhaven National Laboratory on Long Island, New York by the United States Department of Energy. EIC will provide capabilities of colliding beams of polarized electrons with polarized beams of proton and light ions. EIC will be one of the largest and most sophisticated new accelerator facilities worldwide, and the only new large-scale accelerator facility planned for construction in the United States in the next few decades. The versatility, resolving power and intensity of EIC will present many new opportunities to address some of the crucial and fundamental open scientific questions in particle physics. This document provides an overview of the science case of EIC from the perspective of the high energy physics community.

\newpage
%
\tableofcontents
\newpage
%
%
\section{Introduction} \label{sec:intro}
\begin{figure}[!htp]
    \centering
    \includegraphics[width=0.6\textwidth, angle=-90]{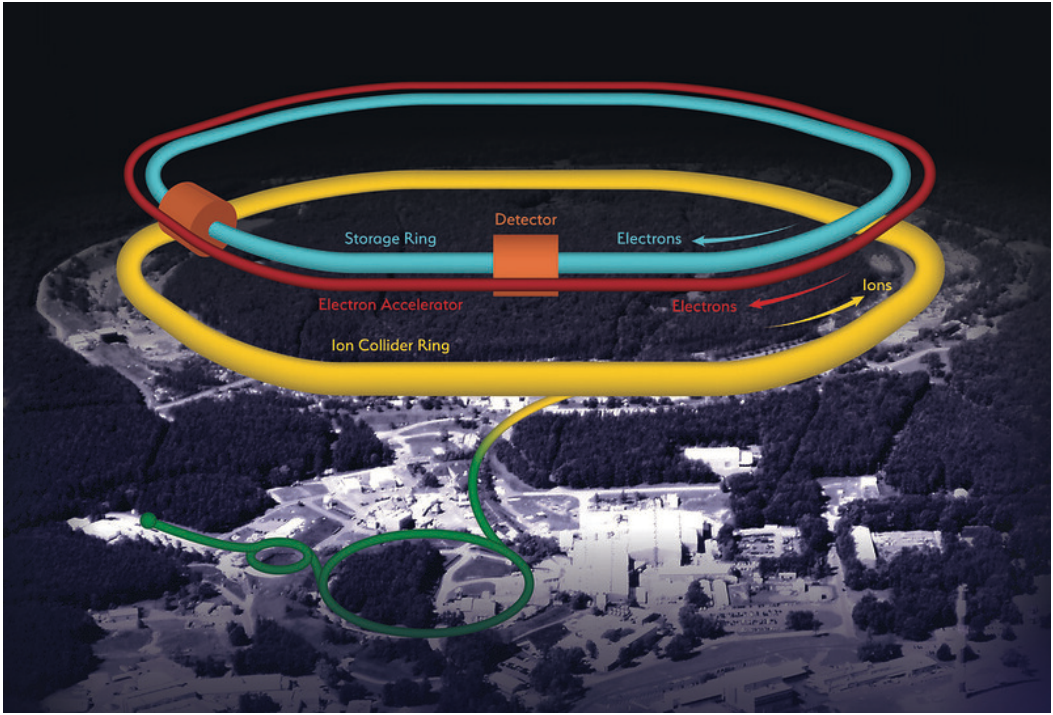}
    \caption{Schematic perspective of the Electon Ion Collider (EIC).}
    \label{fig:intro_eic1}
\end{figure}

The Electron-Ion Collider (EIC) is a particle accelerator facility planned for construction at Brookhaven National Laboratory (BNL) on Long Island, New York by the United States Department of Energy to address the following key science questions:
\begin{itemize}
    \item How do the nucleonic properties such as mass and spin emerge from partons and their underlying interactions?
    \item How are partons inside the nucleon distributed in both momentum and position space?
    \item How do color-charged quarks and gluons, and jets, interact with a nuclear medium? How do the confined hadronic states emerge from these quarks and gluons? How do the quark-gluon interactions create nuclear binding?
\end{itemize}
In addressing these questions, the EIC will undertake a broad program of fundamental QCD measurements, with serious implications for current and future
HEP programs, as we explore in the sections below.

\begin{figure}[!tp]
    \centering
    \includegraphics[width=0.8\textwidth]{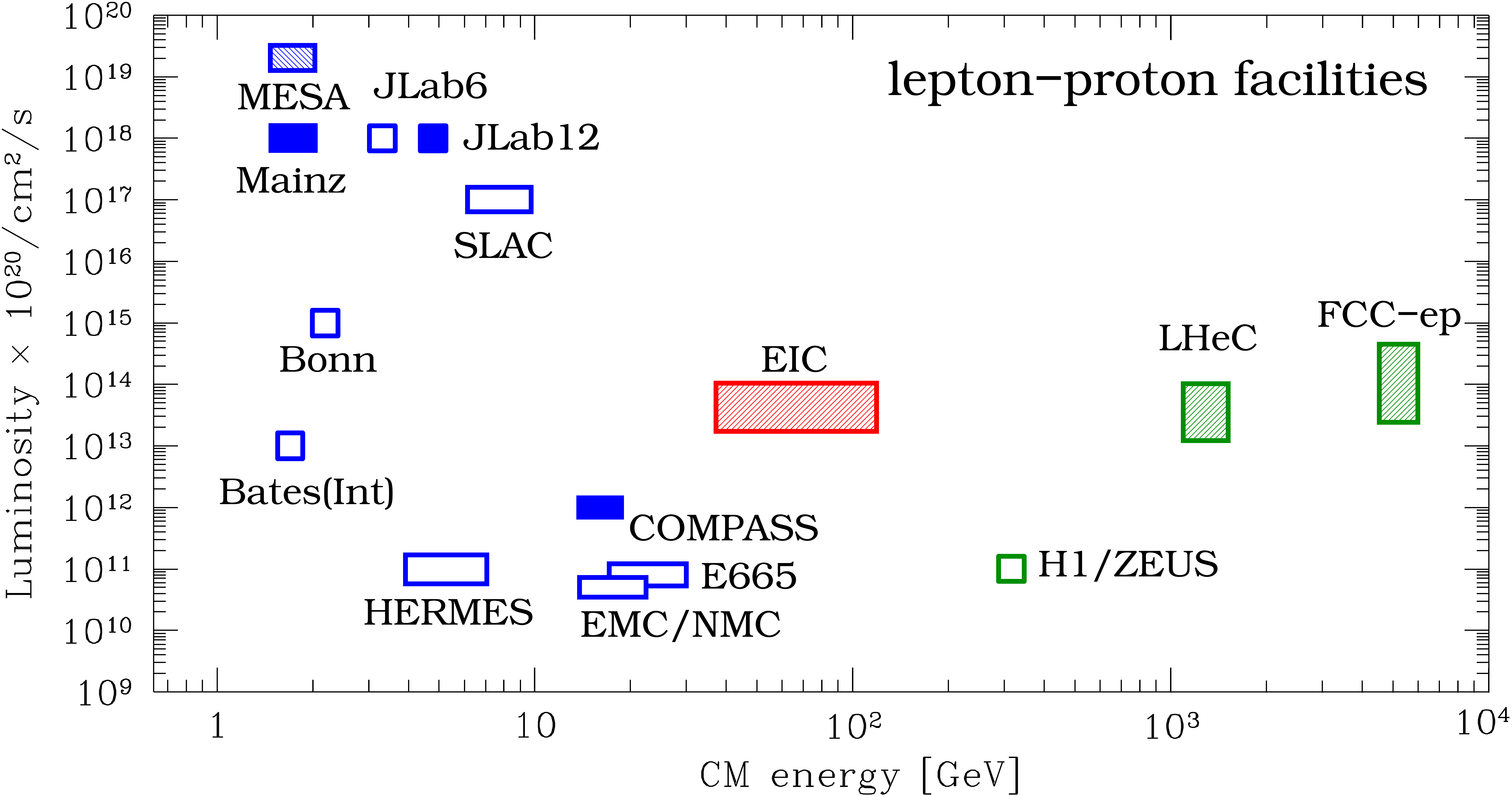}
    \caption{Luminosity vs. center-of-mass energy for the past (open markers), current (solid markers), and some of the planned future lepton-proton or ion collision facilities (shaded markers). Most of the fixed-target facilities (blue) have high luminosity but low energy, while collider facilities (green or red) typically access higher energy but have low luminosity. While all fixed-target facilities can utilize polarized proton (or nuclear) targets, polarized proton or ion beams are only available at EIC and not at HERA, LHeC or FCC-ep.}
    \label{fig:intro_eic2}
\end{figure}

The EIC will provide capabilities of colliding beams of polarized electrons with polarized beams of protons and light ions. A schematic perspective of the planned EIC is shown in \autoref{fig:intro_eic1}. The EIC will collide electrons with protons and nuclei in the intermediate energy range between fixed-target scattering facilities and high energy colliders, see \autoref{fig:intro_eic2} for a comparison with similar facilities globally. The primary design requirements of the planned EIC are:
\begin{itemize}
    \item Capability to achieve high polarization ($\sim70\%$) of both electron and proton beams.
    \item Variable center-of-mass energies $\sqrt{s}=20-100$~GeV, upgradable up to $\sqrt{s}=140$~GeV, for $e+p$ collisions.
    \item Luminosity up to $10^{33}-10^{34}$~cm$^{-2}$s$^{-1}$ for $e+p$ collisions. 
    \item Capability for colliding ion beams from deuterons to heavy nuclei, such as gold, lead or uranium. 
    \item Flexibility to accommodate more than one interaction regions.
\end{itemize}

The EIC is the only new large-scale accelerator facility planned for construction in the United States in the next few decades. It will also be one of the largest and most sophisticated new accelerator facilities worldwide. The versatility, resolving power and intensity of EIC opens up new windows of opportunities to address some of the crucial and fundamental scientific questions in particle physics. This whitepaper provides an overview of the science case of EIC from the perspective of the high-energy physics community. The science case is broadly categorized into and presented in the following sections:
\begin{itemize}
    \item Beyond standard model physics: \autoref{sec:bsm},
    \item Tomogography of hadrons and nuclei: \autoref{sec:tomo},
    \item Jet physics: \autoref{sec:jets},
    \item Physics of heavy flavors: \autoref{sec:hf},
    \item Physics at small Bjoeken-$x$: \autoref{sec:smx}.
\end{itemize}

%
%
%
\newpage
\section{Beyond Standard Model Physics at EIC} \label{sec:bsm} 
\vspace{-2ex}
\centerline{\textit{Editors:} 
\href{mailto:yulia@jlab.org}{\texttt{Yulia Furletova}},
\href{mailto:ciprian.gal@stonybrook.edu}{\texttt{Ciprian Gal}},
\href{mailto:claire.gwenlan@physics.ox.ac.uk}{\texttt{Claire Gwenlan}}.
}
\vspace{3ex}
%


EIC will provide orders of magnitude higher luminosity compared to HERA, the only electron-proton collider operated to date. It will be the first lepton-ion collider that polarizes both the electron and the proton (ion) beams, and the first collider with the capacity to flip its electron beam helicity at the source level. These unique design features will allow a direct extraction of parity-violating components of the electroweak (EW) neutral-current (NC) scattering cross section and the first (and possibly the only) measurement of the polarized NC structure functions. The high luminosity of the EIC will also provide better sensitivity to charged-current EW physics, rare processes such as charged lepton flavor violation, heavy photons, and searches for Lorentz- and CPT-violating effects. In addition to providing measurements in an energy and luminosity range hardly explored before, 
the energy range of the EIC will provide a unique opportunity for the nuclear and high-energy physics communities to collaborate closely and to advance EW and beyond-the-Standard-Model (BSM) physics together. 


\subsection{Neutral Current Electro-Weak and BSM Physics}
\subsubsection{Deep Inelastic Scattering Cross Sections} 
The cross section of Deep Inelastic Scattering (DIS) can be divided into four contributions that depend on the electron and ion polarization: the unpolarized cross section $\sigma_0$, the single-spin polarized cross sections $\sigma_e$ and $\sigma_p$, and the double-spin polarized cross section $\sigma_{ep}$: 
  \begin{eqnarray}
   \frac{d^2\sigma}{dxdy} &=& \frac{d^2\sigma_0}{dxdy} + P_e\frac{d^2\sigma_e}{dxdy} + P_p\frac{d^2\sigma_p}{dxdy} + P_e P_p \frac{d^2\sigma_{ep}}{dxdy}~,
  \end{eqnarray}
  where $P_e,P_p$ are the polarizations of the electron and the proton beams, respectively, with their signs defined by the particle helicity. These cross sections are determined by the structure functions $F_{1,2,3}^{\gamma, \gamma Z, Z}$ and $g_{1,4,5}^{\gamma, \gamma Z, Z}$, the electromagnetic fine-structure constant $\alpha$, and the electron NC couplings $g_V^e$ and $g_A^e$. 
  In the parton model (leading-order or LO), one has for the structure functions $F_2^i=2xF_1^i$ and $g_4^i=2xg_5^i$ with $i=\gamma, \gamma Z, Z$.  The parton-model interpretation of the remaining structure functions $F_{1,2,3}^{\gamma,\gamma Z,Z}$ and $g_{1,4,5}^{\gamma,\gamma Z,Z}$ can be found in the PDG~\cite{PDG:2020} Eq.(18.18):
  \begin{eqnarray}
   \left[F_2^\gamma, F_2^{\gamma Z}, F_2^Z\right] &=& x\sum_f \left[Q_f^2, 2Q_f g_V^f, {g_V^f}^2+{g_A^f}^2\right] (q_f+\bar q_f),\label{eq:SF_F2}\\
   \left[F_3^\gamma, F_3^{\gamma Z}, F_3^Z\right] &=& \sum_f \left[0, 2Q_f g_A^f, 2{g_V^f}{g_A^f}\right] (q_f-\bar q_f),\label{eq:SF_F3}\\
   \left[g_1^\gamma, g_1^{\gamma Z}, g_1^Z\right] &=& \frac{1}{2} \sum_f \left[Q_f^2, 2Q_f g_V^f, {g_V^f}^2+{g_A^f}^2\right] (\Delta q_f+\Delta\bar q_f),\label{eq:SF_g1}\\
   \left[g_5^\gamma, g_5^{\gamma Z}, g_5^Z\right] &=& \sum_f \left[0, Q_f g_A^f, {g_V^f}{g_A^f}\right] (\Delta q_f-\Delta \bar q_f),\label{eq:SF_g5}
  \end{eqnarray}
where $q_f(x,Q^2)$ and $\Delta q_f(x,Q^2)$ are unpolarized and polarized parton distribution functions (PDFs) of quark flavor $f$, respectively; EIC implications for these quantities and HEP are discussed in greater detail in Sec.~\ref{sec:tomo}. The quantity $Q_f$ is the electric charge, and $g_{V,A}^f$ are the vector and axial-vector couplings of quarks. New methods to access these quantities have been proposed in Refs.~\cite{Li:2021uww, Yan:2021htf}.

In all cross section terms, the $\gamma Z$ structure functions are suppressed by 
  \begin{eqnarray}
  \eta_{\gamma Z} &=& \left(\frac{G_F M_Z^2}{2\sqrt{2}\pi\alpha}\right)\left(\frac{Q^2}{Q^2+M_Z^2}\right)~, \label{eq:eta_gZ}
  \end{eqnarray}
  and the $Z$ terms by $\eta_{Z}=\eta_{\gamma Z}^2$.
  Here, $G_F=1.1663787(6) \times 10^{-5}$~GeV$^{-2}$ is the Fermi constant and $M_Z=91.1876 \pm 0.0021$~GeV~\cite{PDG:2020} is the mass of the $Z$ boson. The value of $\eta_{\gamma Z}$ can be tens of percents at EIC energies. 
  
  \subsubsection{Parity Violation Asymmetries and Projections for EIC}~\label{sec:sim}
The parity-violating DIS (PVDIS) asymmetry can be formed by flipping either the electron helicity or the ion helicity. If the electron helicity is flipped, one forms 
\begin{equation}
    A_\mathrm{PV}^{(e)} = \frac{\sigma^{R}-\sigma^{L}}{\sigma^{R}+\sigma^{L}}=\frac{\sigma_e}{\sigma_0}. 
\label{eq:APV}
\end{equation}
Because all lepton and quark NC couplings depend on the weak mixing angle $\sin^2\theta_W$, measurements of $A_{PV}^{(e)}$ will allow the value of $\sin^2\theta_W$ to be determined. 
 Similarly, if the electron beam is unpolarized and the proton (or ion) beam is longitudinally polarized, one forms the so-called {\it polarized} PV asymmetry
  \begin{eqnarray}
   A_{PV}^\mathrm{p (d)}&\equiv&\frac{\sigma_p}{\sigma_0},
   \end{eqnarray}
   which can provide constraints on polarized PDFs via the determination of the $g_{1,4,5}^{\gamma Z}$ structure functions, or a simultaneous determination of the PDF and EW NC couplings.

Projections for $A_{PV}^{(e)}$ were performed using the ECCE detector configuration~\footnote{\href{https://www.ecce-eic.org/}{https://www.ecce-eic.org/}} and the luminosity from Table 10.1 of the EIC Yellow Report~\cite{AbdulKhalek:2021gbh}. Details of the analysis -- asymmetry projection and extractions of $\sin^2\theta_W$ 
-- can be found in~\cite{ecce-note-phys-2021-14-pub} and a brief description is given here. For electroweak physics studies, we assumed the annual luminosity -- ten times the ``High divergence configuration'' value -- 15.4, 100, 44.8, and 36.8~fb$^{-1}$ for $18\times 275(137)$, $10\times 275(137)$, $10\times 100$, and $5\times 100$~GeV for $ep$ ($eD$), respectively. For the deuterium ion beam, the energy specified is per nucleon. We used the Djangoh generator~\cite{Charchula:1994kf} (version 4.6.16~\footnote{\href{https://github.com/spiesber/DJANGOH}{https://github.com/spiesber/DJANGOH}}) that includes full electromagnetic and electroweak radiative effects to generate 20-million (20~M) events for each beam type and energy combination. In lieu of a full GEANT-based simulation, a fast smearing method was used because of the need of high statistics for EW physics study. Events were selected based on DIS condition, to limit photoproduction background, and to ensure high purity of electron samples. Events were unfolded to correct for bin migration. 
  
\subsubsection{Extraction of the Weak Mixing Angle} 
\begin{figure}[!h]
      \includegraphics[width=0.45\textwidth]{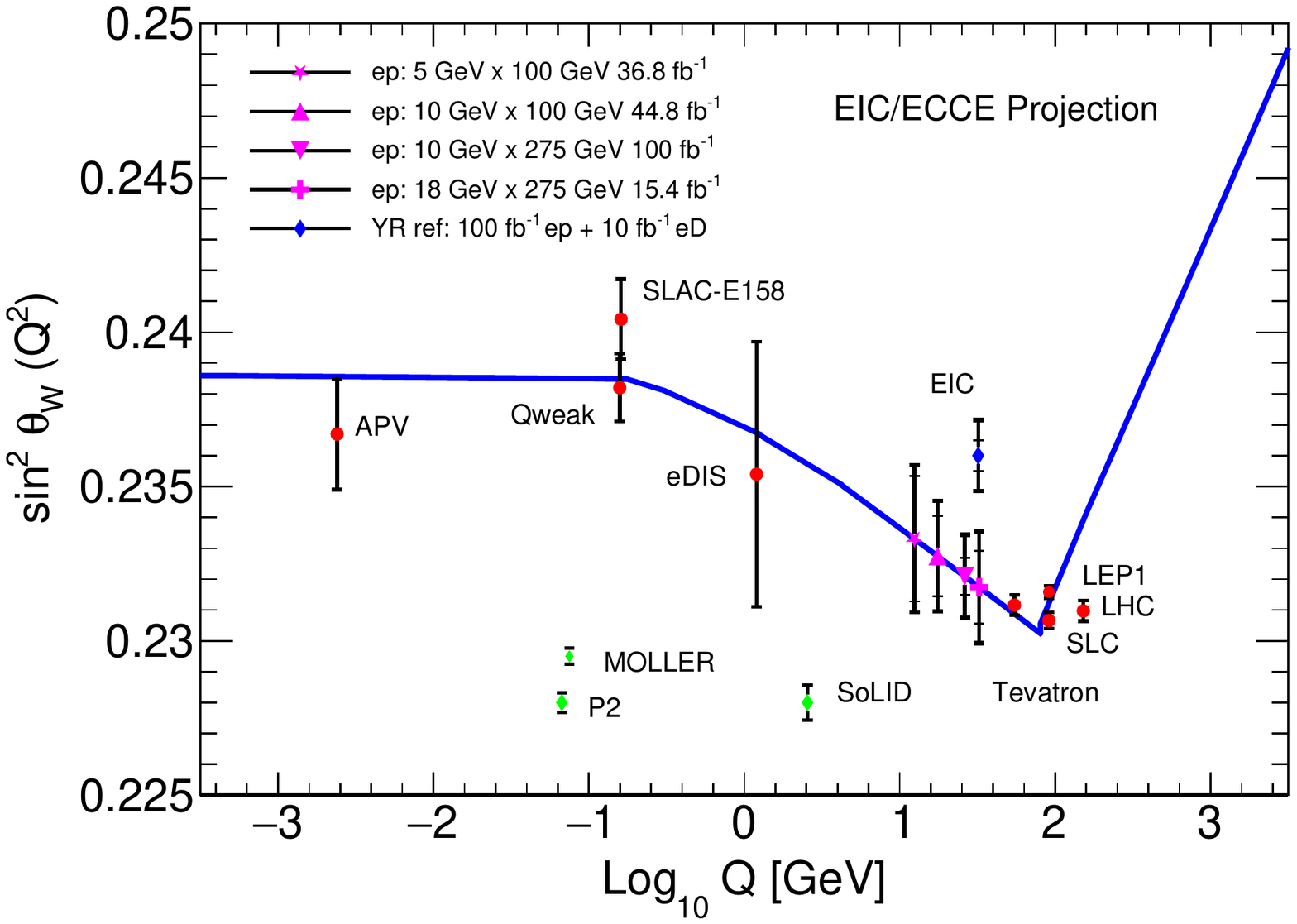}
      \includegraphics[width=0.45\textwidth]{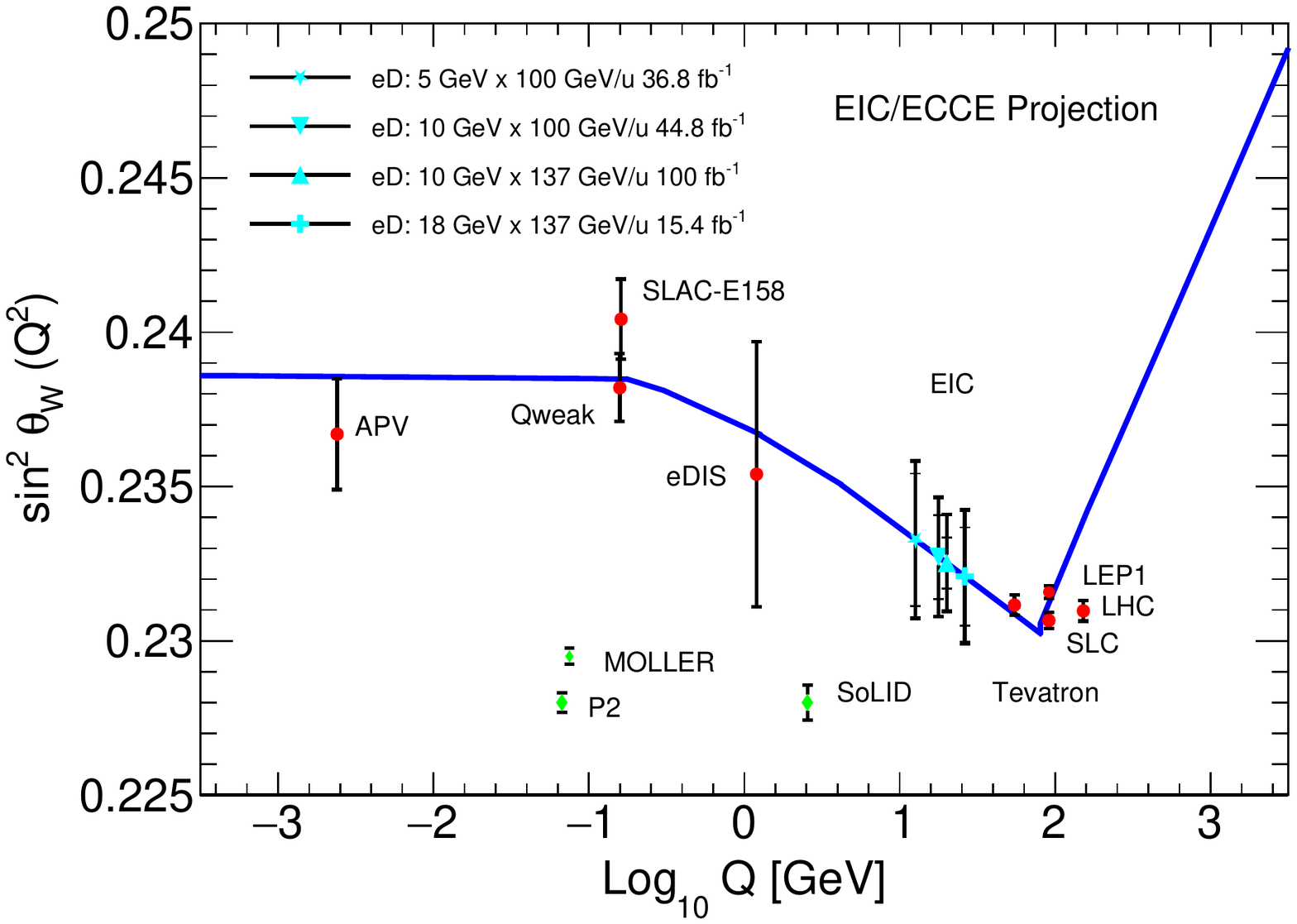}
      \caption{From~\cite{ecce-note-phys-2021-14-pub}: Projected results for $\sin^2\theta_W$ using $ep$ (left, solid magenta markers) and $eD$ (right, solid cyan markers) collision data and the nominal annual luminosity given in Table 10.1 of the Yellow Report~\cite{AbdulKhalek:2021gbh}, along with existing world data (red solid circles). Data points for Tevatron and LHC are shifted horizontally for clarity. Result from combining 100~fb$^{-1}$ $ep$ $18\times 275$~GeV and 10~fb$^{-1}$ $eD$ $18\times 137$~GeV is also shown and is called the "YR reference point" (blue diamond). The PDF uncertainty is from CT18NLO. The electron beam polarization is assumed to be 80\% with a relative 1\% uncertainty. The inner error bars show the combined uncertainty from statistical and 1\% uncorrelated background effect; the median error bars show the experimental uncertainty that includes statistical, 1\% uncorrelated background, and 1\% electron polarimetry. The outer-most error bars (which almost coincide with the median error bars) include all the above and the PDF uncertainty evaluated using the CT18NLO sets. Results evaluated with the MMHT2014 and NNPDF31NLO sets are similar. 
      Also shown are the expected precision from P2~\cite{Becker:2018ggl}, MOLLER~\cite{MOLLERCDR} and SoLID~\cite{Chen:2014psa} PVDIS~\cite{PVDIS,Erler:2014fqa}, respectively.
      The script used to produce this plot was inherited from~\cite{Zhao:2016rfu}, and the scale dependence of the weak mixing angle expected in the SM (blue curve) is defined in the modified minimal subtraction scheme ($\overline{\mathrm{MS}}$ scheme)~\cite{Erler:2004in}.}
      \label{fig:sin2th}
  \end{figure}

  From the projected size and statistical uncertainty of the asymmetry $A_{PV}^{(e)}$, we used the parton-model structure functions of Eqs.~(\ref{eq:SF_F2}-\ref{eq:SF_F3}) to fit the weak mixing angle $\sin^2\theta_W$. The main systematic uncertainties included particle background (1\% relative in each bin, completely uncorrelated among all $(x,Q^2)$ bins), and electron polarization (1\% relative in each bin, completely correlated among all $(x,Q^2)$ bins of data of the same $\sqrt{s}$). The PDF uncertainty was evaluated using the CT18NLO~\cite{Hou:2019qau} (LHAPDF~\cite{Buckley:2014ana} ID 14400--14458), MMHT2014nlo\_68cl~\cite{Harland-Lang:2014zoa} (ID 25100--25150) and NNPDF31\_nlo\_as\_0118~\cite{NNPDF:2017mvq} (ID 303400--303500) PDF sets. The analysis in~\cite{ecce-note-phys-2021-14-pub} treated the PDF uncertainties as completely uncorrelated in $(x,Q^2)$ bins, but we have updated the analysis and have accounted for correlations in the PDF uncertainty here. 
  We found that PDF uncertainties are likely not a dominant uncertainty for EIC projections, but the electron polarization is, for high luminosity settings. 
  Our results for $\sin^2\theta_W$ are 
  shown in Figs.~\ref{fig:sin2th} for five energy and luminosity combinations for $ep$ and $eD$ collisions, respectively, along with existing and near-future world data.
  
  
  Our results show that the EIC will provide determination of  $\sin^2\theta_W$ in an energy scale that bridges higher energy colliders with low-energy tests. 
  Additionally, data points of different $\sqrt{s}$ values of EIC can be combined, or the $Q^2$-dependence of the EW parameter can be explored, depending on the runplan of the EIC. Furthermore, one should explore the exploratory potential of the EIC measurements beyond the scope of a single parameter (the weak mixing angle).  One such framework is given in the next section.

\subsection{Complementarity of EIC and LHC in the SMEFT Framework}

The Standard Model has so far been remarkably successful in describing all observed laboratory phenomena.  No new particles beyond those
present in the SM have been discovered so far, and no appreciable deviation
from SM predictions has been conclusively observed. Given this situation it is increasingly important to understand how indirect signatures of new physics can be probed and constrained by data.  This effort will help guide future searches for new physics by suggesting in what channels measurable deviations from SM predictions may occur given the current bounds.

A convenient theoretical framework for investigating indirect
signatures of heavy new physics without associated new particles is SM
effective field theory (SMEFT), which extends the SM Lagrangian to include terms suppressed by an energy scale $\Lambda$ at which ultraviolet completion becomes important and BSM particles appear (for a review of the SMEFT see Ref.~\cite{Brivio:2017vri}). Truncating the expansion in $1/\Lambda$ at dimension-6, and ignoring operators of odd-dimension, which violate lepton number, we have
\begin{eqnarray}
{\cal L} = {\cal L}_{\rm SM}+ \sum_i C_{i} {\cal
  O}_{i} + \ldots,
\end{eqnarray}
where the ellipsis denotes operators of higher dimensions.  The Wilson
coefficients defined above have dimensions of $1/\Lambda^2$.  When computing cross sections and other observables, we consider only the leading
interference of the SM amplitude with the dimension-6 contribution.
This is consistent with our truncation of the SMEFT expansion above,
since the dimension-6 squared contributions are formally the
same order in the $1/\Lambda$ expansion as the dimension-8 terms which we neglect.
The
following four-fermion operators in Table~\ref{tab:ffops} can affect the DIS cross section
at leading-order in the coupling constants for massless fermions,
which we assume here.
\begin{table}[h!]
\centering
\begin{tabular}{|c|c||c|c|}
\hline
${\cal O}_{ \ell q}^{(1)}$ & $(\bar{ \ell}\gamma^{\mu}  \ell) (\bar{q}\gamma_{\mu} q)$ &  ${\cal O}_{ \ell u}$ & $(\bar{ \ell}\gamma^{\mu}  \ell) (\bar{u}\gamma_{\mu}  u)$ 
  \\
  ${\cal O}_{ \ell q}^{(3)}$ & $(\bar{ \ell}\gamma^{\mu} \tau^I \ell)
                     (\bar{q}\gamma_{\mu} \tau^I q)$ & ${\cal O}_{ \ell d}$ & $(\bar{ \ell}\gamma^{\mu}  \ell) (\bar{d}\gamma_{\mu}  d)$   
  \\
  ${\cal O}_{eu}$ & $(\bar{e}\gamma^{\mu} e) (\bar{u}\gamma_{\mu}  u)$
                                                                     & ${\cal O}_{qe}$ & $(\bar{q}\gamma^{\mu} q) (\bar{e}\gamma_{\mu}  e)$             
  \\
  ${\cal O}_{ed}$ & $(\bar{e}\gamma^{\mu} e) (\bar{d}\gamma_{\mu}  d)$
                                                                     & &
  \\
  \hline
\end{tabular}
\caption{Dimension-6 four-fermion operators contributing to DIS at leading order in the coupling constants. Here, $q$ and {$ \ell $} denote left-handed quark and lepton doublets, while $u$, $d$ and $e$ denote right-handed singlets for the up quarks, down quarks and leptons, respectively. {The} $\tau^I$ denote the SU(2) Pauli matrices. \label{tab:ffops}}
\end{table}
\par
In Table~\ref{tab:ffops}, 
we have suppressed flavor indices
for these operators, and in our analysis, we assume flavor universality
for simplicity.  We note that the overall
electroweak couplings that govern lepton-pair production are also
shifted in the SMEFT by operators other than those considered above.  Such contributions are far better bounded
through other data sets such as precision $Z$-pole
observables, and we neglect them here.  The
above assumptions leave us with the seven Wilson coefficients
associated with the operators in Table~\ref{tab:ffops} entering
the predictions for our observables.

\begin{figure}[!h]
\includegraphics[width=0.48\textwidth]{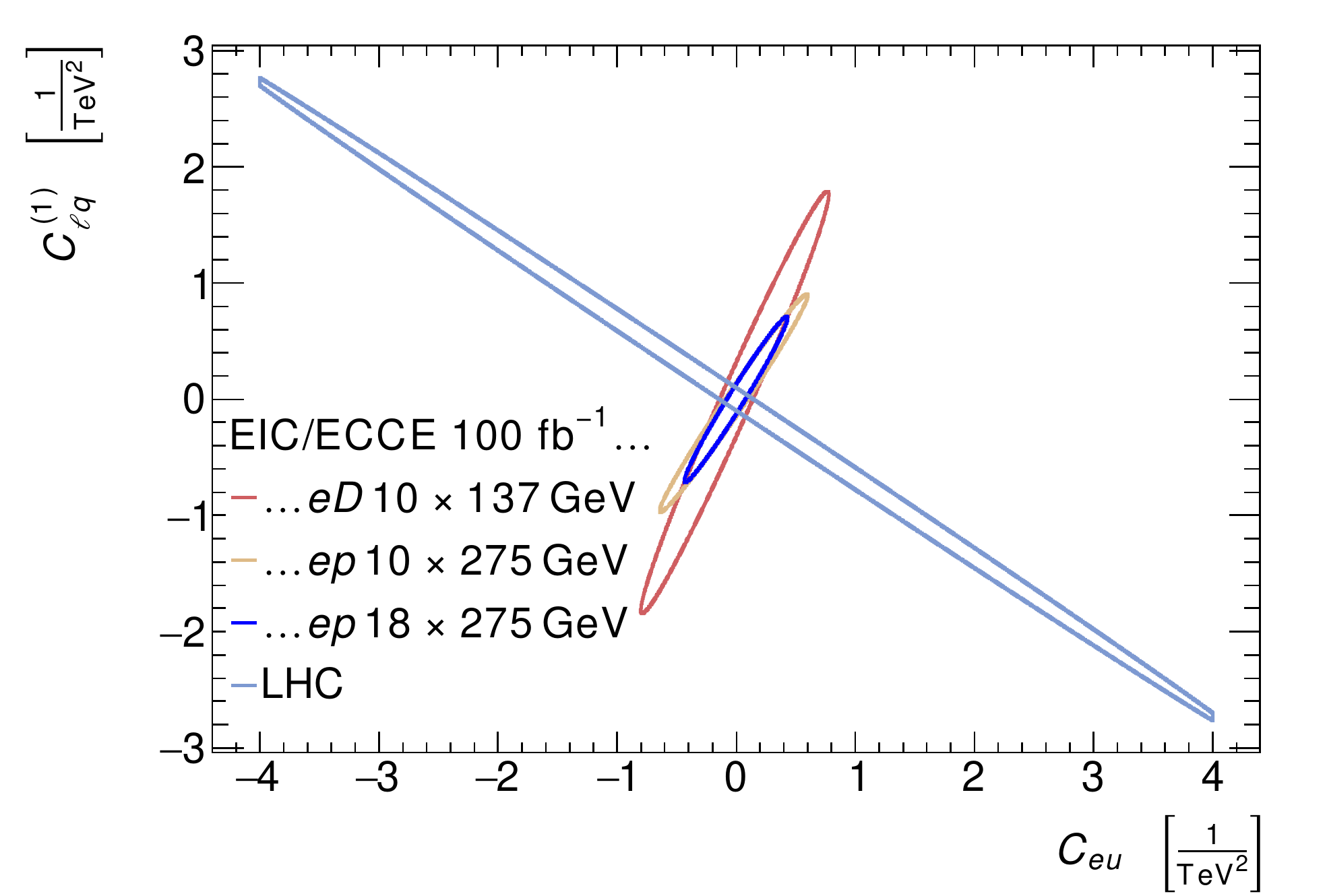}
\includegraphics[width=0.48\textwidth]{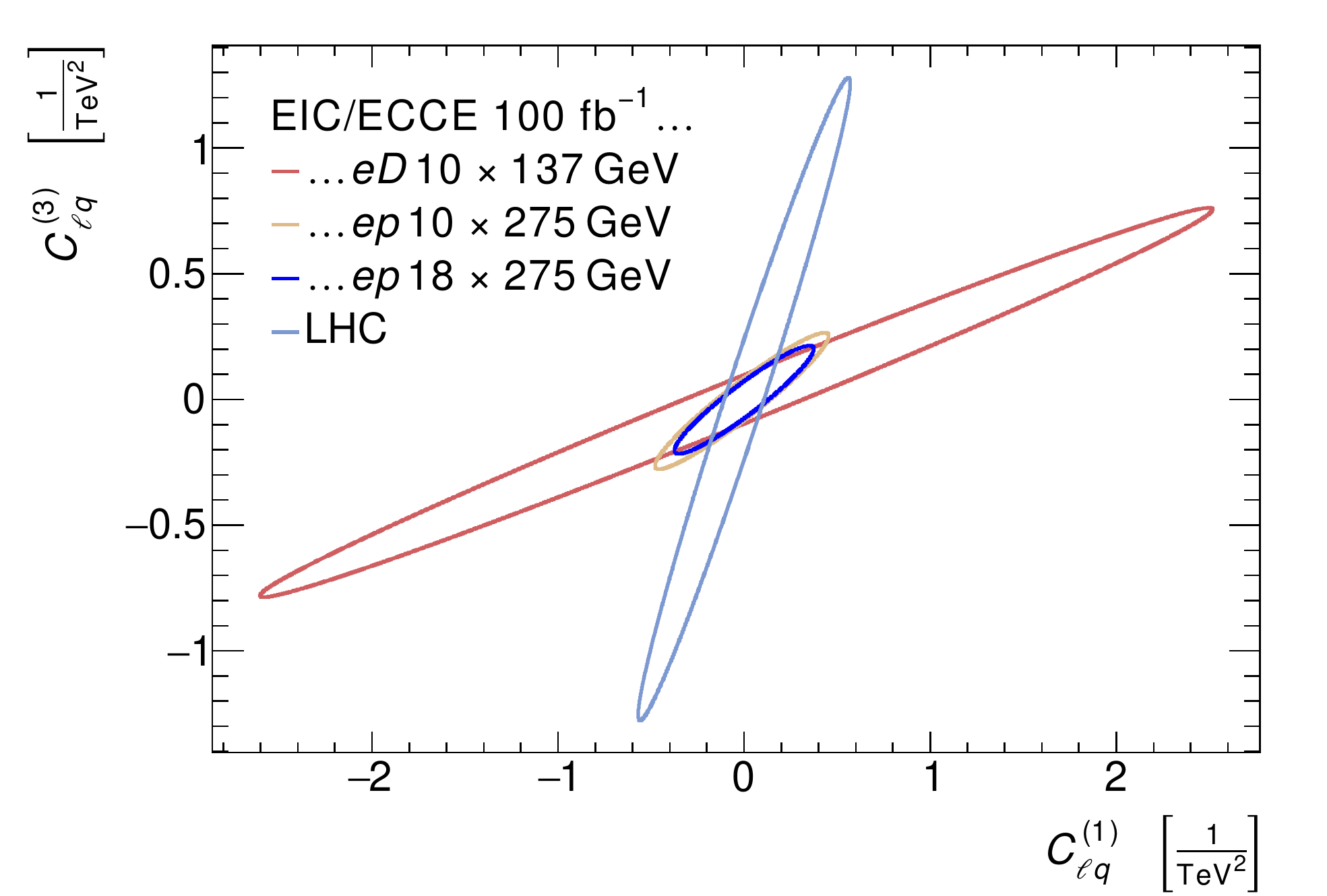}
\caption{From~\cite{ecce-note-phys-2021-14-pub}: Examples of 'flat directions' when using only LHC data to constrain Wilson coefficients. The Wilson coefficients plotted correspond to the operators defined in Table~\ref{tab:ffops}. The inclusion of high precision $A_{PV}^{(e)}$ data from EIC (projected here using the ECCE detector) would provide strong, complementary constraints on the parameter space.
}
\label{fig:smeft}
\end{figure}

The SMEFT framework provides a mechanism to conduct global
analyses of world data across all energy scales. An issue 
that arises in such global fits is the appearance of flat directions that occur when 
the available experimental measurements cannot disentangle the contributions from different EFT operators.
Figure~\ref{fig:smeft} shows two Wilson coefficients and the available 68\% confidence level ellipse for the case where only LHC data is used in the fits (blue curve), only EIC data (yellow curve) and the combination of the two data sets. The EIC data included in this projection are the two (one) highest energy settings of the $ep$ ($eD$) $A_{PV}^{(e)}$ data sets described in Section~\ref{sec:sim}. The LHC data set used in this fit is from the high invariant mass Drell-Yan process measured by ATLAS in Ref.~\cite{ATLAS:2016gic}. More details regarding the analysis of this data set can be found in Ref.~\cite{Boughezal:2020uwq}.
It is apparent that when only LHC data are included, there is a degeneracy in the space of 2-lepton, 2-quark Wilson coefficients. This can be seen from the elongated nature of the LHC constraint ellipses. The EIC will play a crucial role in resolving this flat direction, and combined fits of LHC and projected EIC data lead to much stronger constraints than either experiment alone.  Analysis is ongoing to extract SMEFT constraints using both unpolarized and polarized PV asymmetries projected for EIC. The polarized PV asymmetry in fact will provide further complementarity to unpolarized asymmetries.

\subsection{PDF Fits and Flavour Decomposition}

An impact study of these $\gamma Z$ interference structure functions on the unpolarized proton PDFs has been performed by the JAM 
collaboration [Yellow Report, Chapter 7]. This work represents a specialized study of the EIC's PDF impacts as broadly considered in
Sec.~\ref{sec:tomo} below.
Figure~\ref{fig:impact_APV_JAM} (left panel) shows the impact of $A_\mathrm{PV}^\mathrm{(e)}$ measurements at the EIC, including both proton and deuteron beams with integrated luminosities of 100 fb$^{-1}$ and 10 fb$^{-1}$, respectively, at the energies $\sqrt{s} = 29, 45, 63$ and 141~GeV for the proton and $\sqrt{s} = 29, 66$ and 89~GeV for the deuteron. 
In the case of a longitudinally polarized proton beam, similar enhancement of the sensitivity to various quark flavors has also been observed, 
especially for the strange quark.
Fig.~\ref{fig:impact_APV_JAM} (right panel) shows the impact of $A_{\rm PV}^{\rm p}$ at the EIC on the truncated moments of $\Delta\Sigma$ and $\Delta g$, assuming an integrated luminosity of 100~fb$^{-1}$. During the helicity impact studies, it was found
that the outcome of the impact of EIC parity-violating data has a dependence on the triplet and octet axial charges, $g_A$ and $a_8$. Therefore,
different values were used either from JAM17 collaboration or from Hyperon Decays under SU(3) symmetry, as labeled by different color
in the plot. On the other hand, please be noted that $\gamma-Z$ interference structure functions will reduce the dependence of 
Hyperon decay data in the future global data analysis, since they provide additional inputs.

In short, by taking advantage of parity-violating asymmetry measurements at the EIC, the $\gamma-Z$ interference structure functions can provide unique combinations of unpolarized and polarized PDFs in the parton model. It can enhance the sensitivities to different 
quark flavors on top of the traditional pure electromagnetic structure functions.

\begin{figure}[!h]
\includegraphics[width=0.36\textwidth]{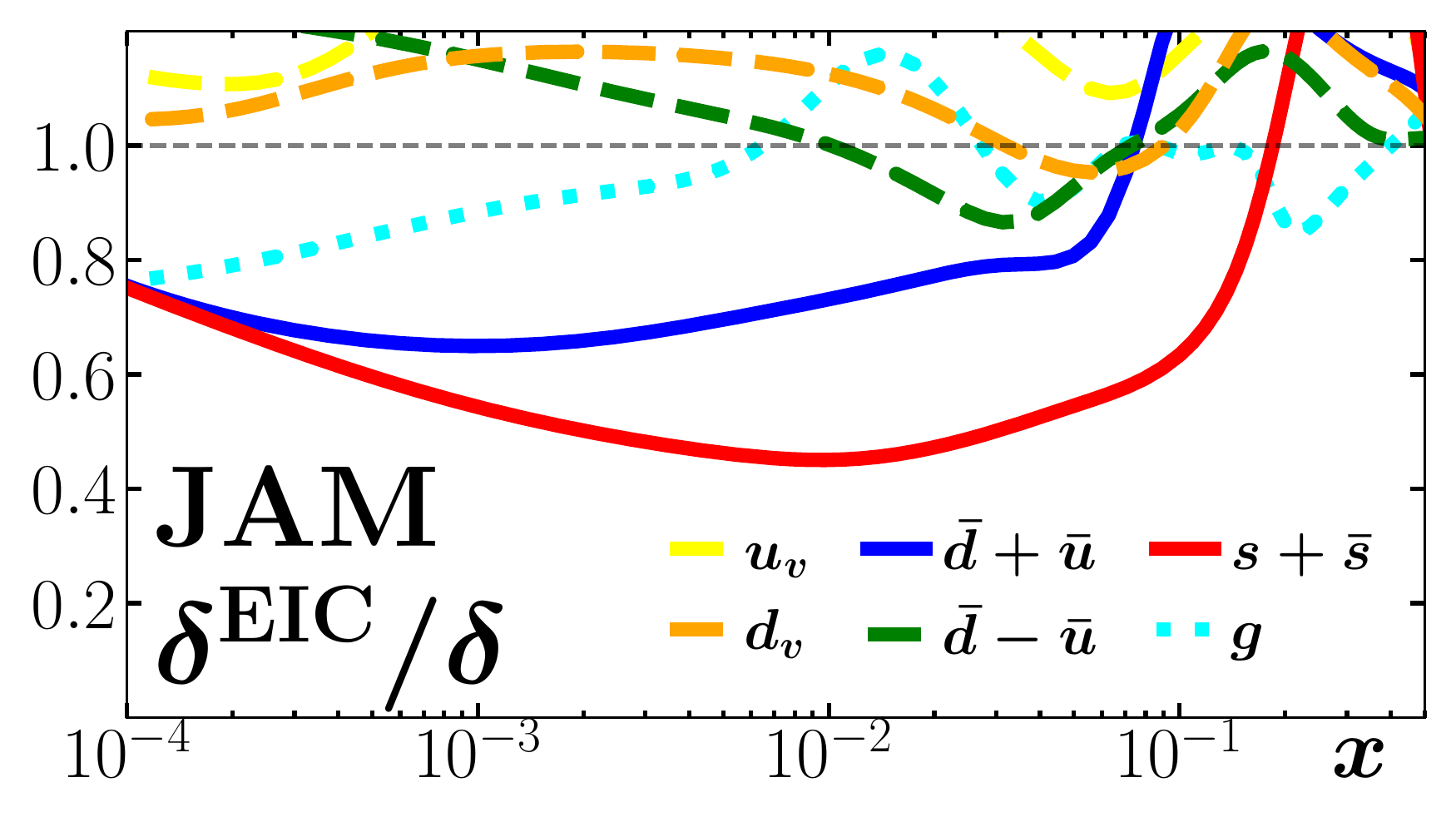}
\includegraphics[width=0.3\textwidth]{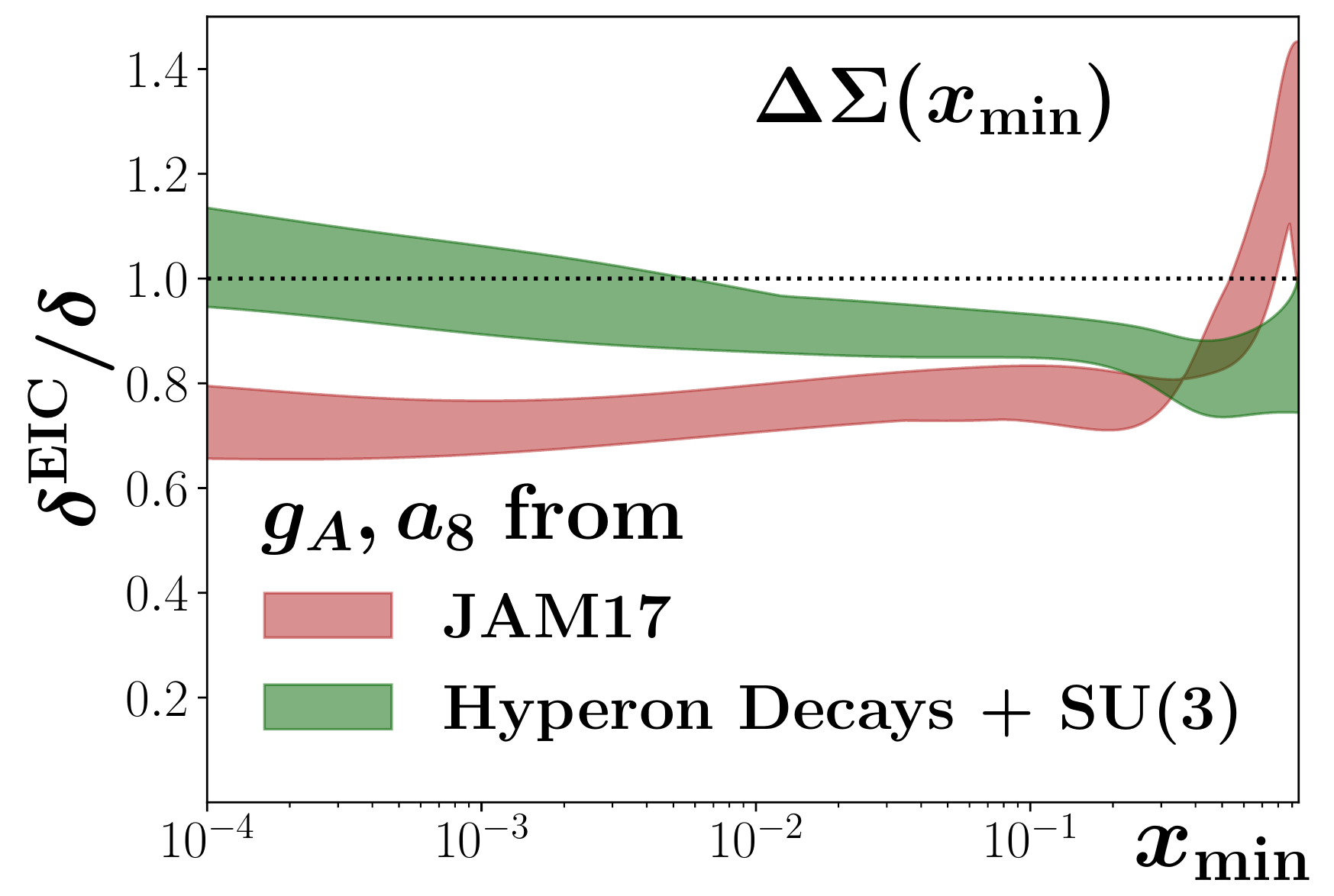}
\includegraphics[width=0.32\textwidth]{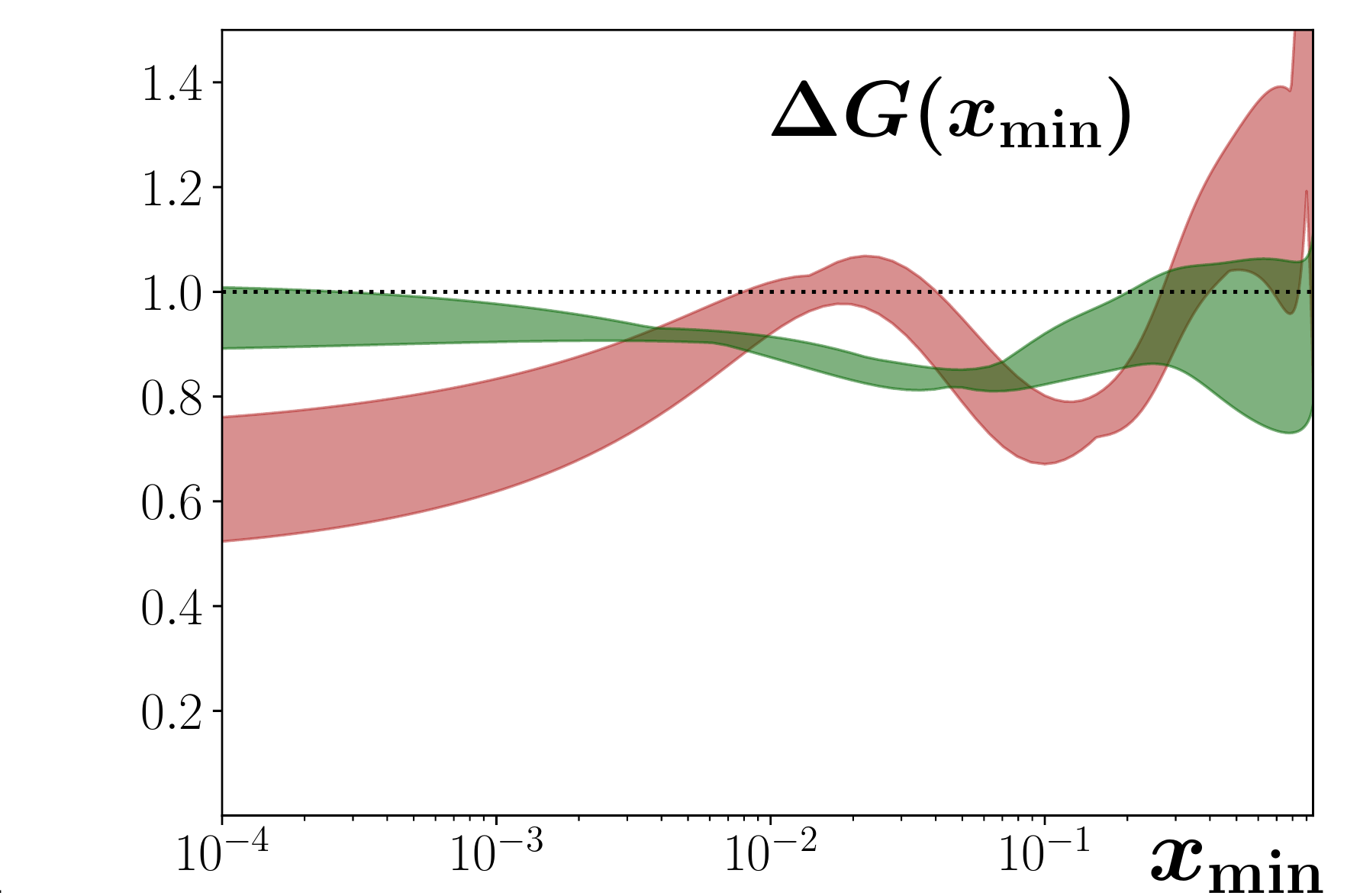}
\caption{{\it Left:} Ratio of uncertainties on the unpolarized PDFs as functions of $x$, including EIC $A_\mathrm{PV}^\mathrm{(e)}$ 
data to those without EIC data, at $Q^2=10$~GeV$^2$. One can see that parity-violating asymmetry measurements
with a longitudinally polarized electron beam can enhance the sensitivity of $\bar{d}+\bar{u}$ and  $s+\bar{s}$ apparently. {\it Middle and Right:}
Ratio of uncertainties on the truncated moments of the quark singlet (top) and gluon (bottom) PDFs as functions of $x_{\rm min}$, including EIC $A_{\rm PV}^{\rm (p)}$ data to those without EIC data, at $Q^2=10$~GeV$^2$.  
The results are sensitive to the triplet and octet axial charges, $g_A$ and $a_8$. Different color in the plot shows the 
results with values of $g_A$ and $a_8$ taken from JAM17~[cite JAM17] (red) or hyperon decays and SU(3) (green).
}
\label{fig:impact_APV_JAM}
\end{figure}

\subsection{Charged Lepton Flavor Violation}
\label{sec:CLFV}

\subsubsection{CLFV and Leptoquarks}
The discovery of neutrino oscillations provided conclusive evidence of lepton flavor violation. Lepton flavor violation in the neutrino sector also results in charged lepton flavor violation (CLFV) through loop-suppressed processes such as $\mu \to e \gamma$. However, the resulting predicted rates are highly suppressed -- Br$(\mu \to e  \gamma) < 10^{-54}$ -- due to the small neutrino masses  and are far beyond the reach of any current of planned experiments. On the other hand, many Beyond Standard Models (BSM) scenarios predict CLFV rates that are both much larger and within reach of present or future experiments. For example, SUSY-based models predict rates as high as Br$(\mu \to e  \gamma) \sim 10^{-15}$~\cite{Barbieri:1995tw}.

There have been extensive searches for such CLFV processes between the first and second lepton generations, denoted as CLFV(1,2) for brevity, and are tightly constrained. For example, the current limit on the $\mu \to e \gamma$ process is Br$(\mu \to e  \gamma) < 4.2\times  10^{-13}$~\cite{MEG:2016leq}. On the other hand, the constraints on CLFV(1,3) processes that involve $e\leftrightarrow \tau$ transitions are weaker by several orders of magnitude~\cite{Gonderinger:2010yn,Cirigliano:2021img}, obtained through searches for $e +  p\to \tau +X$, $\tau\to e\gamma$, and $p+p\to e+\tau + X$ at HERA~\cite{ZEUS:2005nsy,H1:2007dum,ZEUS2012,ZEUS2019}, BaBar~\cite{BaBar:2009hkt}, and the LHC~\cite{ATLAS:2018mrn} respectively. Earlier phenomenological studies~\cite{Gonderinger2010} have shown that even with its comparatively lower center of mass energy, the EIC has the potential to improve upon the HERA limits, owing to the much higher luminosity, and provide complementary information~\cite{Cirigliano:2021img} to the constraints from BaBar and the LHC.

\subsubsection{Event Simulation and Selection}
We have studied the potential of searching for $e^- \rightarrow \tau^-$ events at the EIC based on the ECCE detector configuration and with various SM background processes included to some extent. The leptoquark generator LQGENEP~\cite{Bellagamba:2001fk} (version 1.0) with a default 1.9~TeV leptoquark mass, the Djangoh generator, and the Pythia generator were used to produce the leptoquark signal, background DIS NC and CC, and background photoproduction Monte-Carlo events, respectively. The leptoquark candidate events should contain a high $p_{\rm T}$ quark initiated jet along with an isolated and high-$p_{\rm T}$ $\tau$ which replaces the scattered electron in the NC DIS events. After being produced, the $\tau$ will decay into stable particles after flying a short distance, of the order of millimeters. 
Based on the number of charged particles in the final state, the $\tau$ decay modes are categorized as ``1-prong", ``3-prong", and ``5-prong" modes. This study focuses on only the 3-prong mode that contains ($\pi^-\pi^+\pi^-$) in the final state. 
\begin{figure}[!h]
    \includegraphics[width=0.25\textwidth]{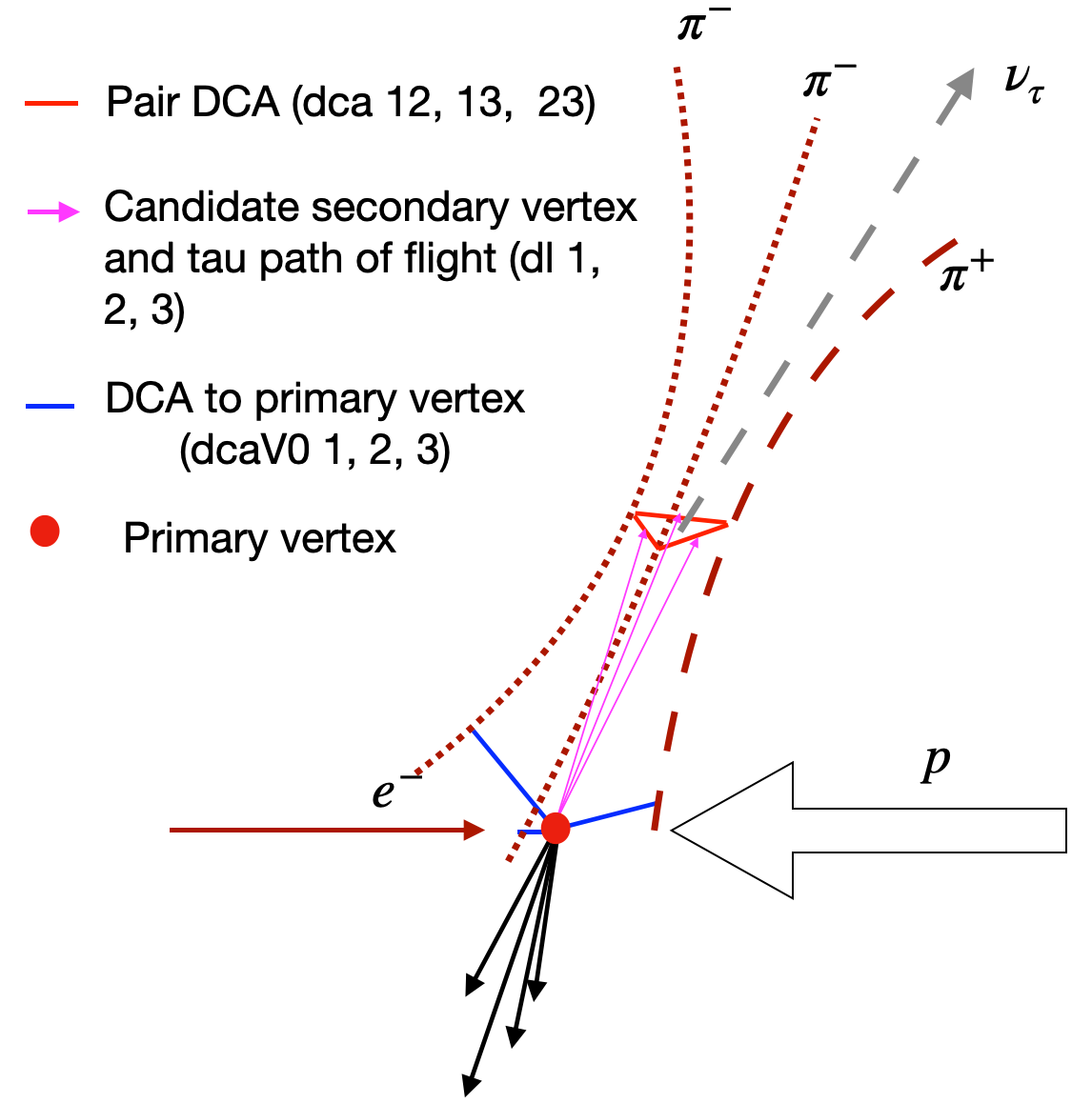}
    \includegraphics[width=0.7\textwidth]{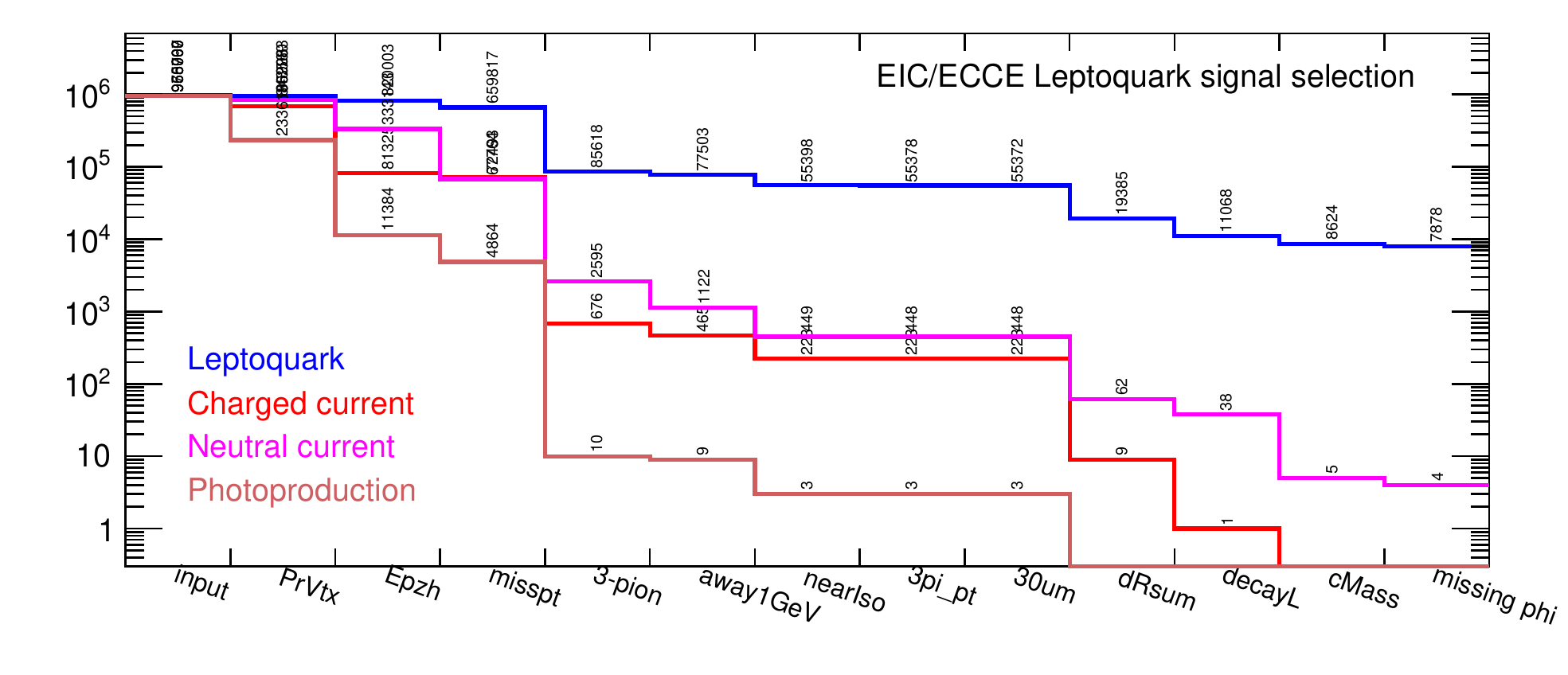} 
    \caption{{\it Left:} Event topology for a ``3-prong" $\tau$-jet event. DCA stands for `distance at the closest approach" of two tracks (e.g. dca 12) or between a track and a point (e.g. dcaV0). {\it Right:} From~\cite{ecce-note-phys-2021-14-pub}: MC statistics of leptoquark (blue), DIS CC (red), DIS NC (magenta), and photoproduction (orange) events, as ten selection criteria are progressively applied on 1\,M input events for each channel. See the text for details.}
    \label{fig:LQ3}
\end{figure}
%
Figure~\ref{fig:LQ3} (left panel) shows a typical ``3-prong" $\tau$-jet event with the secondary vertex reconstruction highlighted. 

After passing all MC events through the ECCE simulation, 
ten selection criteria were used to identify $e\to\tau$ events in the ``3-prong" mode and reject SM backgrounds. Their effects are shown in the right panel of Fig.~\ref{fig:LQ3}, where the vertical axis show how many sample events pass each of the selection criteria, and the horizontal axis are the progressive selections, see~\cite{ecce-note-phys-2021-14-pub} for details. 
It is evident that $e\to\tau$ events can be effectively selected with this set of preliminary cuts.  

\subsubsection{Sensitivity to Leptoquarks}
We can now deduce the sensitivity to the leptoquark signal cross section based on simulations of the 3-prong decay mode (15\% branching ratio) of the $\tau$ lepton. 
One caveat that we can see from 
Fig.~\ref{fig:LQ3} is that while some DIS CC events survive all ten cuts and can provide a reasonable estimate of the background effect, zero DIS NC and photoproduction events passed the cuts in the current simulation. Since DIS NC and photoproduction events could pass all cuts once a simulation is done with a larger sample size, it is not currently possible to project the leptoquark limit with all background effects fully taken into account. In addition, simulation studies of the detection efficiency of the other $\tau$ decay modes remain to be done. With these considerations in mind, we provide the potential for leptoquark exclusion limits under different possible scenarios for the detection efficiency of the  $\tau$ decay channels that are not in the the ``3-prong" mode. We estimate the 3-sigma exclusion limit on leptoquark cross sections to be 11.4\,fb and 1.7\,fb for the case where the decay channels not in the 3-prong mode are not detected and when they are detected with the same efficiency as the ``3-prong" mode presented here, respectively.  

\begin{figure}[!h]
    \includegraphics[width=0.45\textwidth]{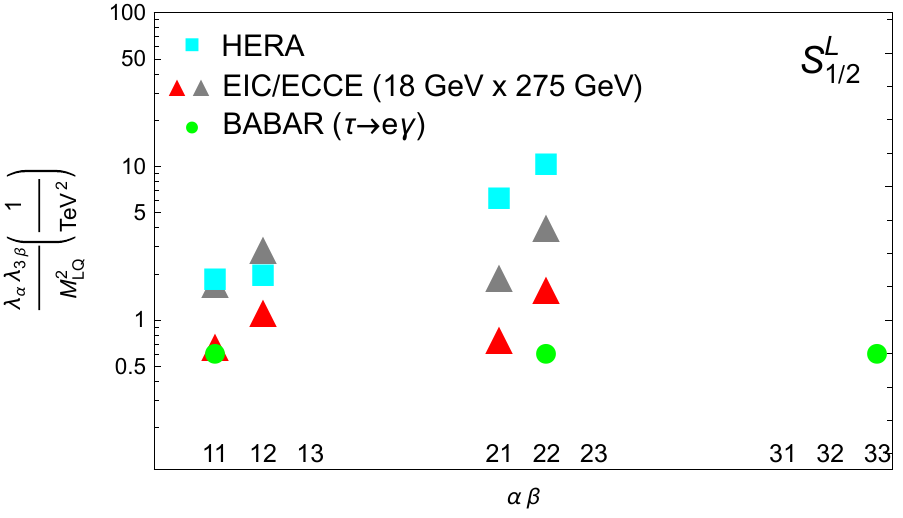}
    \includegraphics[width=0.45\textwidth]{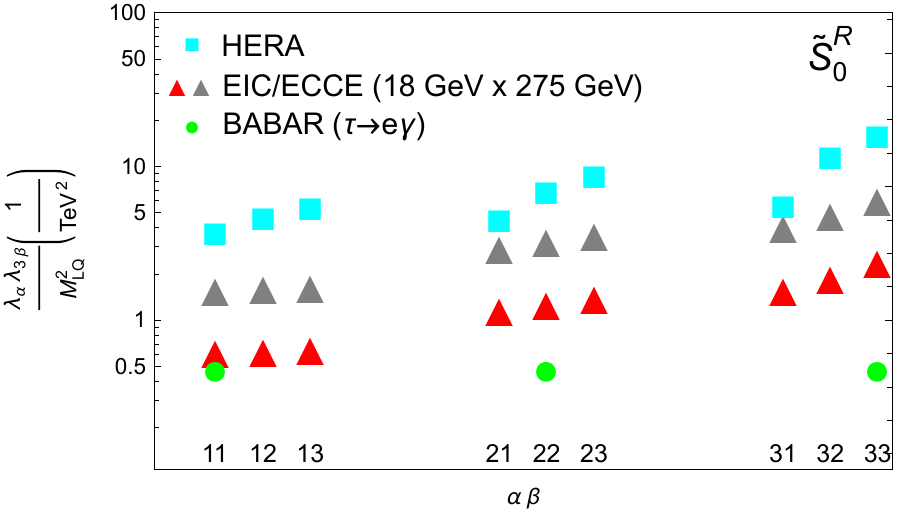}
    \caption{From~\cite{ecce-note-phys-2021-14-pub}: Limits on the scalar leptoquarks with $F=0$ $S_{1/2}^L$ (left) and $|F|=2$ $\tilde{S}_0^R$ (right) from 100~fb$^{-1}$ of $ep$ $18\times 275$~GeV data, based on a sensitivity to leptoquark-mediated $ep\to\tau X$ cross section of size 1.7~fb (red triangles) or 11.4~fb (grey triangles) with ECCE. Note that due to small value of $\sqrt{s}$, EIC cannot constraint the third generation couplings of $S_{1/2}^L$ to top quarks. Limits from HERA~\cite{ZEUS:2005nsy,H1:2007dum,ZEUS:2002dnh,H1:1999dil} are shown as cyan solid squares, and limits from $\tau\to e\gamma$ decays~\cite{Gonderinger:2010yn} are shown as green solid circles.}
    \label{fig:LQ_Limits_Scalar}
\end{figure}

We show in Fig.~\ref{fig:LQ_Limits_Scalar} the leptoquark limits, expressed in terms of $\lambda_{1\alpha}\lambda_{3\beta}/M_{LQ}^2$. The quantity $\lambda_{1\alpha}\lambda_{3\beta}/(M_{LQ}^2)$ characterizes the strength of the leptoquark-mediated contact interaction. The $\lambda_{ij}$ parameters, assumed to real for this analysis, denote the leptoquark couplings between the $i$-th lepton generation and $j$-th quark generation and $M_{LQ}$  denotes the leptoquark mass.
The left and right panels in Fig.~\ref{fig:LQ_Limits_Scalar} correspond to exclusion limits on the $S_{1/2}^L$ ($F=0$) and  $\tilde{S}_0^R$ ($|F|=2$) leptoquarks, respectively, assuming 100~fb$^{-1}$ of luminosity for the $18\times 275$~GeV energy configuration. Here the leptoquark fermion number $F=3B+L$, where $B$ and $L$ denote the baryon and lepton number, respectively. We see that the EIC can improve upon existing HERA limits and complement limits from $\tau\to e\gamma$, since the latter only impact the contact interactions in the quark flavor-diagonal channels ($\alpha=\beta$).
Once again, we note that while the effect of DIS CC background is accounted for, the DIS NC and photoproduction backgrounds are not,  since no such events survived the selection cuts for the sample size used in our study.

\subsubsection{CLFV Mediated by Axion-Like Particles: a Golden Opportunity at the EIC}

Another context in which $e^- \rightarrow \tau^-$ events can naturally appear at the EIC is in the presence of axion-like particles (ALPs) with CLFV couplings, in the form $e A_Z \rightarrow \tau A_Z a$, where $A_Z$ is an ion with charge $Z$ and $a$ is an ALP.  ALPs appear in a wide variety of extensions of the SM, over a broad range of parameters.  Some of these models involve CLFV in the interactions of the ALPs, for example, as a consequence of non-trivial dynamics in a ``hidden" or ``dark'' sector (see Ref.~\cite{Davoudiasl:2017zws}), or as a ``familon'' \cite{Calibbi:2020jvd} or ``Majoron'' \cite{Chikashige:1980ui,Schechter:1981cv} associated with spontaneous breaking of global symmetries.  Signals of this general possibility have been studied for a variety experimental settings in Refs.~\cite{Calibbi:2020jvd,Cornella:2019uxs,Bauer:2021mvw}.  

\begin{figure*}[t!]
     \centering
         \includegraphics[width=\textwidth]{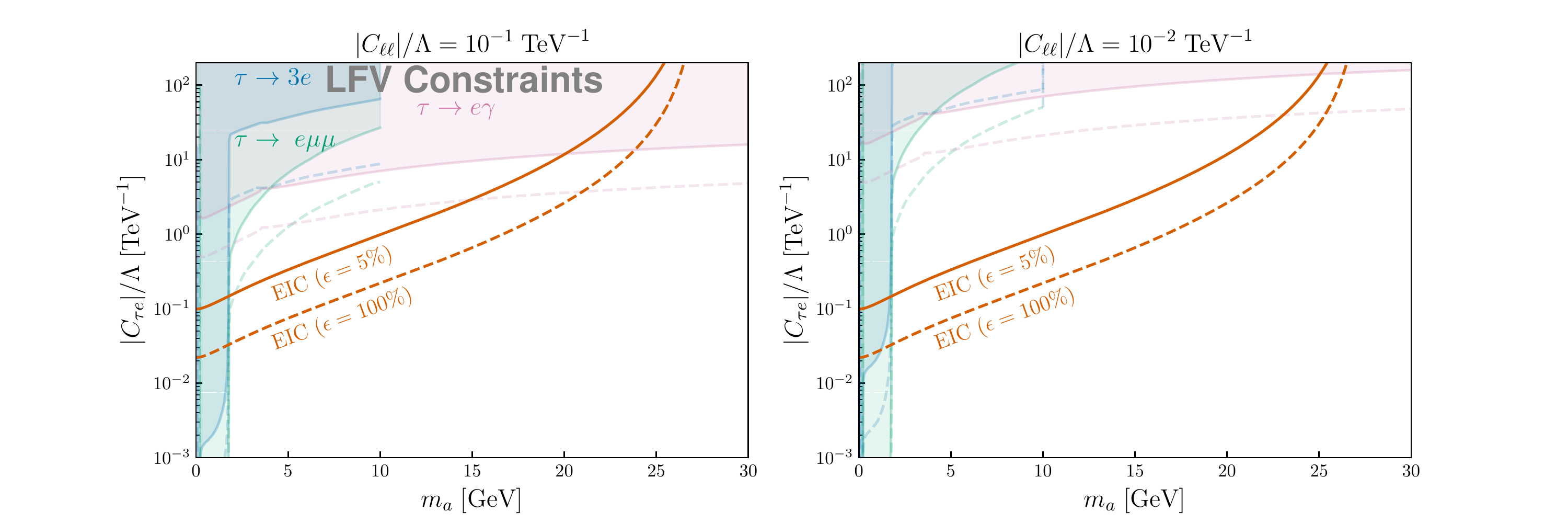}
     \caption{
      Estimated 90\% CL constraints on the ALP interaction strength $C_{\tau e}$, in TeV$^{-1}$ units, from future measurements of electron-gold collisions at the EIC. Other CLFV constraints from BABAR (solid lines) and projections from Belle II (dashed lines) for diagonal couplings  $|C_{\ell\ell}|/\Lambda = 10^{-1}$~TeV$^{-1}$ (left) and $|C_{\ell\ell}|/\Lambda = 10^{-2}$~TeV$^{-1}$ (right). The CLFV limits (90\% CL) are from Ref.~\cite{Cornella:2019uxs} for $m_a \leq 10$~GeV.  The $\tau \rightarrow e\gamma$ constraint for $m_a > 10$~GeV is calculated using expressions from Ref.~\cite{Cornella:2019uxs}, as it is dominant in that regime.  The CLFV constraints are evaluated using  their dependence on $|C_{\ell \ell}| / \Lambda$; the contribution from the $a\gamma \gamma$ coupling is assumed to be negligible.  These plots assume zero background, and acceptance $\times$ efficiency of 100\% (lower curve) or 5\% (upper curve), with the latter corresponding to a rough estimate for efficiency of identifying the conversion of the beam electron to final-state $\tau$, following  \cite{Zhang:2022zuz}.  The plots are from Ref.~\cite{Davoudiasl:2021mjy} (where LFV denotes CLFV).}
     \label{fig:Ctaue}
\end{figure*}

In Ref.~\cite{Davoudiasl:2021mjy}, it was pointed out that electron-ion collisions at the future EIC can provide an excellent opportunity to probe such interactions, in particular for $e-\tau$ CLFV mediated by ALPs in the $\sim$~GeV or higher mass range.  Coherent electromagnetic scattering at low $q^2$ from a high-$Z$ ion -- like the gold nucleus with $Z=79$ -- can lead to significant production of ALPs from the electron beam,  accompanied by a $\tau$ lepton in the final state.  In addition, electron beam polarization can be used to probe the parity properties of the ALP leptonic couplings.  The relatively high center of mass energy $\sqrt{s} \sim 100$~GeV achievable at the EIC \cite{AbdulKhalek:2021gbh} allows it to probe the ALP $e-\tau$ CLFV coupling, $C_{\tau e}/\Lambda$, well below other projected limits, for ALP masses up to $\sim \text{few}\times 10$~GeV, with 100 fb$^{-1}$ of integrated luminosity.  These results are shown in Fig.~\ref{fig:Ctaue}.  Between the left panel and the right panel, the flavor diagonal ALP couplings are reduced, which makes other probes less effective but does not change the EIC reach at all.  This makes the EIC a unique and complementary probe of this flavor-violating coupling.  More information regarding the plots are provided in the figure caption; for a detailed account of the assumptions and calculations relevant to these results, see Ref.~\cite{Davoudiasl:2021mjy}.           

\subsection{Other opportunities}
\subsubsection{Testing Charged Current Chiral Structure}
Polarized electron and positron beams at the EIC would allow measurements of the charged current cross sections for a few different values of the beam polarization $P_e$. 
The cross sections for the charged current processes $e^- + p \to \nu_e + X$ and $e^+ + p \to \bar{\nu}_e +X$  have a linear dependence on the lepton beam polarization: 
\begin{equation}
\sigma_{\rm SM}^{e^{\pm}p} (P_e) = (1\pm P_e) \sigma_{\rm SM}^{e^{\pm}p} (P_e=0), 
\end{equation}
where $P_e$ denotes the lepton beam polarization. Polarized electron and positron beams allow measurements of the charged current cross sections for a few different values of the beam polarization $P_e$. A straight-line fit of this polarization dependence can be  extrapolated to obtain the charged current cross section at $P_e=\mp 1$ and can be compared to the SM prediction $\sigma_{\rm SM}^{e^{\pm}p} (P_e=\mp 1)=0$. Any observed deviation could indicate the presence of new physics that couples to right-handed electrons and left-handed positrons. A non-zero value for $\sigma^{e^{\pm}p} (P_e=\mp 1)$ could arise from the coupling of a right-handed $W$-boson ($W_R$) of mass $M_R$ to a right-handed electron and a right-handed neutrino.

Although the EIC will have smaller cross sections compared to HERA,  due to its lower center of mass energy, its higher luminosity and  degree of lepton beam polarization can allow for a more precise extraction of $\sigma^{e^{\pm}p}_{\rm upper \>bound} (P_e=\mp 1)$,  allowing for the possibility of stronger limits. The limits obtained using positron beams are expected to be stronger, due to the correspondingly smaller SM charged current cross section, allowing for enhanced sensitivity to new physics. For a 10 GeV positron beam colliding with a 100 GeV proton beam ($\sqrt{s}=63.25$ GeV), with a cut of $Q^2 > 100$ GeV$^2$, a 95\% CL upper bound of $\sigma^{e^{+}p} (P_e=- 1) < 0.0207$pb is obtained corresponding to the limit  $M_R > 270$ GeV~\cite{Furletova:2018nci}, assuming a luminosity of 100 fb$^{-1}$.  For an energy setting of $\sqrt{s}=109.5$ GeV the limit improves to $M_R > 285$.  Although more detailed studies are required, these preliminary results indicate that the EIC can compete and make modest improvements on the HERA limits \cite{Aktas:2005ju} which require $M_R > 208$ GeV.  While the Tevatron and the LHC have already set more stringent limits on $M_R$ (in the TeV range) by looking for deviations in the transverse mass distribution of the Drell-Yan process $pp\to W \to \ell \nu_\ell$, the observed distribution is sensitive to a time-like charged boson and in general can be affected by physics that involves a different combination of chiral and flavor structures.  Thus,  new limits from the EIC can provide complementary information to limits from colliders.

\subsubsection{Heavy Photons}

\begin{figure}[!h]
\begin{center}
\includegraphics[width=0.5\textwidth]{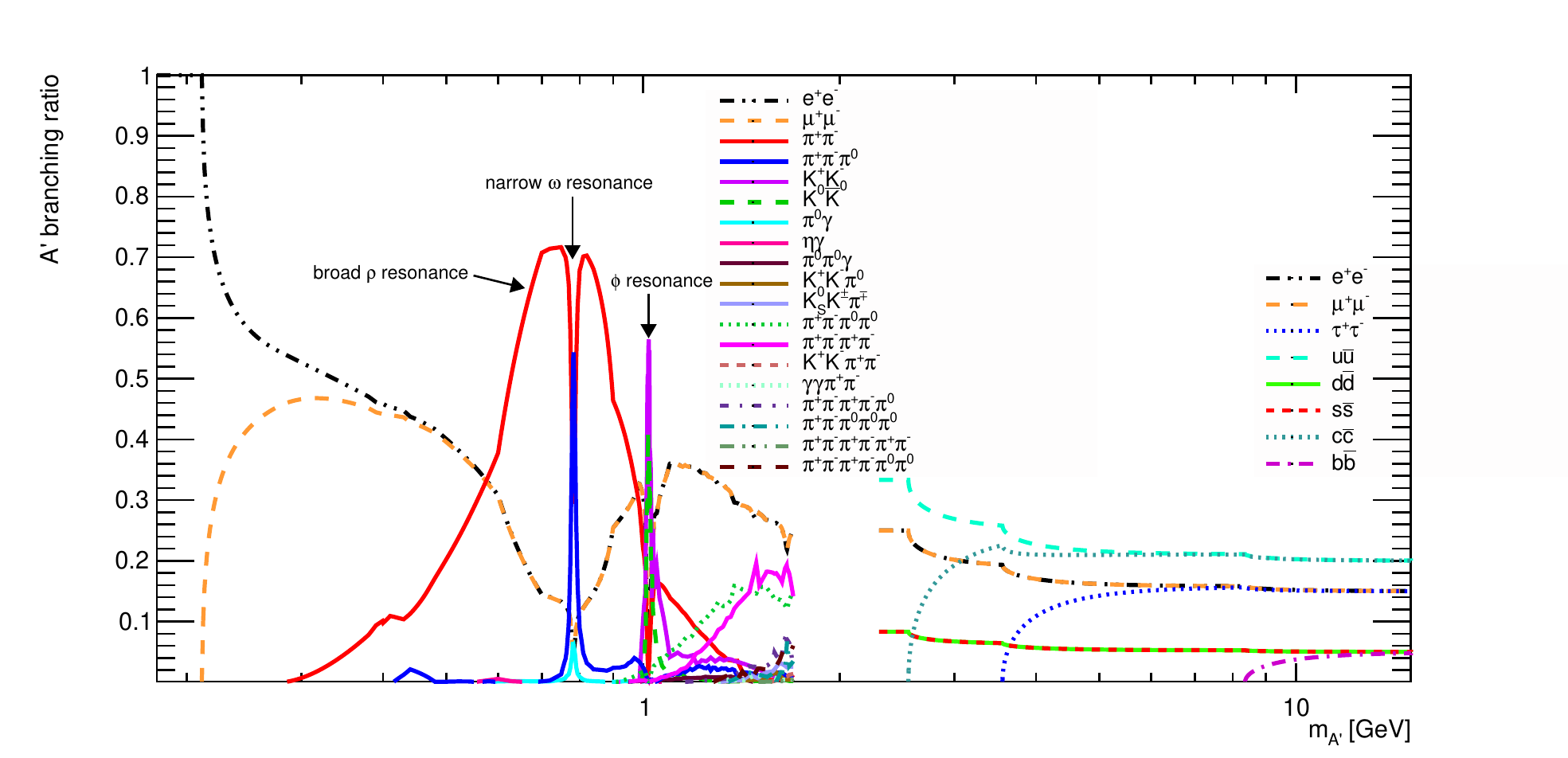}
\includegraphics[width=0.4\textwidth]{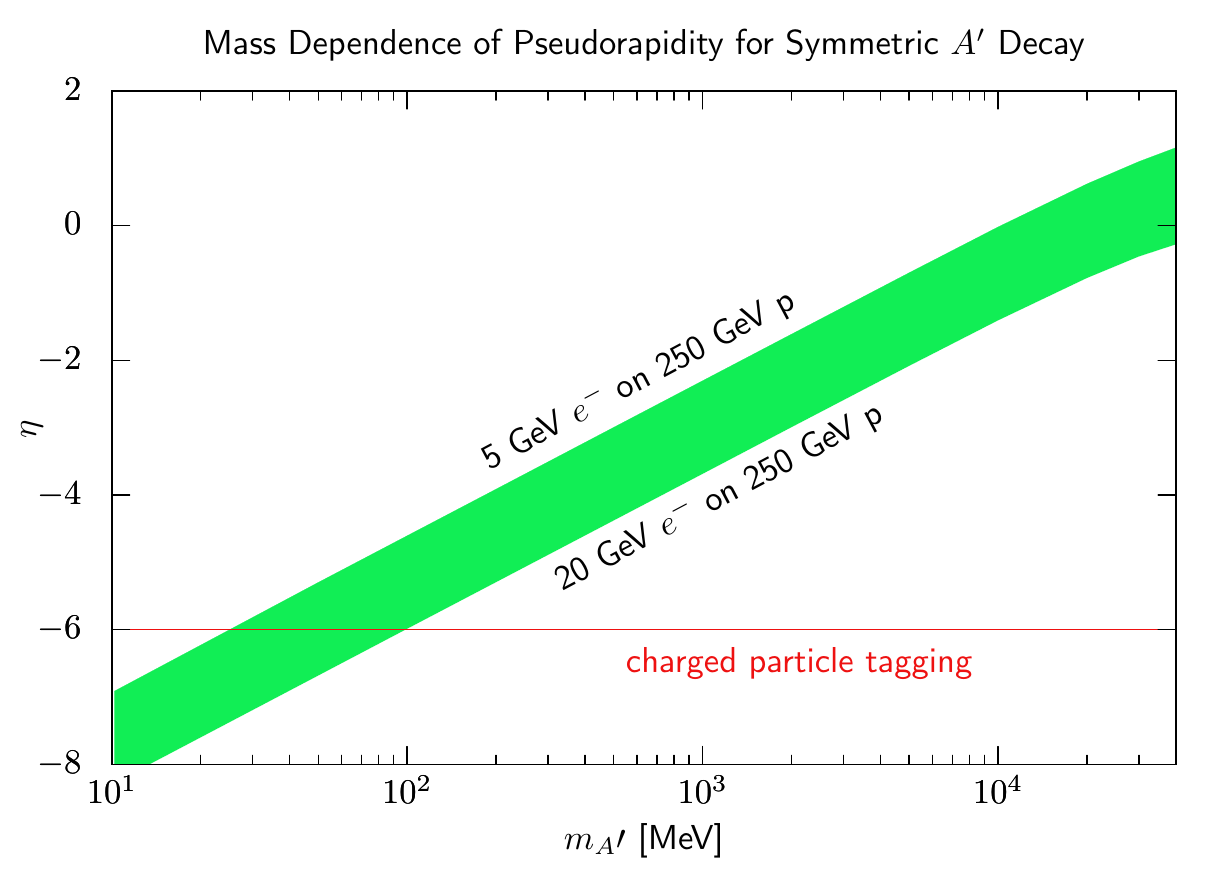}
\caption{Left: Branching ratios for a dark photon to various standard model particles as a function of the dark photon mass, assuming a simple mixing model.  Dilepton final states dominate at lower energies, and hadronic final states dominate at higher energy.  Several resonances are also present.  The relative cleanliness of leptonic final states make these an attractive search channel.  Plot from \cite{Buschmann2015}. Right: Detector pseudorapidity of symmetric-angle decay leptons.  Largest and smallest relative boosts are shown, along with the electron-going bound on particle tagging in the current reference detector design.  The boost of the collision frame with respect to the detector allows access to masses much lower than the center-of-mass energy.}
\label{fig:darkphoton_branching}
\end{center}
\end{figure}

In various SM extensions, an additional U(1) gauge field boson mixes weakly with the SM photon.
\begin{equation}
\mathcal{L}\supset-\frac{\epsilon}{2} F_{\mu\nu}^\prime F^{\mu\nu}.
\end{equation}
Through this mixing, the dark photon ($A'$) inherits a small coupling to SM charge, and can hence be produced through any interaction that could produce a photon in the final state.  These models have two parameters:  the mass of the dark photon, $m_{A'}$, and the strength of the coupling, $\epsilon$.  In the absence of additional dark-sector particles, the branching ratios in the kinetic mixing model can be derived from hadron production measurements at $e^+e^-$ colliders (Figure \ref{fig:darkphoton_branching} Left \cite{Buschmann2015}).  More generic fifth-forces invoke different mechanisms to produce small couplings to SM particles, permitting species-dependent couplings.  At the highest masses, current exclusion limits are dependent on particular Higgs production models.  Direct measurement of leptonic or hadronic couplings in these mass ranges would provide a model-independent constraint on the existence of new interactions.

The EIC offers several opportunities to mount such searches.  For masses above $\sim$100~MeV, up to the center-of-mass energy of the collisions, dark photons can be sought through radiative production in diffractive events and subsequent decay to lepton pairs, $ep\rightarrow epA'\rightarrow epl^+l^-$.  This production mechanism is used in fixed target experiments, but no such experiment can approach the $>$100~GeV center-of-mass energy available at the EIC, the equivalent of a $>$10~TeV fixed-target electron beam.

In the fixed target frame, an $A'$ emitted via initial state radiation (ISR) prefers to carry the majority of the beam momentum, produced at $\theta\sim(m_A/E)^{3/2}$.  Symmetric decays in this frame have an opening angle $\theta_d\sim m_A/E$, making it challenging to separate decay leptons from the outgoing beam for $m_A\ll E_\mathrm{beam}$.  In contrast, the asymmetric beam energies of the EIC drastically reduces the boost of the $A'$, resulting in large opening angles that are compatible with current proposed designs for EIC detectors, in terms of tracking and calorimetry, for masses above $m_{A'}\sim 100\mathrm{MeV}$ (Figure \ref{fig:darkphoton_branching} Right).  

In this search, an $A'$ would manifest as a resonant peak in the invariant mass of dilepton decay products, above a smooth background expected from SM processes. In most models, the intrinsic width of the resonance is expected to be narrow, so angular and energy resolutions in the electron-going direction dominate the reach of this approach.
This process can be used to probe the traditional dark photon landscape parameterized by the particle's mass, $m_{A'}$, and a single coupling strength, $\epsilon$, but also allows separate measurement of the couplings to electrons and muons in more general models.



\begin{figure}
\begin{center}
\includegraphics[width=0.4\textwidth]{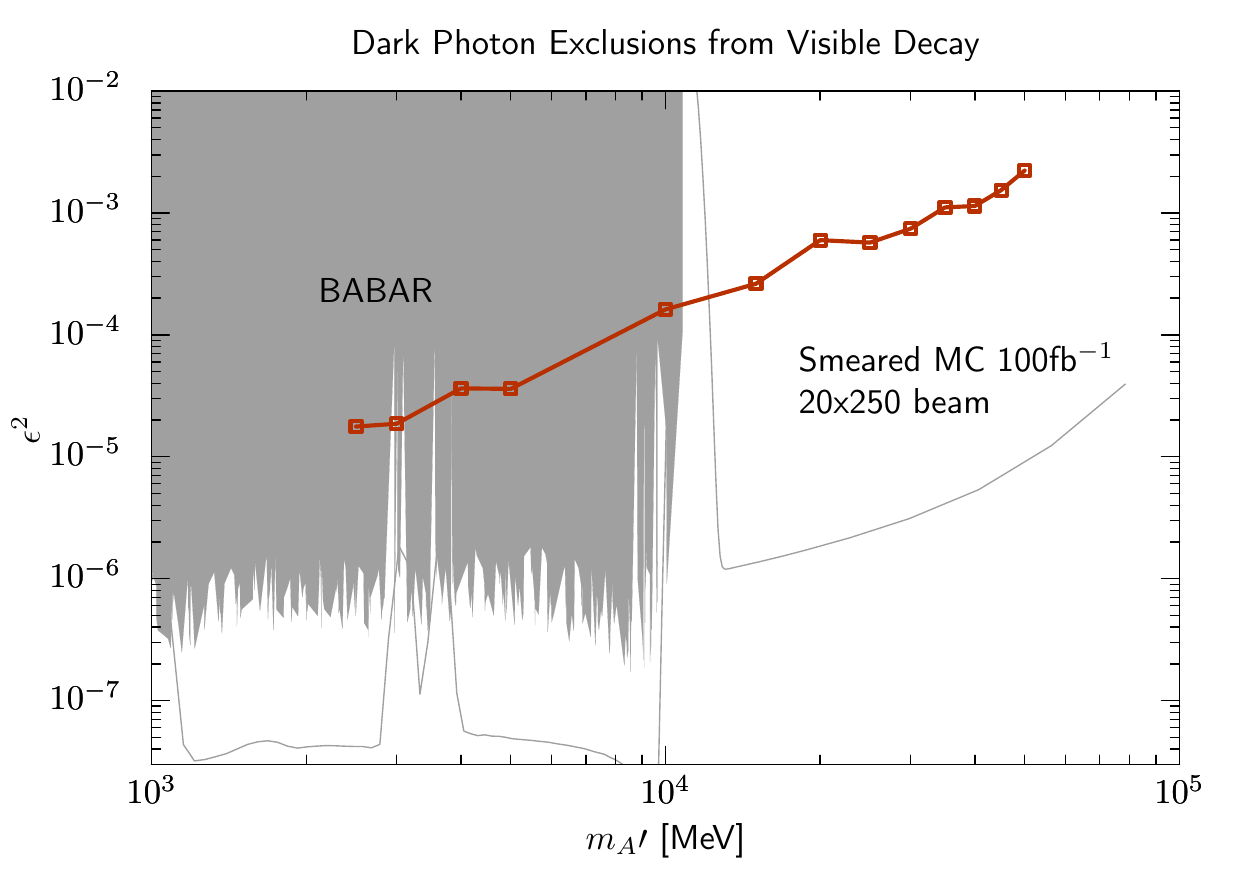}
\caption{Dark photon mass and coupling strength for a benchmark $ep\rightarrow epA'\rightarrow epee^+$ search.  The solid red line reflects the curve above which a dark photon could be excluded at $2\sigma$ with 100~fb$^{-1}$ of data.  Performance is shown without cuts to enhance signal, and without hadronic background contributions.  Effects due to structures in the branching ratio are not represented, but will follow similar structures in existing measurements.  The solid gray area represents the region excluded by BABAR's existing search,  while the thin gray lines represent projections for Belle II (leptonic, left) and LHCb (hadronic, right) regions excluded through existing leptonic searches. Further details are in the text.}
\label{fig:darkphoton_reach}
\end{center}
\end{figure}

A benchmark study has been performed using MadGraph Monte Carlo samples of $ep\rightarrow epA'\rightarrow epee^+$ for a range of dark photon masses, along with QED background processes with the same final state.  The projected sensitivities shown in Figure~\ref{fig:darkphoton_reach} consider only signal and irreducible QED background, assuming 100\% branching to electrons and using a benchmark detector configuration.  With 100~fb$^{-1}$, the higher energy settings of the EIC allow access beyond the upper limit probed by BABAR, and provide a leptonic-generation counterpart to projections from LHCb.  The effect of known resonances in these mass regions are not represented in this plot, but are expected to produce `blind spots' similar to those in Belle II and others.

Other backgrounds such as hadrons mis-identified as leptons, or combinatoric backgrounds from multiple events, require further study.  Studies currently underway suggest that the distinct event topology of the signal, two high-energy leptons back-to-back in azimuth and with correlated pseudorapidity, will offer additional ways to discriminate against these backgrounds.  Alternative Monte Carlo generators are also being explored to natively incorporate proton form factor effects and radiative corrections (the latter of which is not treated in the current approach).

In addition to the leptonic decays described here, dijet final states could also provide a probe of quark couplings to a generalized dark photon produced via ISR.  The EIC aims at relatively precise (and comparably ``clean'') measurements of the hadronic final state, allowing more sophisticated testing of models with leptophobic mediator couplings.  At lower dark photon masses, Dalitz decays of $\pi^0$ and $\eta$ to $\gamma l^+l^-$ final states provide additional probes up to the mass of the parent meson and could provide substantially larger samples than in the high-mass study presented here. 

Additional dark photon studies have been recently performed for HERA and the EIC~\cite{Yan:2022npz}.

\subsubsection{Lorentz- and CPT-Violating Effects}

The high precision of the data that is expected from the EIC opens up the possibility to test both Lorentz and CPT symmetry at new levels. While these symmetries are so far very well established, BSM theories exist that have one or both of these symmetries be spontaneously broken. This summary is based on the work of E. Lunghi, V. A. Kostelecky and collaborators. For specific details please see \cite{Lunghi:2018uwj, Kostelecky:2019fse,Lunghi:2020hxn} and references therein.  

Low-energy tests of Lorentz and CPT symmetry can be performed using the effective field theory framework known as the Standard-Model Extension (SME). To date, SME operators describing Lorentz- and CPT-violating effects on QCD degrees of freedom have been largely unconstrained.
The studies cited above suggest differential cross section measurements at the EIC will allow 
for precision tests of Lorentz and CPT symmetry in the quark sector. Specifically, data for unpolarized inclusive deep inelastic scattering at 100~fb$^{-1}$ luminosity can increase bounds on quark sector coefficients by two orders of magnitude over data taken at HERA. 

Symmetry violations would be visible as  variations in the cross section as a function of sidereal time. Additional processes, including those with polarization effects, charged-current exchange, and QCD corrections, have the ability with the EIC to place first constraints on a number of completely unexplored effects stemming from Lorentz and CPT violation.

\subsection{Summary}
\begin{itemize}
    \item The Electron Ion Collider data will give access to EW and BSM observables that are of interest to the entire particle physics community;
    \item The EIC will be able to make extractions of the weak mixing angle using both polarized electron-proton and electron-deuteron scattering in a phase space between the $Z$-pole and the upcoming SoLID PVDIS determination; 
    \item The high precision polarized data from the EIC provides unique constraints within the SMEFT framework;
    \item We expect that with the data collected at the EIC we will be able to increase the sensitivity to CLFV in the $e$ to $\tau$ sector, through leptoquark exchange or ALP production;
    \item The EIC data will extend the reach of dark-photon searches beyond what BABAR has determined and provide a leptonic-generation counterpart to the projections from LHCb.
\end{itemize}


%
\def\head#1{\vspace{6pt} \noindent \textbf{\textcolor{black}{ #1:}}\ }  
\newpage
\section{Tomography of Hadrons and Nuclei at the EIC} \label{sec:tomo}
\vspace{-2ex}
\centerline{\textit{Editors:} 
\href{mailto:thobbs@fnal.gov}{\texttt{Timothy Hobbs}},
\href{mailto:avp5627@psu.edu}{\texttt{Alexei Prokudin}},
\href{mailto:alessandro.vicini@mi.infn.it}{\texttt{Alessandro Vicini}}.
}
\vspace{3ex}
The future program of the High-Luminosity LHC (HL-LHC)~\cite{Apollinari:2015bam} is premised on the realization of
next-generation sensitivity to a wide variety of both standard model (SM) and beyond
standard model (BSM) processes. The success of this program in stringently
testing the SM and performing impactful measurements at the TeV-scale is
dependent upon improvements to modern knowledge of the internal structure of hadrons.
This dependence is realized in numerous circumstances, including the substantial
PDF limitations to determinations of standard-candle measurements at $pp$ colliders,
including the Higgs cross section, gauge-boson masses, and electroweak couplings and
mixing angles.
Simultaneously, a number of HEP activities on the Intensity and Neutrino Frontiers, including
the upcoming program at LBNF/DUNE, are similarly precision-limited by both
single-nucleon and nuclear uncertainties. In the case of LBNF/DUNE, these
uncertainties come primarily in the form of incomplete knowledge of neutrino-nuclear
($\nu A$) interactions for neutrino energies falling within the few-GeV region,
$E_\nu\!\sim$ few GeV.
Related considerations apply
to the proposed Forward Physics Facility (FPF)~\cite{Anchordoqui:2021ghd}
at the HL-LHC exploiting the intense
flux of neutrinos produced
in $pp$ collisions in the forward direction.

In parallel with these future experiments, the Electron-Ion Collider (EIC) envisions a program dedicated
to unraveling the details of the strong interaction through an array of
precision DIS measurements.
While the strongly-interacting character of QCD makes it especially challenging to compute,
it is exactly this complexity that leads to fascinating phenomena such as 
confinement at large distance scales and asymptotic freedom at short distance scales.
The ultimate goal of the EIC program is a more
systematic understanding of these challenging features of QCD,
including a campaign to probe the transition separating perturbative
QCD dynamics from nonperturbative phenomena.

\subsection{Tomography and HEP}
\label{sec:tomography}

In exploring this physics, the EIC will for the first time comprehensively
map the internal landscape of hadrons --- foremost, the proton --- as well as
both light and heavy nuclei. These multi-dimensional maps amount to an unraveling of
the {\it tomography} of these QCD bound states, an undertaking which will provide
detailed information on nonperturbative distributions like nucleon and nuclear PDFs,
transverse-momentum dependent (TMD) distributions, and generalized parton distributions
(GPDs). The EIC will therefore specialize in constraining the very objects
that so frequently represent systematic limitations to HEP precision. For this reason,
the EIC is distinguished as a facility for both precision QCD as well as a facility
that will critically advance key objectives of HEP by enhancing precision at
the Energy, Intensity, and Neutrino Frontiers.

In this Section, we first briefly review the conceptual basis for hadronic and 
nuclear tomography in Sec.~\ref{sec:tomography} below, before providing a brief survey in
Sec.~\ref{sec:QCD} of a number of representative HEP areas which will be particularly impacted by the EIC's precision tomography
program. Thereafter, we highlight several areas subsumed by tomography, including
nucleon and nuclear (un)poloarized PDFs, TMDs, and GPDs. In a number of instances,
we anticipate projected impacts of the EIC program on these quantities, and note the
possible HEP phenomenological implications for these improvements.

\subsubsection{Hadronic and nuclear tomography}
\label{sec:tomography_nphad}
With growing precision in both Energy and Intensity Frontier activities,
an improved understanding of the subtleties of the QCD transition from
nonperturbative to perturbative dynamics will be essential
to continued progress. The signatures of this transition are imprinted
on the partonic quark-gluon distributions of hadrons, including the proton,
as well as of nuclei. 
In fullest generality, these distributions may assume the form of unintegrated
quantities such as the 5-dimensional {\it Wigner distribution}~\cite{Ji:2003ak,Belitsky:2003nz,Lorce:2011kd,Lorce:2015sqe},
\begin{equation}
    W^{\pm}_q(x,b_\perp,k_\perp) = \int \frac{d^2\Delta_\perp}{(2\pi)^2}\,e^{-i\Delta_\perp\cdot b_\perp}\frac{1}{2}\int \frac{dz^- d^2z_\perp}{(2\pi)^3} \,e^{ixp^+z^- -ik_\perp\cdot z_\perp} 
    \left\langle p+\tfrac{\Delta_\perp}{2} \right|\bar{q}(-\tfrac{z}{2})\Gamma\mathcal{U}^{\pm} q(\tfrac{z}{2})\left|p-\tfrac{\Delta_\perp}{2}\right\rangle\ ,
\end{equation}
which describes the quark-level substructure of a hadron simultaneously in terms of its
localization in longitudinal momentum ($x$), transverse momentum ($k_\perp$), and
impact-parameter space ($b_\perp$). The extraction of this multidimensional
information on hadron structure amounts to a systematic probe of the
hadron's tomography --- a chief objective of the EIC. As we elaborate
below, constraining the tomographic structure of the proton will be closely
bound up with enhanced precision at the LHC, LBNF/DUNE, and other HEP facilities.

While unintegrated quantities like the Wigner function provide a holistic
picture of hadronic wave functions, a crucial aspect is the possibility of
projecting other distributions more familiar to HEP studies. For instance,
from a generic Wigner function, one might obtain (schematically) from successive
integrations
\begin{equation}
    f(x) = \int d^2k_\perp f(x,k_\perp) = \int d^2b_\perp \int d^2k_\perp W(x,b_\perp,k_\perp)\ ,
\end{equation}
the collinear parton distribution function (PDF) of the nucleon, $f(x)$, in which we have
suppressed explicit factorization scales and other features. This collinear
PDF can in turn be obtained from a more general transverse momentum dependent (TMD)
PDF; meanwhile, distinct integrations yield the generalized parton distributions
(GPDs), which encode impact-parameter dependence.
Notably, these distributions are of a fundamentally nonperturbative
nature, being defined as matrix elements of partonic currents within the
hadron at characteristic scale(s) comparable to the hadronic mass. At
softer scales typical of QCD bound-state masses, the properties of QCD
as a non-Abelian gauge theory are such that the strong interaction $\alpha_s$
is too strong to allow standard small-coupling perturbation theory.
While this is the case, nonperturbative methods like lattice QCD may provide
useful information as we discuss further in the sections below.
In the meantime, the predominant means of extracting information on hadron
structure is the technique of QCD global analysis, which combines parametrizations
of the nonpertubative quantities like the PDFs with perturbatively-calculable
Wilson coefficients or parton-level matrix elements.
These calculations are embodied by QCD factorization theorems. For inclusive
quantities like the DIS structure functions, these assume a relatively simple
form.

The LHC and EIC together will provide a combined data set with
unprecedented precision and reach in energy across a diverse
range of targets.
However, to obtain the optimal benefit from this combined data, it will
take coordination and communication between the LHC and EIC projects.
Therefore, it is entirely appropriate that the Snowmass planning
process is bringing these communities together.
In a broader sense, the study of hadronic structure will provide a
deeper comprehension of both the QCD theory and the encompassing
Standard Model (SM); hence, this endeavor is synergistic with a broad
spectrum of research as diverse as dark matter searches and
electroweak (EW) symmetry breaking.

In these Snowmass proceedings, we consider this broader connection,
detailing several main areas related to the physics reach of the
EIC to HEP via improvements to the knowledge of hadronic structure.
We start in Sec.~\ref{sec:QCD} by further expounding upon the phenomenological
implications of the EIC measurements, particularly at the LHC, but also in connection
to precision measurements in upcoming $\nu A$ programs.
We then build the connection on EIC-driven improvements to knowledge
of the collinear structure of hadrons and nuclei in Sec.~\ref{sec:collinear},
moving through a discussion of the nucleon's unpolarized and
spin-polarized PDFs in Sec.~\ref{sec:unPDFs} and~\ref{sec:spinPDFs}, respectively,
before discussing nuclear PDFs in Sec.~\ref{sec:nPDFs} thereafter.
Following this discussion, we move on to TMD quantities in Sec.~\ref{sec:TMDs}, considering
implications related to  TMD PDFs (Sec.~\ref{sec:TMDPDF}) and TMD fragmentation functions
(\ref{sec:TMDfrag}). Sec.~\ref{sec:GPDs} develops the role of GPDs, including discussions
of both phenomenological and lattice QCD studies overlapping EIC science. 

\subsubsection{Precision in Perturbative QCD and the EW Sector}
\label{sec:QCD}

\begin{figure}[th]
\centering
\includegraphics[width=0.48\textwidth]{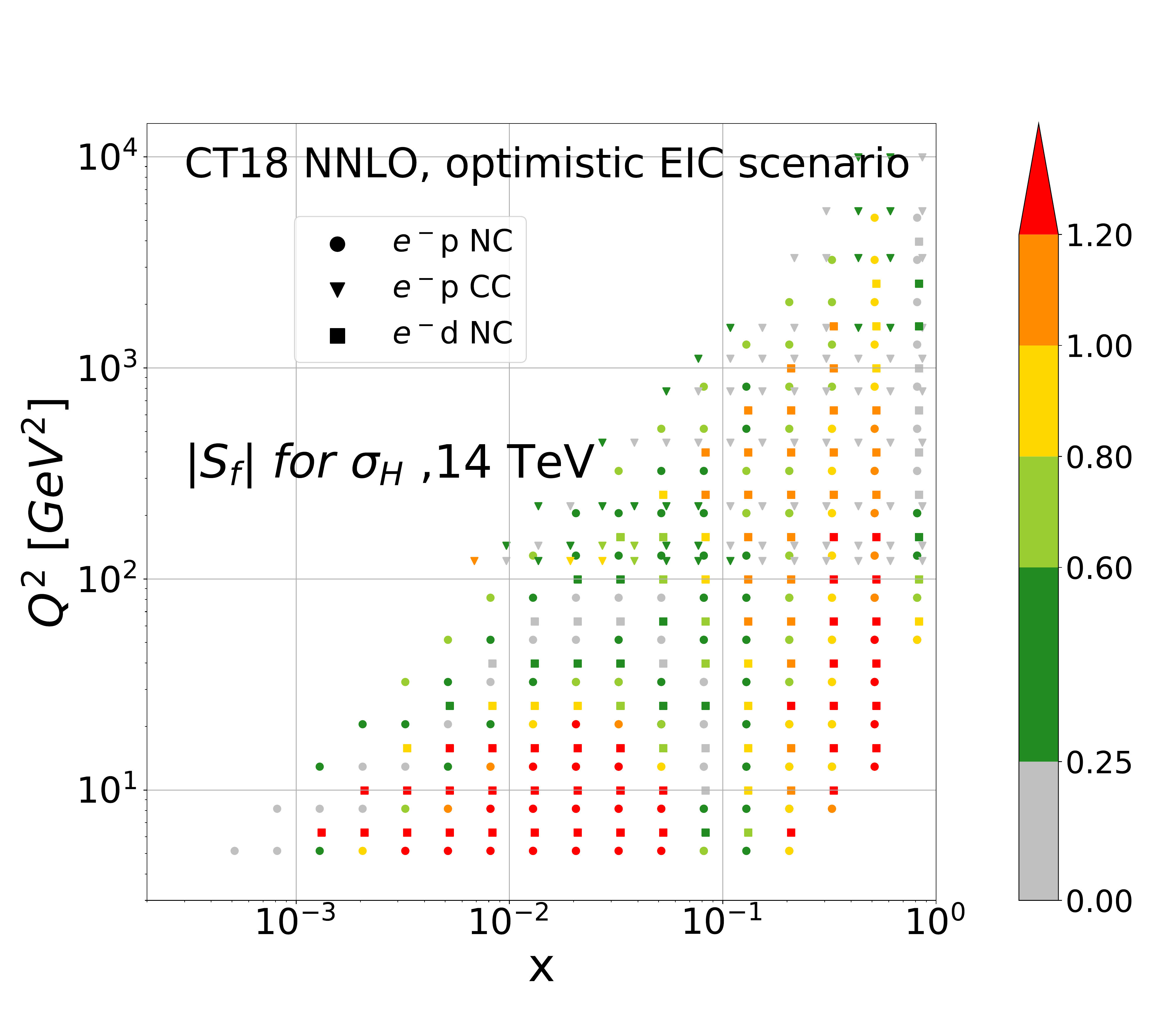} \ \
	\raisebox{0.1cm}{\includegraphics[width=0.46\textwidth]{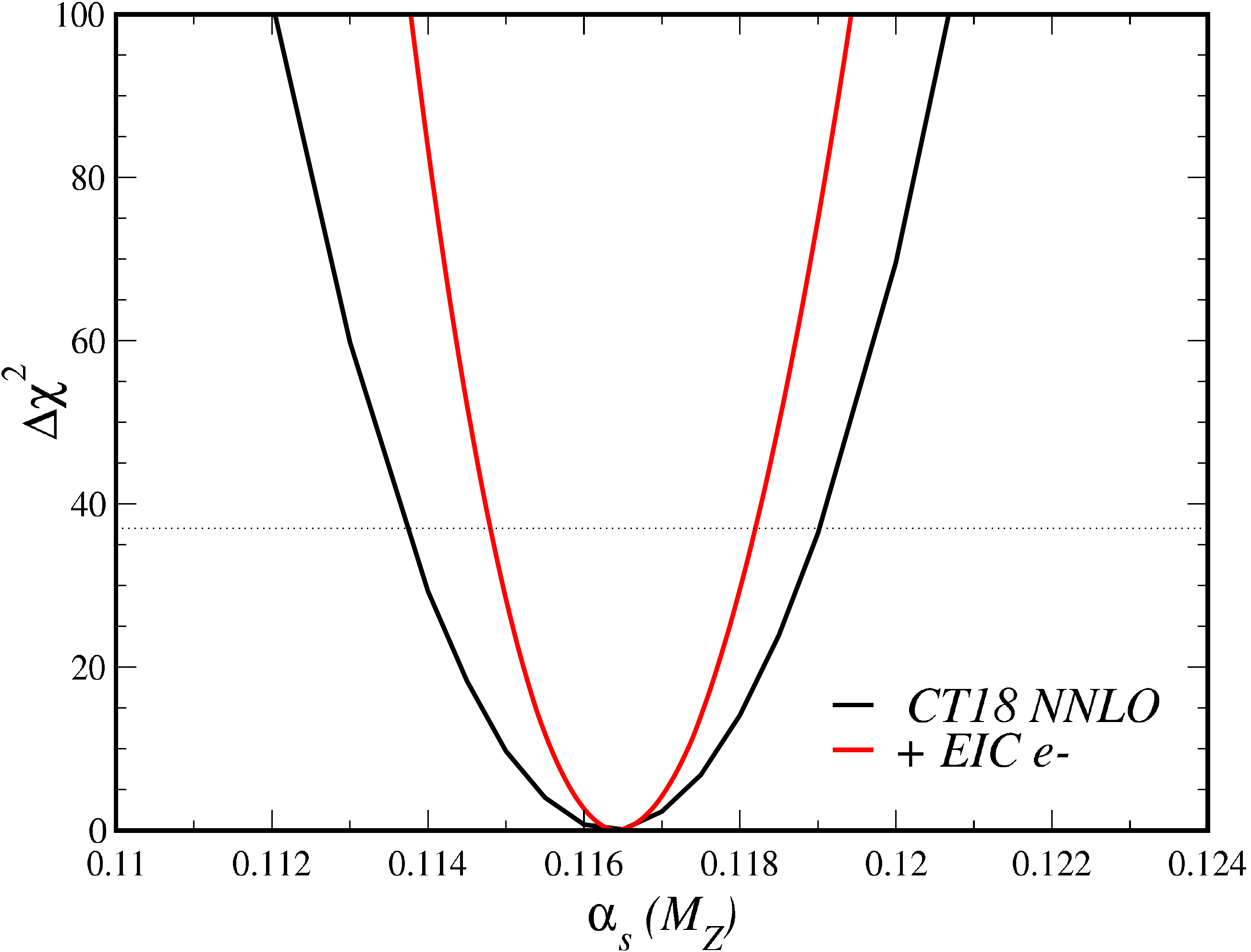}}
\caption{The PDF sensitivity of the $100$ fb$^{-1}$ EIC pseudodata explored in the EIC Yellow Report translates into a
	substantial point-by-point impact on the SM Higgs production cross section (left), as visualized using the
	\texttt{PDFSense} methodology~\cite{Wang:2018heo}. Similarly, extensive DIS collisions covering a range of scales
	and $x$ values results in important potential constraints on the strong coupling, $\alpha_s$ (right); the precision
	on $\alpha_s$ can improve by $\sim 40\%$ over the corresponding uncertainty in CT18 NNLO following the inclusion
	and analysis of $100$ fb$^{-1}$ of electron-scattering data. Both panels are taken from Ref.~\cite{AbdulKhalek:2021gbh}.
	}
\label{fig:EIC}
\end{figure}

\begin{figure}[th]
\centering
\includegraphics[width=0.46\textwidth]{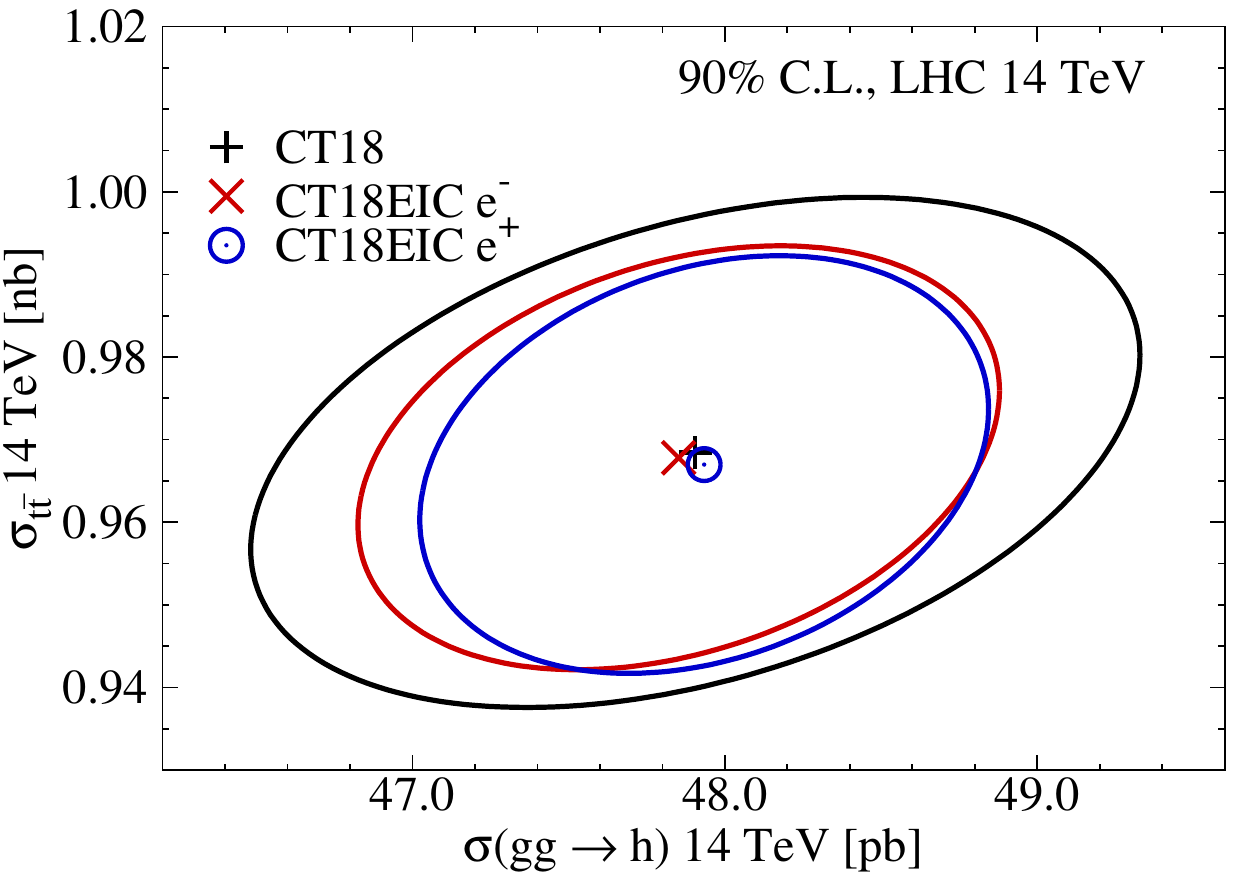} \ \
\includegraphics[width=0.46\textwidth]{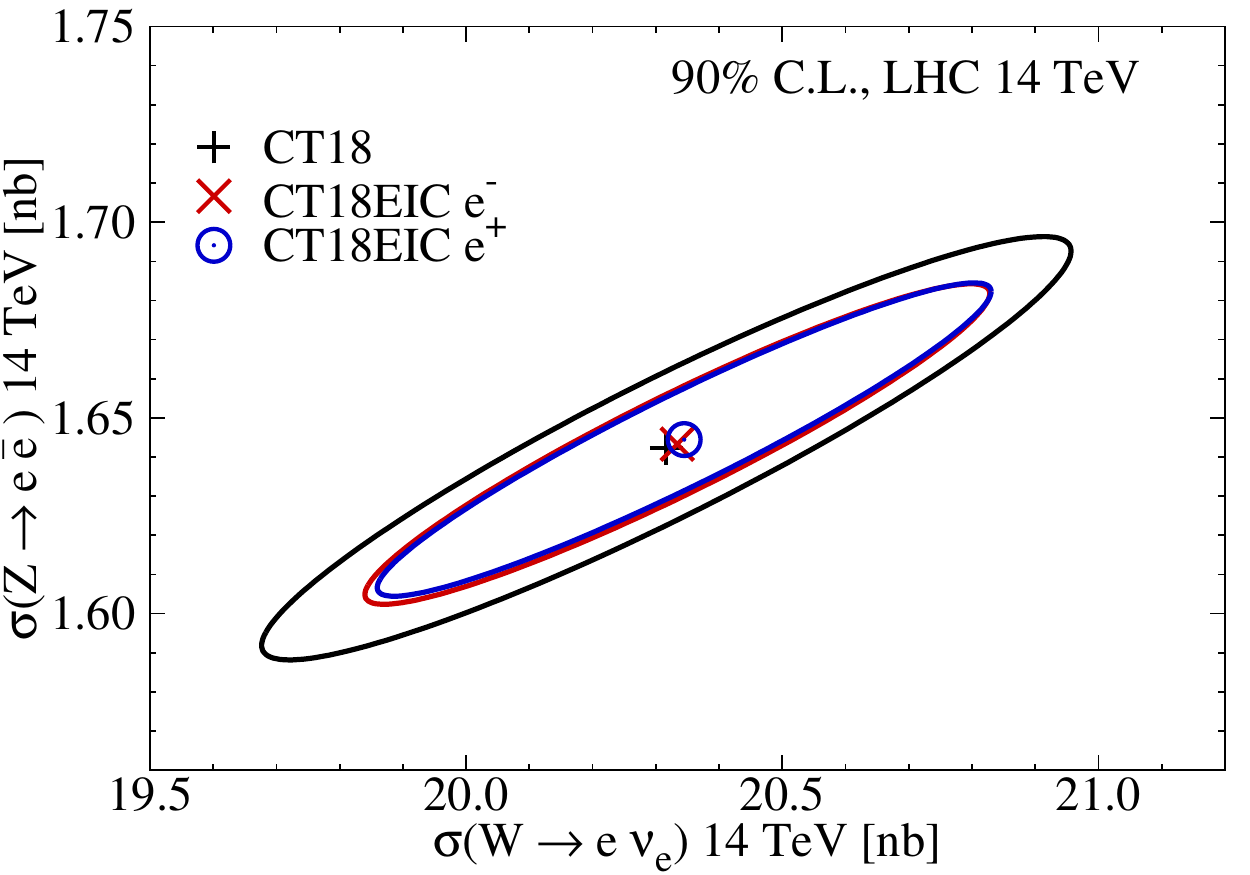}
\caption{Through its PDF sensitivity, the EIC can drive improvements in the uncertainties of
LHC processes, including Higgs, $t\bar{t}$, and electroweak-boson production cross sections. In this
case, we plot the improvements over CT18 NNLO to the 14 TeV cross sections for $t\bar{t}$ vs.~Higgs production (left)
and $Z$ vs.~$W$ production (right), due to the optimistic electron (red ellipses) and electron + positron
apseudodata (blue ellipses).
	}
\label{fig:EIC-LHC-XS}
\end{figure}

\noindent
{\bf High-energy QCD and perturbative QCD developments.}
By measuring a range of interactions up to $\sqrt{s}\! =\! 140\,\mathrm{GeV}$
in comparatively clean $ep$ and $eA$ DIS processes, the EIC will probe fundamental
aspects of QCD that are otherwise challenging to disentangle via hadronic
collisions alone. These include the DIS production and dynamics of heavy quarks and
QCD jets (discussed at length in companion studies --- Refs.~\cite{EICHFLOI,EICjetLOI},
respectively). In turn, such processes in the unpolarized case offer an array of
complementary channels to inclusive neutral- and charged-current structure function
measurements. When taken together, these data can drive future QCD global analyses by
providing a comprehensive flavor separation of (un)polarized PDFs and enhancing
precision determinations of QCD sector SM quantities, including $\alpha_s$
and the heavy-quark masses (see Refs.~\cite{LOITheory,LOIN3LO,HeavyQuarkMassLOI}).
These PDF improvements suggest that the EIC will have a considerable impact upon a range of observables
of high interest within HEP, including the inclusive Higgs-production cross section. For example,
we can evaluate the PDF sensitivity of a typical set of EIC pseudodata (specifically,
the {\it optimistic} scenario reduced cross sections examined at PDF-level in Sec.~\ref{sec:unPDFs}
below) to the 14 TeV $gg \to H$ cross section, represented in terms of the {\it sensitivity} metric
of Ref.~\cite{Wang:2018heo}, $|S_f|$. In Fig.~\ref{fig:EIC} (left), we map the EIC pseudodata in the space of $(x,Q^2)$ probed by
the EIC, with the plotted colors indicating very pronounced sensitivity to the standard model
Higss cross section at the LHC --- sensitivity driven by potential EIC improvements to the gluon PDF. This PDF sensitivity translates into potential
for significant improvements to the Higgs cross section, such as that
shown together with the $t\bar{t}$ production cross section in the left
panel of Fig.~\ref{fig:EIC-LHC-XS}.
Similarly, the $\mathcal{L}\! =\! 100\, \mathrm{fb}^{-1}$ electron data set has the potential to markedly improve
extractions of the QCD coupling, $\alpha_s$, as we show in Fig.~\ref{fig:EIC} (right), based on an
analysis in the CT18 NNLO framework.

In order to facilitate the unambiguous extraction and interpretation of multi-dimensional distributions, it will be
necessary to further develop the QCD factorization theorems which allow the separation
of soft, non-perturbative matrix elements from hard partonic sub-processes. Extension to
higher orders in perturbative QCD will also be necessary for higher precision. The delineation
of leading, twist-$2$ from higher-twist matrix elements will be enabled by high-precision
EIC data, and the accompanying theory must be developed to extract higher-twist effects.
These developments can be expected to benefit the understanding of multi-parton interactions
at hadron colliders as well as QCD processes like jet-$p_T$ broadening in nuclear
scattering.

Forward physics is becoming more and more important at the LHC, HL-LHC and future hadron colliders, not only for QCD physics, {\it e.g.}, central exclusive production, but also for EW measurements, {\it e.g.}, Higgs production in diffractive mode or multi-boson production with associated forward jet/proton tagging. It has also been demonstrated that measurements in this phase-space corner can provide powerful constraints on BSM searches.

At jet $p_T$ values on the order of 10-20 GeV (or lower), hadronization, missing higher-order corrections, parton shower (PS), ME-PS matching schemes as well as underlying event (UE)/multi-parton interactions (MPIs) play significant roles in uncertainties. In several of these measurements at the LHC, where jet vetoes are applied on jets with $p_T> 20$ or 30 GeV  the theory/model uncertainties are already by far dominating the experimental results, limiting the theoretical interpretation of the data in search for BSM anomalies.
The clean EIC environment, where QCD processes can be studied using an electromagnetic probe and where factorisation holds, will help test the calculations and models. 
For instance, we can test the MC models for fragmentation, PS, ME-PS matching, 
including whenever possible perturbative higher-order corrections,
in events where low-$p_T$ jets are produced centrally, together with some forward activity, {\it e.g.}, a forward jet or an intact proton/ion. Studies of cross sections as a function of the gap size, for instance, will allow tests of the modeling of color-flow in MCs. By using the tracking system in EIC detectors, very low-$p_T$ jets or hadrons can be identified and studied as a function of the event topology and event activity, providing stringent tests to MCs.
Such studies will be an important starting point for a further validation phase at hadron colliders.

Furthermore,
since jet sub-structure has been extensively studied at the LHC and has been found to be an extremely powerful tool to improve searches for BSM physics,
similar investigations can be carried on at the EIC on lower $p_T$ jets, 
composed of far fewer particles than at the LHC. 
Such studies will surely provide interesting results on jet properties that will also help HEP studies, {\it e.g.}, in low-$p_T$ jets appearing in EW analyses.\\

\noindent
{\bf QED interaction of the partons and photon parton density.}
The advent of high-precision measurements at hadron colliders necessitates the systematic and consistent inclusion
of EW corrections for many processes under study. For instance, the appearance of photonic initial-state collinear
divergences naturally leads to a treatment analogous to the one adopted in QCD for their factorization and reabsorption
in the physical PDFs; similarly, a photon density in the proton must be taken into account.
\begin{table}[!h]
\begin{center}
\begin{tabular}{|c|c|c|c|c|c|c|}
\hline
  collider & PDF set &  $\sigma_{QCD}$  &  $\sigma_{QCD\times EW}$ & $\delta_{QCD\times EW} $ & $\Delta_{env} $ & $\delta_{PDF}$\\
\hline\hline
$p\bar p$ 1.96 TeV &  {\tt NNPDF3.1} & 7710.0 & 7649.5 & -0.8 & 0.3\% & $^{+1.7\%}_{-1.7\%}$ \\
\hline
           &  {\tt CT18}     & 7683.8 & 7640.7 & -0.6 & & $^{+1.6\%}_{-2.3\%}$\\
\hline
           &  {\tt MMHT2015} & 7701.1 & 7625.8 & -1.0 & & $^{+2.4\%}_{-2.4\%}$\\
\hline
LHC 7 TeV &  {\tt NNPDF3.1} & 29356.2 & 29120.4& -0.8 &  1.8\% & $^{+0.9\%}_{-0.9\%}$\\
\hline
           &  {\tt CT18}     & 28836.9 & 28702.4 & -0.5 &  & $^{+1.5\%}_{-2.4\%}$ \\
\hline
           &  {\tt MMHT2015} & 29023.0 & 28709.1 & -1.1  &  & $^{+2.0\%}_{-2.1\%}$\\
\hline
LHC 8 TeV &  {\tt NNPDF3.1} & 34116.0 & 33840.2 & -0.8 & 1.6\% & $^{+0.8\%}_{-0.8\%}$\\
\hline
           &  {\tt CT18}     & 33562.2 & 33407.5 & -0.5 & & $^{+1.6\%}_{-2.4\%}$\\
\hline
           &  {\tt MMHT2015} & 33792.4 & 33420.8 & -1.1  &  & $^{+2.0\%}_{-2.1\%}$\\
\hline
LHC 13 TeV &  {\tt NNPDF3.1} & 57769.1 & 57287.6 & -0.8 & 1.1\% & $^{+0.8\%}_{-0.8\%}$ \\
\hline
           &  {\tt CT18}     & 57152.1 & 56898.9 & -0.4 &  & $^{+1.9\%}_{-2.5\%}$\\
\hline
           &  {\tt MMHT2015} & 57564.8 & 56899.3 & -1.2  &  &  $^{+2.1\%}_{-2.1\%}$\\
\hline
LHC 14 TeV &  {\tt NNPDF3.1} & 62454.4 & 61931.2 & -0.8 & 1.0\% &  $^{+0.8\%}_{-0.8\%}$\\
\hline
           &  {\tt CT18}     & 61840.8 & 61568.1 & -0.4 &  &  $^{+2.0\%}_{-2.5\%}$\\
\hline
           &  {\tt MMHT2015} & 62278.6 & 61553.7 & -1.2 &  &  $^{+2.2\%}_{-2.2\%}$\\
\hline
LHC 100 TeV &  {\tt NNPDF3.1} & 418617 & 412815 & -1.4 &  2.4\% & $^{+3.1\%}_{-3.1\%}$ \\
\hline
           &  {\tt CT18}     & 420218 & 418344 & -0.4 &  & $^{+5.5\%}_{-3.8\%}$\\
\hline
           &  {\tt MMHT2015} & 410367 & 405238 & -1.2 &  & $^{+6.4\%}_{-4.4\%}$\\
\hline
\end{tabular}
\caption{\label{tab:photonPDF}
Cross sections for on-shell $Z$ production, expressed in picobarns and computed with different PDF sets at different collider types and energies. The two columns show the results obtained with PDF parameterisations determined with a QCD-only analysis or including also EW effects. We define $\delta_{QCD\times EW}=100\, (\sigma_{QCD\times EW}/\sigma_{QCD} -1)$, while $\Delta_{env}$ is the percentage width of the envelope of the three PDF sets predictions in the QCD model, with respect to their mean value. The experimental PDF uncertainty $\delta_{PDF}$ in the QCD model is computed according to the definitions of each group.
} 
\end{center}
\end{table}
The photon parton density has been related, in the LUX-QED formulation \cite{Manohar:2016nzj}, to the structure functions $F_{2,L}$
describing the proton structure.
The perturbative evolution,
ruled by the DGLAP equations with QCD+QED kernels,
dynamically leads to a non-vanishing photon density, relevant for precision tests of the SM
and for the study of gauge-boson scattering at large energy scales.
While the evolution depends solely on the splitting functions,
the photon parameterisation at low scales
has been implemented in different ways by the CT~\cite{Xie:2021equ}, MSHT~\cite{Cridge:2021pxm} and NNPDF~\cite{Ball:2013hta,NNPDF:2017mvq,Bertone:2017bme} collaborations.

A natural comparison of the impact of these choices
can be found by computing a standard candle like the total cross section
for on-shell $Z$ production in hadron-hadron collisions, at different hadronic center-of-mass energies.
The results discussed in \cite{Bonciani:2019nuy,Bonciani:2020tvf,Bonciani:2021iis}, including radiative corrections up to NNLO QCD-EW,
are presented for the ease of discussion in Table \ref{tab:photonPDF}.
In order to perform a fair evaluation, for each collaboration two proton PDF sets,
one analysed in pure QCD and the other  in presence of QCD and QED effects, are considered.
The difference of the predictions in the two models
can be taken as an estimate of the role of the EW and QCD-EW corrections,
including the photon-induced contributions.
Such effects are different for the three collaborations and show a moderate but different evolution
with respect to the collider energy.
We have considered three pairs of PDF sets:
{\tt NNPDF31\_nnlo\_as\_0118} and
{\tt NNPDF31\_nnlo\_as\_0118\_luxqed}~\cite{NNPDF:2017mvq},
{\tt MMHT2015\_nnlo} and
{\tt MMHT2015qed\_nnlo}~\cite{Harland-Lang:2019pla}, and
{\tt CT18NNLO} \cite{Hou:2019efy} and
{\tt CT18qed} \cite{Xie:2021equ}.
The column $\delta_{QCD\times EW}$ gives the percentage difference between the total cross section computed in the two models, one including only QCD effects in the partonic cross section and in the PDF evolution, the other instead
consistently including all the QCD and EW SM corrections. The differences emerging from this comparison appear on top of the methodological differences in the QCD sector and are not due to different data sets, since the percentage is computed separately for each collaboration, considering two fits of the same data set.

The EIC provides a very interesting opportunity to test and validate
the proton model that includes a photon density and obeys a DGLAP QCD+QED evolution.
A precise determination of the
photon density over an extended range of partonic $x$ and its interplay with quark-gluon degrees-of-freedom will be informed
by the broad range of observables available at the EIC. Ultimately, an upgraded fitting framework will be needed to exploit
this potential: a systematic collection of partonic matrix elements with NLO-EW effects as well as control over the scale dependence
of the resulting PDFs with combined N$^3$LO (QCD) + NLO (QED) accuracy.
The possibility to link in a coherent way inelastic and elastic phenomena, with the two regimes of QED interaction,
is an important topic, both from the theoretical and experimental points of view.
The description of the transition between the two regimes and
the overall global intensity of the electromagnetic interaction
are important features of the proton structure.\\

\noindent
{\bf Electroweak boson production and $p_T$ spectra at $pp$ colliders}.
The production of electroweak bosons at the LHC is a fertile ground for
tests of the standard model. These tests substantially depend on precise
control over the standard model cross section, along the lines of the discussion
above for Higgs production. As with extractions of electroweak parameters
like $M_W$, the $W, Z$ cross sections are themselves PDF limited; as a result,
even the wide-coverage reduced cross sections the EIC will measure using
both neutral- and charged-current interaction can produce significant
improvements to the predicted cross section. We illustrate this effect based
on a CT18 NNLO analysis in the error ellipses of Fig.~\ref{fig:EIC-LHC-XS}
(right).

Precise measurements of $M_W$ at the LHC heavily rely on Z-boson differential distributions, such as the transverse momentum $p^Z_T$ spectrum (or the $\phi^*_{\eta}$) spectrum. An accurate theoretical prediction for $p^Z_T$, together with a reliable estimate of the associated perturbative and non-perturbative uncertainties, is thus mandatory for the precision physics programme.

It is well-known that fixed-order predictions for $p^Z_T$ show a divergent behavior for $p^Z_T \ll m_Z$ due to the presence of large logarithmic terms of the Sudakov type (double logs). An all-order resummation of these divergent contributions is thus required to high logarithmic accuracy in order to match the extremely high accuracy reached by experimental collaborations at the LHC.
Many mature formalisms have been developed by different communities and implemented in numerical codes: arTeMiDe \cite{Scimemi:2017etj}, NangaParbat \cite{Bacchetta:2019sam} and ResBos2 \cite{Wang:2012xs} solving evolution equations for the TMD parton distribution functions, DYRes/DYTurbo \cite{Camarda:2019zyx} and reSolve \cite{Coradeschi:2017zzw} implementing soft-gluon resummation in perturbative QCD, CuTe-MCFM \cite{Becher:2020ugp} and SCETlib \cite{scetlib} making use of Soft-Collinear Effective Theory, RadISH \cite{Bizon:2018foh} and PB-TMD \cite{BermudezMartinez:2018fsv} implementing coherent parton branching, or resummation in the direct $k_T$ space.
One of the goals of the EW Working Group at CERN is the benchmarking of all these different codes: to this extent, Drell-Yan pair production is a perfect playground since for this process all previous frameworks prove to be equivalent and should thus provide very similar results.
The benchmarking exercise aims at comparing predictions and understanding their differences, also trying to give a careful estimate of theoretical uncertainties. While a general agreement on central predictions has been reached, establishing a one-to-one correspondence between theoretical uncertainties stemming from different formalisms has proved to be non-trivial. 
In the very low $p^Z_T$ region ($\cal{O}$ 1 GeV), a potentially relevant source of uncertainty is the intrinsic-$k_T$ of the incoming partons. Global fits of the non-perturbative structure of TMD PDFs have been performed in recent years \cite{Bertone:2019nxa,Bacchetta:2019sam} from Drell-Yan and SIDIS data, pointing to a non-trivial flavor and $x$-dependence of these proton structure functions.
This emerging field suffers at present from the limited availability of Drell-Yan data with fine binning at very low $p^Z_T$ and the restricted kinematical region of the currently available SIDIS data. With its very large kinematical coverage for SIDIS processes at unprecedented luminosity, the EIC will surely play a relevant role in the 3D mapping of the proton, allowing us to enter a precision era for TMD PDFs.

\subsection{Collinear Structure of Hadrons and Nuclei}
\label{sec:collinear}

\subsubsection{Unpolarized nucleon PDFs}
\label{sec:unPDFs}

\noindent
{\bf Detailed quark-gluon structure of the proton}.
The EIC will pursue a dedicated program to constrain the unpolarized quark-gluon PDFs 
of the proton down to $x\!\! \sim\! 10^{-4}$, with strong sensitivity
to standard-candle measurements at the energy frontier like the
inclusive Higgs cross section highlighted in Fig.~\ref{fig:EIC} (left).
By performing a wide array of DIS measurements, the EIC will help relieve PDF limitations on various HEP observables by resolving a number
of issues in PDF phenomenology. These include significant sensitivity
to high-$x$ physics, including through investigations of phenomena like higher-twist (HT) and target mass 
corrections (TMCs). It should be stressed that high-$x$ phenomenology is especially important for  
constraining signatures of BSM physics in the rapidity or invariant-mass distributions commonly measured
in Drell-Yan processes with large mass scales at the LHC.
The EIC will also have important capabilities with respect to dynamics at small-$x$. These include a program
to investigate saturation, recombination, and low-$x$ resummation, potentially opening the door to more
detailed studies of the effects of BFKL and possible DGLAP violations, which have been challenging for
theoretical calculations. 
Also, as discussed in Sec.~\ref{sec:QCD} above, electroweak accuracy at the EIC will connect to PDF improvements
and higher precision: as QCD accuracy increases, new effects must be considered.  For example,
at sufficient precision (${\sim} 1\%$) QED effects enter, making necessary the consideration of
dedicated photon PDFs and isospin-violating effects.
In terms of heavy flavor, heavy-quark masses introduce an additional scale in the DIS process
which complicates the calculations. Using the combined energy reach of both the EIC and LHC we
can explore the full range of scales from the low-energy decoupling region to the high-energy (massless)
limit. Systematic treatments of heavy-quarks in DIS at the EIC, including charged-current DIS~\cite{Gao:2021fle} which is
important for PDF flavor separation, will both benefit from and be essential to next-generation
PDFs at the EIC. We also note that EIC measurements involving heavy flavor can be very consequential
for constraints to nucleon PDFs, as explored in more detail in Sec.~\ref{sec:hf}.
In addition to these PDF-related lines of investigation at the EIC, a primary future direction
for this field will be the use of EIC and other hadronic data
to explore phenomenological intersections between collinear PDFs and multi-dimensional distributions;
this will eventually entail expanding collinear PDF global fits to include larger simultaneous analyses of
the TMDs and GPDs discussed in greater detail in Sec.~\ref{sec:TMDs} and~\ref{sec:GPDs} below.
Leveraging the EIC to make progress on these issues will require 
a multi-disciplinary effort drawing upon expertise from many
subfields throughout the physics community. 
However, the likely reward for a concerted attack on these problems 
will be a revolutionary understanding of the fundamental dynamics of the
QCD dynamics within hadronic structure.\\
%

\begin{figure}[h]
  \centering
  \includegraphics[width=\textwidth]{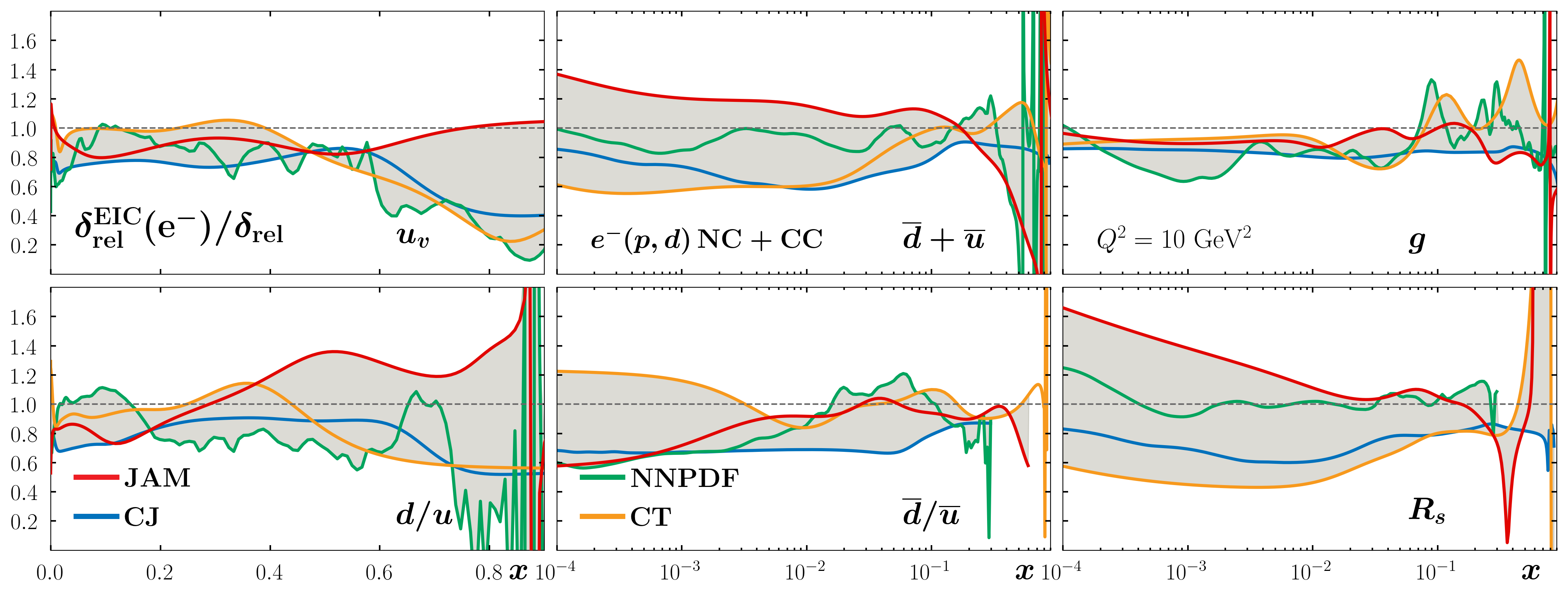}
  \caption{
  The EIC electron-scattering data have the potential to broadly improve
  the unpolarized PDFs of the proton. Here we show the result of the EIC YR~\cite{AbdulKhalek:2021gbh} analysis based on the {\it optimistic} systematic
  error performance scenario. Results were determined by four main QCD analysis
  groups, such that the relative improvement in the PDF uncertainty has some
  methodological dependence.
}
  \label{fig:ppdfs}
\end{figure}


\noindent
{\bf Proton PDF impact studies}.
The excellent potential of the EIC to improve our knowledge of the proton's unpolarized PDFs~\cite{Gao:2017yyd}
has been examined in a series of impact studies, particularly those carried out in the
context of the recent EIC Yellow Report~\cite{AbdulKhalek:2021gbh}. The most systematic examination of PDF
impacts involved studies of inclusive DIS cross sections, which might be measured on
both the proton and deuteron to afford greater flavor separation, as well as through
DIS interactions mediated by both neutral- and charged-current exchanges. Due to the
copious statistics expected when running at nominal EIC luminosities, a clear message
was the fact that efforts to unravel unpolarized PDFs will be limited by systematics,
which were in turn explored and estimated in a series of detector performance scenarios;
these were designated ``pessimistic'' and ``optimistic'' scenarios, corresponding to
comparatively conservative and aggressive systematic error projections, respectively. It is worth
noting that detector performances in these studies were significantly informed by
previous experience at HERA, such that further PDF impact studies associated with ongoing
detector concept development is to be expected over the coming years.
We also note that the default operating configuration of the EIC assumes access to a
polarized electron beam, with positron scattering as a possible upgrade. For this reason,
the EIC YR studies also explored this possibility as an independent scenario. We refer the interested reader to 
Sec.~7.1.1~of Ref.~\cite{AbdulKhalek:2021gbh} for details
related to the potential PDF implications of positron
beams at the EIC; we note that the additional flavor separation
afforded by positron-initiated charged-current DIS complements
the sensitivity to certain BSM possibilities, such as
charged-lepton flavor violating interactions as discussed
in Sec.~\ref{sec:CLFV}.

To explore the potential impact of inclusive DIS measurements at the EIC,
impact studies for the EIC YR took a set of pseudodata corresponding
to a broad range of reduced cross sections involving neutral-current electron scattering from the proton and deuteron as well as charged-current
scattering from the proton. Several QCD analysis groups --- the CTEQ-JLab (CJ), CTEQ-TEA (CT), Jefferson Lab Angular Momentum (JAM), and NNPDF Collaborations --- fitted these pseudodata in their respective frameworks, resulting in the results displayed in Fig.~\ref{fig:ppdfs}.
In Fig.~\ref{fig:ppdfs}, for each PDF global analysis, we plot the relative improvement in the PDF uncertainty after including the inclusive EIC pseudodata, showing a range of improvements, depending upon methodological
and other aspects of each fitting framework.
Fig.~\ref{fig:ppdfs} allows us to make a number of observations. First, the impact of the
EIC pseudodata can occur in various regions of $x$, depending upon the specific PDF flavor, as can be expected from the wide kinematical coverage of the EIC, which extends from low to very high $x$. Impacts at high $x$ can be especially significant for, {\it e.g.}, the $u_v$ PDF, for which PDF uncertainties could be reduced by up to
a factor of five for $x\gtrsim 0.8$. Significant impacts are also possible at large $x$ for the
$d/u$ PDF ratio (for which uncertainty reductions of as much as a factor of two for $0.5\lesssim x \lesssim 0.6$) and for the strange-suppression factor, $R_s$ (for which uncertainty reductions of $\sim\! 60\%$ may be achievable at $x \lesssim 0.01$). The relative uncertainty of the gluon PDF, is more modestly reduced at low-to-intermediate $x$, but with important implications for the Higgs cross sections as shown in the left panel of Fig.~\ref{fig:EIC}.
These features rely in part on the unique ability of the EIC to perform precise CC DIS
measurements at large $x$ and large $Q^2$: their theoretical interpretation
remains particularly clean, as any non-perturbative large-$x$ contamination
due, {\it e.g.}, to higher-twist effects, is suppressed. This possibility
distinguishes the EIC from HERA, which had a similar reach at high $Q^2$ but a
more limited access at large-$x$, and from fixed-target experiments
(including the recent JLab-12 upgrade~\cite{Dudek:2012vr}), which can access the
high-$x$ region only at small $Q^2$.

We note that complementary proton PDF impact studies have been conducted in parallel with the EIC Yellow Report analyses.
For example, a dedicated  NNPDF impact study~\cite{Khalek:2021ulf} examined the relative uncertainty of the proton PDFs
in an variant of the NNPDF3.1 global analysis~\cite{Faura:2020oom,Ball:2017nwa} once 
EIC pseudodata are included in the fit both for the optimistic and pessimistic scenarios.
The FONLL general-mass variable-flavor-number scheme~\cite{Forte:2010ta}
as implemented in {\tt APFEL}~\cite{Bertone:2013vaa} was used
for the consistent generation of pseudo-data.
A subset of flavors (or flavor combinations) was found to be most affected by the EIC pseudodata: $u$, $d/u$, $s$, and $g$.
The impact of the EIC pseudodata
was not found to depend on the scenario considered: the reduction of PDF
uncertainties remains comparable irrespective of whether optimistic or
pessimistic pseudodata projections are included in the fits. Because the two
scenarios only differ in systematic uncertainties, we conclude that it may be
sufficient to control these to the level of precision forecast in the pessimistic scenario.

\begin{figure}[h]
  \centering
  \includegraphics[scale=0.7]{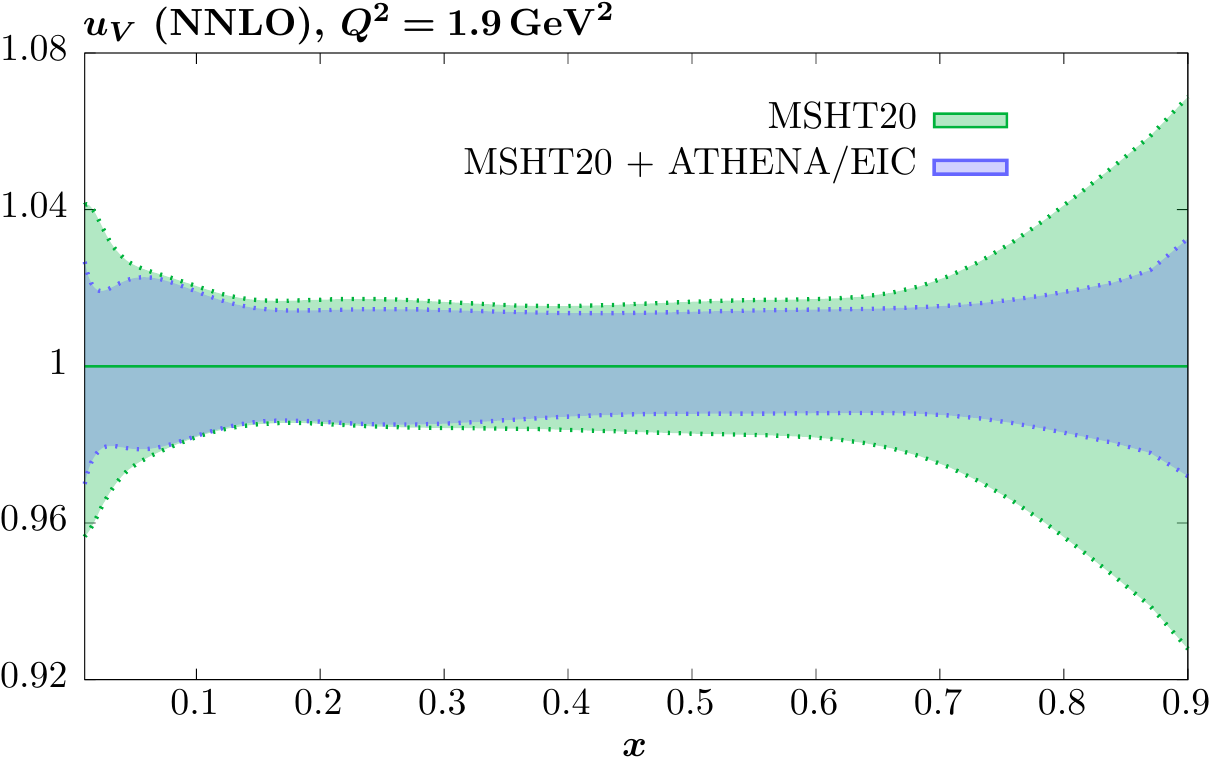}
  \includegraphics[scale=0.7]{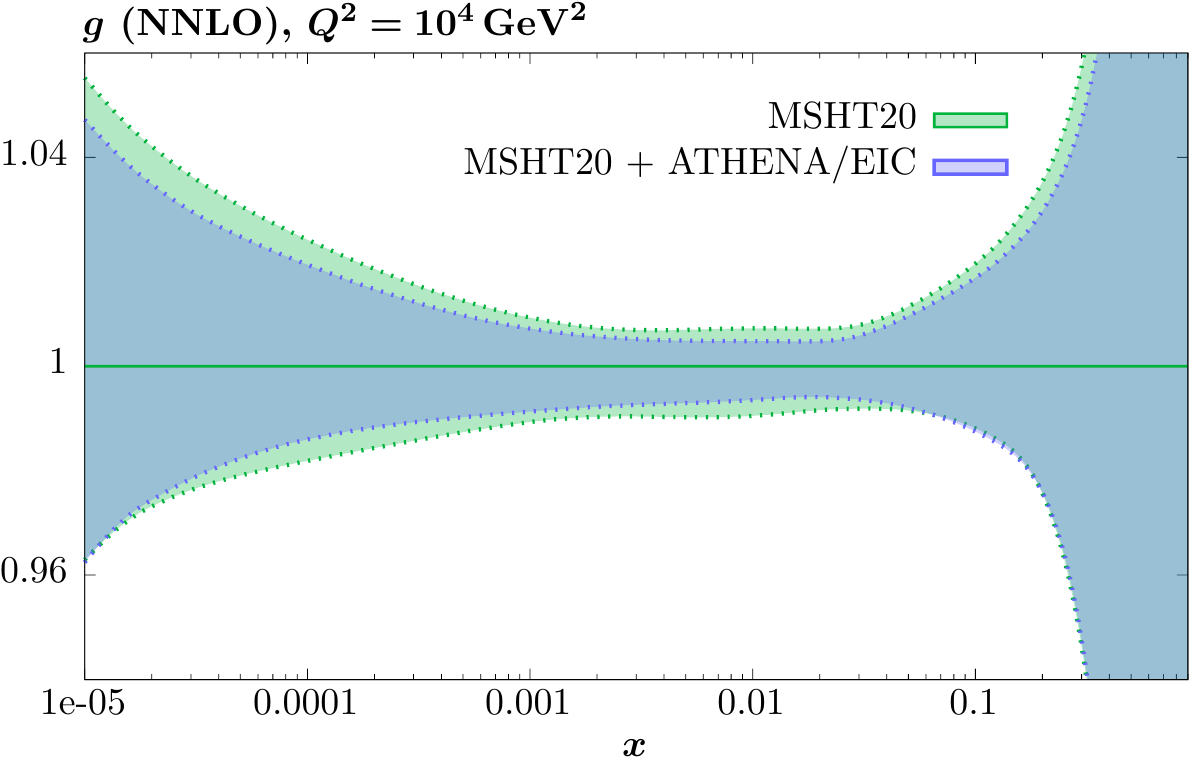}
  \caption{ Impact on the up valence (left) and gluon (right) PDF errors of ATHENA/EIC pseudodata, with the corresponding factorization scale $Q$ indicated.
}
  \label{fig:MSHT}
\end{figure}

Similarly, in Fig.~\ref{fig:MSHT} we show a small selection of results of an impact study within the MSHT20~\cite{Bailey:2020ooq} framework. In particular, $ep$ inclusive NC and CC pseudodata have been generated for a range of collider energies, and accounting for the expected $x$ and $Q^2$ binning (see~\cite{athena} for further information). These are then included in addition to the default MSHT20 global dataset, and a refit is performed. The pseudodata are produced using NLO QCD theory and consistently MSHT20NLO PDFs, while the fit is performed at NNLO, in order to effectively inject some amount of inconsistency between theory and (pseudo) data, as one might expect to occur in a real data/theory comparison, although the final result will not be too sensitive to this. These results have been produced as part of the ATHENA detector proposal, but are consistent with expectations for any EIC multi-purpose detector.

The PDFs of the valence $u$-quark, focused at high $x$, and the gluon across a range of $x$ are shown to give a representative picture of the impact of these pseudodata where it is greatest; other partons, for example the down quark, are less affected as we would expect from DIS data. We can see that the expected impact on the up valence at high $x$ is significant, with a relative reduction of up to $\sim 50\%$ in the highest $x$ region. This demonstrates the clear potential for the EIC to impact on the high $x$ region, as described in more detail above. We can also see a moderate but non--negligible impact on the gluon (shown at higher scale for clarity, in order to avoid any negativity at low $x$) across a range of $x$, due to the constraints from scaling violations at higher--order gluon--initiated production in the DIS process. Therefore, the potential for the EIC to constrain PDFs away from the high $x$ region, at least at higher scales, is also clear.

\subsubsection{Spin-Dependent nucleon PDFs}
\label{sec:spinPDFs}

A precise determination of spin-dependent PDFs is essential for understanding the spin structure of the proton, and the decomposition of the total spin into its quark and gluon helicity and orbital angular momentum components.
To this end, knowledge of the spin PDFs in the small-$x$ region, and in particular, the polarized gluon PDF $\Delta g$, is vital for the reconstruction of the moments.

\begin{figure}[t]
    \centering
    \includegraphics[width = 0.44\textwidth]{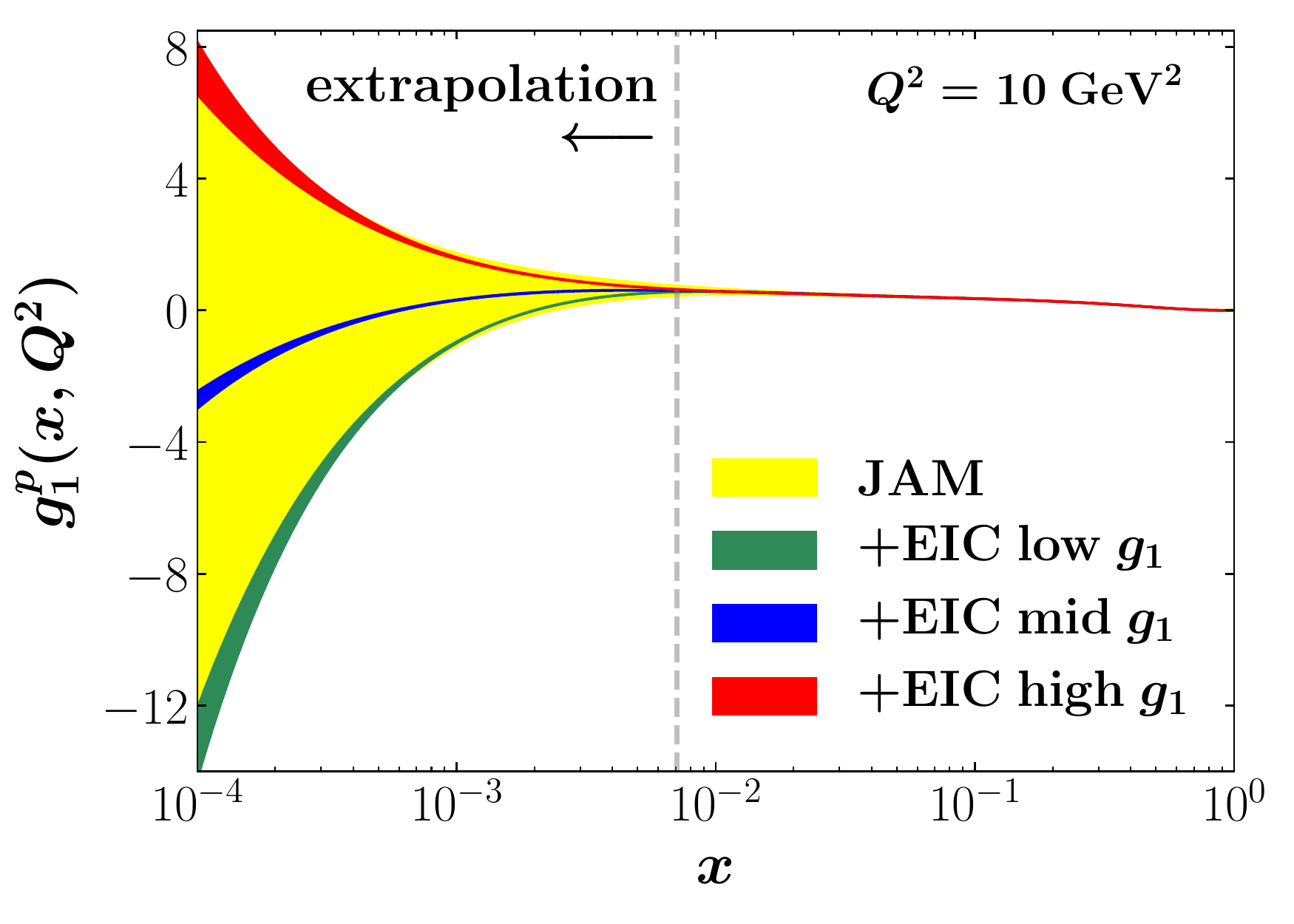}
    \includegraphics[width = 0.44\textwidth]{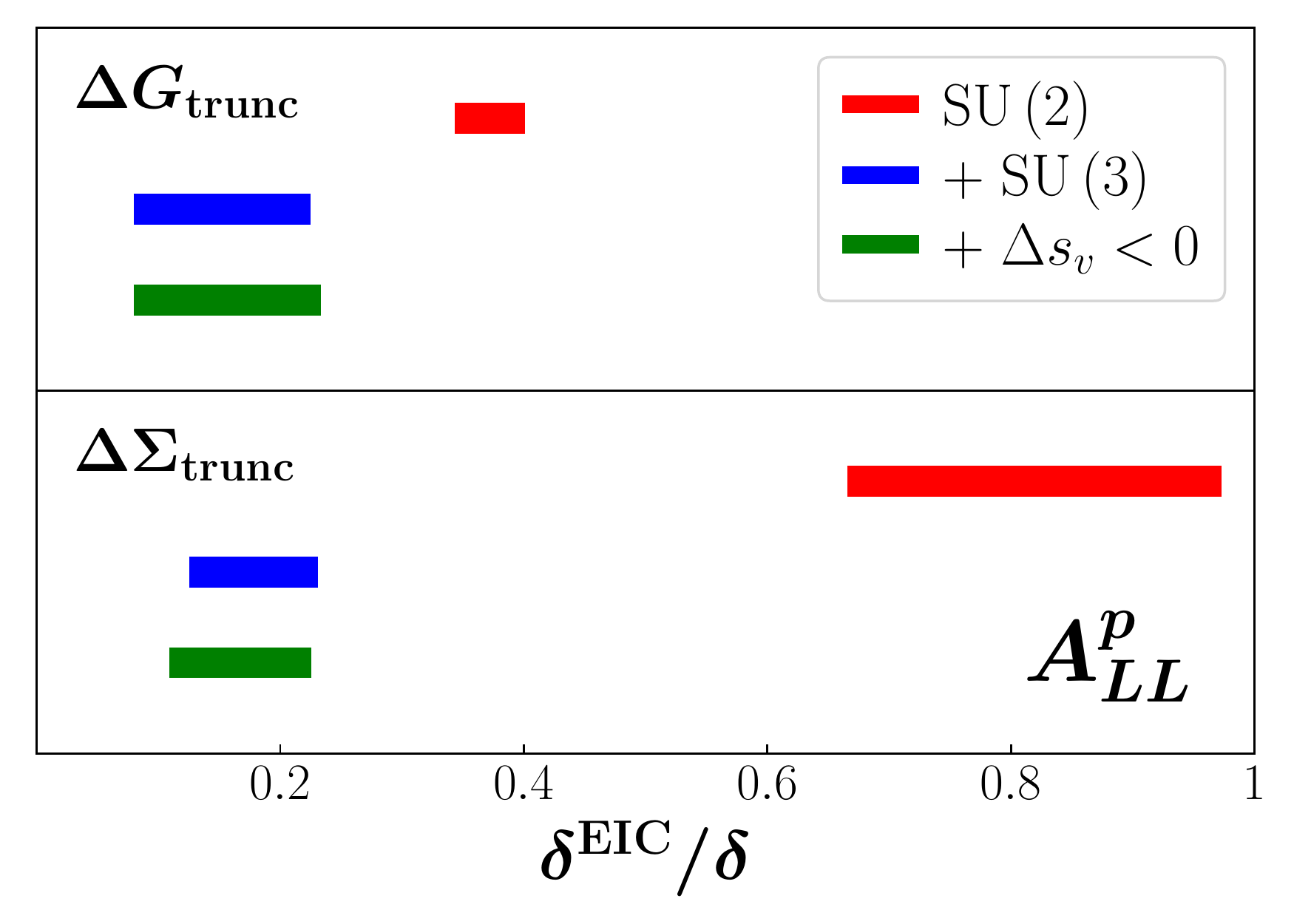}
    \caption{
        {
        {\bf (Left)} Impact of projected $A_{LL}^p$ data at EIC kinematics on $g_1^p$, relative to the JAM global QCD analysis~\cite{Zhou:2021llj} (yellow band), taking $+1\sigma$ (``high $g_1$'', red band),  $-1\sigma$ (``low $g_1$'', green band) and central (``mid $g_1$'', blue band) uncertainties of $A_{LL}^p$.
        {\bf (Right)} Uncertainty on the quark singlet ($\Delta G_{\rm trunc}$) and gluon ($\Delta \Sigma_{\rm trunc}$) truncated moments from $x_{\rm min}=10^{-4}$ to 1 with EIC data ($\delta^{\rm EIC}$) normalized to the baseline PDFs uncertainties ($\delta$)~\cite{Zhou:2021llj}, covering the ``low'', ``mid'' and ``high'' scenarios, for the case of no SU(3) symmetry (red lines) and with SU(3) symmetry (blue lines).}
        }
    \label{fig:A_LL_p}
\end{figure}

The impact of future EIC data on longitudinal double-spin asymmetries, $A_{LL}$, on spin-dependent PDFs was explored recently in Ref.~\cite{AbdulKhalek:2021gbh}.
One of the challenges in assessing the impact of the inclusive $A_{LL}$ measurements at the EIC is that the predictions for rates are based on extrapolation from existing measurements that extend only down to $x \sim 0.01$. 
A study of the uncertainty on the helicity distributions associated with the extrapolation of $A_{LL}$ for the EIC pseudodata is shown in Fig.~\ref{fig:A_LL_p}.
The analysis is performed within the JAM global QCD analysis framework at NLO, including all existing data on $A_{LL}$ and inclusive jet production from polarized $pp$ scattering at RHIC~\cite{Zhou:2022wzm}, along with $A_{LL}$ from EIC proton pseudodata (simulated with ${\cal L} = 100~\mathrm{fb}^{-1}$, 2.3\% normalization uncertainty, and 2\% point-by-point uncorrelated systematic uncertainties).

To explore the impact of the extrapolation region, three sets of pseudodata were generated by shifting the unmeasured region at low $x$ with $\pm 1\sigma$~CL using existing helicity PDF uncertainties along with the central predictions. 
In Fig.~\ref{fig:A_LL_p} the uncertainty bands for $g_1^p$ before and after the three scenarios ($\pm 1\sigma$~CL and central) at the EIC are shown, along with the ratios $\delta^{\rm EIC}/\delta$ of uncertainties on the truncated moments of the quark singlet and gluon PDFs, $\Delta\Sigma_{\rm trunc}$ and $\Delta G_{\rm trunc}$, integrated between $x_{\rm min}\!=\!10^{-4}$ and 1, with EIC data to the baseline JAM results with existing data.
The results indicate that if one assumes SU(3) symmetry for the axial vector charges, the uncertainty on $\Delta G_{\rm trunc}$ can improve by $80\%-90\%$, depending on the behavior of the low-$x$ extrapolation of $g_1^p$, with an $\approx 80\%$ reduction in the uncertainty on $\Delta\Sigma_{\rm trunc}$. 
The reduction is more modest, however, if one does not impose SU(3) symmetry, in which case the gluon moment uncertainty decreases by $\approx\! 60\%$, but no clear reduction in the quark singlet uncertainty is apparent from proton EIC data alone.

Overall, the $A_{LL}$ impact study suggests that the effect of new EIC data on spin-dependent PDFs does depend on the theory choices, and additional channels may be needed.
These include polarized parity-violating DIS for better determination of the polarized quark singlet distribution, and open charm production for the polarized gluon PDF.

\subsubsection{Nuclear PDFs}
\label{sec:nPDFs}

Although hadron colliders like the LHC primarily
examine proton-proton events, in many cases there remains an indirect dependence on
nuclear-scattering information --- entering, for instance, via PDFs constrained with nuclear data.
The treatment of nuclear DIS, however, involves various ambiguities associated with
corrections due to the nuclear environment. While the EIC will help dissect nuclear-medium
effects on various tomographic distributions, it will also measure hadronic structure with
high precision in a way that avoids excessive dependence on nuclear targets through
combinations of EW currents and other dedicated processes.

In a complementary direction, ultra-peripheral photonuclear collisions at hadron colliders \cite{UPCLOI} are also sensitive to the internal structure of both the proton and of nuclei. Coherent photoproduction of vector mesons and other final states is sensitive to the internal structure of the target; the Fourier transform of $d\sigma/dt$ gives access to the transverse density distribution --- essentially the GPD for nuclei.   In the Good-Walker paradigm, $d\sigma/dt$ for incoherent photoproduction is sensitive to event-by-event internal fluctuations of the target structure, notably including gluonic hot spots \cite{Klein:2019qfb}. These studies will also be pursued with increased precision at the EIC \cite{Accardi:2012qut}, where they will provide synergies with corresponding DIS measurements.

Leveraging the full breadth of the $\mathcal{O}(1\,\mathrm{ab}^{-1})$ data set expected
from the EIC will require a combination of theoretical, computational, and experimental
advances. We identify and briefly describe several leading examples below, concentrating on EIC implications for nuclear PDFs (nPDFs)
and, ultimately, precision at the HL-LHC and future $\nu A$ programs.\\

\noindent
{\bf Challenges for nPDFs.}
\label{sec:challenges}
Below, we outline a representative sampling of contemporary issues
associated with the partonic structure of nuclei; for these,
we expect an interplay of EIC and HL-LHC measurements to make
important progress.
With copious high-precision data from the EIC, a number of exciting possibilities and challenges await studies of nuclear PDFs. For years, the relationship between free-nucleon and nPDF global analyses has been developed in the direction of greater reciprocity. As such, improved understanding of nuclear effects in nPDF analyses will
influence proton PDF phenomenology, benefiting precision at the HL-LHC. 
Nuclear data from the EIC invites further extensions of these efforts, including investigations of whether the precision tools developed for proton PDFs might be applied to nPDFs.
This program might result in nPDF analyses of greater parametric flexibility with the aim of enhancing nPDF precision to more closely resemble that of the proton PDFs.
Augmented precision could also witness parametrizations not just in $A$ but also in $Z$, allowing studies to move away from the nuclear stability line; access to mirror nuclei species could be instrumental to this activity.

Moreover, determining consistent nPDFs for few-baryon ({\it i.e.}, light nuclear) systems
in global analyses with fully general $A$ dependence can be challenging. In this light,
the EIC program to more precisely determine the $Z$ and $A$ dependence of nPDFs
would complement data at RHIC on isobars and JLab measurements on $^3$He to move toward
more reliable extractions of the $A$ dependence of nPDFs from small to large $A$.
Exploring the nPDF parametrization uncertainty through high-quality EIC data
will be valuable to dissecting the underlying QCD physics in parallel with
lattice QCD calculation, hydrodynamic simulations, and (possibly)
Machine Learning (ML) approaches.
Similarly, in this scenario, next-generation nPDF studies facilitated by the EIC could be the vehicle for obtaining greater control over various questions in nuclear structure, including the quest to characterize the EMC effect across a wider range of $A$ and energy scales.

The EIC would also be well-positioned to resolve an issue of
central importance to $\nu A$ and nPDF phenomenology: the question
of whether there are systematic differences between charge- and
neutral-current nuclear DIS processes. This issue is particularly relevant to the interpretation of neutrino DIS data, which can be challenging to simultaneously describe in conjunction with charge-lepton scattering information as some data sets have suggested possible
tensions between these two processes.
the implications of this interplay will be consequential to attempts to quantify the DIS
contributions to the total $\nu A$ cross section at future neutrino facilities like DUNE/LBNF. The EIC can inform this subject by performing detailed comparisons of charged-current ($W^\pm$) and neutral
current ($\gamma,Z$) induced processes. The appropriate setting to explore this would be nPDF global QCD analyses.\\

\begin{figure}[th]
  \centering
  \includegraphics[width=\textwidth]{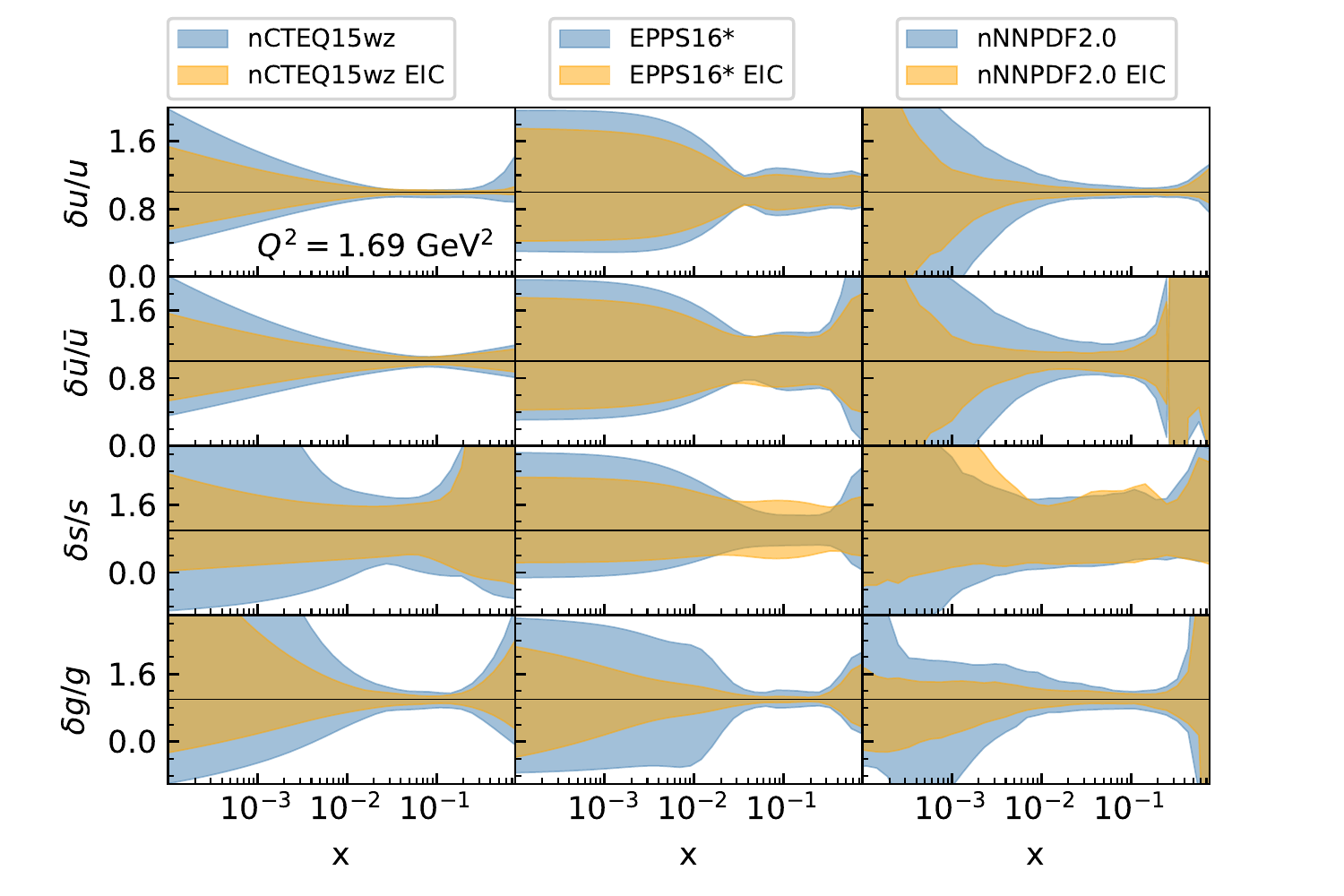}
  \caption{The relative uncertainty, based on one standard deviation, of the nuclear PDFs determined in the context of the EIC Yellow-Report studies (see Ref.~\cite{AbdulKhalek:2021gbh}, Sec.~7.3.3), in this case
  for $\mathrm{Au}$ and at a low scale, $Q^2\! =\! 1.69$ GeV$^2$,
  near the boundary for perturbative QCD evolution in typical
  nPDF global analyses.
  The projected impact shown here corresponds to the optimistic
  scenario for systematic uncertainties and detector effects.
  The results shown here for $\mathrm{Au}$ were founds to be
  generally representative of other nuclei.
  We plot PDF flavors particularly affected by the EIC pseudodata: $u$, $\bar{u}$,
  $s$ and $g$. Note the use of a log scale on the $x$ axis.}
  \label{fig:npdfs}
\end{figure}

%

\noindent
{\bf nPDF Impact studies.}
As with the proton PDFs discused in Sec.~\ref{sec:unPDFs}, a number of dedicated investigations were
carried out within the EIC Yellow Report to assess the potential 
implications of the EIC program for nPDFs.
In Fig.~\ref{fig:npdfs}, we illustrate the result of a PDF
impact study carried out in the context of the EIC Yellow Report
(see Sec.~7.3.3 of Ref.~\cite{AbdulKhalek:2021gbh}).
We note that the pseudodata on which these studies were based involved
an assumed $\mathcal{L}\!=\! 10\, \mathrm{fb}^{-1}$ of accumulated statistics
on $^{197}$Au; these data were further generated at three selected
scattering energies: 5 GeV electrons on 41 GeV$/A$ ions, as well
as 10 GeV $e^-$ on 110 GeV ions and 18 GeV  $e^-$ on 110 GeV ions. Systematic
uncertainties were again informed by detector studies and fast Monte Carlo
runs.
As a characteristic example, we plot the impact of EIC inclusive
nuclear pseudodata on the relative nPDF uncertainty for the
$\mathrm{Au}$ nucleus. We observe a reduction of nuclear
PDF uncertainties, due to EIC pseudodata, that varies with the
region of $x$ considered as well as the flavor of the PDF. 
The plotted uncertainties correspond to one standard deviation, and
are computed as a function of $x$ at $Q^2$=1.69~GeV$^2$, close
to the perturbative evolution boundary. Results are displayed for
$^{197}$Au and for a selection of PDF flavors to which the examined
pseudodata demonstrate particular sensitivity: $u$, $\bar{u}$, 
$s$ and $g$.
The more substantial uncertainty reductions are comparatively
localized to the small-$x$ region, where little or no data are
currently available, and in the large-$x$ region, where nuclear
PDFs benefit from the increased precision of the baseline proton
PDFs. In the case of the gluon PDF, the reduction of uncertainties
is seen for the whole range in $x$. This is a consequence of the
extended data coverage in $Q^2$, which allows one to constrain
the gluon PDF even further via perturbative evolution.

\begin{figure}[t]
\centering
\includegraphics[width=0.65\textwidth]{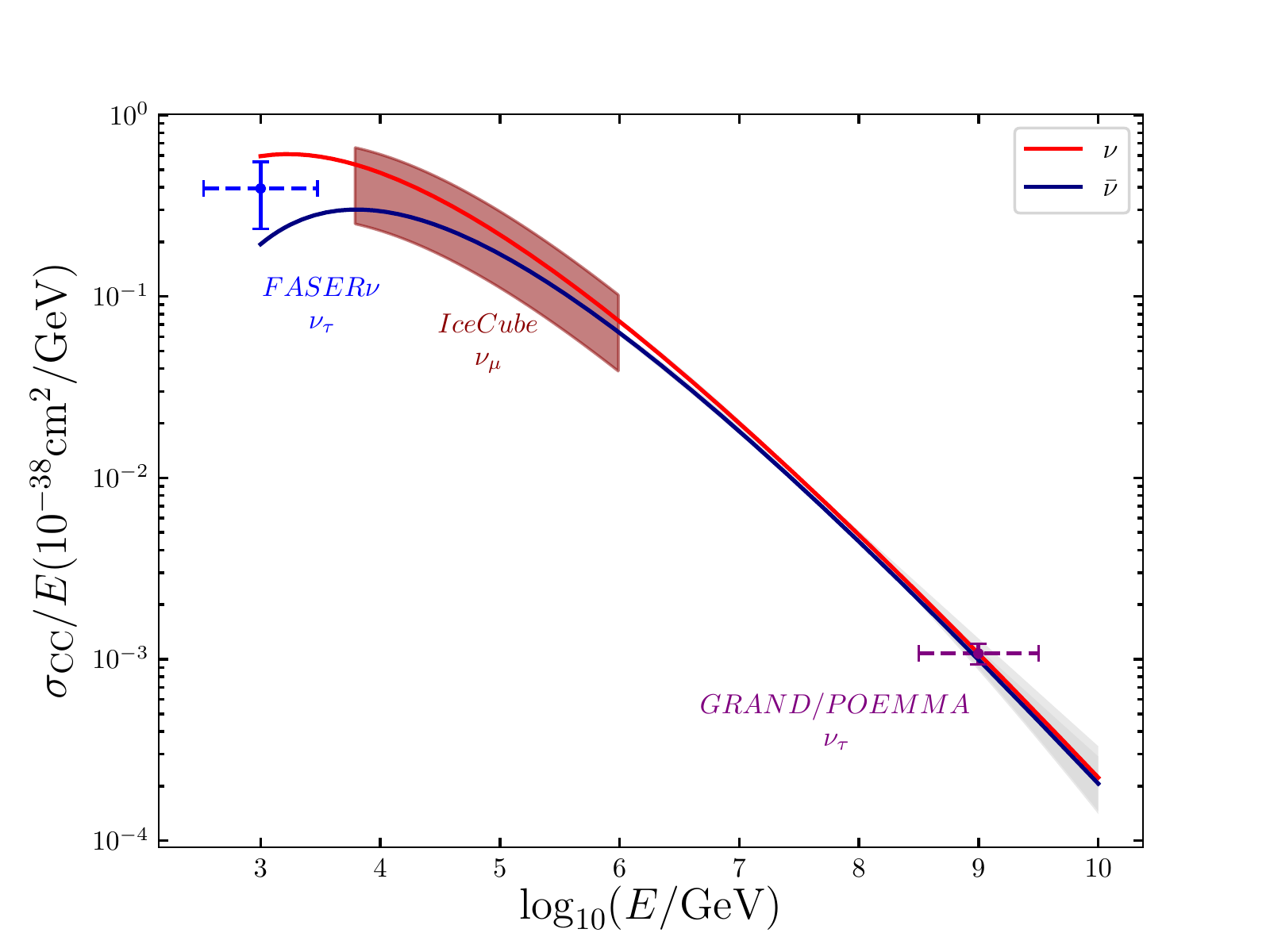}
\caption{The high- and ultra-high energy neutrino-nucleon cross section and the PDF uncertainty, shown in gray, as computed in Ref.~\cite{1102.0691}. Existing constraints from IceCube~\cite{IceCube:2017roe} and expected future sensitivities from FASER$\nu$~\cite{FASER:2019dxq} and next-generation astrophysical experiments \cite{2007.10334} are also shown.}
\label{fig:HEnu}
\end{figure}

In addition to the EIC Yellow Report studies, other groups have carried out informative and complementary analyses. For instance, the dedicated NNPDF impact study mentioned in the proton PDF discussion~\cite{Khalek:2021ulf} also compared the relative uncertainty of the
nuclear PDFs in the nNNPDF2.0 analysis~\cite{AbdulKhalek:2019mzd, AbdulKhalek:2020yuc} used to generate pseudodata in the pessimistic and in the optimistic scenarios EIC performance scenarios with the results of
nNNPDF2.0+EIC fits.
Overall, the heavier the nucleus,
the largest the reduction of PDF uncertainties. This is a consequence of the
fact that nuclear PDFs are customarily parametrized as continuous functions
of the nucleon number $A$: nuclear PDFs for $^4$He, which differ from the proton
PDF boundary by a small correction, are better constrained than nuclear PDFs
for $^{197}$Au because proton data are more abundant than data for nuclei.
In this respect, the EIC will allow one to perform a comparatively accurate
scan of the kinematic space for each nucleus individually, and, as shown in
Fig.~\ref{fig:npdfs}, to determine the PDFs of all ions with similar precision. 
As observed in the case of
the proton PDFs, the fits obtained upon inclusion of the EIC pseudodata do not
significantly depend upon whether the optimistic or the pessimistic scenario is
considered, except at very small values of $x$. In this case, the optimistic
scenario leads to a significantly more marked reduction in the PDF uncertainties.

It is also worth noting in this context that the EIC measurements benefit from several 
important advantages as compared
to measurements from pPb collisions at the LHC
used to constrain the small-$x$ nPDFs,
such as the $D$-meson and dijet production data included
in the nNNPDF3.0~\cite{Khalek:2022zqe} and EPPS21~\cite{Eskola:2021nhw} global fits.
In particular, the much
cleaner environment of lepton-nuclei collisions 
facilitates disentangling
cold nuclear matter contributions from other
possible nuclear effects taking place in pPb collisions.
Furthermore, the availability of data for different $A$
will pin down the $A$-dependence of the small-$x$ quark
and gluon shadowing, which in current fits suffers from
a sizable model dependence.\\

\begin{figure}[!t]
  \centering
  \includegraphics[width=\textwidth]{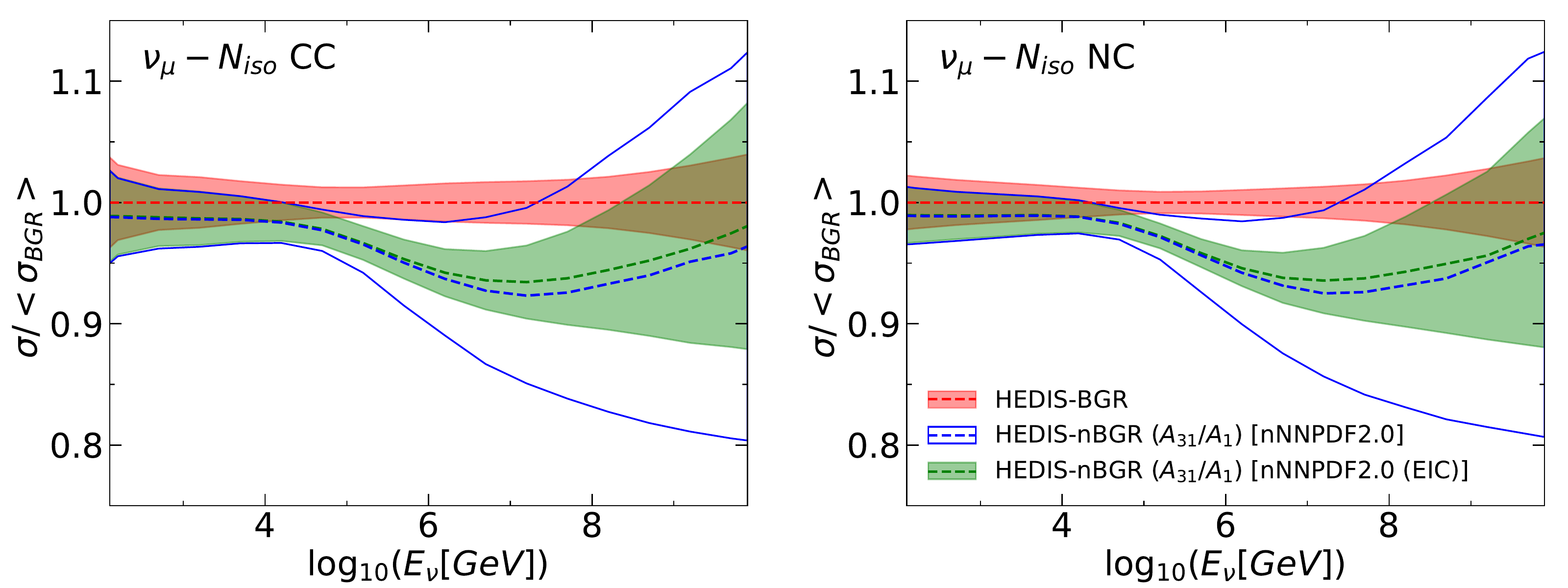}
  \caption{The CC (left) and NC (right) neutrino-nucleus DIS cross
  sections, with their one-sigma uncertainties, as a function of the neutrino
  energy $E_{\nu}$. Predictions correspond to the HEDIS-BGR
  computation~\cite{Garcia:2020jwr} with the proton PDF of~\cite{Gauld:2016kpd},
  and with the nNNPDF2.0 and nNNPDF2.0+EIC nuclear PDFs. They are all
  normalized to the central value of the proton results. See text for details.}
  \label{fig:UHExsec}
\end{figure}

\begin{figure}[!t]
  \centering
  \includegraphics[width=\textwidth]{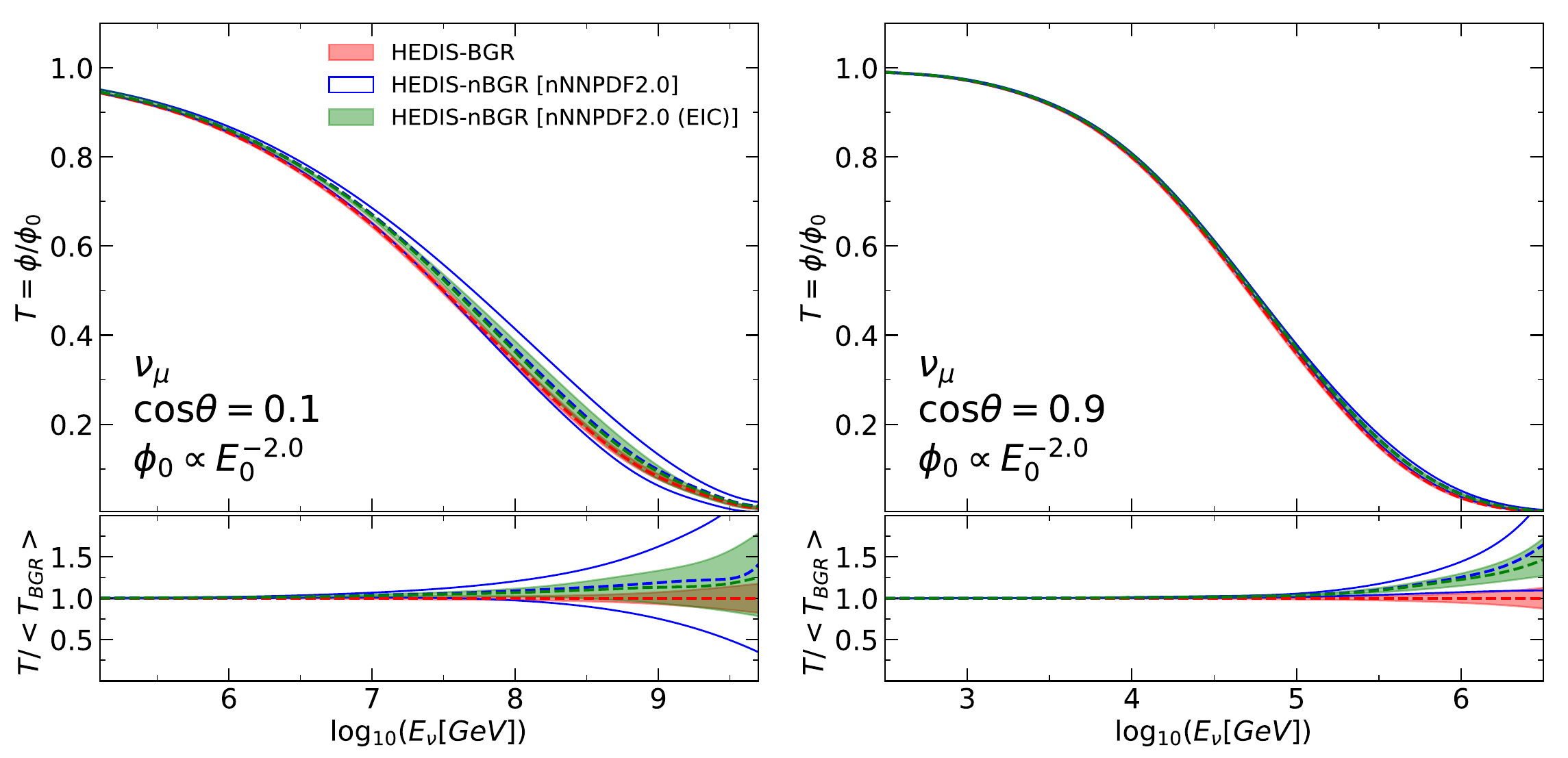}
  \caption{The transmission coefficient $T$ for muonic neutrinos as a function
  of the neutrino energy $E_\nu$ and for two values of the nadir angle $\theta$.
  Predictions correspond to the computation of~\cite{Garcia:2020jwr} with the
  proton PDF of~\cite{Gauld:2016kpd}, and with the nNNPDF2.0 and nNNPDF2.0+EIC
  nuclear PDFs. They are all normalized to the central value of the proton
  results. See text for details.}
  \label{fig:attenuation}
\end{figure}
%

\noindent
{\bf Implications for neutrino scattering.}
The reduction of nuclear PDF uncertainties due to EIC (pseudo)data has important phenomenological implications for a variety of neutrino measurements, particularly, ultra-high energy neutrino experiments such as GRAND~\cite{1810.09994}, POEMMA~\cite{1907.03694}, and PUEO~\cite{2010.02892}, among others --- see reviews \cite{2203.08096,2203.05591}. These experiments can also measure the neutrino-nucleus cross section at the $\mathcal O(20\%)$ level at $E_\nu\sim10^9$ GeV \cite{2007.10334,2205.09763} which would provide an independent cross-check of PDFs at these energies. Corresponding theory predictions for the
ultra-high energy cross section are shown in Fig.~\ref{fig:HEnu}, including an associated nucleon PDF uncertainty based on Ref.~\cite{1102.0691}. In addition,
it was shown in~\cite{Garcia:2020jwr} that the dominant source of
uncertainty in the theoretical predictions for the cross section of
neutrino-matter interactions is represented by nuclear effects. In Fig.~\ref{fig:UHExsec}, we show the CC (left) and NC (right)
neutrino-nucleus inclusive DIS cross sections, with their one-sigma PDF
uncertainties, as a function of the neutrino energy, $E_\nu$. Moreover, in
Fig.~\ref{fig:attenuation}, we show the transmission coefficient, $T$, for muon
neutrinos, defined as the ratio between the incoming neutrino flux $\Phi_0$ and
the flux arriving at the detector volume, $\Phi$ (see Eq.~(3.1), and the ensuing
discussion in Ref.~\cite{Garcia:2020jwr} for details); $T$ is displayed for two
values of the nadir angle, $\theta$, as a function of $E_\nu$.
In both cases, we compare predictions obtained with the calculation presented
in Refs.~\cite{Bertone:2018dse,Garcia:2020jwr} and implemented in
{\sc hedis}~\cite{Brown:1971qr}. For a proton target the prediction is made
with the proton PDF set determined in Ref.~\cite{Gauld:2016kpd},
a variant of the NNDPF3.1 PDF set in which small-$x$ resummation
effects~\cite{Ball:2017otu} and additional constraints from $D$-meson
production measurements in proton-proton collisions at 5,7 and
13~TeV~\cite{Aaij:2013mga,Aaij:2015bpa,Aaij:2016jht} have been included.
This prediction is labeled HEDIS-BGR in
Figs.~\ref{fig:UHExsec}-\ref{fig:attenuation}. For a nuclear
target ($A=31$ is adopted as in Ref.~\cite{Garcia:2020jwr}), the prediction is
made alternatively with the nNNDPF2.0 and the nNNPDF2.0+EIC (optimistic) PDFs.
The corresponding predictions are labeled HEDIS-nBGR [nNNPDF2.0] and HEDIS-nBGR
[nNNPDF2.0 (EIC)] in Figs.~\ref{fig:UHExsec}-\ref{fig:attenuation}. Predictions
are all normalized to the central value of the proton result. In comparison to
nNNPDF2.0, the effect of the EIC pseudodata is seen to reduce the uncertainty
of the prediction for a nuclear target by roughly a factor of two for
$E_\nu\gtrsim 10^6$~GeV. The reduced uncertainty no longer encompasses the
difference between predictions obtained on a proton or on a nuclear target,
except in the case of an attenuation rate computed with a large nadir angle.

\subsection{Transverse momentum dependent distributions}
\label{sec:TMDs}

\subsubsection{TMD PDFs}
\label{sec:TMDPDF}

\noindent
{\bf TMDs for precision electroweak physics.}
The $W$-boson mass, $M_W$, is an important testbed for the SM in the EW sector, being potentially
sensitive to oblique, propagator-level corrections due to insertions of hypothetical BSM
degrees-of-freedom. Like the Higgs cross section, determinations of $M_W$ are limited
by nucleon structure uncertainties that the EIC stands to improve.
Currently, the best determinations of $M_W$ come from global EW fits (with $\delta[{M_W}]$ = 8 MeV). 
Precise extractions have also been obtained by fitting
the transverse mass and transverse-momentum distributions of the decaying leptons in proton-proton collisions
at ATLAS and in proton-antiproton collisions at D0 and CDF. The result of the fit has a total error $\delta[{M_W}]$ = 19 MeV, for which PDF uncertainties are a limitation (see also \cite{Bagnaschi:2019mzi} for a discussion about the ultimate precision achievable).
In the latter approach, an additional nonperturbative effect from the flavor dependence of the intrinsic
transverse momenta of the partons entering the collision has a statistically significant impact on the
extracted values of the $W^{\pm}$ masses \cite{Bacchetta:2018lna,Bozzi:2019vnl}, inducing shifts in $M_W$
comparable to those associated with PDF variations.
By improving knowledge of TMD PDFs as well as TMD fragmentation functions and
hadronization processes through precise measurements of unintegrated SIDIS multiplicities,
the EIC will enhance the accuracy of $M_W$ extracted from hadron collider data and EW/BSM
phenomenology generally \cite{EICBSMEWLOI}.\\

\begin{figure}[t]
\centering
\includegraphics[width=0.45\textwidth]{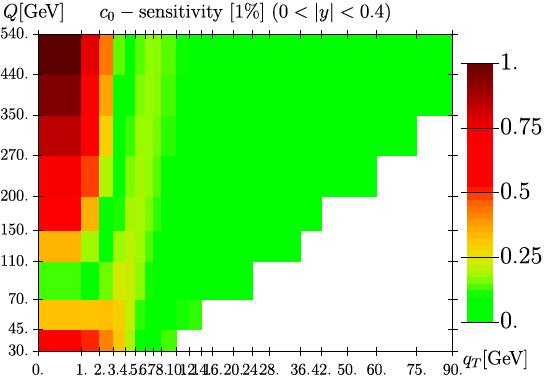}
\includegraphics[width=0.45\textwidth]{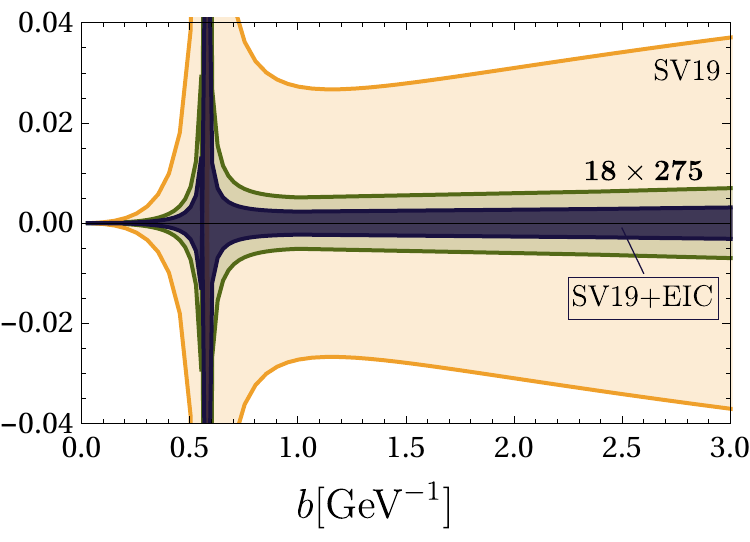}
\caption{Left: sensitivity to nonperturbative Collins-Soper kernel in the kinematics of LHC. The red region indicate the range of utmost impact on determination of Collins-Soper kernel. The sensitivity is calculated assuming 1\% uncertainty of data. Right: the impact of EIC data onto the determination of the Collins-Soper kernel.}
\label{fig:CS-kernel}
\end{figure}

\noindent
{\bf Nonperturbative evolution.}
The nonperturbative evolution kernel is the central feature of TMD distributions. It reflects the sensitivity of temperate transverse motion of partons to the confining forces that bind the nucleon together. \cite{Vladimirov:2020umg}. This information is accumulated in the Collins-Soper kernel, extracted from scattering data alongside TMD distributions. Importantly, the effects of nonperturbative evolution are universal and impact all TMD observables. Currently, its uncertainty is one of the main sources of uncertainties in unpolarized and polarized TMDs \cite{Bury:2021sue, Bury:2022czx}.

The precise extraction of the Collins-Soper kernel requires a broad coverage in the hard scale $Q$, which helps to decorrelate effects of nonperturbative evolution from the nonperturbative transverse momentum. The effects of confining forces are not negligible at high energies contrary to naive expectation \cite{Hautmann:2020cyp}. It is due to the hierarchy of scales within the multi-differential cross section. So, on the one hand, the large $Q$ guarantees the validity of factorization theorems; on the other hand, the small $q_T$ provides sensitivity to nonperturbative physics. Therefore, future EIC and HL-LHC will be an ideal laboratory to study the Collins-Soper kernel and related properties of the QCD vacuum. Herewith, the EIC and HL-LHC will contribute complementary to the determination of Collins-Soper kernel see figure \ref{fig:CS-kernel}. The HL-LHC will provide very precise data with an enormous coverage in $Q$, with relatively wide bins in $q_T$. The EIC will measure a small $q_T$ preciser in a smaller range of $Q$. Together these sources of information will pin down the values of the Collins-Soper kernel.\\

\noindent
{\bf Phenomenological studies of TMDs.}
With the advent of the EIC, systematic studies of observables sensitive to quark-gluon dynamics and hadronization like $J/\psi$ production will become
feasible, offering unique opportunities to deepen knowledge, including of the gluon TMDs --- particularly at low $x$. Given the
need for flexible models suited for phenomenology, common frameworks for all T-even and T-odd gluon TMDs at twist-2 must be produced as in
Ref.~\cite{Bacchetta:2020vty}, which calculated a spectator model for
the parent nucleon, encoding effective small-$x$ effects through BFKL resummation.
Reliable models are important inputs for the generation of pseudodata
allowing the identification of key processes at the EIC sensitive to the details of gluon
dynamics.
Additionally, TMD fits based on Drell-Yan data are now available at N$^3$LL accuracy
\cite{Scimemi:2019cmh, Bacchetta:2019sam}, and the flavor dependence of the intrinsic-$k_T$ has
been recently included in numerical codes widely used at LHC experiments
\cite{Camarda:2019zyx}. Ensuring that TMD improvements from the EIC complement
analogous measurements at hadron colliders will depend upon the continued
development of this work.
Similar arguments apply to modeling and simulation of the nucleon GPDs, providing
novel insights into the spatial distribution of quark-gluon degrees-of-freedom.

The complete set of twist-2 gluon TMDs, accounting for the polarization both of the parent nucleon and of the extracted gluon, was determined for the first time in Ref.~\cite{Mulders:2000sh}. Tab.~\ref{tab:gluon_TMDs} shows the eight twist-2 gluon TMDs for a spin-1/2 target, using the nomenclature as in Ref.~\cite{Meissner:2007rx}, which fairly corresponds to quark-TMD one (see also Ref.~\cite{Lorce:2013pza}). The definition of gluon TMDs for a spin-1 target was given in Ref.~\cite{Boer:2016xqr}, this leading to the emergence of 11 new distributions.
The two functions on the diagonal in Tab.~\ref{tab:gluon_TMDs} are the density of unpolarized gluons inside an unpolarized nucleon, $f_1^g$, and of circularly polarized gluons inside a longitudinally polarized nuclon, $g_1^g$. They respectively correspond to the well-known unpolarized and helicity gluon PDFs, in the collinear limit.

As for quarks TMDs, gluon TMDs embody terms coming from the resummation of transverse-momentum logarithms in perturbative calculations. They represent the ``perturbative contribution" to TMDs. Much is know about them~\cite{Bozzi:2003jy,Catani:2010pd,Echevarria:2015uaa}, but very little is known about the intrinsic nonperturbative components.

The $h_1^{\perp g}$ is the density of linearly-polarized gluons inside an unpolarized nucleon, and plays a key role in generating spin effects in collisions of unpolarized hadrons~\cite{Boer:2010zf,Sun:2011iw,Boer:2011kf,Pisano:2013cya,Dunnen:2014eta,Lansberg:2017tlc}. They are commonly known as Boer--Mulders effects. Only one part of these contributions are perturbatively generated via the aforementioned resummation.
The remaining part is due to the \emph{intrinsic} motion of gluons inside the hadron, which has a nonperturbative nature and cannot be caught by the resummation, and its weight must be assessed through fits on global data.
The $f_{1T}^{\perp g}$ is the so-called gluon Sivers TMD and carries information on the density of unpolarized gluons in a transversely-polarized nucleon. It plays a very important role in the description of transverse-spin asymmetries.
In the forward limit the gluon Sivers is connected to the QCD Odderon and can be accessed even in unpolarized collisions of leptons and nucleons~\cite{Boussarie:2019vmk}.
The precise definition of gluon TMDs accounts also for their dependence on (at least) two different gauge-link structures. This brings to the distinction between $f$-type and $d$-type gluon TMDs, also known in the context of small-$x$ investigations as Weisz\"acker-Williams (WW) and dipole (DP) TMDs~\cite{Kharzeev:2003wz,Dominguez:2010xd,Dominguez:2011wm}.
In practice, for a given gauge link one has the full set of functions shown in Tab.~\ref{tab:gluon_TMDs}.

{
\renewcommand{\arraystretch}{1.7}
\begin{table}
\centering
 \hspace{1cm} gluon pol. \\ \vspace{0.1cm}
 \rotatebox{90}{\hspace{-1cm} nucleon pol.} \hspace{0.1cm}
 \begin{tabular}[c]{|c|c|c|c|}
 \hline
 & $U$ & circular & linear \\
 \hline
 $U$ & $f_{1}^{g}$ & & \textcolor{blue}{$h_{1}^{\perp g}$} \\
 \hline	
 $L$ & & $g_{1}^{g}$ & \textcolor{red}{$h_{1L}^{\perp g}$} \\
 \hline	
 $T$ & \textcolor{red}{$f_{1T}^{\perp g}$} & \textcolor{blue}{$g_{1T}^{g}$} & \textcolor{red}{$h_{1}^{g}$}, \textcolor{red}{$h_{1T}^{\perp g}$} \\
 \hline
 \end{tabular}
 \caption{Gluon TMD PDFs at twist-2. $U$, $L$, $T$ describe unpolarized, longitudinally polarized and transversely polarized nucleons, respectively. $U$, `circular', `linear' stand for unpolarized, circularly polarized and linearly polarized gluons. Functions in blue are T-even. Functions in black are T-even and survive integration over the transverse momentum. Functions in red are T-odd.}
 \label{tab:gluon_TMDs}
\end{table}
}

The WW TMDs contain either $[+,+]$ or $[-,-]$ gauge links, whereas the dipole TMDs contain either $[+,-]$ or $[-,+]$ gauge links. Here $+$ and $-$ stands for the direction for future- and past-pointing  Wilson lines which correspond to final- and initial-state interactions, respectively. The WW structure is probed via processes where the gluon interacts with a color-singlet initial particle (for example, a deep-inelastically scattered photon) and produces two colored objects (for example, a di-jet system). In reactions featuring the interaction between two initial-state gluons (color-octet configuration) and production of a colorless particle (as a Higgs boson), the emerging gauge link is the $[-,-]$ one. TMD factorization is expected to hold in all these reactions, and the relations below follow from time-reversal invariance (T-symmetry):
\begin{align}
  f_1^{g\,[+,+]} &= f_1^{g\,[-,-]} \; && \text{(T-even)}\ , \nonumber \\
 f_{1T}^{g\,\perp[+,+]} &= -f_{1T}^{g\,\perp[-,-]}  && \text{(T-odd)}\ .
\end{align}
DP TMDs emerge when a gluon interacts with a colored initial object and produces a colored final particle (as the hadroproduction of a photon-jet system). Here, TMD factorization has not been proven and may be affected by effects of color entanglement~\cite{Rogers:2013zha}. More complicated gauge links are involved in reactions where multiple color states are present both in the initial and final state~\cite{Bomhof:2006dp}. Here, the TMD factorization can be even more questionable~\cite{Rogers:2010dm}.

\begin{figure}[htbp!]
\centering
\includegraphics[scale=0.300]{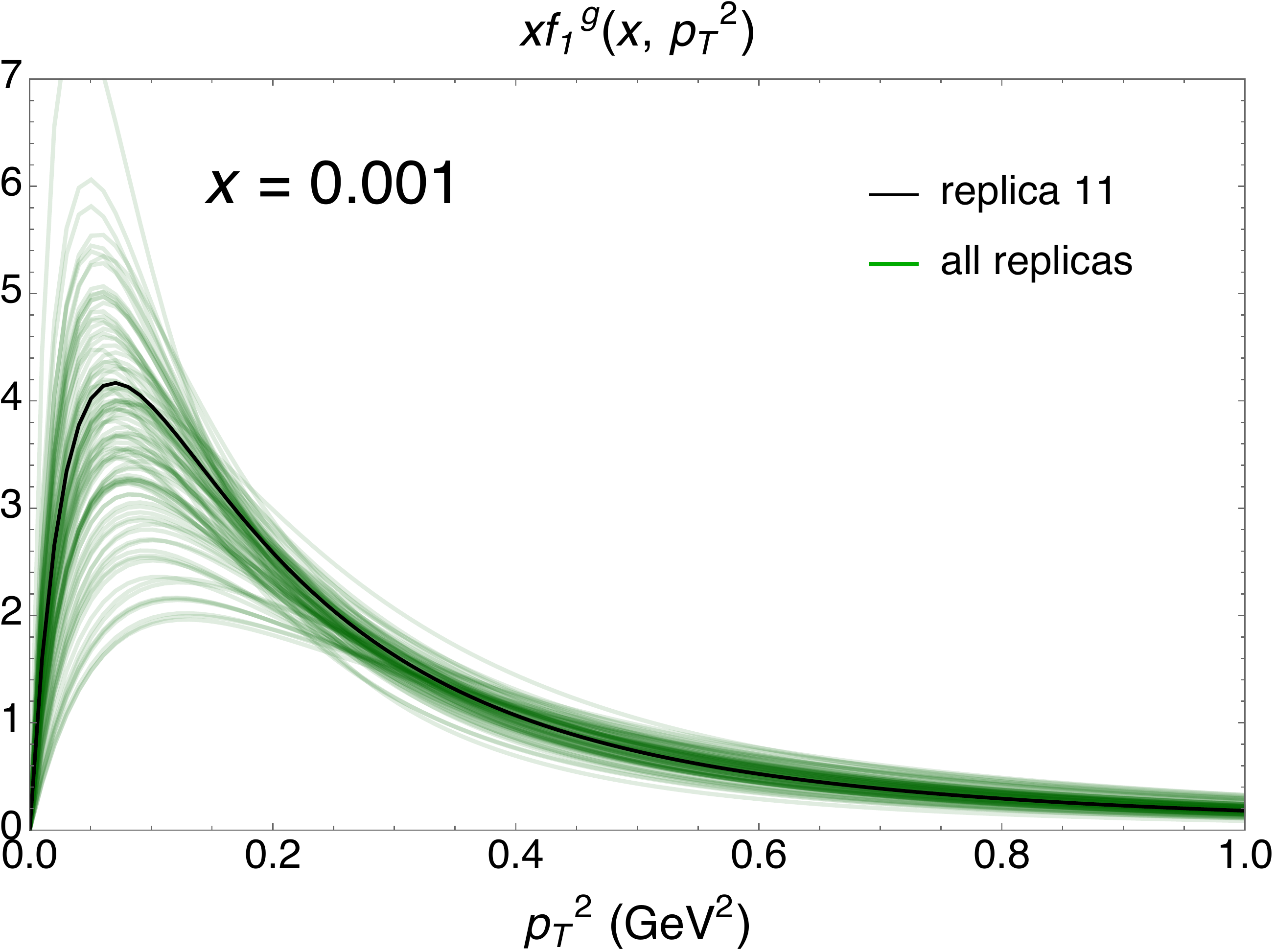} \hspace{0.50cm}
\includegraphics[scale=0.310]{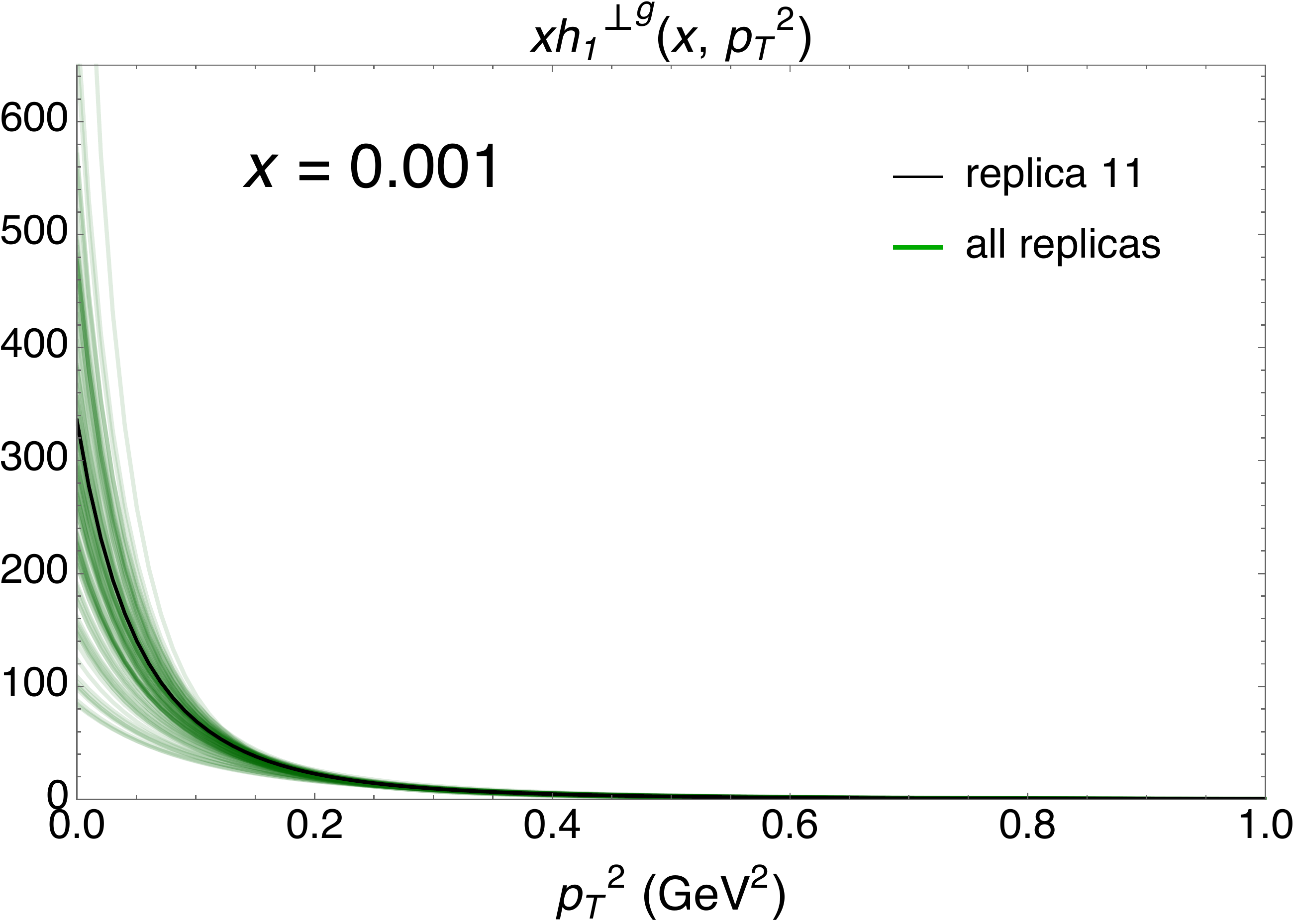} 

\vspace{0.55cm}

\hspace{-0.65cm}
\includegraphics[scale=0.320]{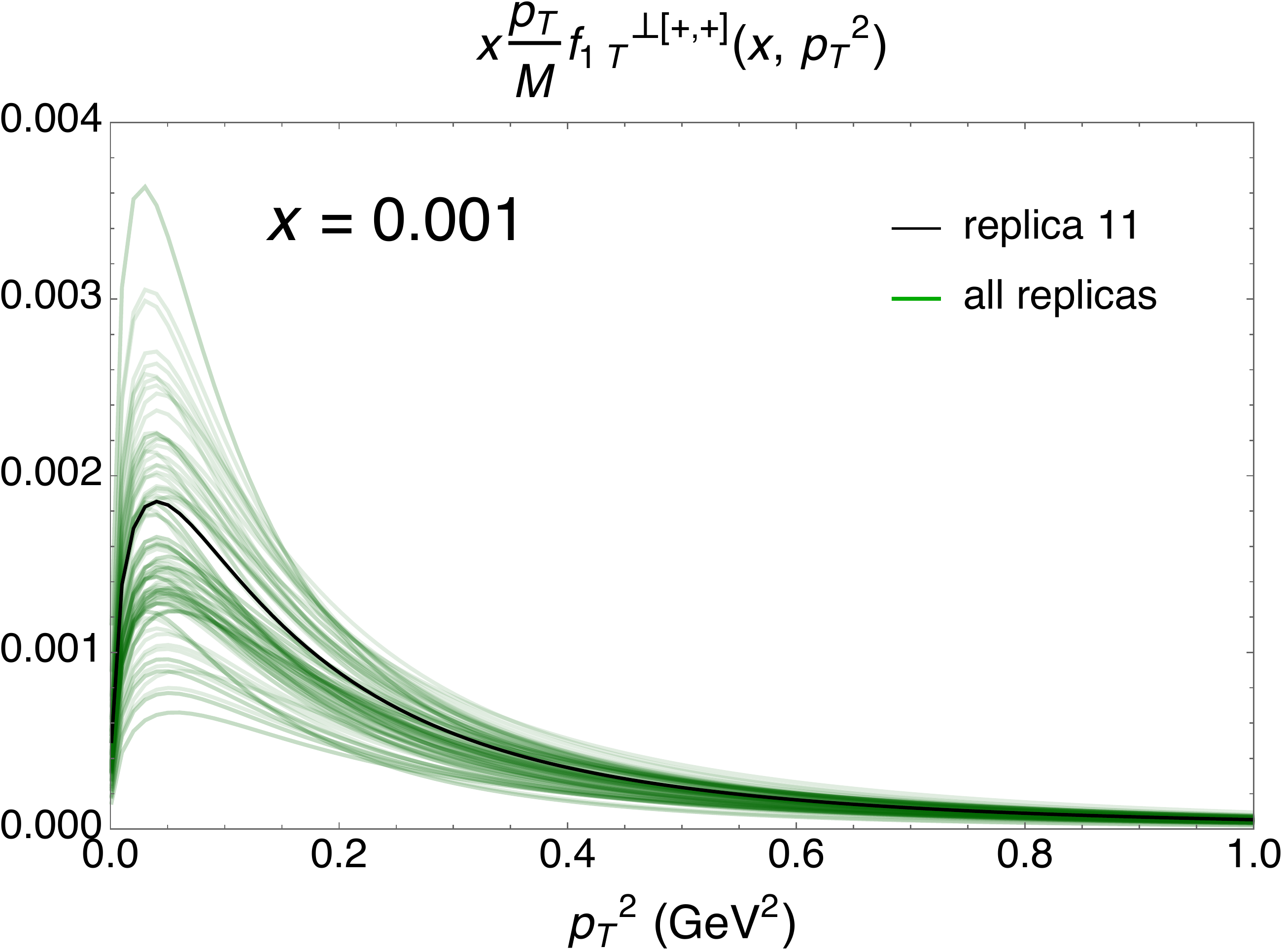} \hspace{0.00cm}
\includegraphics[scale=0.3275]{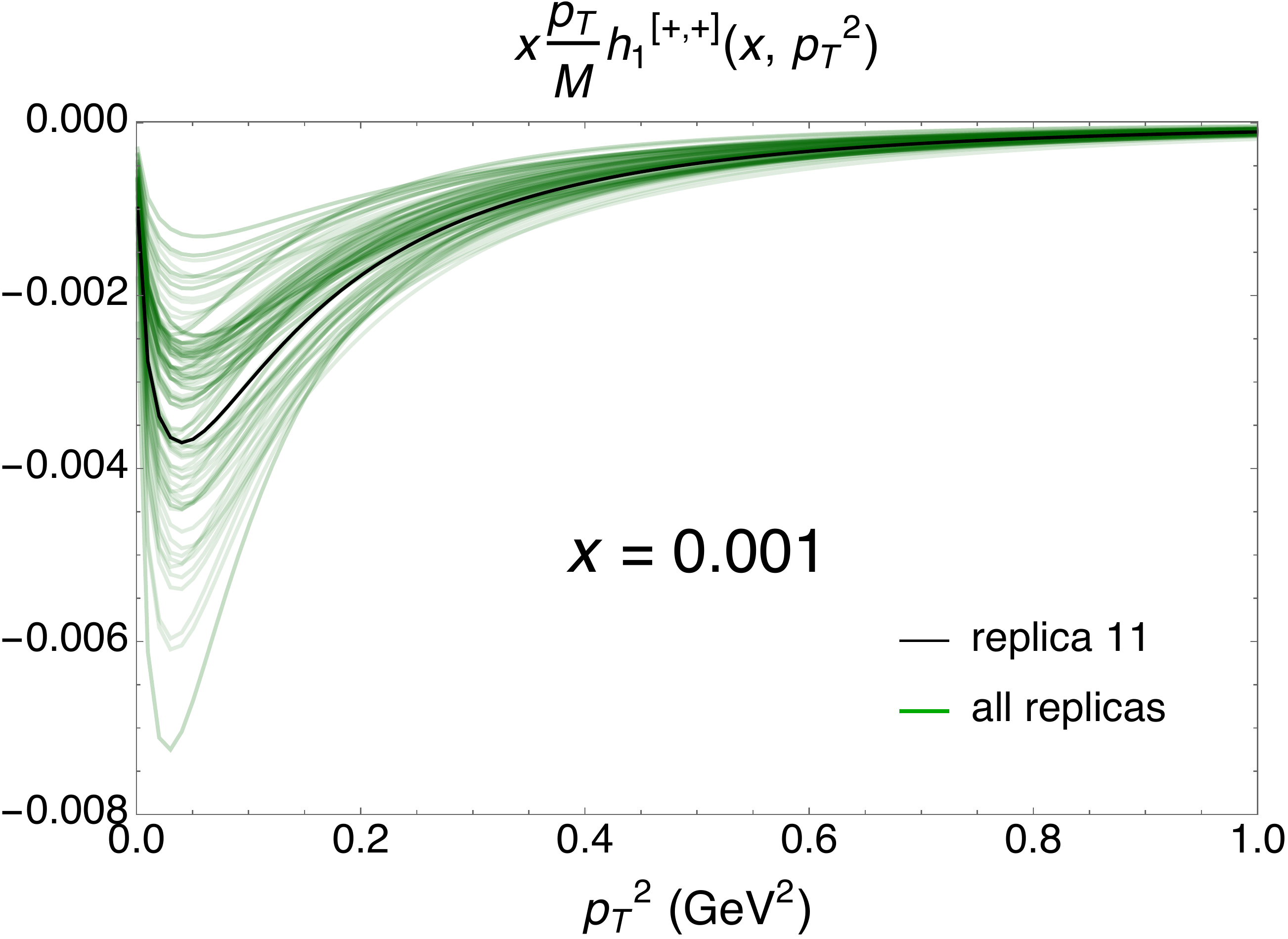} 
\caption{Examples of model calculations of four leading-twist gluon TMDs as functions of $\boldsymbol{p}_T^2$, at $x=10^{-3}$ and for the initial scale $Q_0 = 1.64$ GeV. Figures adapted from Refs.~\cite{Bacchetta:2020vty,Bacchetta:2021lvw,Bacchetta:2021twk,Bacchetta:2022esb}.}
\label{fig:twist-2_gluon_TMDs}
\end{figure}

\vspace{0.25cm}

At high transverse momentum and at small-$x$, namely, in the domain of the \emph{high-energy} factorization, the unpolarized and linearly-polarized gluon distributions, $f^g_1$ and $h^{\perp g}_1$, are connected~\cite{Dominguez:2011wm,Nefedov:2021vvy,Hentschinski:2021lsh} to the BFKL \emph{unintegrated gluon distribution} (UGD)~\cite{Fadin:1975cb,Kuraev:1976ge,Kuraev:1977fs,Balitsky:1978ic} (see Refs.~\cite{Hentschinski:2012kr,Anikin:2009bf,Anikin:2011sa,Besse:2013muy,Bolognino:2018rhb,Bolognino:2018mlw,Bolognino:2019bko,Bolognino:2019pba,Celiberto:2019slj,Brzeminski:2016lwh,Garcia:2019tne,Celiberto:2018muu,Celiberto:2022fam} for recent applications).
A study on the inclusion of gluon-TMD inputs within high-energy factorization was recently done in the context of the exclusive electroproduction of polarized $\rho$ mesons at the EIC~\cite{Bolognino:2021niq,Bolognino:2021gjm,Bolognino:2022uty}.
From the experimental point of view the sector of gluon TMDs as a largely unexplored field. 
First attempts to get a phenomenological description of the unpolarized gluon TMD and the Sivers one were done in Refs.~\cite{Lansberg:2017dzg,Gutierrez-Reyes:2019rug,Scarpa:2019fol} and~\cite{Adolph:2017pgv, DAlesio:2017rzj,DAlesio:2018rnv,DAlesio:2019qpk}, respectively.
Since our present knowledge of gluon TMDs is very limited, exploratory studies by means of phe\-no\-me\-no\-lo\-gi\-cal models are needed. Significant progresses in this respect were obtained by making use of the so-called \emph{spectator-model} framework~\cite{Bacchetta:2008af,Bacchetta:2010si,Gamberg:2005ip,Gamberg:2007wm,Jakob:1997wg,Meissner:2007rx,Lu:2016vqu,Mulders:2000sh,Pereira-Resina-Rodrigues:2001eda,Goldstein:2010gu,Goldstein:2013gra}.
Within this model it is possible to calculate all the leading-twist gluon TMD functions in Tab.~\ref{tab:gluon_TMDs}.
Small-$x$ improved T-even gluon distributions at twist-2 were recently calculated in Refs.~\cite{Bacchetta:2020vty} (see also Refs.~\cite{Celiberto:2021zww,Bacchetta:2021oht,Celiberto:2022fam}). Preliminary results for the T-odd counterparts were presented in Refs.~\cite{Bacchetta:2021lvw,Bacchetta:2021twk,Bacchetta:2022esb}. 

Results obtained in Ref.~\cite{Bacchetta:2020vty} for the two T-even functions, the unpolarized gluon TMD, $x f_1^g$, and the linearly polarized gluon TMD, $x h_1^{\perp g}$, are shown in upper panels of Fig.~\ref{fig:twist-2_gluon_TMDs} as functions of the gluon transverse momentum squared, $\boldsymbol{p}_T^2$, at $x=10^{-3}$ and for the initial energy scale $Q_0 = 1.64$ GeV (namely without the applying the standard TMD evolution). 
Preliminary results~\cite{Bacchetta:2021lvw,Bacchetta:2021twk,Bacchetta:2022esb} for two T-odd WW functions, the Sivers TMD, $x \,|\boldsymbol{p}_T|/M \, f_{1T}^{g [+,+]}$, and the linearity, $x \, |\boldsymbol{p}_T|/M \, h_{1}^{g [+,+]}$, are shown in lower panels of Fig.~\ref{fig:twist-2_gluon_TMDs} for the same kinematic configuration.
Our plots show a set of 100 replicas, which are statistically equivalent and reproduce the $x f_1^g$ and $x g_1^g$ collinear PDFs. Each line in the plot refers to a single replica, and the black solid line corresponds to the most representative replica (n.~11). Although all the given replicas reproduce collinear PDFs well, they predict different results for TMDs. Furthermore, each TMD presents a peculiar pattern both in $\boldsymbol{p}_T^2$ and $x$. Our TMDs exhibit non-Gaussian $\boldsymbol{p}_T^2$-behaviors, three of them going to small but non-vanishing values in the $\boldsymbol{p}_T^2 \to 0$ limit. The linearly-polarized gluon TMD is large at small $\boldsymbol{p}_T^2$ and fall off very fast fast.
The ratio $f_1^g / h_1^{\perp g}$ behaves as a constant in the asymptotic limit $x \to 0$. This is in line with the BFKL evolution of the low-$x$ UGD, which predicts an ``equal number" of unpolarized and the linearly-polarized gluons up to effects beyond the leading-twist accuracy. This represents a contact point between the TMD and the high-energy dynamics that could be unveiled via analyses on reactions featuring a \emph{natural stability} of the high-energy resummation~\cite{Celiberto:2017ius,Celiberto:2015yba,Celiberto:2020wpk,Celiberto:2015mpa,Celiberto:2016ygs,Mueller:2012uf,Mueller:2015ael,Xiao:2018esv,Celiberto:2016hae,Celiberto:2017ptm,Caporale:2016zkc,Celiberto:2016vhn,Bolognino:2018oth,Bolognino:2019yqj,Bolognino:2019cac,Celiberto:2017nyx,Bolognino:2019yls,Bolognino:2019ccd,Golec-Biernat:2018kem,Celiberto:2020tmb,Celiberto:2021fjf,Celiberto:2021txb,Celiberto:2021xpm,Celiberto:2021tky,Bolognino:2021mrc,Bolognino:2021hxx,Celiberto:2020rxb,Celiberto:2021dzy,Celiberto:2021fdp,Celiberto:2022dyf}.
Future data on gluon TMDs are expected to exclude many replicas and to constrain parameters which are not yet well constrained by studies on collinear PDFs.

Studies at the EIC will represent a key ingredient to shed light of the gluon dynamics inside nucleons and nuclei via a 3D tomographic imaging.
The information coming from EIC will be complemented by the one collected in hadronic collisions at large $x$ from NICA-SPD~\cite{Arbuzov:2020cqg,Abazov:2021hku} and at high energies from the HL-LHC~\cite{Chapon:2020heu} and the Forward Physics Facility (FPF)~\cite{Anchordoqui:2021ghd}.\\

\begin{figure}
    \centering
    \includegraphics[width=0.85\textwidth]{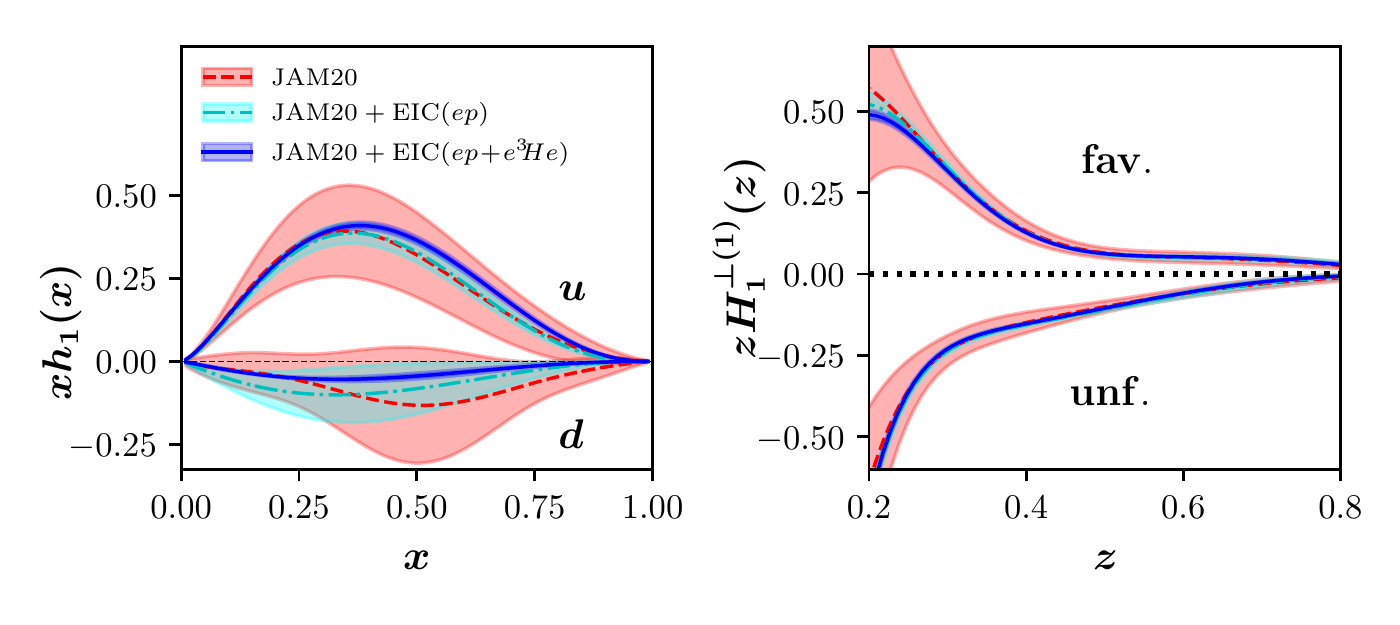}
    \includegraphics[width=0.5\textwidth]{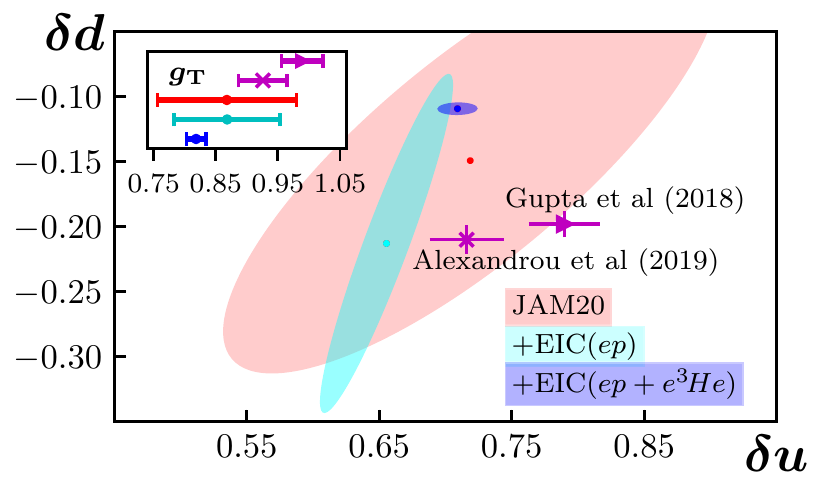}
    \caption{Top:~Plot of the transversity function for up and down quarks as well as the favored and unfavored Collins function first moment from the JAM20 global analysis~\cite{Cammarota:2020qcw} (light red band with the dashed red line for the central value) as well as a re-fit that includes EIC Collins effect pion production pseudo-data for a proton beam only (cyan band with the  dot-dashed cyan line for the central value) and for both proton and $^3$He beams together (blue band with the solid blue line for the central value). Bottom:~Individual flavor tensor charges $\delta u$, $\delta d$ as well as the isovector charge $g_T$ for the same scenarios.  Also shown are the results from two recent lattice QCD calculations~\cite{Gupta:2018qil,Alexandrou:2019brg} (purple).  All results are at $Q^2=4\,{\rm GeV^2}$ with error bands at $1$-$\sigma$ CL.}
    \label{fig:h1_Collins_gT}
\end{figure}

\noindent
{\bf Nulceon tensor-charge studies.}
The tensor charge $g_T$ is a fundamental charge of the nucleon and sits at the intersection of three keys area of nuclear physics:~TMD phenomenology~(see, e.g.,~\cite{Anselmino:2013vqa,Goldstein:2014aja,Radici:2015mwa,Kang:2015msa,Radici:2018iag,Benel:2019mcq,DAlesio:2020vtw,Cammarota:2020qcw}), searches for beyond the Standard Model (BSM) physics~(see, e.g.,~\cite{Courtoy:2015haa,Yamanaka:2017mef,Gao:2017ade,Gonzalez-Alonso:2018omy}), and {\it ab initio} approaches like lattice QCD or Dyson-Schwinger Equations~(see, e.g.,~\cite{Gupta:2018qil,Yamanaka:2018uud,Hasan:2019noy,Alexandrou:2019brg,Pitschmann:2014jxa}).  The recent lattice calculations at the physical point~\cite{Alexandrou:2019brg,Gupta:2018qil} have provided the most precise determinations of $g_T$ as well as the individual flavor (up and down) tensor charges $\delta u, \delta d$. Phenomenological approaches rely on extractions of the transversity PDF $h_1(x)$, from which $g_T$ can then be calculated as
\begin{equation}
    g_T = \delta u - \delta d\quad {\rm where} \quad \delta u = \int_0^1 \! dx\,(h_1^u(x)-h_1^{\bar{u}}(x))\,, \quad \delta d = \int_0^1 \! dx\,(h_1^d(x)-h_1^{\bar{d}}(x))\,. \label{e:gT}
\end{equation}
Analyses of this type have suffered from large uncertainties due to unmeasured regions at large and small $x$, which nonetheless enter the integrals in Eq.~(\ref{e:gT}).  Future data from the EIC will greatly improve the phenomenological situation.
In a recent work~\cite{Gamberg:2021lgx}, an impact study was performed using EIC pseudodata for the Collins asymmetry $A_{UT}^{\sin(\phi_h+\phi_S)}$ in semi-inclusive deep inelastic scattering (SIDIS) for both proton and $^3$He beams at various center-of-mass energies.  Including this pseudodata in the QCD global analysis of single transverse-spin asymmetries (SSAs) conducted in Ref.~\cite{Cammarota:2020qcw} (JAM20) allowed for an assessment of the reduction in the uncertainties for both the up and down transversity PDFs and the tensor charges $\delta u, \delta d$, and $g_T$.  The results are shown in Fig.~\ref{fig:h1_Collins_gT}. 

The EIC clearly will  significantly reduce the uncertainties of these quantities.  We find all relative errors are now  $\lesssim 5\%$: $\delta u = 0.709(15), \delta d  = -0.109(5), g_T= 0.818(16)$.  One can see the increase in precision of the extracted up quark transversity function and $\delta u$ due to the proton EIC data and further dramatic reduction of errors, in particular for $\delta d$ and $g_T$, in a combined analysis of proton and $^3$He EIC data.  The polarized $^3$He data is of particular importance for the down quark transversity and $\delta d$ tensor charge as well as decorrelating $\delta u$ from $\delta d$.  We also mention that, because the transversity TMD couples to the Collins fragmentation function (FF), there is a significant reduction in the error for the Collins FF, which will allow for an important test of universality with results from $e^+e^-$ annihilation.  With EIC data, uncertainties for phenomenological extractions of the tensor charges will become as precise as current lattice calculations --- see the bottom of Fig.~\ref{fig:h1_Collins_gT}. This expectation is confirmed also by simulations of the extraction of transversity as a collinear PDF through the inclusive di-hadron production channel~\cite{AbdulKhalek:2021gbh}. Thus, the EIC will provide a unique opportunity to rigorously explore the compatibility of lattice QCD and phenomenology/experiment. Any possible differences may become indications of BSM physics since they are not included in lattice QCD simulations~\cite{Courtoy:2015haa,Gao:2017ade}.   


\subsubsection{TMD Fragmentation}
\label{sec:TMDfrag}

To capitalize on a new era of experiments like the EIC
and HL-LHC, sound predictions for all parton distributions are necessary. TMD studies bring an
additional complication: namely, TMD extraction from data requires knowledge of fragmentation functions
(FFs), for which realistic calculations are limited. The future of momentum imaging depends critically
on progress with the calculation of FFs; hence, this will be an effort focus.  Links will be forged
between experiment, phenomenology and lattice.
The TMD FFs are one of the key tools necessary for investigating the mechanism of hadronization in a 3D-picture.
Historically, they have been accessed through Semi-Inclusive DIS (SIDIS) and $e^+e^-$ annihilation in two almost back-to-back hadrons. 
However, phenomenological extractions based on such processes are complicated by the fact that, in the cross section, the TMD FF does not appears on its own, but it is always convoluted to another TMD (two TMD FFs in $e^+e^-$ annihilations, one TMD PDF and one TMD FF in SIDIS). Disentangling these functions is usually a rather difficult task.
This problem can be bypassed provided we can extend the TMD factorization scheme to cross sections that involve only one TMD FF. 
In this sense, the cleanest process that allows to access one TMD FF is the single hadron production in $e^+e^-$ annihilations, $e^+e^- \to h\,X$, which involves a single TMD FF associated to the final detected hadron.
A factorization theorem was recently derived in Ref.~\cite{Boglione:2021wov} for $e^+e^- \to HX$, where the transverse momentum $P_T$ of the detected hadron is measured with respect to the thrust axis. Under certain approximations, this cross section can be written as a
convolution of a TMD FF with a coefficient that is totally predicted by perturbative QCD and can be interpreted
as a partonic cross section~\cite{Boglione:2020auc, Boglione:2020cwn}.
For spinless hadron with fractional energy $z_h$ and for a $2$-jet like topology of the final state, {\it i.e.}, $T \sim 1$, we have:
\begin{align}
\frac{d \sigma}{dz_h \, dT \, dP_T^2} = 
\pi \sum_j \, \int_{z_h}^1 \,
\frac{d z}{z} \, 
\frac{d \widehat{\sigma}_j}{d {z_h}/{z} \, dT} \,
D_{1,\,h/j}(z,\,P_T)
\left[
1+\mathcal{O}\left( \frac{P_T^2}{Q^2}\right) 
\right]
\,
\left[
1+\mathcal{O}\left( \frac{M_H^2}{Q^2}\right)
\right],
\label{eq:xs_e+e-single}
\end{align}
were $D_{1,\,h/j}$ is the unpolarized fragmentation function of the parton $j$ into  hadron $h$.
Since this process is more inclusive than SIDIS and $e^+e^-$ annihilation in two hadrons, the role of the soft gluons is different and the usual definition of TMDs~\cite{Collins:2011book,Aybat:2011zv,Aybat:2011vb}, where a square root of the soft factor describing the soft radiation is absorbed in the TMD definition itself, cannot be used. 
In this case, the contribution of the soft radiation is implicitly written inside the partonic cross section, being totally predicted by perturbative QCD. Therefore, the TMD FF appearing in Eq.~\eqref{eq:xs_e+e-single} is simply the operator obtained from the factorization procedure, without any further contamination from the interplay with the soft factor; therefore it describes only genuinely collinear particles. This definition is referred to as the ``factorization definition", in contrast to the ``square root definition" used in the past.
An exact relation relating the two definitions restores the possibility to perform global phenomenological analyses,  including single-hadron production. 
It is interesting to point out that the differences between these two definition of TMDs arise only in their non-perturbative behavior, namely they differ by the square root of a function, called the soft model $M_S$, i.e. schematically $\widetilde{D}^{\text{\small sqrt}} = \widetilde{D} \times \sqrt{M_S}$ (in the Fourier conjugate space of the transverse momentum of the fragmenting parton).
The soft model is a new, purely non-perturbative object that needs to be extracted from experimental data.
The advantage of this formulation is that, by neatly disentangling the soft non-perturbative part of the TMD 
at the root of the cross section, it is possible to define a phenomenology work plan that involves a much larger number of different processes by dealing with one single unknown at a time:
\begin{enumerate}
    \item Extract the TMD FF $D_{h/j}$, defined through the factorization definition, from $e^+e^- \to h\,X$ (see Eq.~\eqref{eq:xs_e+e-single}). 
    \item Apply the $D_{h/j}$ TMD as extracted above to $e^+e^- \to h_1\,h_2$ processes in order to extract  $M_S$ which, at this stage, is the only unknown in  this cross section:
    \begin{align}
        \frac{d \sigma}{dz_A dz_B d q_T} = 
        \sum_j
        \mathcal{H}^{e^+e^- \to h_A h_B}_{j\,\Bar{j}}
        \int \frac{d^2\vec{b}_T}{(2\pi)^2}
        e^{-i \vec{b}_T \cdot \vec{k}_T}
        \widetilde{D}_{h_A/j}(b_T)
        \widetilde{D}_{h_B/\Bar{j}}(b_T)
        M_S(b_T)
        \label{eq:xs_e+e-double}
    \end{align}
    \item Exploit SIDIS processes to extract the TMD PDF $f_{j/h_A}$, in the factorization definition. This is now the only remaining unknown in the SIDIS  cross section:
    \begin{align}
        \frac{d \sigma}{dz_A dz_B d q_T} = 
        \sum_j
        \mathcal{H}^{\text{sidis}}_{j\,j}
        \int \frac{d^2\vec{b}_T}{(2\pi)^2}
        e^{-i \vec{b}_T \cdot \vec{k}_T}
        \widetilde{F}_{j/h_A}(b_T)
        \widetilde{D}_{h_B/j}(b_T)
        M_S(b_T)
        \label{eq:xs_sidis}
    \end{align}
    At this stage, all necessary building blocks are available to construct and predict the cross sections involving up to two TMDs.
\end{enumerate}
Within this framework, the future EIC, which will explore a very  broad kinematical region, will provide solid ground for such phenomenological analyses.

In addition, it is important to emphasize that studies of fragmentation will be crucial for understanding the complete mechanism that includes interplay of the TMD and collinear factorization regions. By matching these regions, one will be able to reveal a consistent picture of factorization that includes both collinear and TMD FFs. In this respect, the future data from the EIC and the HL-LHC will play a crucial role.


\subsection{Generalized Parton Distributions}
\label{sec:GPDs}

Quantum chromodynamics (QCD) provides the commonly accepted description of hadrons. In this fundamental theory, quarks and gluons appear as the basic building blocks of hadron structure. The attributes of partons are therefore the essential degrees of freedom from which the macroscopic properties of hadrons --- like charge and spin --- emerge. The question naturally arises of how partons are distributed in hadrons: what are their longitudinal and transverse momenta, position in 3D space; finally, what are the resulting characteristics of the strongly-interacting matter they build, like, {\it e.g.}, the pressure and shear force at a given point. Despite very active progress towards the understanding of hadronic structure, in particular, that of the proton, finding answers to these questions still remains among the main challenges faced by studies of QCD and particle physics.

 QCD factorisation theorems provide us with tools to answer all of the above-mentioned questions. In particular, generalised parton distributions (GPDs) offer a rigorous theoretical framework that can be used to study the 3D structure of nucleons \cite{Muller:1994ses, Ji:1996ek, Radyushkin:1996ru}. These objects describe the partonic structure of hadrons and, to some extent, may be considered a unifying basis subsuming the collinear PDFs and elastic form factors. This non-trivial connection is used to obtain tomographic pictures of the nucleon, where the spatial distributions of partons carrying a fraction of the nucleon’s momentum are projected onto the plane perpendicular to the direction of the nucleon’s motion~\cite{Burkardt:2000za}. These pictures reveal the true nature of the nucleon, as GPDs allow us to depict hadrons as extended objects composed of quarks and gluons. Another exciting feature of GPDs is their relation to the QCD energy-momentum tensor, providing its only experimental connection practicable today. This unique relation allows us to evaluate the total angular momentum carried by gluons or quarks of a given flavor~\cite{Ji:1996ek}, which is essential to solve the long-standing problem of the nucleon spin decomposition. The relation between GPDs and energy-momentum tensor can be also used to access information on ``mechanical'' properties of partonic systems, like the distribution of pressure inside the nucleon \cite{Polyakov:2002yz,Goeke:2007fp,Polyakov:2018zvc,Lorce:2018egm,Shanahan:2018nnv,Freese:2021czn,Freese:2021qtb}. This may help understand the properties of partonic media. A comprehensive review of the GPD formalism and the notations used in the present document can be found in Ref.~\cite{Diehl:2003ny}.

GPDs are studied via exclusive processes, where the properties of all particles in both initial and final states are reconstructed. A distinctive feature of processes used to study GPDs is the requirement that the probed hadron not break in the interaction, allowing one to study its transition from one state to another at a partonic level. Nowadays, measurements of exclusive reactions are among the main goals of a new generation of experiments. This includes JLab and the LHC (including the possible LHeC~\cite{LHeCStudyGroup:2012zhm}) as well as the EIC.

\subsubsection{Phenomenological extraction of GPDs from DVCS}

Nowadays, the main source of information on GPDs comes from deeply virtual Compton scattering (DVCS). 
This process is relatively simple in its interpretation, as the only non-perturbative object entering its amplitudes, which are referred to as Compton form factor (CFFs), are GPDs. 
DVCS has been extensively studied during the last two decades and we refer to Ref.~\cite{Kumericki:2016ehc} for putting these developments into a phenomenological context. 
In view of the perspective open by future colliders and the lattice QCD progress, we focus here on GPD or CFF extractions from DVCS measurements.
So far most of the phenomenological effort in the GPD community has followed three directions:
\begin{enumerate}
    \item Consider data sets as samples of DVCS observables over many kinematic bins. Perform independent fits on kinematic bins (as many fits as kinematic bins).
    
    \item Build GPD or CFF models and compare them to existing data, possibly adjusting some of their parameters to obtain better agreement between model expectations and measurements.  
    
    \item Build GPD or CFF parameterizations and adjust all their parameters to existing data. 
\end{enumerate}

Approach~1 is commonly referred to as \emph{local fits}. It was first presented in Ref.~\cite{Guidal:2008ie} and was used in numerous subsequent studies \cite{Guidal:2009aa, Guidal:2010de, Guidal:2010ig, Boer:2014kya, Kumericki:2013br}. 
The advantage of this method is that it is quite easy to implement and has a low-model dependence. 
Unfortunately, it provides information only on the kinematic bins where data are available, i.e. it does not have any predictive power. 
Interpretation of results obtained in local fits is also difficult, as mixing information coming from independent kinematic bins introduces unknown systematic uncertainty. 
In addition, local fits do not allow to incorporate theoretical constraints on GPDs, suggesting for instance the continuity of CFFs. 
Consequently more sophisticated extractions of CFFs recently combined local fits with kinematic corrections \cite{Defurne:2017paw, Benali:2020vma}, Rosenbluth separation \cite{Kriesten:2020apm} or even para\-me\-tric global fits \cite{Moutarde:2009fg, Dupre:2016mai, Dupre:2017hfs, Burkert:2018bqq, Burkert:2021ith}. 

Approach~2 corresponds to, {\it e.g.}, the GK model~\cite{Goloskokov:2005sd, Goloskokov:2007nt, Goloskokov:2009ia}, the VGG model \cite{Vanderhaeghen:1998uc, Vanderhaeghen:1999xj, Goeke:2001tz, Guidal:2004nd}, or the reggeized diquark model \cite{Ahmad:2006gn, Ahmad:2007vw, Goldstein:2010gu, GonzalezHernandez:2012jv}. 
Even if these models provide GPDs defined over the whole $(x_B, t, Q^2)$ kinematic domain, they usually do not provide a good agreement to the existing DVCS data. 
The lack of flexibility of these models if often signaled by a $\chi^2/\mathrm{ndf} \gg 1$. 
However, these physically-motivated parameterizations are valuable to making forecasting studies or to designing future experiments, since they can be used in event generation to obtain sensible mock data.
An on-going theoretical effort now provides wide families of GPD models which consistently address polynomiality and positivity properties. 
This allows robust predictions of new processes in as yet unexplored kinematic domains, like, {\it e.g.}, DVCS off a pion target at the EIC~\cite{Chavez:2021koz, Chavez:2021llq}.

\begin{figure*}[!ht]
\begin{center}
\includegraphics[width=0.95\textwidth]{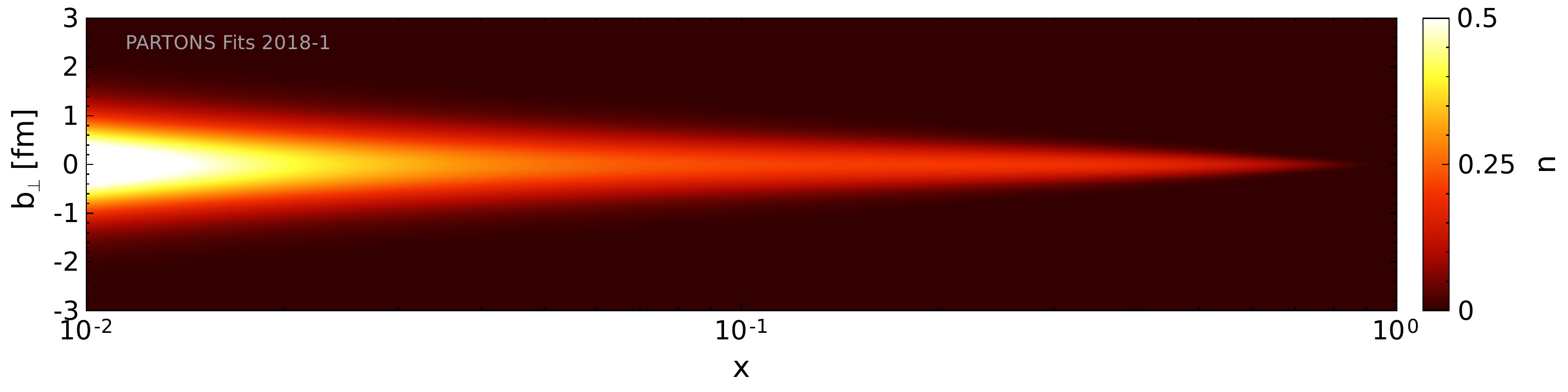}
\caption{Transverse position of up quarks in an unpolarized proton as a function of the longitudinal momentum fraction $x$ \cite{Moutarde:2018kwr}.}
\label{fig:NT:2D}
\end{center}
\end{figure*}

Approach~3, commonly referred to as \emph{global fits}, has been achieved by the KM and PARTONS groups \cite{Kumericki:2007sa, Kumericki:2009uq, Kumericki:2013br, Moutarde:2018kwr}, however only for CFFs. As a representation of typical results, the tomography of the nucleon for up quarks, obtained with a specific choice of CFF parameterization and constrained in a global fit to world data, is shown in Fig.~\ref{fig:NT:2D}. The issue of the potential parameterization bias was addressed by both groups utilising artificial neural-network techniques~\cite{Kumericki:2011rz, Moutarde:2019tqa}. 
These parameterizations are flexible enough to reproduce the existing DVCS data with $\chi^2/\mathrm{ndf} \simeq 1$.  
As for today, there are no global fits of GPDs including gluons, sea and valence quarks.
Only the KM model performed fits of gluon and sea quark GPDs in the small-$x_B$ region.
The extension to the case of GPDs of the artificial neural-network techniques involved in CFF fitting was recently considered~\cite{Dutrieux:2021wll} and it is hoped that neural network-based GPD fits will be obtained in the next few years.

\subsubsection{Hard exclusive processes and multi-channel analysis}

Multi-channel analyses are necessary because they constitute universality checks of GPDs, provide different handles on the partonic content of the nucleon, and may avoid the aforementioned uniqueness issue of the DVCS deconvolution problem~\cite{Bertone:2021yyz}. 
Most of the existing data come either from DVCS or DVMP. Since these two (families of) processes have different coefficient functions, a study at the GPD-level is mandatory. 
On the contrary, TCS occupies a special place in multi-channel analyses. 
Its intimate relation to DVCS makes possible a joint study of DVCS and TCS at the CFF-level, see {\it e.g.}, Ref.~\cite{Grocholski:2019pqj}.
The first TCS measurements were made by CLAS~\cite{CLAS:2021lky}.

However, most of the multi-channel studies made so far for DVCS, DVMP, or TCS and in various kinematic domains were produced using models of approach~2 above.
There are fewer studies with the flexible fit parameterizations of approach~3 (KM and PARTONS).
The feasibility of a simultaneous global GPD fit to collider measurements (H1 and ZEUS) to DVMP ($\rho^0$ and $\phi$ mesons) and DVCS was shown in Ref.~\cite{Meskauskas:2011aa}.
Nevertheless the fit, performed at LO, achieved a $\chi^2/\mathrm{ndf} \simeq 2$ and an extension of this study to NLO was advocated. 
The theoretical description of DVMP at NLO was introduced in Ref.~\cite{Mueller:2013caa} and applied in a pioneering unpublished study \cite{Lautenschlager:2013uya} for DIS, DVCS, and $\rho^0$ and $\phi$ electroproduction, but still for H1 and ZEUS data. 

In that respect, the future EIC represents a challenge for modern GPD phenomenology.
Indeed, a large amount of precise data is expected, for DVCS, TCS, and DVMP over a wide kinematic range. 
Such a situation remains uncharted for GPD physics, and phenomenological software such as {\sc partons} \cite{Berthou:2015oaw} and {\sc gepard} \cite{Kumericki:2015lhb} will need to be upgraded to fully exploit the vast amount of forthcoming data.
But the situation also carries the promise of potential breakthroughs in exclusive physics. 
It is expected to offer a unique opportunity to shed light on the properties of the nucleon, from its 3D structure up to its mechanical properties through the connection with the energy-momentum tensor. 

\subsubsection{Energy-momentum tensor form factors}

For a spin-$1/2$ hadron, the matrix elements of a general asymmetric energy-momentum tensor (EMT) can be parametrized in terms of five form factors \cite{Bakker:2004ib, Leader:2013jra}

\begin{align}
\label{Eq:def-EMT-kinetic}
    \langle p^\prime, s^{\,\prime}|  T_a^{\mu\nu}(0) |p, s\rangle
    = \bar u(p^\prime, s^{\,\prime})\biggl[A_a(t)\,\frac{P^\mu P^\nu}{M_N} &+
    J_a(t)\ \frac{i(P^{\mu}\sigma^{\nu\rho} + P^{\nu}\sigma^{\mu\rho})\Delta_\rho}{2M_N}
    + D_a(t)\,
    \frac{\Delta^\mu\Delta^\nu-g^{\mu\nu}\Delta^2}{4M_N} \nonumber \\
    &-S_a(t)\ \frac{i(P^{\mu}\sigma^{\nu\rho} - P^{\nu}\sigma^{\mu\rho})\Delta_\rho}{2M_N} + \bar{C}_a(t)\,M_N\,g^{\mu\nu}\biggr]u(p, s),
\end{align}
where $p$ and $p^\prime$ denote the incoming and outgoing momenta, while $s$ and $s^\prime$ denote the incoming and outgoing polarizations, respectively. Despite the EMT form factors for each parton type $a=q,g$ being renormalisation scale dependent, the total EMT form factors are renormalisation scale independent. Furthermore, the conservation of the EMT, i.e. $\partial_{\mu}T^{\mu\nu} = 0$, implies that $\bar C\equiv\sum_{a} \bar{C}_{a} = 0$. EMT form factors encode essential information about the fundamental properties of the nucleon. When there is no invariant momentum transfer to the system, i.e. at $t=0$, it follows from the Poincar\'e symmetry that $A(0) = 1$ and $J(0)=1/2$ \cite{Ji:1996ek,Lowdon:2017idv,Cotogno:2019xcl}. The constraints on $A$ and $J$ reflect the fact that the total momentum and angular momentum of the hadron are carried by their constituents, respectively. On the other hand, $D(0)$, also known as the D-term \cite{Polyakov:1999gs}, is unconstrained and must be determined through experiments. There have already been several phenomenological efforts on the extraction of the D-term \cite{Burkert:2018bqq,Kumericki:2019ddg,Dutrieux:2021nlz}, and the subject is gaining growing interest because it provides an insight into the pressure and shear forces inside the proton \cite{Polyakov:2002yz,Goeke:2007fp,Polyakov:2018zvc,Lorce:2018egm,Shanahan:2018nnv,Freese:2021czn,Freese:2021qtb}. The form factor $S_q(t)$, on the other hand, is associated with the anti-symmetric part of the EMT and its quark part is related to the axial form factor $G_A(t)$ by $S_q(t) = -G_A(t)$ \cite{Bakker:2004ib,Lorce:2017wkb}.  

Accessing to the EMT form factors is in principle possible through exclusive processes. For $a\!=\!q$ (see the analogous expression for gluons in \cite{Diehl:2003ny}), the second Mellin moments of GPDs $H^{a}(x,\xi,t)$ and $E^{a}(x,\xi,t)$ are related to the three EMT form factors $A_a(t), J_a(t),$ and $D_a(t)$ as follows \cite{Ji:1996ek,Ji:1998pc}
\begin{align}
    \int_{-1}^{1} dx\;x\;H^{a}(x,\xi,t) &= A_{a}(t) + \xi^2\;D_{a}(t) \\
    \int_{-1}^{1} dx\;x\;E^{a}(x,\xi,t) &= 2J_{a}(t) - A_{a}(t) - \xi^2\;D_{a}(t)
\end{align}
or, 
\begin{equation}\label{Eq:Ji-sum-rule}
    \int_{-1}^{1} dx\;x\;\Big(H^{a}(x,\xi,t) + E^{a}(x,\xi,t)\Big) = 2J_{a}(t)\; .
\end{equation}
Eq. (\ref{Eq:Ji-sum-rule}) is known as the Ji sum rule and, in the limit of $t\rightarrow 0$, it allows one to determine the individual contributions of each parton type to the total longitudinal angular momentum of the nucleon directly from the GPDs $H$ and $E$. For the transverse angular momentum, the combination of $H$ and $E$ will usually depend on the frame~\cite{Leader:2011cr,Ji:2020hii,Lorce:2021gxs}. Therefore, the future EIC data will not only enable us to narrow down uncertainties in GPDs but also, subsequently, enhance our ability to access fundamental properties of the nucleon at much higher precision.

\subsubsection{Connections to lattice QCD}
\begin{figure*}[tb]
\centering
\includegraphics[width=0.42\textwidth]{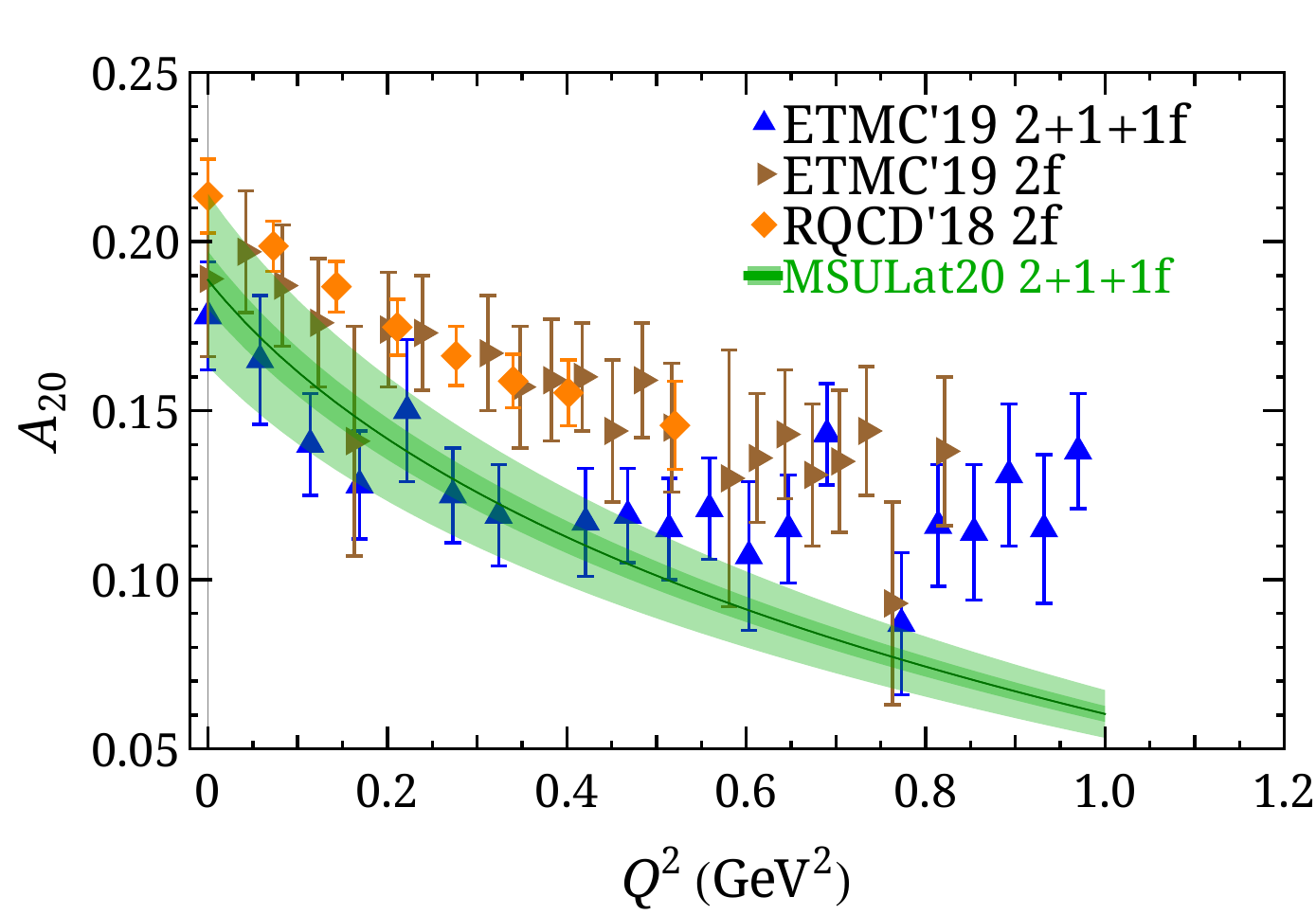} 
\includegraphics[width=0.42\textwidth]{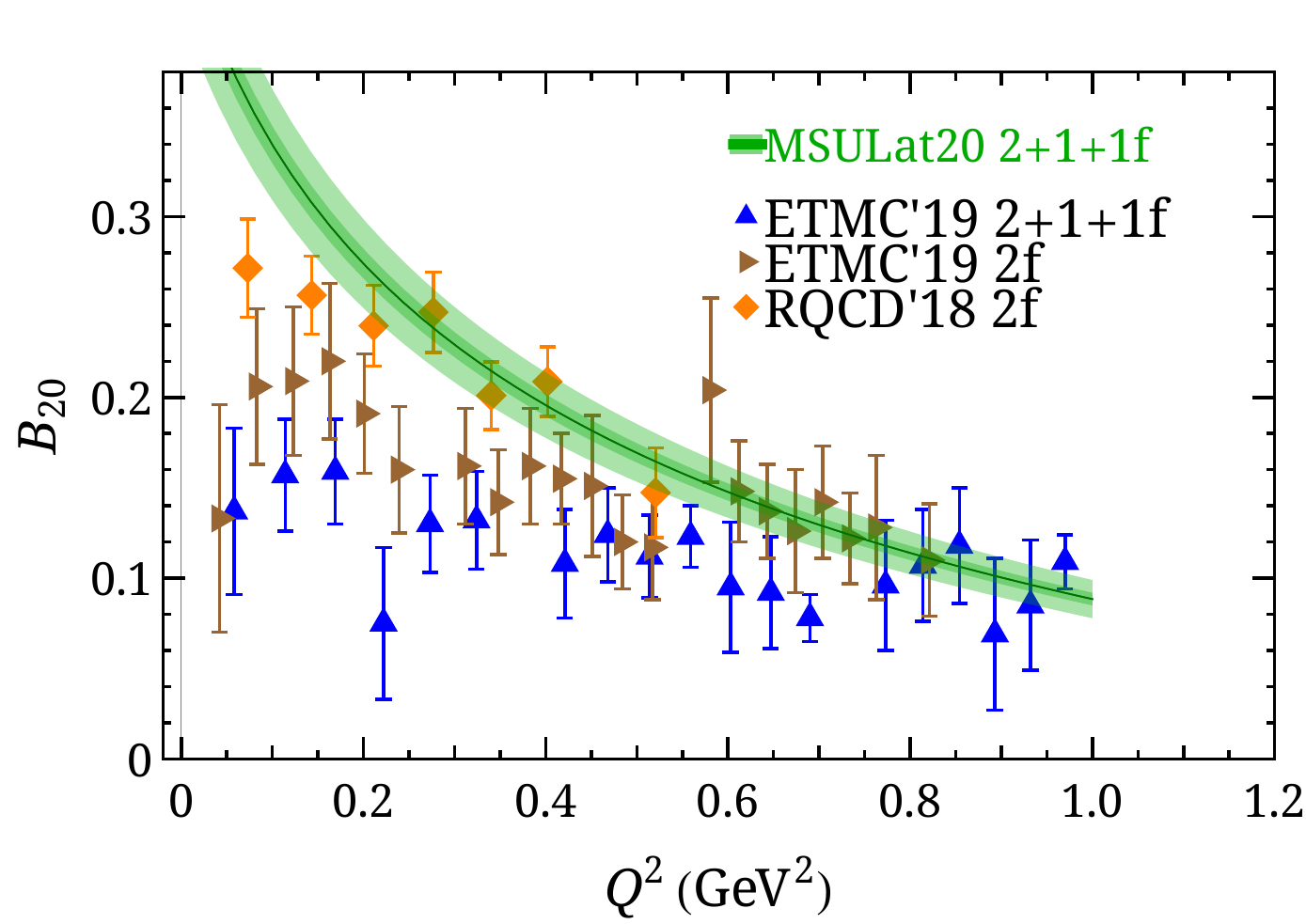}
\includegraphics[width=0.42\textwidth]{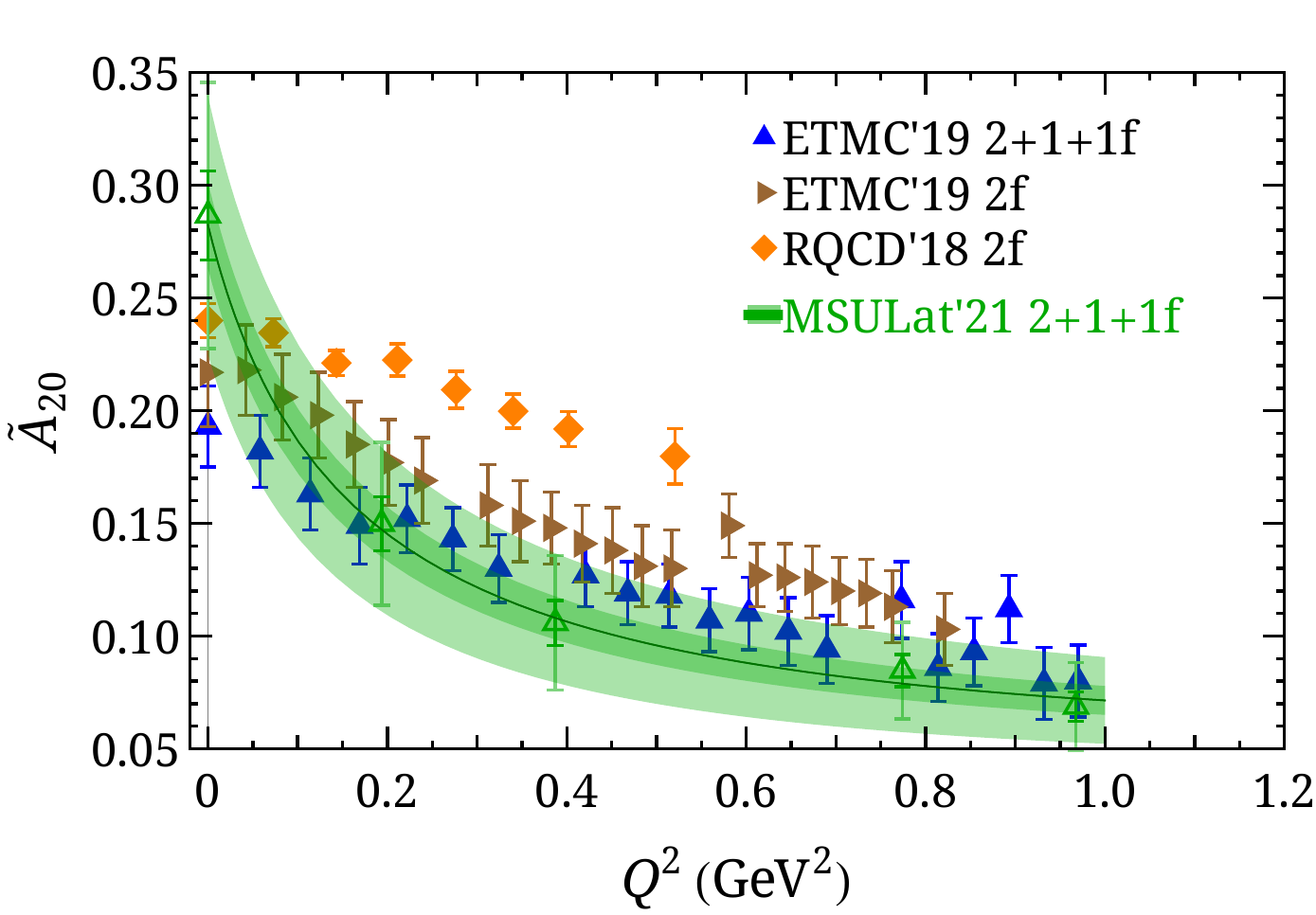} 
\includegraphics[width=0.42\textwidth]{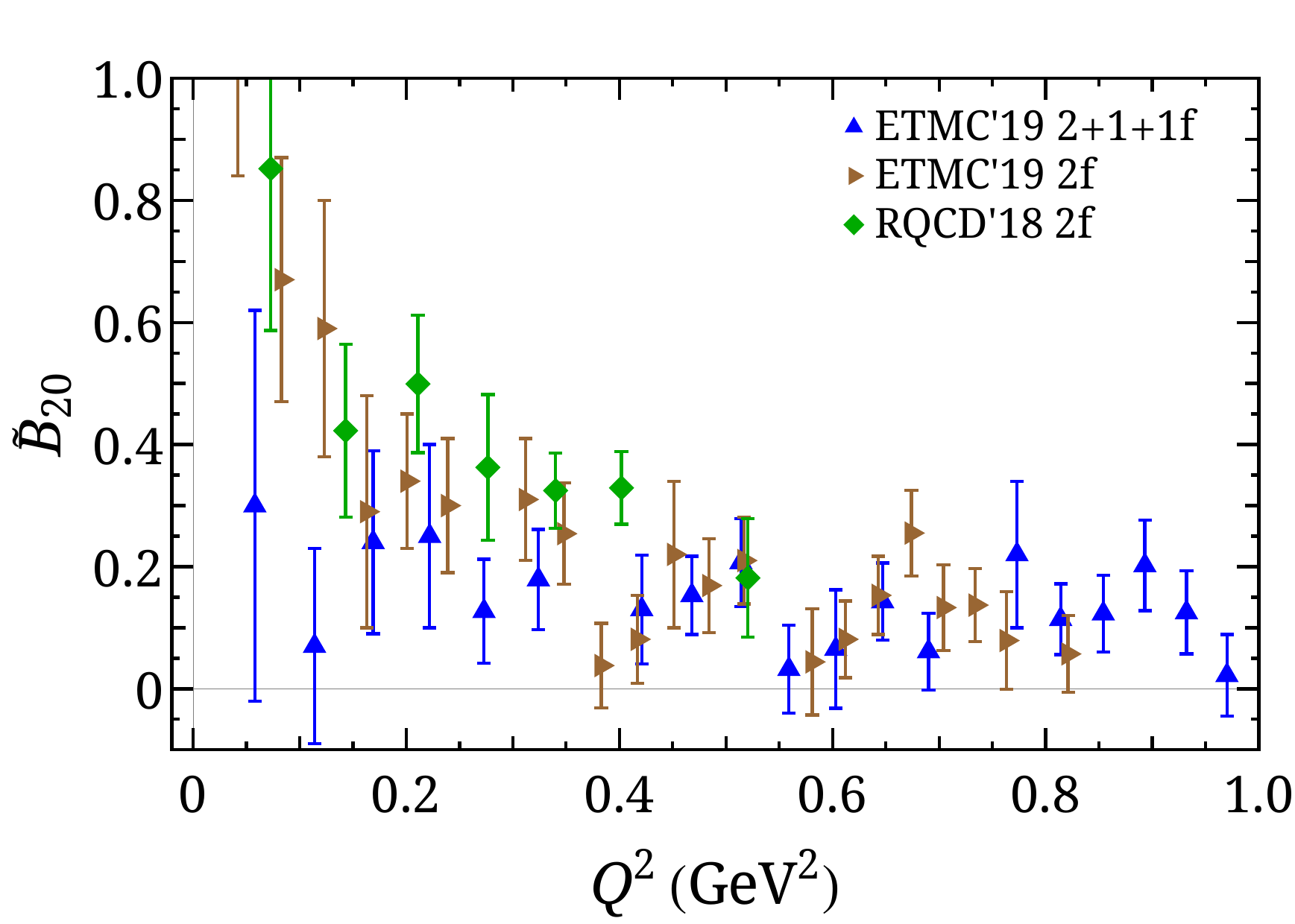}
\caption{\label{fig:LatGFF}
Unpolarized (left) and linearly polarized (right) nucleon isovector GFFs from near physical pion mass
as functions of transferred momentum $Q^2$. 
The references corresponding to the above works are:
LaMET method by MSULat20/21 2+1+1f~\cite{Lin:2020rxa,Lin:2021brq} by integrating over $x$-dependent GPDs,
OPE methods from
2+1+1f and 2f (only the larger-volume results are quoted here) from
ETMC19~\cite{Alexandrou:2019ali} and
2f RQCD19~\cite{Bali:2018zgl}.
}
\end{figure*}
Recent progress in lattice QCD has led to many exciting
possibilities, including the direct lattice calculation of
collinear PDFs (unpolarized as well as the helicity and transversity distributions) and
extension to GPDs \cite{Constantinou:2022yye,Constantinou:2020hdm,GPDsLOI}.  Synergies with QCD global analyses in the EIC
era will require further theoretical refinements. Opportunities include studies
of the gluonic structure of the proton and of light nuclei, in which a synthesis of EIC
measurements and lattice calculations may enlighten a range of issues including the
double helicity flip gluon structure functions of light nuclei and the gluon PDFs,
GPDs (including the $D$-term, pressure and shear), and TMDs of the proton and light
nuclei.

For a long time, lattice-QCD has been limited to calculate the  Mellin moments of the GPDs  using local matrix elements through the operator product expansion (OPE):
\begin{subequations}
\begin{align}
\label{eq:GFFs}
\int_{-1}^{+1}\!\!dx \, x^{n-1} \, H(x, \xi, -Q^2) =
\sum\limits_{i=0,\text{ even}}^{n-1} (-2\xi)^i A_{ni}(Q^2) &+ (-2\xi)^{n} \,  C_{n0}(Q^2)|_{n\text{ even}}, \\
\int_{-1}^{+1}\!\!dx \, x^{n-1} \, E(x, \xi, -Q^2) =
\sum    \limits_{i=0,\text{ even}}^{n-1}	(-2\xi)^i B_{ni}(Q^2)	 &- (-2\xi)^{n} \,  C_{n0}(Q^2)|_{n\text{ even}}\,,
\end{align}
\end{subequations}
where the generalized form factors (GFFs) $A_{ni}(Q^2)$, $B_{ni}(Q^2)$ and $C_{ni}(Q^2)$ in the $\xi$-expansion on the right-hand side are real functions.
When $n=1$, we get the Dirac and Pauli electromagnetic form factors $F_1(Q^2) = A_{10}(Q^2)$ and $F_2(Q^2) = B_{10}(Q^2)$. 
The unpolarized GFFs ($A_{20}$, $B_{20}$) and the linearly polarized GFFs ($\tilde A_{20}$, $\tilde B_{20}$) using the $x$-dependent GPD functions~\cite{Lin:2020rxa,Lin:2021brq} and Eq.\ref{eq:GFFs} are shown in Fig.\ref{fig:LatGFF}, along with the simulations at/near the physical point by ETMC~\cite{Alexandrou:2019ali} and RQCD~\cite{Bali:2018zgl} using the traditional moment method.  It is interesting to observe that the two data sets for $A_{20}$ in the ETMC calculation exhibit some tension.
This is an indication of systematic uncertainties.
Given that the blue points correspond to finer lattice spacing, larger volume and larger $m_\pi L$, we expect that the blue points have smaller systematic uncertainties.
Both ensembles used by ETMC lead to compatible results for the other GFFs.
A comparison between the $N_f=2$ ETMC data and $N_f=2$ RQCD data reveals agreement for $A_{20}$, $B_{20}$ and $\tilde B_{20}$.
However, the RQCD data have a different slope than the ETMC data, attributed to the different analysis methods and systematic uncertainties.
The integral of $\tilde{H}$ for $n=2$ from the LaMET method providing $\tilde{A}_{20}$ also shows results consistent within two sigma with most past work from various actions and other lattice parameters.
This seems to verify that LaMET GPD calculations give reasonable results for pursuing precision calculations in the future.

\begin{figure*}[tb]
\includegraphics[width=0.8\textwidth]{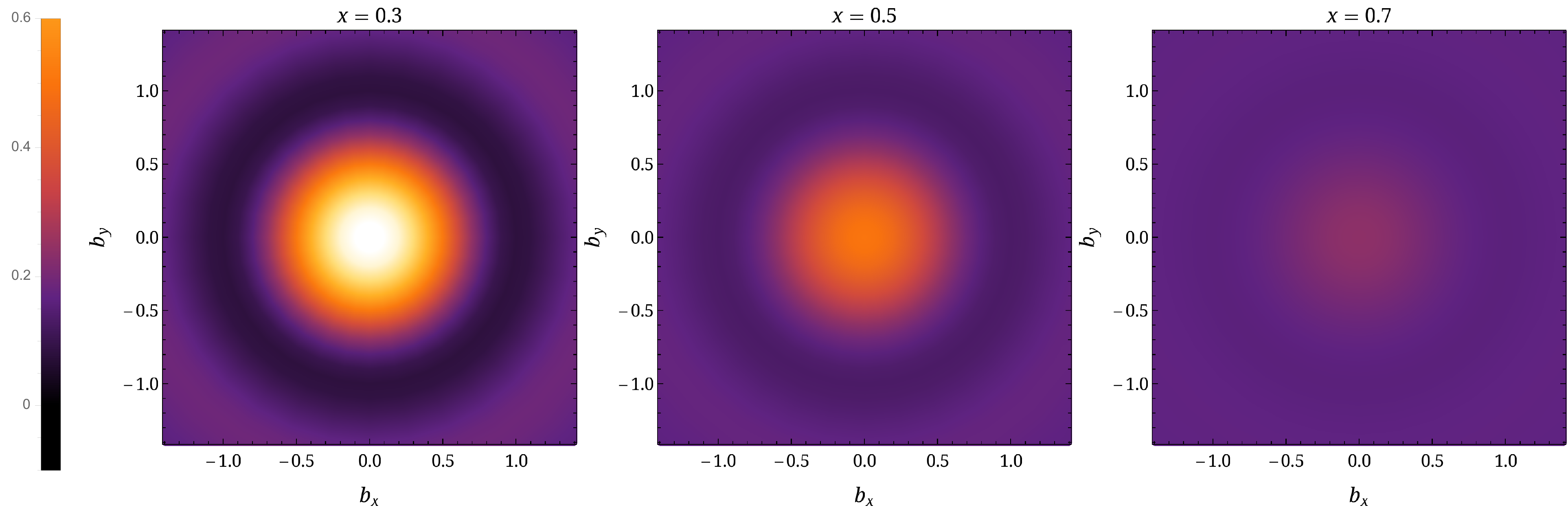}
\caption{
The two-dimensional impact-parameter--dependent distribution of isovector nucleon GPDs for $x=0.3$, 0.5 and 0.7 from lattice QCD calculation at physical pion mass~\cite{Lin:2020rxa}.
\label{fig:lat-impact-distribution}}
\end{figure*}

Recently, there has been a couple of new calculations~\cite{Chen:2019lcm,Lin:2020rxa,Alexandrou:2020zbe,Lin:2021brq} exploring $x$-dependent GPD structure using quasi-GPDs. 
On the lattice, calculations are done in the Breit frame starting with the calculating lattice matrix element
\begin{equation}
\langle N(P_z+{Q}/2)|\bar\psi\left(0\right)\Gamma W(0,z)\psi\left(z\right)|N(P_z-{Q}/2)\rangle.
\end{equation}
Both calculations use a $N_f=2+1+1$ (degenerate up/down, strange and charm QCD vacuum) lattice at lattice spacing around 0.09~fm. 
ETMC~\cite{Alexandrou:2020zbe} reports results at a single $Q^2$ for both unpolarized and polarized GPDs with a 260-MeV pion with skewness of 0 and 0.3, while MSULat~\cite{Lin:2020rxa} focuses on unpolarized zero-skewness GPDs with 135-MeV pion at multiple $Q^2$. 
One can now actually take the moments of MSULat's $x$-dependent GPDs  and compare with lattice results using transitional moment local-operator via the OPE method as shown in Fig.~\ref{fig:LatGFF}.
With multiple $Q^2$ values, MSULat~\cite{Lin:2020rxa} can then take Fourier transforms of the non--spin-flip GPD $H(x,\xi=0,-Q^2)$ to calculate the impact-parameter--dependent distribution $\mathsf{q}(x,b)$~\cite{Burkardt:2002hr}
\begin{equation}\label{eq:impact-dist}
\mathsf{q}(x,b) = \int \frac{ d^2 \Delta_\perp}{(2\pi)^2}\,e^{-i\Delta_\perp\cdot b_\perp }\, H(x,\xi=0,t=-\Delta^2_\perp) ,
\end{equation}
where $b=|b_\perp|$ is the transverse distance from the center of momentum.
Figure~\ref{fig:lat-impact-distribution} shows the first results of impact-parameter- dependent  two-dimensional distributions at $x=0.3$, 0.5 and 0.7
from lattice QCD. 
The impact-parameter--dependent distribution describes the probability density for a parton with momentum fraction $x$ at distance $b$ in the transverse plane. 
Using Eq.~(\ref{eq:impact-dist}) and the $H(x,\xi=0,-Q^2)$ obtained from the lattice calculation at the physical pion mass, we can take a snapshot of the nucleon in the transverse plane to perform $x$-dependent nucleon tomography using lattice QCD for the first time.



\subsection{Summary}
\label{sec:tomo_summary}
The EIC is a machine for precision QCD targeted at unfolding the multi-dimensional structure of the proton and other
hadrons, and of the nucleus. As such, it will have a number of far-reaching implications at the Energy and Intensity
Frontiers.
While the issues discussed in this report present formidable challenges, 
the HL-LHC and EIC projects will assemble a world-wide collaboration of
scientists, all working together to solve these puzzles. 
When combined, the complementary information from the EIC and
LHC will provide an unparalleled description of precision QCD and of how the strong force
generates the structure of hadrons.
Continued communication between these projects beyond the Snowmass study
will help ensure optimal use of relevant data sets and serve as   
the keystone to a deeper understanding of QCD.
\begin{itemize}
    \item EIC data will provide information of interest to the entire particle physics community on various aspects of hadron structure encoded in collinear PDFs, GPDs, and TMDs; these improvements to PDF precision will significantly impact LHC measurements of, {\it e.g.}, the Higgs cross section, $M_W$, and various electroweak couplings;
        \item A unique capability of the EIC is to provide comprehensive flavor separation of (un)polarized PDFs to enhance precision determinations of QCD sector of SM quantities, including the running coupling and the heavy quark masses;
    \item The EIC will be allow for precise extraction of the tensor charge of the nucleon and its decomposition in partonic degrees of freedom. This information is synergistic with both lattice QCD studies and the searches of BSM physics; 
    \item Recent progress in lattice QCD has led to many exciting
possibilities, including the direct lattice calculation of
collinear PDFs  and
extension to GPDs and TMDs.  Synergies with QCD global analyses that will include EIC and LHC data will allow for further theoretical refinements and better understanding of QCD and SM;
    \item The EIC data together with the data from high-energy experiments such as those at the LHC will provide the necessary information for unraveling the novel three-dimensional structure of the nucleon that will allow nuclear, lattice QCD, and HEP communities to answer fundamental questions on the origin of the spin and the mass of the nucleon.
\end{itemize}
%
%
%
\newpage
\section{Jets at EIC} \label{sec:jets}
\vspace{-2ex}
\centerline{\textit{Editors:} 
\href{mailto:miguel.arratia@ucr.edu}{\texttt{Miguel Arratia}},
\href{mailto:zkang@ucla.edu}{\texttt{Zhong-Bo Kang}},
\href{mailto:stefan.prestel@thep.lu.se}{\texttt{Stefan Prestel}}.
}
\vspace{3ex}

The advent of the ElC with its high luminosity ($\sim 1000$ times higher than HERA) and polarized hadron beams will produce the first-ever jets in polarized electron-hadron scattering, and will unlock the full potential of jets as novel tools for probing the structure of nucleon and nuclei. Jet studies have played a key role in the exploration of QCD since its conception~\cite{Ali:2010tw}. With the advances in experimental techniques, and theory development over time, jets have become powerful tools for exploring the fundamental properties and regimes of QCD, and when searching for unexpected phenomena in high-energy collisions~\cite{Larkoski:2017jix,Asquith:2018igt,Marzani:2019hun}. This has pushed jet physics to the forefront of phenomenology at the LHC and RHIC.

While jets are familiar in high-energy physics analyses, and appear in many different guises, jets at the EIC can add important pieces of the puzzle on top of insights gained at hadron-hadron machines: Jets at EIC are naively expected to be very “clean”, i.e. little energy is not associated with the jets. However, the jets themselves contain relatively few particles, and the particles have moderate energies. This offers unique challenges and opportunities: Every particle is precious and differences between jet algorithms or substructure methods can become very apparent, while at the same time, underlying event contamination (that continues to be a major challenge at the LHC) will be much smaller. This already makes an assessment of jet properties at EIC an exciting theoretical and experimental prospect. On top of that, non-perturbative contributions from fragmentation are more pronounced at lower jet masses, which makes EIC jets a stress test for the universality of jet-based methods in high-energy physics. In addition, because of its polarization capability and unmatched versatility, the EIC will surely serve as a venue for developing new jet-based observables that exploit the spin. Thus, dedicated studies of jet substructure at the EIC are critical in order to realize the full potential of the EIC in jet physics. Jets are guaranteed to contribute significantly at the EIC to a variety of key electron-nucleus and electron-hadron physics topics, which we study in detail in this section.

\begin{figure}
    \centering
    \includegraphics[width=0.49\textwidth]{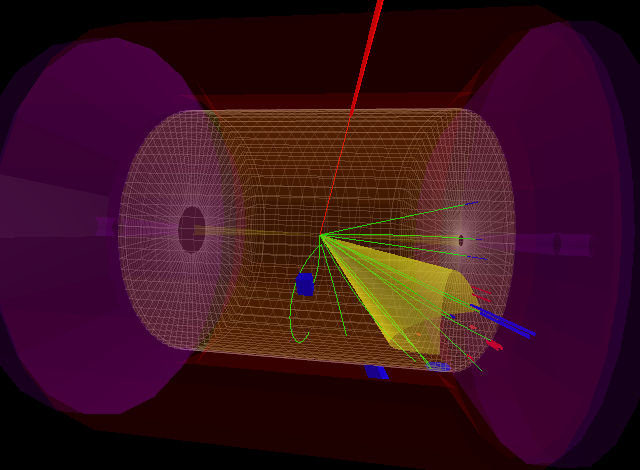}
    \includegraphics[width=0.49\textwidth]{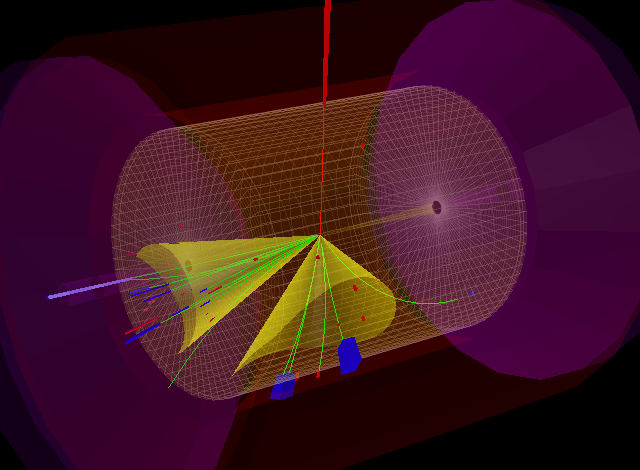}\\
        \includegraphics[width=0.49\textwidth]{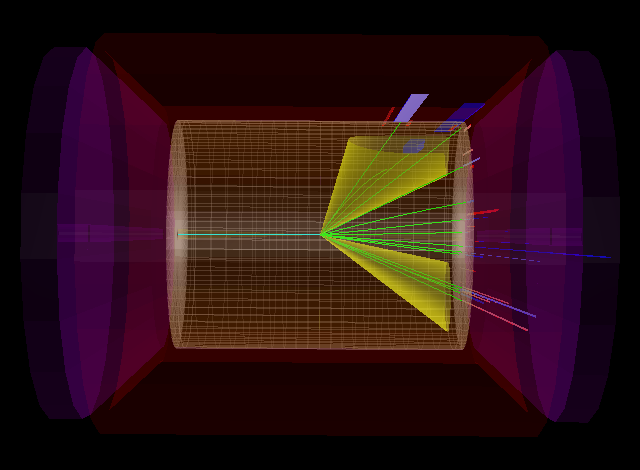}
     \includegraphics[width=0.49\textwidth]{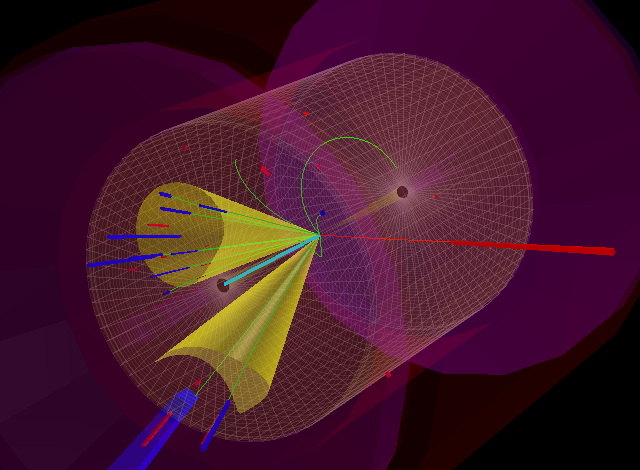}

    \caption{Event displays for simulated EIC events. Single-jet in from Born-level DIS event (upper left); di-jet in photon-gluon fusion event (upper right); di-jet in photo-production event (bottom left); di-jet in hard-diffractive DIS event (bottom right). }
    \label{fig:eventdisplays}
\end{figure}

\subsection{Jets for flavor and spin structure of hadron}
The femtoscale structure of the nucleon is one of the major scientific pillars of the EIC~\cite{Accardi:2012qut,Proceedings:2020eah}. In particular, the flavor and spin structure of the nucleon in terms of both one dimensional~(1D) and three-dimensional (3D) imaging provides fascinating glimpses into the non-perturbative QCD dynamics. Production of jets and jet substructure have shown their great power in providing important information and constraints on partonic structure of the hadron.

\subsubsection{1D spin structure}
In the unpolarized and longitudinal polarized lepton and nucleon scattering, one would be able to learn 1D unpolarized collinear parton distribution functions (PDFs) and helicity distribution functions. While the collinear PDFs are crucial ingredient for the prediction of any cross section at the LHC, the helicity distribution functions enables us to extract important information on the spin decomposition of the proton. 

There has been tremendous progress in the past a few years in advancing the perturbative computations of jet cross sections. In~\cite{Abelof:2016pby,Currie:2017tpe}, the authors performed the next-to-next-to-leading order (NNLO) calculation for single inclusive jet production in unpolarized electron+nucleon collisions. A next-to-next-to-next-to leading (N$^3$LO) calculation has been performed in~\cite{Currie:2018fgr}. The NNLO computations in longitudinally electron+nucleon collisions have been worked out in~\cite{Borsa:2020ulb}.

Among the most intriguing aspects of hadronic physics is the spin decomposition of the proton in terms of its partonic constituents. This has remained an outstanding puzzle for decades and is one a primary motivation for the EIC. To determine the contribution of quarks and gluons to the spin of the proton one needs to extract the helicity-dependent parton distribution functions (PDFs). A standard way to approach this goal is to perform a global QCD analysis of all available data taken in spin-dependent scattering. The accuracy of these global fits relies upon the validity of QCD factorization and the high precision computation of the perturbative hard coefficients. 

\begin{figure}[h!]
\centering
\includegraphics[width=0.5\textwidth]{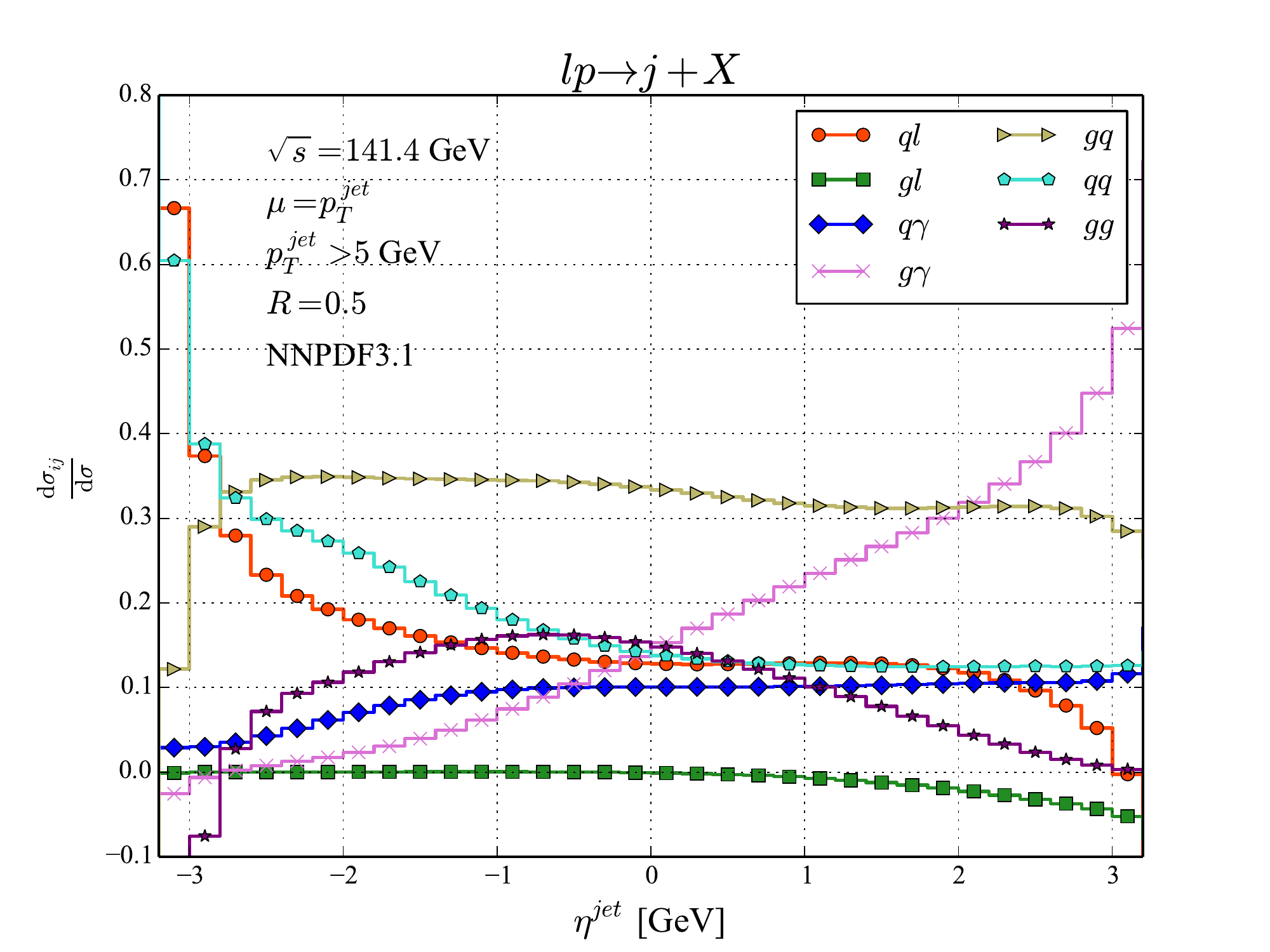}
\caption{Decomposition of inclusive jet production at an EIC into its partonic constituents as a function of jet pseudorapidity. 
\label{fig:kin}}
\end{figure}

\begin{figure}[h!]
\centering
\includegraphics[width=0.49\textwidth]{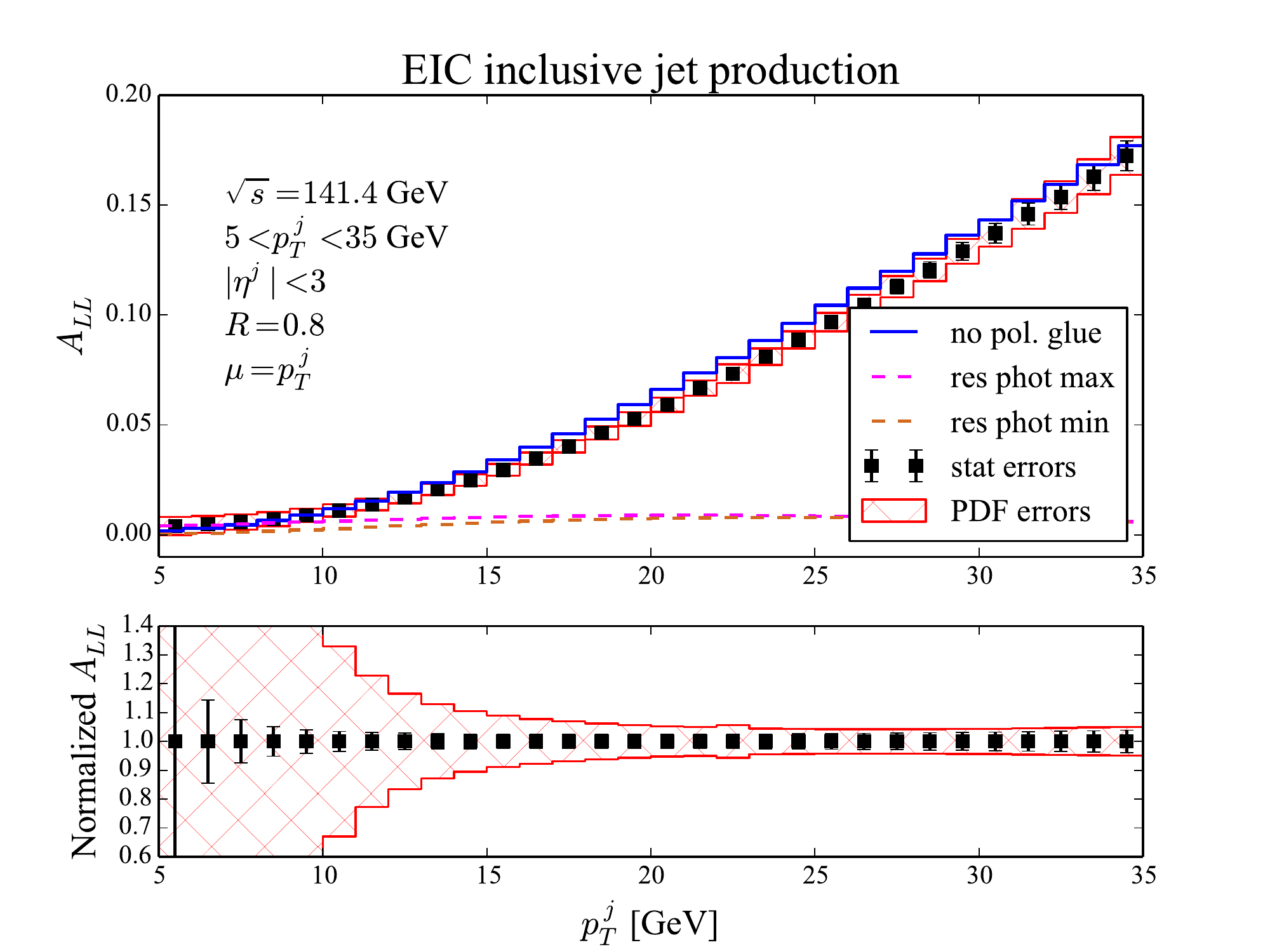}
\caption{Double-longitudinal spin asymmetry in inclusive jet production at an EIC as a function of jet transverse momentum. Figure taken from~\cite{Boughezal:2018azh}.\label{fig:AllpT}}
\end{figure}

Jets are one of the main probes of the unpolarized partonic structure of the nucleon in current global fits, and it is worthwhile to explore their impact on helicity-dependent PDFs as well. Several detailed studies of the impact of EIC jet data on helicity-dependent PDFs have been performed~\cite{Chu:2017mnm,Boughezal:2018azh,Page:2019gbf}.  One advantage of jets is that measurements of their transverse momentum and pseudorapidity give kinematic handles that allow the effects of different distributions to be disentangled.  This is demonstrated in Fig.~\ref{fig:kin}, where inclusive jet production is split into its partonic constituents. Both Weizacker-Williams and resolved photonic processes are included. At high lab-frame pseudorapidities the gluon-photon scattering process dominates, while resolved-photon processes become important at intermediate and backward pseudorapidities. More details on this result are given in Ref.~\cite{Abelof:2016pby}. Studies of the double longitudinal spin asymmetry in jet production at an EIC indicate that polarized PDF errors are much larger than anticipated experimental errors, as shown in Fig.~\ref{fig:AllpT}. It is expected that information from jet production at an EIC will complement inclusive DIS measurements, similarly to the case at HERA where it was useful in reducing uncertainties in gluon PDFs~\cite{ZEUS:2021sqd}. Moreover, dijet production at the EIC will offer a unique opportunity to determine the polarized PDFs of the photon~\cite{Chu:2017mnm}, and diffractive PDFs~\cite{Guzey:2020gkk}.

\begin{figure}[h!]
\centering
\includegraphics[width=0.49\textwidth]{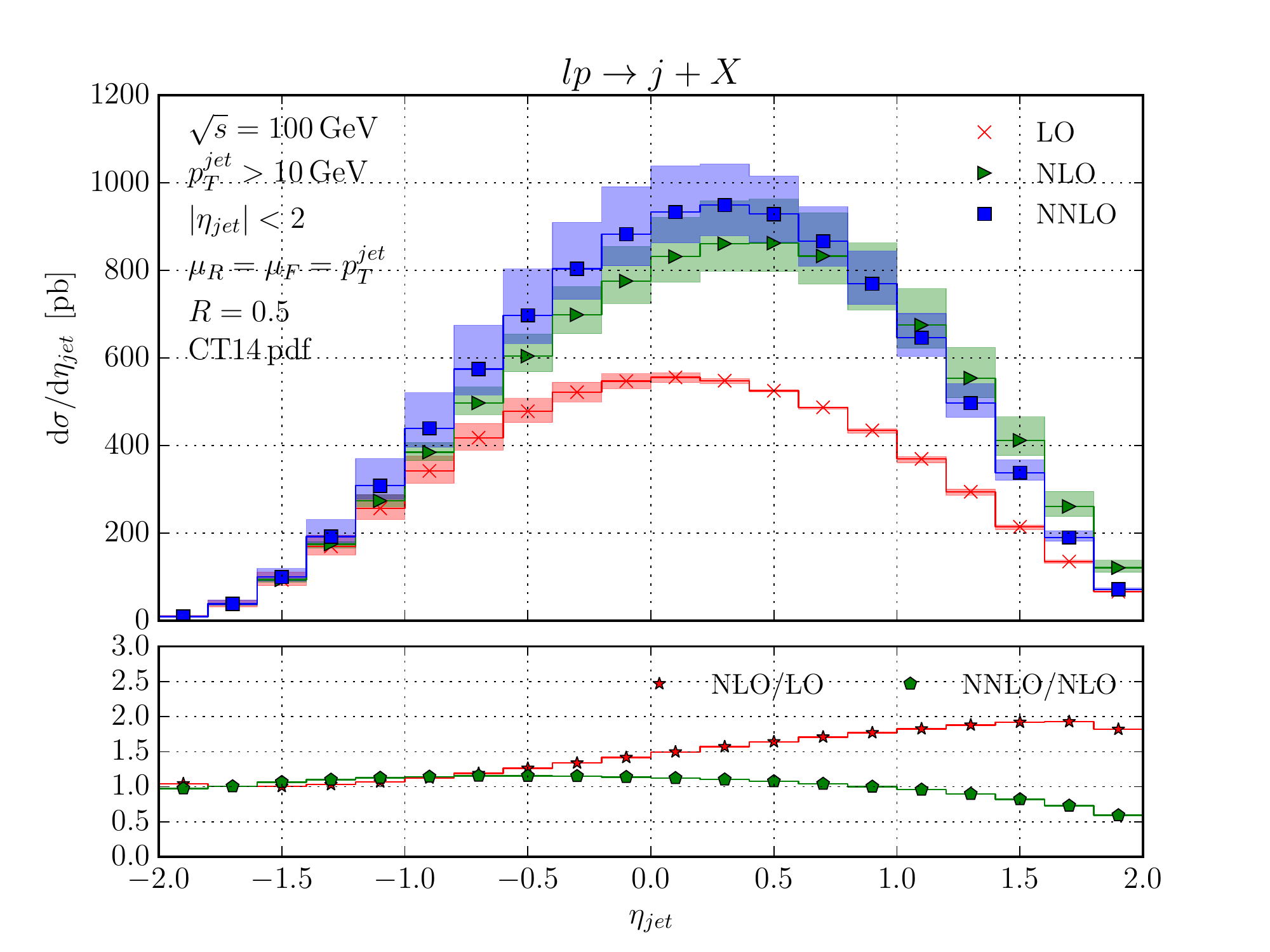}
\caption{Inclusive-jet pseudorapidity distribution at LO, NLO and NNLO in QCD perturbation theory.  The upper panel shows the distributions with scale uncertainties, while the lower panel shows the $K$-factors for the central scale choice. Figure from~\cite{Abelof:2016pby}. \label{fig:NNLO}}
\end{figure}

The eventual goal of this program is to use jet production along with DIS to determine the helicity-dependent PDFs at the next-to-next-to-leading order (NNLO) in perturbative QCD. This would match the precision obtained for unpolarized PDFs; their determination to this level of accuracy has had a profound impact on the physics program of the LHC, and a similarly precise determination of polarized PDFs is expected to greatly influence the EIC physics program. One step toward achieving this goal is the calculation of the NNLO corrections to the required hard-scattering cross sections. The full calculation of the ${\cal O}(\alpha^2 \alpha_s^2)$ perturbative corrections to jet production in electron-nucleus collisions, including all photon-initiated processes, is available~\cite{Abelof:2016pby}. The behavior of the perturbative corrections in perturbation theory is shown in Fig.~\ref{fig:NNLO}. While the NLO corrections are large, the perturbative expansion shows good convergence once NNLO is included. We note that the total NNLO correction comes from an intricate interplay between all contributing channels, with different ones dominating in different $\eta_{jet}$ regions.  Only the gluon-lepton partonic process is negligible over all of phase space.  For negative $\eta_{jet}$, the dominant contribution is given by the quark-quark process.  This appears first at ${\cal O}(\alpha^2\alpha_s^2)$.  It is therefore effectively treated at leading-order in our calculation, and consequently has a large scale dependence.  At high $\eta_{jet}$, the distribution receives sizable contributions from the gluon-photon process.  No single partonic channel furnishes a good approximation to the shape of the full NNLO correction. Recently, the NNLO corrections to the spin asymmetry in polarized collisions have become available~\cite{Borsa:2020ulb,Borsa:2020yxh,Borsa:2021afb}, albeit without photon-initiated processes.

To conclude several salient features of the impact of jet production at an EIC on polarized PDF determinations are summarized below.
\begin{itemize}

\item Collisions at the highest center-of-mass energies offer the broadest sensitivity to polarized hadronic structure.  Both the resolved photon distributions and the polarized gluons and quarks can be probed by selecting appropriate regions of jet transverse momentum and pseudorapidity.  Low transverse-momentum inclusive jet production and dijet production provide access to the polarized photon PDFs, while intermediate-to-high transverse momenta are sensitive to the polarized gluon.

\item The estimated polarized PDF errors are much larger than the expected EIC statistical errors.  The theoretical scale uncertainties are small once NNLO corrections are included. 

\end{itemize}

\subsubsection{3D imaging}

\begin{figure}[t]
  \centerline{\includegraphics[width = .48\textwidth]{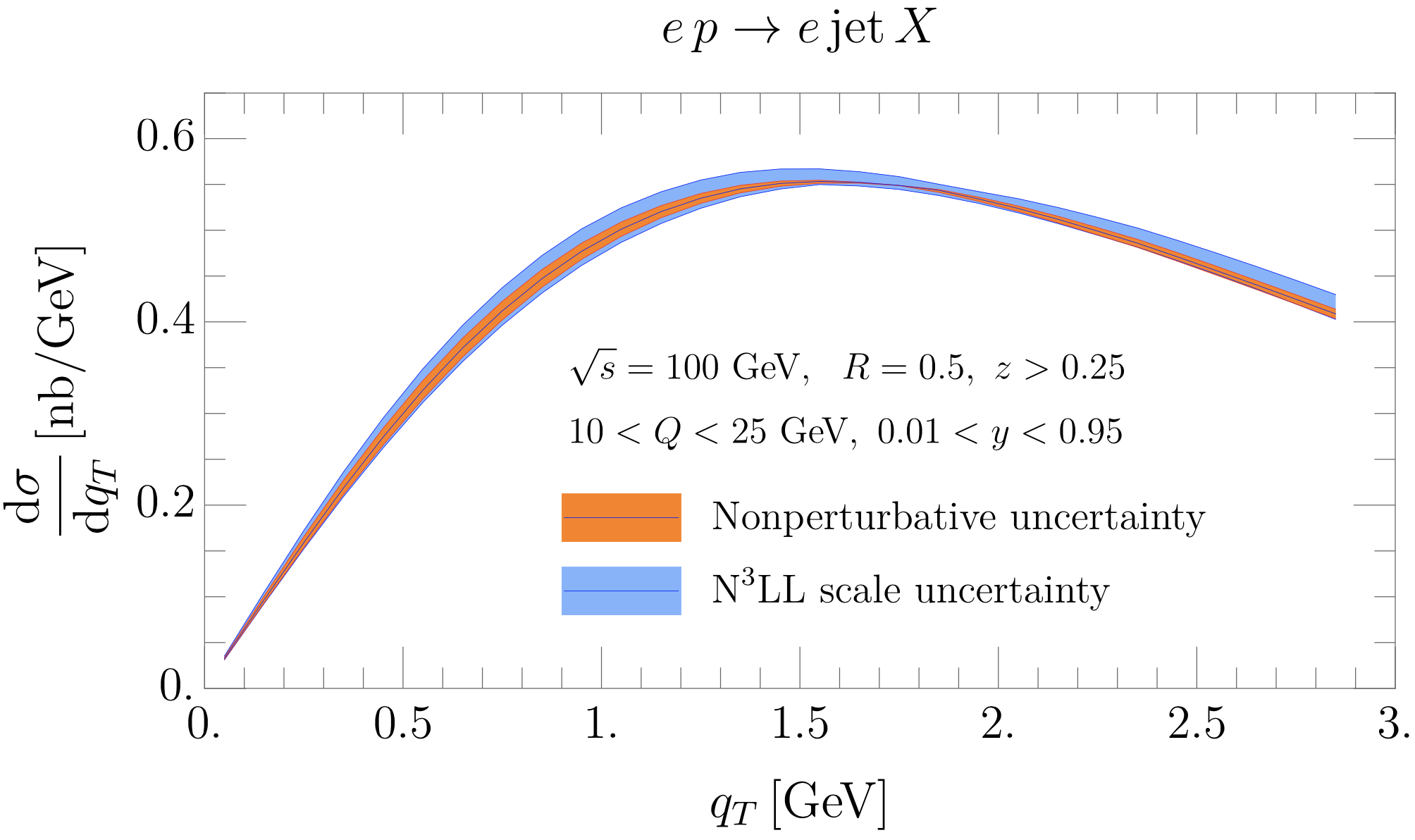}
  \hspace*{1cm}
  \includegraphics[width = .4\textwidth]{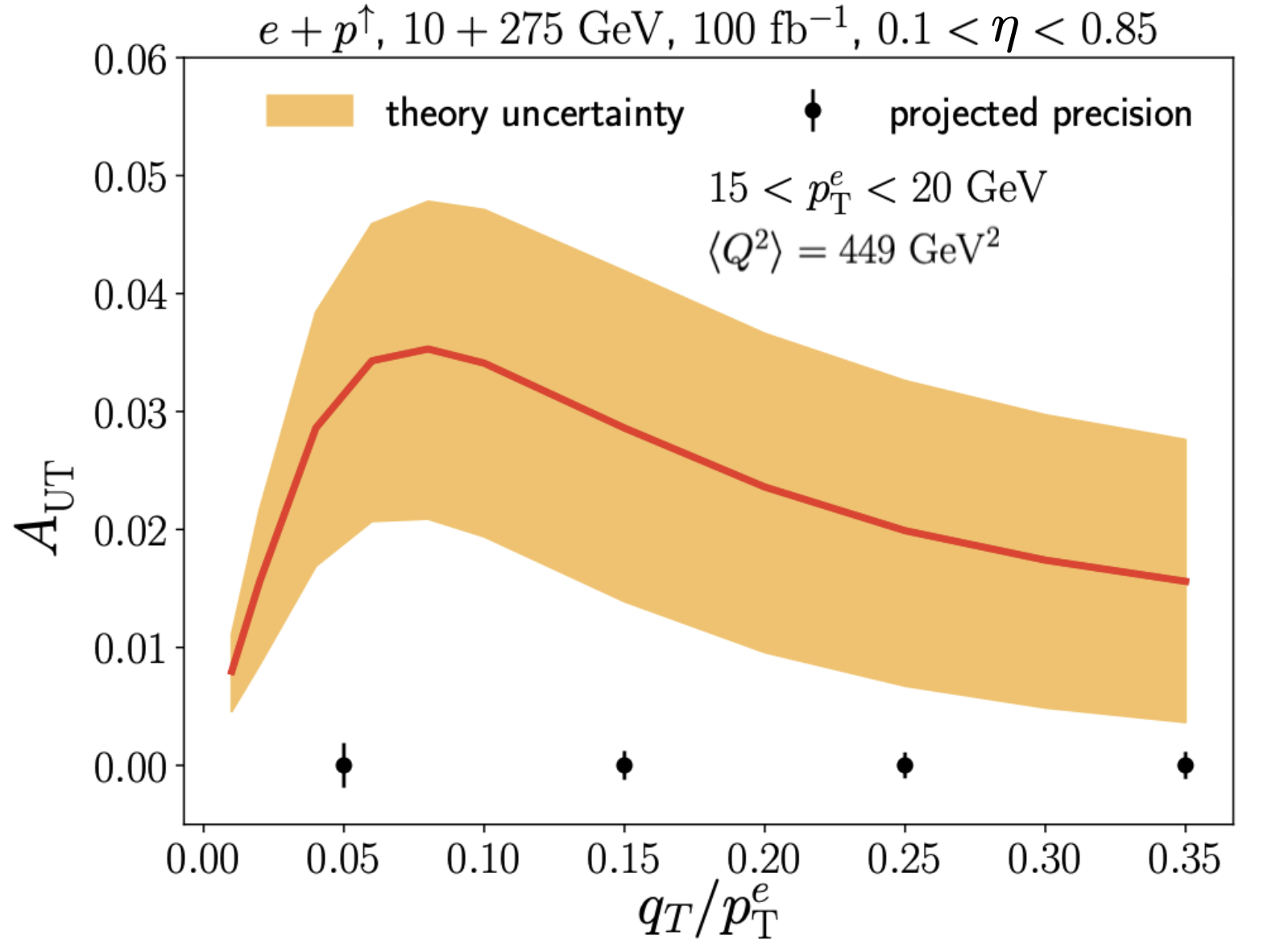}}
  \caption{Left: The unpolarized transverse momentum distribution of jets in the Breit frame for EIC kinematics~\cite{Gutierrez-Reyes:2018qez,Gutierrez-Reyes:2019vbx}. Right: The transversely polarized spin asymmetry $A_{\rm UT}$ for jets recoiling against the scattered lepton and reconstructed in the lab frame. The current theory uncertainty is shown as well as projections for the EIC~\cite{Liu:2018trl,Liu:2020dct,Arratia:2020nxw}.~\label{fig:TMDs}}
\end{figure}

Jets are expected to play an important role at the future EIC which are relevant for all aspects of the EIC science program. Basic aspects and kinematics of jets have been explored in Refs.~\cite{Page:2019gbf,Arratia:2019vju}. Here we will focus specifically on Transverse Momentum Dependent (TMD) studies with jets and jet substructure observables.

The standard benchmark process of TMDs at the EIC is Semi-Inclusive Deep Inelastic Scattering (SIDIS) where a hadron is measured in the final state and which depends both on TMD PDFs and TMD FFs. One of the key aspects of TMD studies with jets is that we can construct observables that are sensitive only to TMD PDFs or FFs alone. 

Jets can be reconstructed in different frames and with different jet algorithms, which is important for the physics we are trying to probe. For example, jets can be identified in the laboratory frame similar to $pp$ collisions~\cite{Cacciari:2008gp}. However, for several TMD studies it is necessary to reconstruct jets in the Breit frame. In Ref.~\cite{Arratia:2020ssx} a spherically invariant algorithm and the Centauro distance metric were proposed for the Breit frame. This allows for the measurement of the jet energy relative to the hard scale of the process $Q^2=-q^2$. In this case, jets are not required to have a large transverse momentum in the Breit frame. This allows for the measurement of quark TMD PDFs analogous to SIDIS with hadrons. In the corresponding factorization theorem, the final TMD FF in SIDIS is replaced by a TMD jet function, which can be calculated purely perturbatively~\cite{Gutierrez-Reyes:2018qez,Gutierrez-Reyes:2019vbx}. Thus, this process is only sensitive to TMD PDFs, which can provide important constraints in global analyses. Numerical predictions for this process are shown in the left panel of Fig.~\ref{fig:TMDs}. Recent developments for Breit frame measurements are also presented in Ref.~\cite{Liu:2021ewb}. 

An alternative observable, which is sensitive to quark TMD PDFs is the lepton-jet decorrelation in the laboratory frame. Also in this case, only the TMD PDF appears in the factorization and the final state jet with large transverse momentum can be calculated perturbatively~\cite{Liu:2018trl,Liu:2020dct,Arratia:2020nxw}. In the transversely polarized case, we are sensitive to the TMD Sivers function. Numerical results for the asymmetry $A_{\rm UT}$ are shown in the right panel of Fig.~\ref{fig:TMDs} along with projections for the EIC. Furthermore, in Ref.~\cite{Kang:2020fka}, the jet charge~\cite{Krohn:2012fg} was proposed to introduce a flavor sensitivity to the jet observables discussed here. The flavor tagging of jets is particularly important for spin asymmetries where the contributions of different quark flavors often have opposite sign, which can lead to large cancellations. Analogous studies that provide access to gluon TMDs can be performed with dijets in the Breit frame. See Refs.~\cite{Zheng:2018ssm,delCastillo:2020omr,Kang:2020xgk,delCastillo:2021znl,Boer:2021upt} for recent theoretical calculations relevant for the EIC. In addition, TMD fragmentation functions can be measured in isolation using jet substructure observables. See Refs.~\cite{Bain:2016rrv,Neill:2016vbi,Kang:2017glf,Arratia:2020nxw} for more details.

Understanding of transverse momentum dependent functions will elucidate our understanding of the internal structure of hadron in terms of elementary quarks and gluons and provide us with a deep insight into the elusive mechanism of hadronization. The TMD functions appearing in relation to the initial and final state hadrons are referred to as TMD distributions and TMD fragmentation functions, respectively. Traditionally, a limited number of processes have been used to constrain such TMD functions\ \cite{Collins:2011zzd}. We briefly discuss three processes involving jets that are useful to studying TMD functions at the EIC, which is by no means an exhaustive list:
\begin{enumerate}
\item back-to-back production of $e\ +\ $jet: $e+p \to e\ +\ $jet$\ +\ X$ with $q_T \ll p_T$.
\item TMD hadron distribution inside an inclusive jet production process: $e+p \to\ $jet (h)$\ +\ X$ with $j_\perp \ll p_T R$.
\item TMD hadron distribution inside an inclusive jet produced back-to-back with $e$: $e+p \to e\ +\ $jet (h)$\ +\ X$ with $q_T \ll p_T$, $j_\perp \ll p_T R$.\end{enumerate}

Here, $q_T$ is the imbalance momentum between the transverse momenta of the final lepton and jet, which needs to be much smaller than their individual transverse momenta $\approx p_T$ to be produced in the back-to-back configuration. On the other hand, $j_\perp$ is the transverse momenta of the final hadron inside the jet with respect to the jet axis\footnote{we use $T$ and $\perp$ to distinguish quantities measured with respect to the beam and jet axes, respectively.}, which needs to be much smaller than the jet scale $p_T R$ to be sensitive to the standard TMD fragmentation functions. The first process only involves a single TMD distribution, as the process only involves a single incoming hadron\ \cite{Arratia:2020nxw}. This is to be contrasted with the standard SIDIS, which involves an additional TMD fragmentation function dependence coming from the measurement of a hadron in the final state\ \cite{Bacchetta:2006tn}. The second process is inclusive on the events outside the measured jet, which suppresses the dependence on the transverse momentum structure from the incoming hadron. On the other hand, one can still probe the transverse momentum distribution of the final state hadron inside the jet with respect to the jet axis, which becomes sensitive to the various standard TMD fragmentation functions for $j_\perp \ll p_T R$\ \cite{Kang:2017glf,Kang:2019ahe,Kang:2020xyq,Kang:2021ffh}. 

\begin{figure}[hbt]
\centering
\includegraphics[width = 0.6\textwidth]{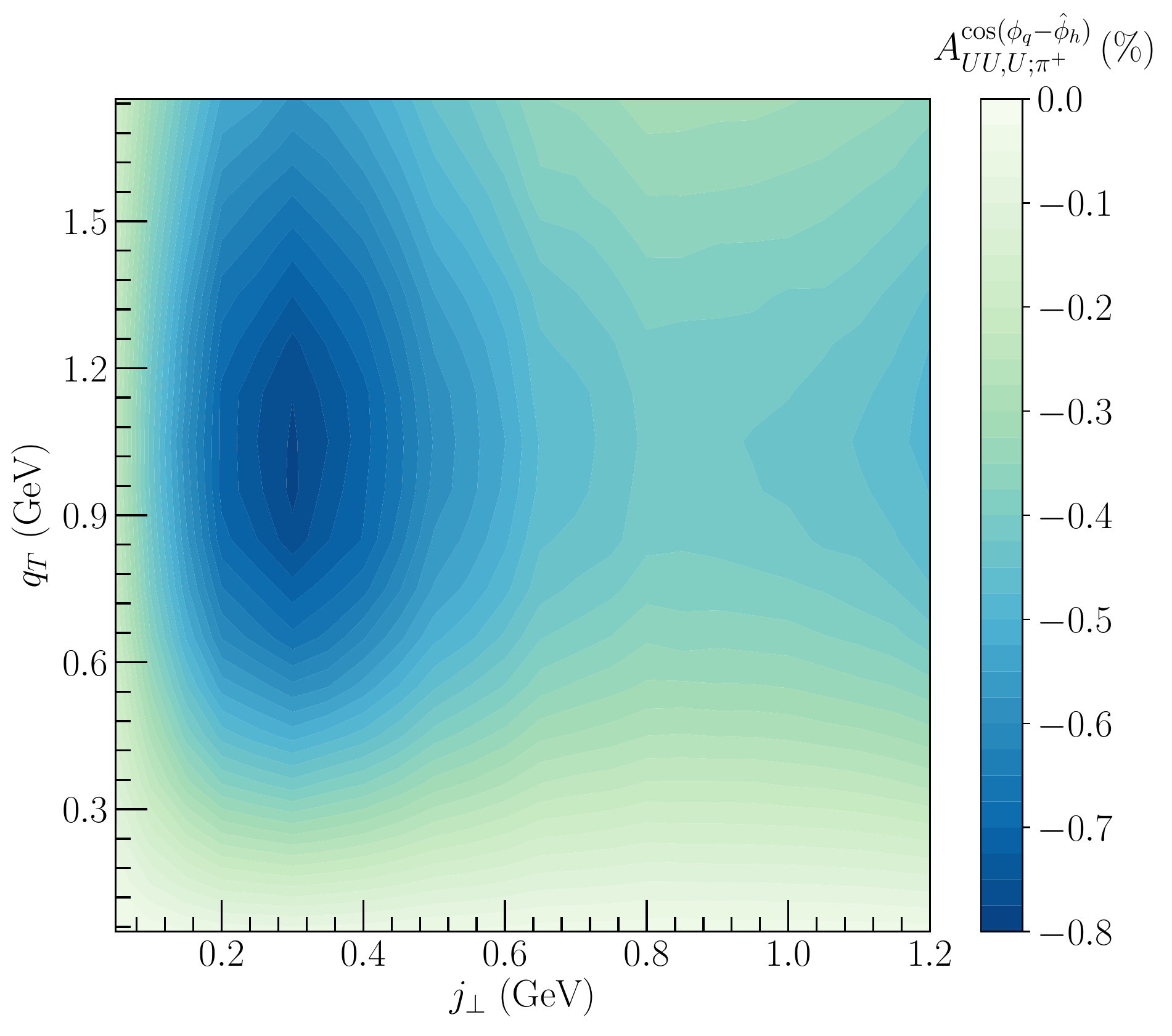}
\caption{The contour plot of azimuthal asymmetry of the third process sensitive to the Boer-Mulders TMD distribution and the Collins TMD fragmentation function at the EIC kinematics. The usage of jets decouple the two TMD functions functional dependence, allowing such contour plot.}
\label{fig:third}
\end{figure}

The third and last process can be seen as a combination of the first two processes, where the lepton and jet are produced in back-to-back configuration and the hadron distributions inside the final jet are measured. This process gives sensitivity to both TMD distributions and fragmentations simultaneously. However, usage of jets decouples and isolate the functional dependence of the TMD distribution and fragmentation functions to only $q_T$ and $j_\perp$, respectively. This separation is useful to separately constrain the TMD distribution and fragmentation function. For different asymmetries of the azimuthal angles involved in the process, one can study different combinations of the TMD distribution and fragmentation functions. In Fig.\ \ref{fig:third}, we see an example~\cite{Kang:2021ffh} of a contour plot produced from the asymmetry related to the Boer-Mulders TMD distribution and the Collins TMD fragmentation function, each separately sensitive to $q_T$ and $j_\perp$ spectrum, respectively. Another example that has been in active discussion is to utilize hadron-in-jet for probing the so-called quark polarizing fragmentation functions (PFFs) for $\Lambda$ hyperons, $D_{1T\,\Lambda/q}^{\perp}$, giving the probability that an unpolarized quark fragments into a transversely polarized spin-1/2 hadron. Recently the Belle collaboration measured the transverse polarization of $\Lambda$ in the production of back-to-back $\Lambda$+h in $e^+e^-$ collisions~\cite{Guan:2018ckx}. Two groups have performed phenomenological analysis of the data for the extraction of the PFFs~\cite{DAlesio:2020wjq,Callos:2020qtu}. The authors of Ref.~\cite{Kang:2021kpt} provided projections for the statistical uncertainties in the corresponding spin observables at the future EIC. In particular, they studied the transverse polarization of $\Lambda$ hyperons inside the jet produced in the scattering of an electron with either an unpolarized or transversely polarized proton beam. 

Jets might also be used to probe the Wigner distribution, which encodes the quantum phase density of the proton in terms of quarks and gluons. In particular, it was argued in Ref.~\cite{Hatta:2016dxp} that using dijets in diffractive scattering (see Figure~\ref{fig:eventdisplays}) is sensitive to the Wigner distribution and parton orbital-angular momentum~\cite{Boussarie:2021ybe}. The proposed observable is the correlation between the dijet system and the scattered proton. The measurement of the proton is needed to limit the influence of soft-gluon radiation in the observable, which can be sizable~\cite{Hatta:2020bgy,Hatta:2021jcd,Hatta:2019ixj}.

\subsection{Jet substructure and precision QCD}
\subsubsection{Precision QCD}
\begin{figure}[t]
  \centerline{\includegraphics[width = .5\textwidth]{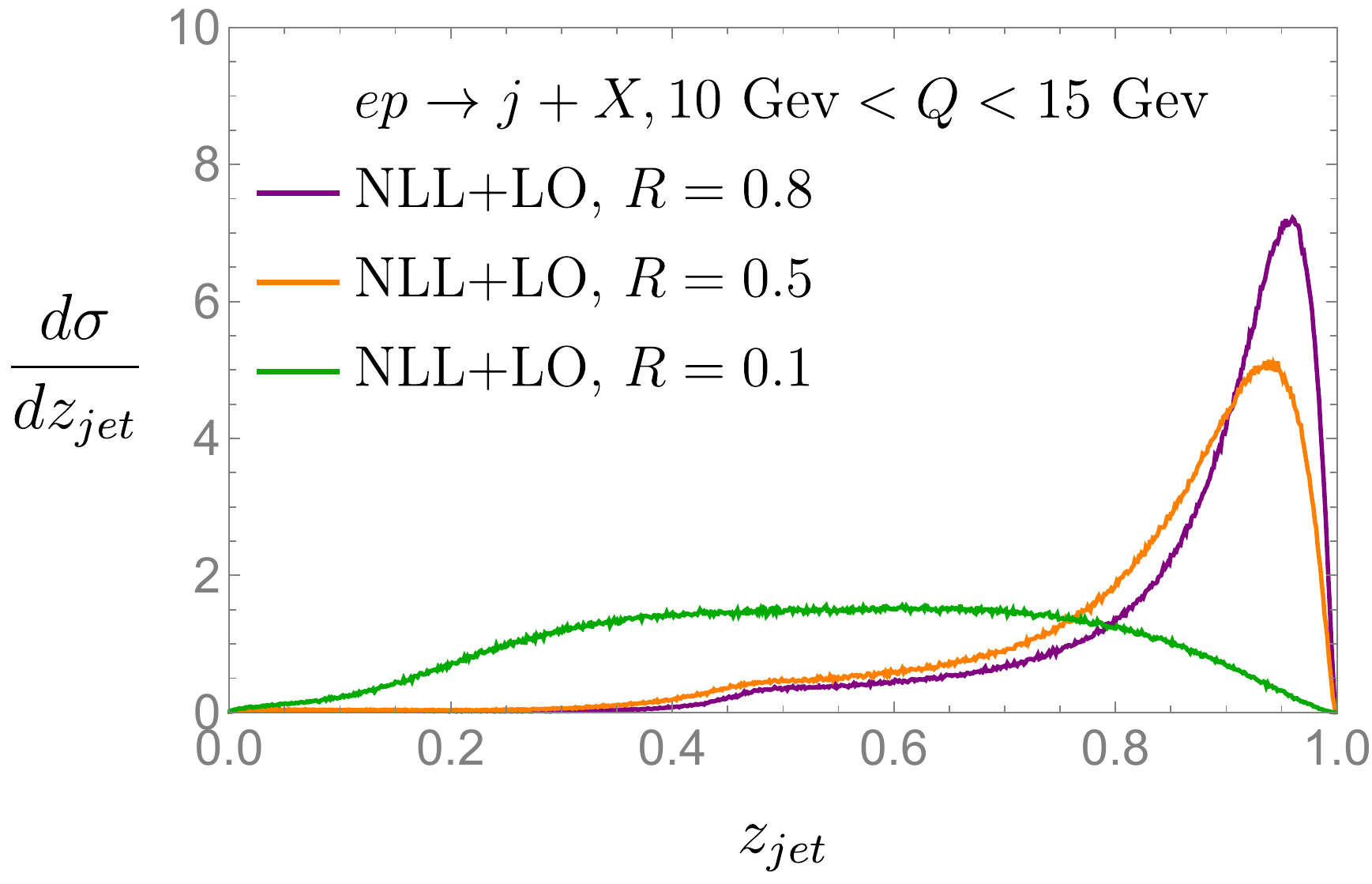}
  \includegraphics[width = .5\textwidth]{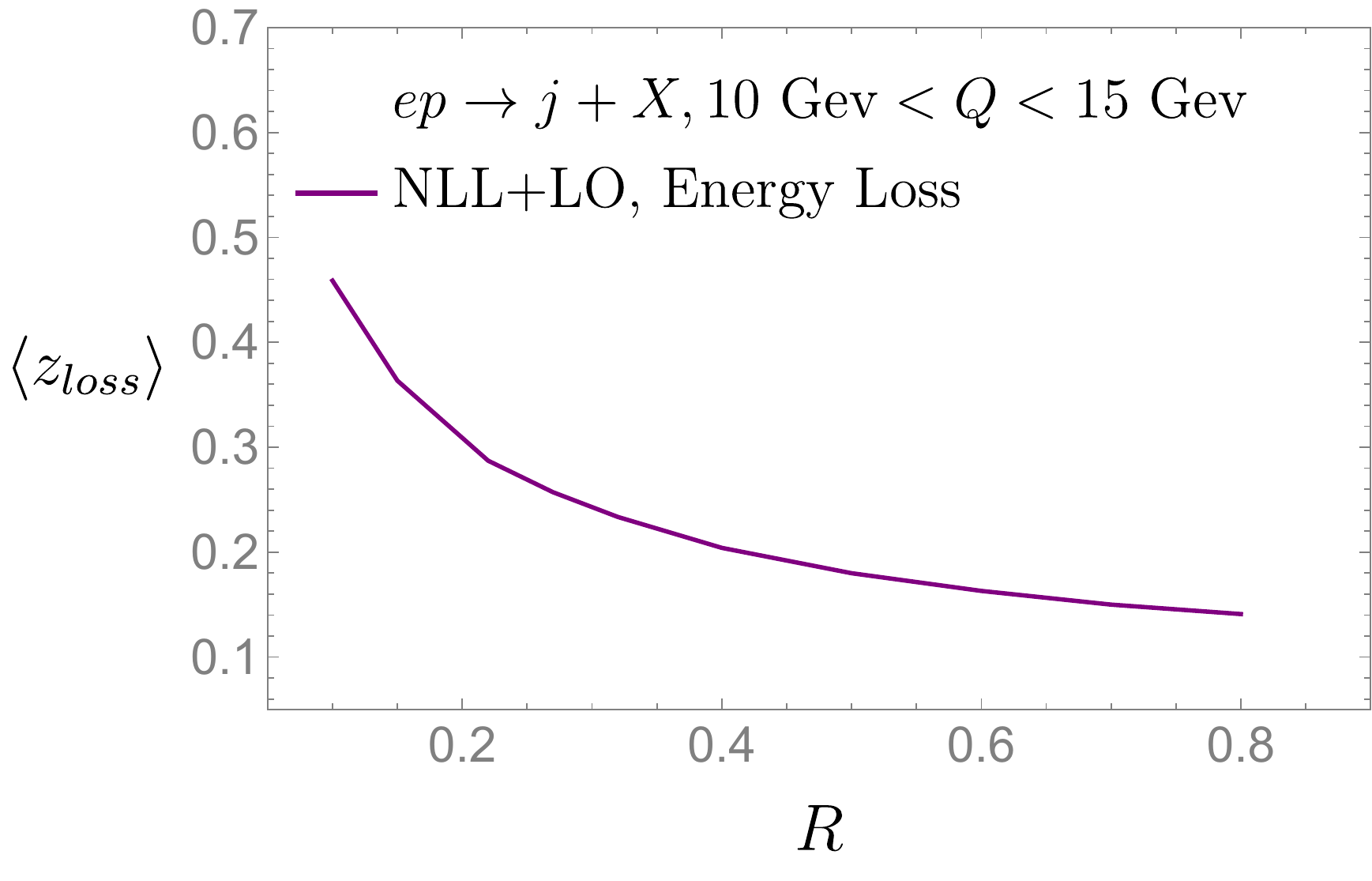}}
  \caption{Left: The energy spectrum $z_{\rm jet}$ of leading jets at NLL$'$ for EIC kinematics for different values of the jet radius $R$. Right: The average energy loss of leading jets $\langle z_{\rm loss}\rangle=1-\langle z_{\rm jet}\rangle$. See Ref.~\cite{Neill:2021std}.~\label{fig:eloss_qg}}
\end{figure}

At the future EIC, jets will be important tools since they are excellent proxies of parton level dynamics. A central question is how quarks and gluons fragment into the observed jets which can be quantified by studying the jet energy fraction $z_{\rm jet}$ relative to the initial hard reference scale $Q^2=-q^2$ of the virtual photon. The asymmetric inital state of $ep$ scattering at the EIC is ideal to study this quantity when suitable jet algorithms are used in the Breit frame~\cite{Arratia:2020ssx}. Instead of inclusive jets, which are frequently studied in the literature~\cite{Dasgupta:2014yra,Kaufmann:2015hma,Kang:2016mcy,Dai:2016hzf}, leading jets require non-trivial factorization theorems and involve non-linear evolution equations~\cite{Neill:2021std,Dasgupta:2014yra,Scott:2019wlk}. However, they have the advantage that we are able to measure observables not only on a sample of (inclusive) jets but on a well-defined single object -- the leading jet. The leading jet spectrum $z_{\rm jet}$ is shown in the left panel of Fig.~\ref{fig:eloss_qg}. The results are obtained using a parton shower approach at next-to-leading (NLL$'$) accuracy where threshold resummation is included~\cite{Neill:2021std}. For large values of the jet radius $R$, the spectrum peaks close to $z_{\rm jet}=1$, which means that a large fraction of the initial parton momentum is contained in the leading jet. This peak structure, different than for hadron cross sections, is an illustration that jets are indeed excellent proxies of parton level dynamics, as advertised above. For lower values of $R$, the distribution becomes broader, and it eventually turns into the leading hadron spectrum. Given the angular-ordered evolution of jets, which can be captured as snap-shots in ``time'' by the jet radius $R$, we can dynamically observe the fragmentation process and probe the transition to non-perturbative processes.

Leading jets allow for a well-defined notion of ``energy loss'', which we define as all the energy which is not contained in the leading jet $z_{\rm loss}=1-z_{\rm jet}$. Distinct from inclusive jets, the leading jet cross section is normalized to unity and can be considered as a probability density. Therefore, we can calculate statistical quantities such as the average energy loss of leading jets $\langle z_{\rm loss}\rangle$. At leading-logarithmic order, the average energy loss of leading jets agrees with parton energy loss. Numerical results, are shown in the right panel of Fig.~\ref{fig:eloss_qg} as a function of the jet radius $R$. As $R$ becomes smaller, less energy is captured in leading jet and, hence, the average energy loss becomes larger. At the EIC it will be possible to perform corresponding measurements both in $ep$ and $eA$ collisions to study cold nuclear matter effects. The proposed measurements can also build an important bridge to observables in $pp$ and especially heavy-ion collisions, where the energy loss mechanism plays an important role.

\subsubsection{Jet substructure}
\begin{figure}[t]
  \centerline{\includegraphics[width = .47\textwidth]{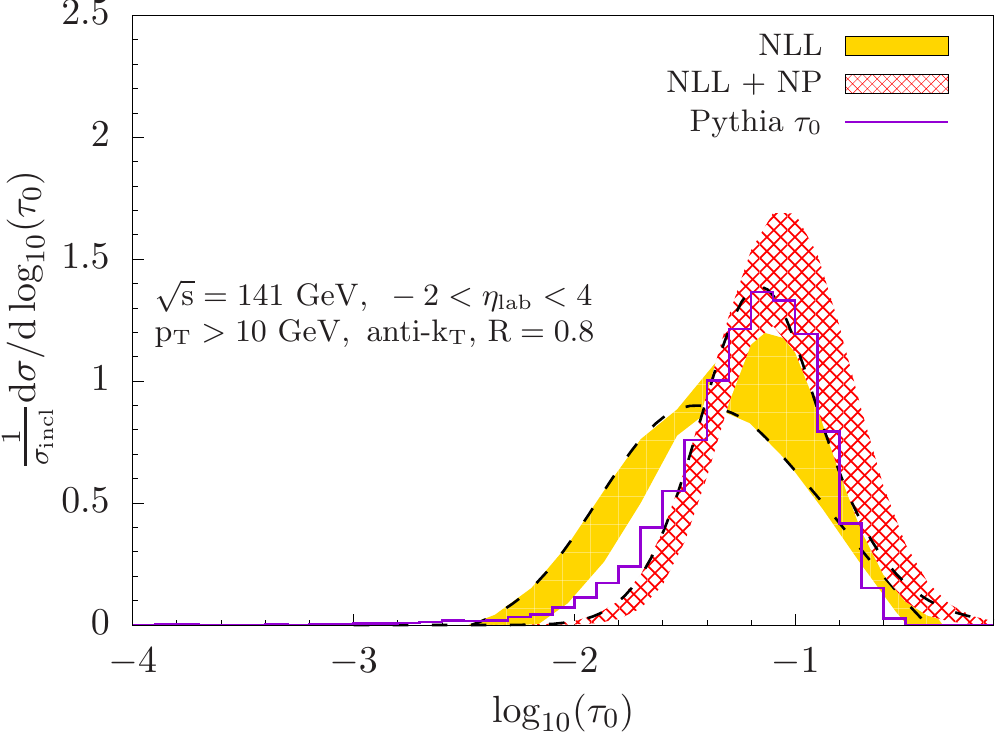}
  \hspace*{1cm}
  \includegraphics[width = .4\textwidth]{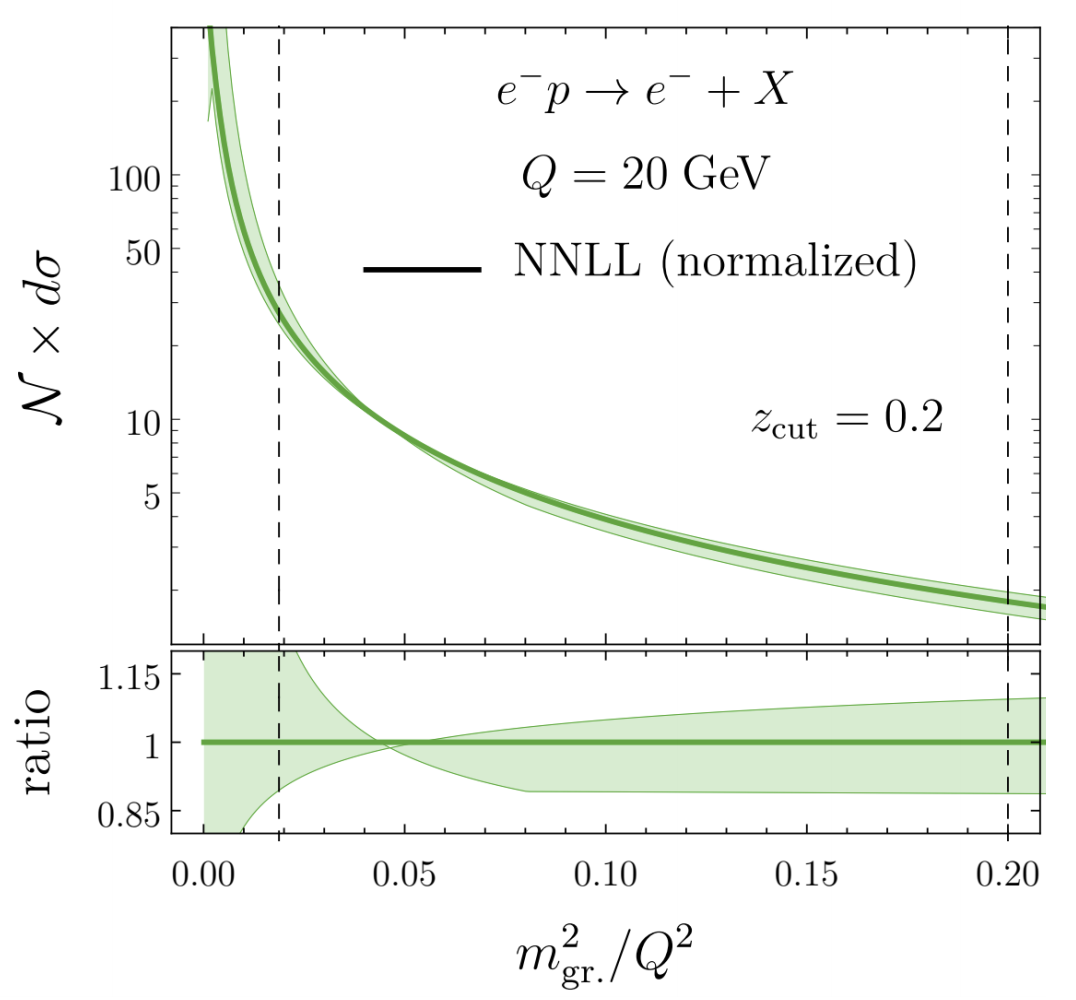}}
  \caption{Left: Jet angularities in photoproduction at the EIC~\cite{Aschenauer:2019uex}. Right: The event-wide groomed jet mass~\cite{Makris:2021drz}.~\label{fig:JSS}}
\end{figure}

Jet substructure observables have received a growing attention at the LHC and RHIC in recent years. See Ref.~\cite{Larkoski:2017jix,Asquith:2018igt} for recent reviews. Related observables have recently been proposed at the EIC to probe interesting aspects of perturbative QCD, hadronization, cold nuclear matter effects and gluon saturation~\cite{Li:2020rqj,Zhang:2019toi,Barata:2020rdn,Chien:2021yol,Li:2021gjw,vanHameren:2021sqc,Iancu:2021rup,Caucal:2021ent,Boussarie:2021ybe,Zhang:2021tcc} in $eA$ collisions. A frequently studied class of observables are jet angularities which characterizes the radiation pattern inside jets~\cite{Berger:2003iw,Ellis:2010rwa}. A continuous parameter is used to interpolate between different traditional jet shape observables. In the left panel of Fig.~\ref{fig:JSS}, we show EIC predictions for an example of the jet angularities which coincides with the jet mass~\cite{Aschenauer:2019uex}. The development of jet grooming techniques~\cite{Dasgupta:2013ihk,Larkoski:2014wba} has lead to a whole range of new jet substructure observables. Initially, jet grooming was introduced as a tool to systematically remove soft wide-angle radiation which is difficult to calculate from first principles in QCD. In addition, jet grooming lead to the development of a whole range of new observables which can only be defined on a groomed jet and which probe interesting aspects of perturbative and nonperturbative physics. Examples include the groomed jet mass, the groomed jet radius~\cite{Larkoski:2014wba,Kang:2019prh} and the groomed momentum sharing fraction which is a direct probe of the QCD splitting function~\cite{Larkoski:2015lea}. At the EIC these observables can be studied in a clean environment which was explored in Refs.~\cite{Arratia:2019vju,Makris:2021drz}. Moreover, in Ref.~\cite{Makris:2021drz} it was proposed to groom not only reconstructed jets but the entire event. In the right panel of Fig.~\ref{fig:JSS}, the theoretical prediction for the event-wide groomed mass is shown.

\subsection{Jets at $e+A$ collision}

\subsubsection{Nuclear PDFs}
Dijet photoproduction in lepton-nucleus ($eA$) scattering at the EIC provides information on the QCD structure of nuclei, in particular, nuclear PDFs, which is complimentary to that in $eA$ DIS. This has been exploited in ultraperipheral collisions (UPCs) of heavy ions at the LHC, see, e.g.~\cite{Guzey:2019kik}.
Using the framework of collinear factorization and next-to-leading order (NLO) perturbative
QCD~\cite{Klasen:2002xb}, we calculated the
cross section of dijet photoproduction in $eA\to e+{\rm 2 jets}+X$ lepton-nucleus
scattering~\cite{Guzey:2020zza} 
\begin{equation}
  d\sigma(eA \to e+{\rm 2 jets}+X)=\sum_{a,b} \int dy \int dx_{\gamma} \int dx_A
  f_{\gamma/e}(y)f_{a/\gamma}(x_{\gamma},\mu^2)f_{b/B}(x_A,\mu^2) d\hat{\sigma}
  (ab \to {\rm jets})\,,
 \label{eq:cs}
\end{equation}
where $a,b$ are parton flavors; $f_{\gamma/e}(y)$ is the flux of equivalent photons of the
electron; $f_{a/\gamma}(x_{\gamma},\mu^2)$ are photon PDFs for the resolved photon contribution~\cite{Gluck:1991jc};
$f_{b/B}(x_A,\mu^2)$ are nuclear PDFs~\cite{Kovarik:2015cma}; $y$, $x_{\gamma}$, and $x_A$ denote the momentum fractions of the photon, partons in the photon, and partons in the nucleus, respectively;
$d\hat{\sigma}(ab \to {\rm jets})$ is the elementary cross section for 
production of jets in hard scattering
of partons $a$ and $b$. 

\begin{figure}[hbt]
\centering
\includegraphics[width = 0.8\textwidth]{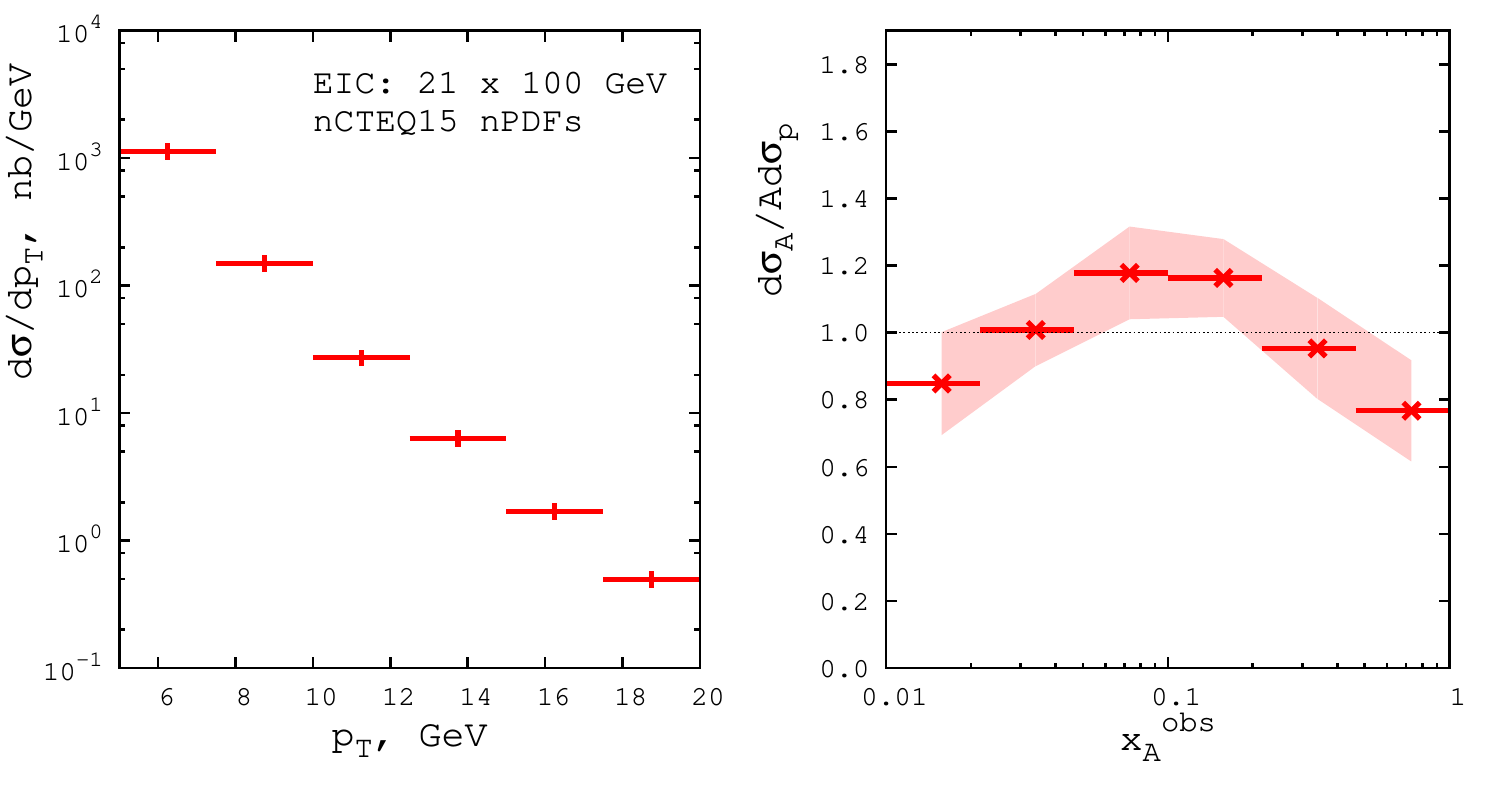}
\caption{NLO QCD predictions for the dijet photoproduction cross section in $eA$ scattering at the EIC.
Left: the dependence on ${\bar p}_T$. Right: the ratio of the dijet cross sections on a nucleus and the proton as a function of 
 $x_A^{\rm obs}$.
}
\label{fig:gk_final}
\end{figure}

An example of our results is presented in Fig.~\ref{fig:gk_final}. The left panel shows the dependence on the
dijet average transverse momentum ${\bar p}_T=(p_{T,1}+p_{T,2})/2$, which extends up to 20 GeV. The kinematic coverage in other
variables spans $-2 < \bar{\eta} \leq 3$,
$0.03 \leq x_{\gamma}^{\rm obs} \leq 1$, and $0.01 \leq x_A^{\rm obs} \leq 1$.
The right panel shows the ratio of the dijet cross sections on a nucleus and the proton, which exhibits the $x_A^{\rm obs}$ 
dependence similar to that of the ratio of the gluon densities $g_A(x,\mu^2)/[Ag_p(x,\mu^2)]$ with $10-20$\% nuclear modifications.

\subsubsection{Nuclear modification of jets and jet substructure}
Because of the much cleaner environment, the uncertainties from background subtraction will be small for jet measurements at the EIC.  For nuclei effects in e+A collisions in relative to the e+p case, both the initial-state effects and final-state jet-medium interactions may be important. It would be crucial to maximize and isolate the effects arising from final-state jet-medium interactions which will benefit both studying parton showers in cold nuclear matter and global nPDF fits. 

The work~\cite{Li:2020rqj} carried out the calculation of inclusive jet cross section and the jet charge in electron-nucleus collisions at the EIC. The inclusive jet cross section was investigated with the  semi-inclusive jet functions,  which can be written as 
 \begin{align}
  E_{J} \frac{d^{3} \sigma^{l N \rightarrow j X}}{d^{3} P_{J}} &=\frac{1}{S} \sum_{i, f} \int_{0}^{1} \frac{d x}{x} \int_{0}^{1} \frac{d z}{z^{2}} f_{i / N}(x, \mu) \hat{\sigma}^{i \rightarrow f}(s, t, u, \mu)J_{f }(z, p_T R, \mu) \;, 
\end{align}
where the hard kernel   $\hat{\sigma}^{i\to f}$  are taken from Ref.~\cite{Hinderer:2015hra} up to NLO. $ f_{i / N}$ is the parton distribution function of parton $i$ from nucleon $N$ where the initial state effects are encoded in global fitted nuclear PDFs.  $J_{f}$ is the semi-inclusive jet functions  which were explored in Refs.~\cite{Kang:2016mcy,Dai:2018ywt}. The first demonstration about the nuclear matter effects was achieved by Ref.~\cite{Kang:2017frl,Li:2018xuv} use of the medium induced splitting kernels  derived in the framework of SCET$_{\rm G}$~\cite{Ovanesyan:2011xy,Ovanesyan:2011kn}.

The average jet charge is defined as 
\begin{align} \label{eq:charge}
    Q_{\kappa, {\rm jet}}  = \frac{1}{\left(p_T^{\rm jet}\right)^\kappa } \sum_{\rm i\in jet} Q_i \left(p_T^{i} \right)^{\kappa } \; ,  \quad \kappa > 0 \; .
\end{align}
with summing over the transverse momentum weighted charge $Q_i$. 
In comparison with inclusive jet cross sections, a smaller intrinsic scale for QCD radiation is expected to study jet substructures, where the phase space is restricted and non-perturbative corrections will be larger. The works~\cite{Krohn:2012fg,Li:2019dre} studied jet charge in proton-proton and heavy-ion collisions. 

\begin{figure}
    \centering
    \includegraphics[width=0.48\textwidth]{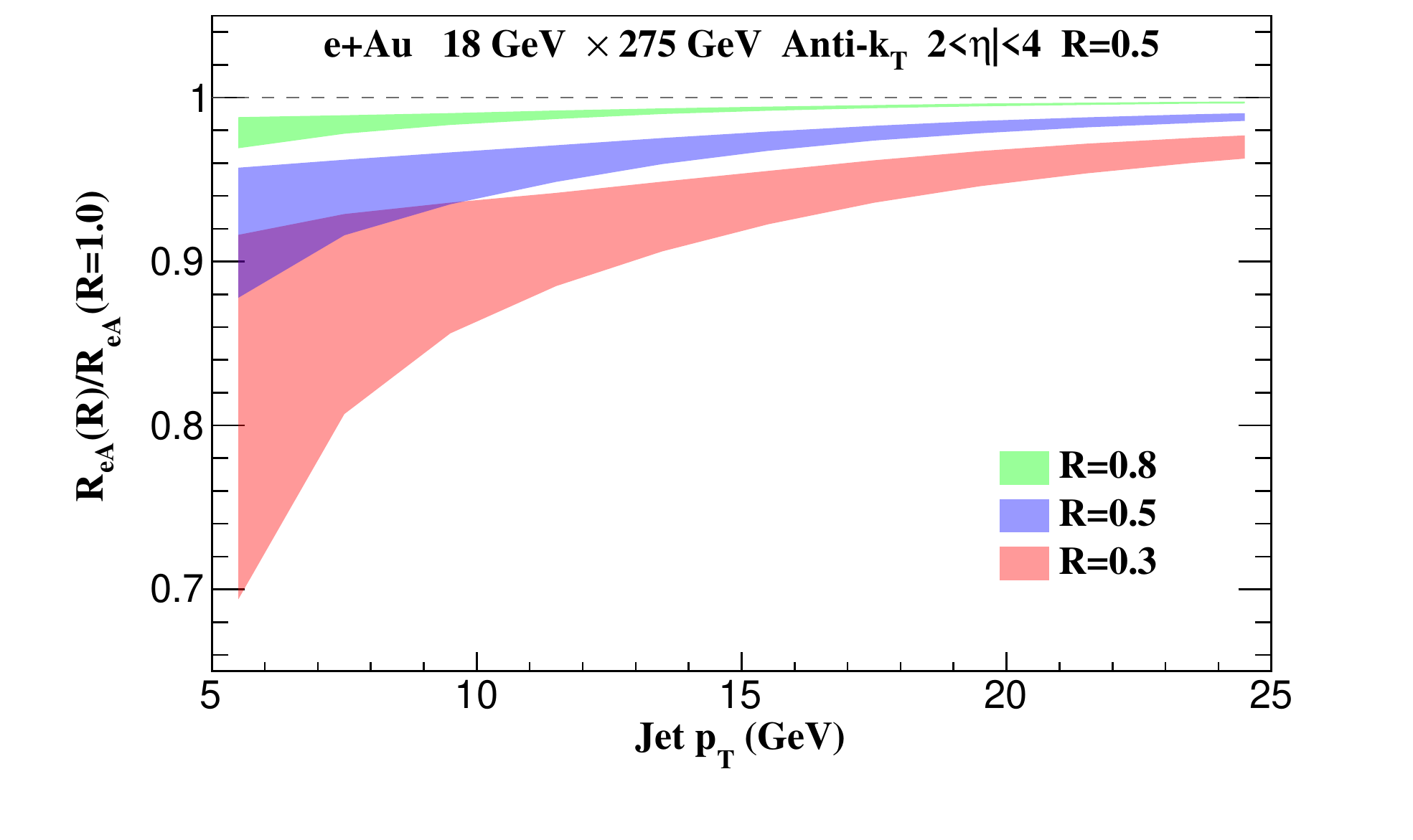}
    \includegraphics[width=0.48\textwidth]{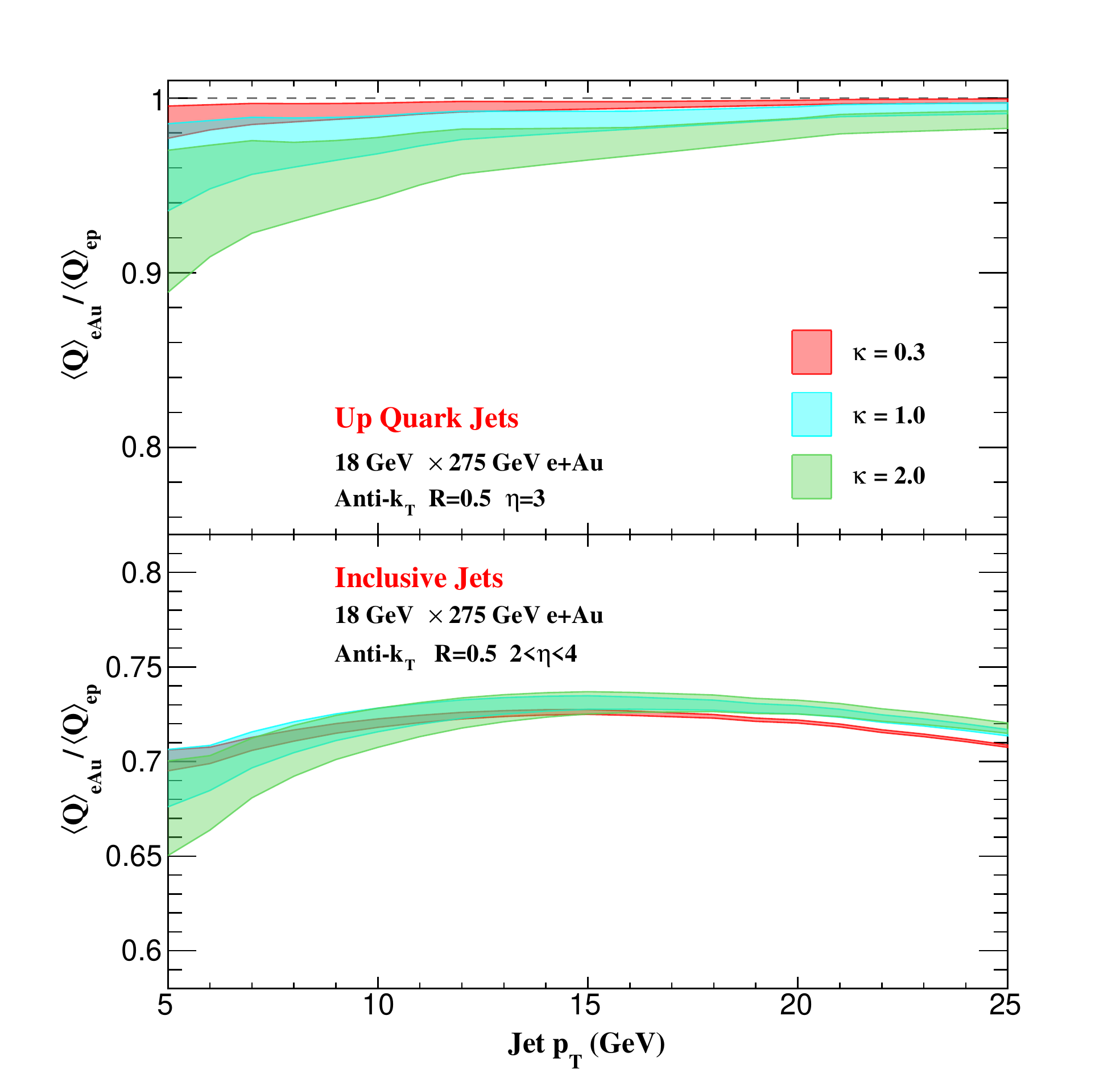}
    \caption{The left plot shows the ratio of jet  cross section modifications for different  radii. The right plot presents the modifications of the jet charge in e+Au collisions.
    }
    \label{fig:ReAto1}
\end{figure}

Figure.~\ref{fig:ReAto1} presents the prediction of the ratio of jet cross section suppression where 
\begin{align}
R_{\rm eA}(R) = \frac{1}{A}  \frac{d\sigma^{eA}/d\eta dp_T}{d\sigma^{ep}/d\eta dp_T}\,. 
\end{align}
The ratio of $R_{\rm eA}(R)$ eliminates initial-state effects. Smaller Jet radii will cause sligtly larger energy loss in cold nuclear matter where  the suppression from medium induced interactions is larger. Jet charge modifications at the EIC are presented in the right plot of Figure.~\ref{fig:ReAto1}. For final state effects (upper panel), the overall modifications decrease with increasing jet $p_T$ which are around 10\%  for small $p_T$. For the charge of inclusive jet in the lower pane, the average charge is sensitive to isospin effects and nuclear PDFs. Precision studies for jet charge will help to constrain isospin effects and  the PDF ratio of up and down quarks in the nucleus.

\subsection{Monte Carlo and Machine Learning}

General-purpose Monte-Carlo event generators are a cornerstone of high-energy physics research, as they allow to simulate events at particle colliders fully differentially in the many-body phase space~\cite{Buckley:2011ms}. These tools simulate the highest-energy scattering that may seed the emergence of jets, the renormalization group evolution of the scattering by means of soft/collinear radiation that leads to the structure of jets, and the fragmentation of low-energy partons that produces the hadrochemistry of the jets and the scattering event as a whole. 
Programs like Herwig~\cite{Bellm:2019zci}, Pythia~\cite{Sjostrand:2014zea} and Sherpa~\cite{Bothmann:2019yzt} have been developed over the past decades (after the end of the main HERA programme) alongside many experimental discoveries, and have been extended and refined to describe the phenomena observed at lepton-lepton, lepton-hadron and hadron-hadron colliders. 
Most existing general-purpose event generators are capable of computing unpolarized deep-inelastic scattering to next-to-leading order in QCD perturbation theory~\cite{Hoche:2010pf,DErrico:2011wfa}, and next-to-next-to-leading order precision has been achieved in some cases~\cite{Hoche:2018gti}.
A large family of simulation programs also exists for heavy-ion physics and air showers~\cite{Gyulassy:1994ew,Engel:1995yda,Ostapchenko:2004ss,Riehn:2015oba,Pierog:2013ria}. The EIC will provide us with the opportunity to test QCD in an entirely new regime~\cite{Accardi:2012qut}, presenting new challenges to the existing numerical calculations. This may also open new research directions at other high-energy colliders, e.g.\ the LHC.

It is expected that the double longitudinal spin asymmetry for jet production with large transverse momentum in DIS will offer excellent sensitivity to the spin-dependent parton distribution functions of the individual quarks and gluons in the proton. The simulation of spin dependent evolution in parton showers, and the proper interface to spin-dependent parton distribution functions~\cite{deFlorian:2014yva,Nocera:2014gqa} will therefore become highly relevant for EIC physics. Spin-dependent splitting functions have been included in some parton shower frameworks~\cite{Larkoski:2013yi,Richardson:2018pvo}, and a general algorithm to tackle the complexity of non-trivial correlations in the evolution exists~\cite{Collins:1987cp,Richardson:2001df}. The modeling of the fragmentation process in the Monte-Carlo simulation must account for the parton spin. The hadronization of partons is usually described either in the Lund string model~\cite{Artru:1974hr,Bowler:1981sb,Andersson:1983jt}, or in the cluster fragmentation model~\cite{Field:1982dg,Webber:1983if}. Practical implementations of these two approaches only include a  limited number of spin effects and will therefore need to be be extended in order to accurately reflect the parton-to-hadron transition in polarized scattering events. First steps towards this goal have been taken in~\cite{Kerbizi:2018qpp,Kerbizi:2021pzn}.

In order to enable precision measurements, it will also be required to match the parton shower to the corresponding higher-order calculations using techniques such as MC@NLO~\cite{Frixione:2002ik} and POWHEG~\cite{Nason:2004rx}. Notable progress on the fixed-order inputs to this matching has been made recently at NNLO~\cite{Abelof:2016pby,Boughezal:2018azh,Boughezal:2021wjw} and N$^3$LO~\cite{Currie:2018fgr,Gehrmann:2018odt}, and it is expected that more calculations of this type will become available in the future. Simultaneously, one can expect that more sophisticated matching procedures~\cite{Bertone:2022hig} will be applied to DIS simulations.

The above described modifications to existing event generators will enable improved simulations in lepton-hadron scattering at the EIC and provide a valuable starting point for the extension of the generators to electron-ion collisions. With an important goal of the EIC being to investigate saturation effects, it becomes mandatory to have reliable numerical predictions for fully differential final states. To provide these, parton showers must be capable of simulating the more general QCD evolution equations for transverse momentum dependent parton densities and fragmentation functions \cite{Collins:1981uw,Collins:1984kg,Collins:2011zzd}. This will require the development of new algorithms, some of which might be inspired by the methods for implementing the CCFM equations~\cite{Ciafaloni:1987ur,Catani:1989yc,Catani:1989sg,Marchesini:1994wr} used in Ref.~\cite{Marchesini:1990zy,Jung:2000hk}. Schematically, the structure of existing parton showers must be adapted to incorporate the additional loss term due to parton recombination, which is not present in the DGLAP equations, and which fundamentally changes the structure of the evolution in the saturation region. It will also be important to quantify the agreement with analytic resummation as obtained from the BK~\cite{Balitsky:1995ub,Balitsky:1998ya,Kovchegov:1999yj,Kovchegov:1999ua} and JIMWLK~\cite{Iancu:2000hn,Iancu:2001ad,JalilianMarian:1998cb,JalilianMarian:1997gr} equations.

Some extensions and modifications of existing event generators are needed that will require substantial effort from the community, but few new algorithmic developments. One such example is the precise simulation of QED radiation in the YFS framework~\cite{Yennie:1961ad}, including interference effects between the leptonic and hadronic radiators.
These topics make the EIC an excellent opportunity to understand QCD at a fully differential level through event-generator development. Such a research program also needs to leverage modern strategies to make high-quality, highly differential measurements, which pose challenges on the experimental side; in particular, on unfolding methods used to correct for detector effects, which we discuss next.

\subsubsection{Unfolding in high number of dimensions}

\begin{figure}[htp]
\centering
\includegraphics[width=0.77\textwidth]{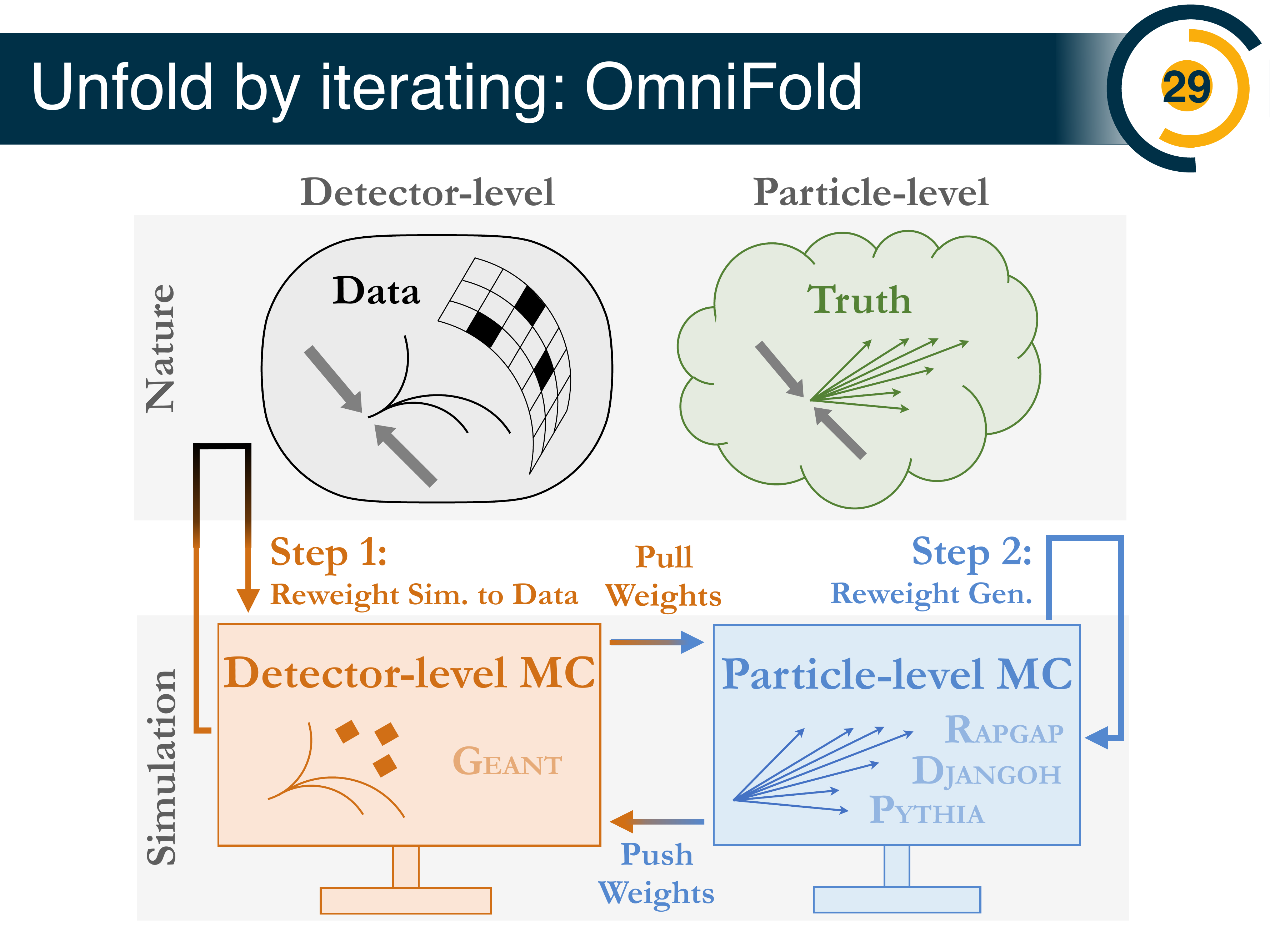}
\caption{An illustration of the OmniFold method.  Adapted from Ref.~\cite{Andreassen:2019cjw}.}
\label{fig:figure1}
\end{figure}

Differential cross section measurements are performed by applying methods to remove detector distortions called \textit{unfolding}.  Traditionally, these methods act on histograms.  A histogram can be represented as a vector and one can then transform the measurement problem into a linear algebra problem: $\vec{T}=R\vec{D}$, where $\vec{T}$ is the target particle-level histogram, $R$ is the response matrix, and $\vec{D}$ is the detector-level histogram. Typically, measurements are limited to a fixed $\mathcal{O}(1)$-$\mathcal{O}(10)$ number of bins. This means that one dimension can be finely measured, or multi-differential observables can be measured only coarsely. Recently, this limitation has been tackled by employing machine-learning techniques: A new method called OmniFold~\cite{Andreassen:2019cjw} used deep learning to achieve unbinned unfolding in any number of dimensions\footnote{Other machine-learning approaches are described in Ref.~\cite{Arratia:2021otl}.}.  

The OmniFold algorithm is presented schematically in Fig.~\ref{fig:figure1}.  The procedure proceeds by iteratively unfolding reweighting various datasets.  The final result is a set of particle-level Monte Carlo (MC) events with a set of weights.  In this way, one can choose any binning \emph{after} the measurement in order to represent the data as a histogram.  See Ref.~\cite{Andreassen:2019cjw} for further details and references.

\begin{figure}[!htp]
\centering
\includegraphics[width=0.75\textwidth]{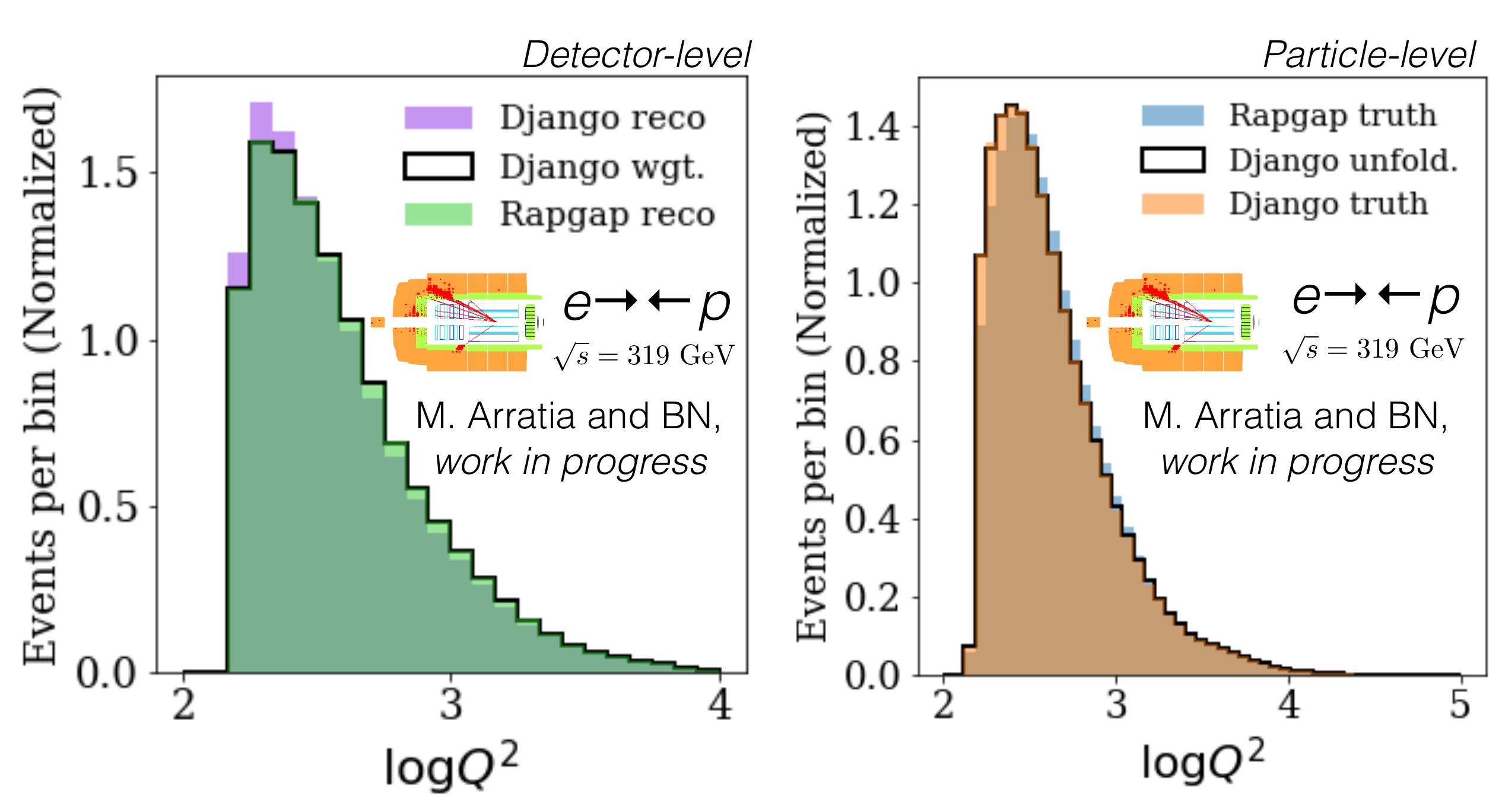}
\caption{An example observable from an eight-dimensional reweighting (left) and unfolding (right).  In the right plot, Django is treated as `data' and Rapgap is the `simulation'.}
\label{fig:figure2}
\end{figure}

The OmniFold approach has been applied to $ep$ data in order to show its utility for studying the physics of TMDs and other QCD phenomena~\cite{H1:2021wkz}.  Figure.~\ref{fig:figure2} illustrates one out of eight dimensions of this procedure.  Note that for OmniFold, adding more dimensions amounts to simply changing one line of python code.

Methods like OmniFold will significantly increase the utility of EIC data and are critical for future-proofing it.  The fact that the data are unbinned means that comparing to other experiments or event-generator calculations is simple: One can pick any observable built from the unfolded phase space to study after and not before the measurement.  However, OmniFold also will require computational innovation and planning.  In particular, we need to rethink how we publish experimental results~\cite{Arratia:2021otl}.  If the data are unbinned, then the usual HEPData repository~\cite{Maguire:2017ypu} for histograms may not suffice.  Furthermore, unbinned data will require more disk space and even though they allow for more flexible interpretation, they also require more complex tools for analysis.  Finally, the unfolding itself will require significant resources for large datasets.  The ML training is fast using Graphical Processing Units (GPUs).  State-of-the-art GPUs have only $O(10)$ GB of memory, so large unbinned unfoldings will require data parallel learning.  While all of these computing challenges can be solved, they require innovation and time so we must begin now if we are going to be ready for large-scale analyses with EIC data.

\subsection{Summary}
Overall, the EIC will be an excellent opportunity for innovation in theory simulation and may set new trends for experimental measurements. A variety of key measurements in jet physics that can be performed at the future EIC include (but are not limited to): 
\begin{itemize}
\item Jets for studies of flavor and spin structure of the nucleon, in particular three-dimensional (3D) imaging of the nucleon and even 5D Wigner distribution
\item Jets for study of 3D evolution equations, which go beyond the well-established DGLAP equations
\item Jet substructure (such as jet shape, jet mass, jet angularity, etc) in electron-proton collisions as powerful probes for QCD dynamics

\item Jet-based observables and event shapes (such as 1-jettiness) as  precision probes for extraction of fundamental QCD parameters, notably the strong coupling constant and its running

\item Modification of jets and jet substructure from $e+p$ to $e+A$ collisions for studying the transport of partons through nuclear matter
\end{itemize}
To tackle these exciting topics, in this section, we have highlighted some recent developments in the following directions:
\begin{itemize}
\item New theory studies and developments for jets and jet substructure at the EIC

\item Monte Carlo simulation

\item Machine learning techniques
\item Areas with high potential of future developments 
\end{itemize}

%
%
%
\newpage
\section{Heavy Flavors at EIC}\label{sec:hf}
\vspace{-2ex}
\centerline{\textit{Editors:} 
\href{mailto:xdong@lbl.gov}{\texttt{Xin Dong}},
\href{mailto:ssekula@smu.edu}{\texttt{Stephen Sekula}},
\href{mailto:ivitev@lanl.gov}{\texttt{Ivan Vitev}}.
}
\vspace{3ex}
%


Open and hidden heavy-flavor production in deep-inelastic scattering benefits from a status review of the tools and underlying theory in order to ensure precise predictions for future EIC experiments. Heavy flavor is a powerful tool in deep inelastic scattering that provides complementary and, in many cases, unique information on the heavy quark and gluon content of individual nucleons and whole nuclei. 

The range of potential measurements, observations, or targets of opportunity is significant in this sector. For example, the EIC will provide the possibility to establish precision information about heavy-flavor content of nucleons/nuclei. For example, the strange-quark parton contribution --- "strangeness" --- can be accessed via charged-current reactions. The production of open heavy flavor and quarkonia has emerged as a set of premier probes of  nuclear matter; these approaches can shed light on the transport properties of large nuclei and the underlying physics of hadronization. At lower transverse momenta and in both polarized and unpolarized reactions at the EIC, heavy quark production has already been shown to be sensitive to TMD physics. To ensure the precise measurement of heavy flavor, new developments in detector technology and analysis techniques are also needed. Highlighted below are some of the theoretical and experimental opportunities on the road to the EIC.

\subsection{Heavy quark production}\label{sec:hq}

Heavy flavor production occurs across a range of $Q^2$ values that can be probed at the EIC. Different quarks, dependent on their masses in relation to $Q^2$, will vary in how their contributions are accounted in processes. It is essential to establish and utilize an appropriate scheme, or schemes, for choosing how and which quark flavors to incorporate in different calculations. This issue is addressed by the development and adoption of different flavor number (FN) schemes, of which multiple have been proposed. 

Heavy-quark PDFs can be uniquely obtained by using the variable-flavor-number (VFN) scheme, represented by the following equations:
\begin{eqnarray}
  \label{eq:VFNS-hq}
  f_{h+\bar h}(n_f+1, \mu^2)
  &=& {A_{hq}^{ps}\Big(n_f, \frac{\mu^2}{m_h^2}\Big)} \otimes {q^{s}(n_f, \mu^2)}
  + {A_{hg}^{s}\Big(n_f, \frac{\mu^2}{m_h^2}\Big)} \otimes {g(n_f, \mu^2)}
  \, ,
\\
  \label{eq:VFNS-lq}
  q^{s}(n_f+1,\mu^2)
  &=& \left[ A_{qq,h}^{ns}\Bigl(n_f, \frac{\mu^2}{m_h^2}\Bigr) + 
    A_{qq,h}^{ps} \Bigl(n_f, \frac{\mu^2}{m_h^2}\Bigr) +
    A_{hq}^{ps} \Bigl(n_f, \frac{\mu^2}{m_h^2}\Bigr) \right] \otimes q^{s}(n_f,\mu^2)
 \nonumber\\ &&
  +\left[A^{s}_{qg,h}\Bigl(n_f, \frac{\mu^2}{m_h^2}\Bigr)
    + A^{s}_{hg}\Bigl(n_f, \frac{\mu^2}{m_h^2}\Bigr) \right] \otimes g(n_f,\mu^2)
  \, ,
  \\
  \label{eq:VFNS-g}
  g(n_f+1, \mu^2)
  &=& A_{gq,h}^{s}\Bigl(n_f,\frac{\mu^2}{m_h^2}\Bigr) \otimes q^{s}(n_f,\mu^2)
  + A_{gg,h}^{s}\Bigl(n_f,\frac{\mu^2}{m_h^2}\Bigr) \otimes g(n_f,\mu^2)
  \, ,
\end{eqnarray}
and one more relation in the non-singlet case, where all contributions are known even to three-loop order. 
The heavy-quark PDFs result from the massless PDFs above a matching scale $\mu \geq m_h$. Furthermore, also the
light-quark PDFs are modified.
The functions $A_{ij}$ denote the different massive operator matrix 
elements (OMEs) which are known at the two-loop level in the unpolarized~\cite{Buza:1995ie,Bierenbaum:2009mv}
and in the polarized case \cite{Buza:1996wv,Blumlein:2019zux}. 
At three-loop order the program of calculating these OMEs is nearly complete, 
cf.~\cite{Bierenbaum:2009mv,Kawamura:2012cr,Ablinger:2014lka,Ablinger:2014nga,Behring:2015zaa,Alekhin:2017kpj,Behring:2021asx}. 
In addition, there are also two-mass corrections, cf. \cite{Alekhin:2020edf}
for an illustration of their size.
The corrections to \autoref{eq:VFNS-hq}--\autoref{eq:VFNS-g} 
currently used in data analysis are at two-loop order.

There are phenomenological aspects to be described in the area of FN schemes, and we follow the example of Ref.~\cite{Alekhin:2020edf}.
In a fixed-flavor number (FFN) scheme, as used for the ABMP16 PDFs~\cite{Alekhin:2017kpj}, 
only light quarks and gluons are considered in the initial state.
Charm and bottom quarks are produced in the final state from the hard scattering of the incoming massless partons.
The available experimental data on heavy-quark DIS production is observed to be well-described with the FFN scheme~\cite{Accardi:2016ndt,H1:2018flt}. 

\begin{figure}[htbp]
\centering 
\includegraphics[width=.6\textwidth,height=0.6\textwidth]{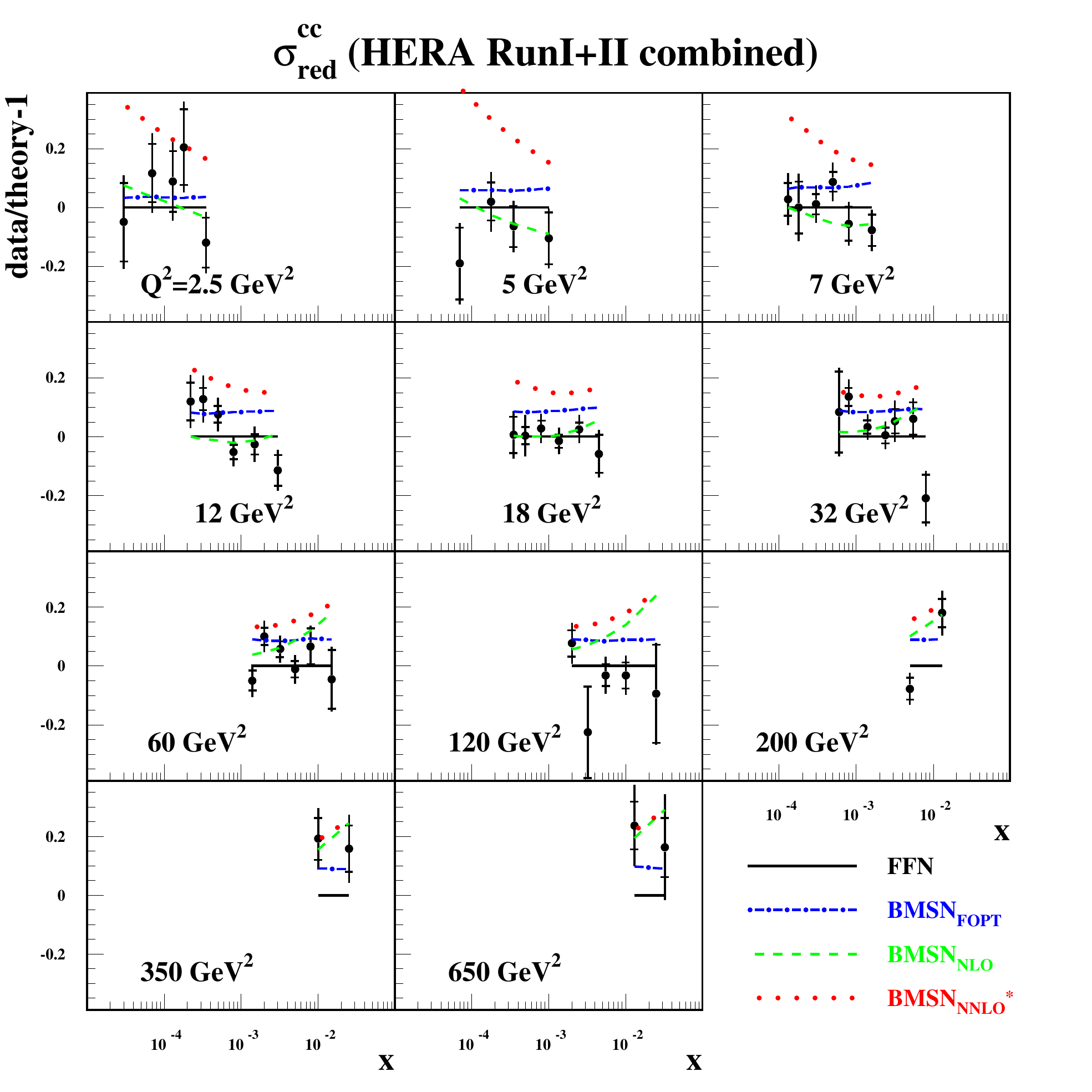}
\hfill
  \caption{\small 
    The pulls obtained for the combined HERA data 
    on DIS charm-quark production~\cite{H1:2018flt} 
    in the a FFN analysis (solid lines) versus $x$ in bins on $Q^2$.
    The predictions obtained using the BMSN prescriptions of the VFN scheme 
    with various versions of the heavy-quark PDFs with respect to 
    the FFN fit are displayed for comparison 
    (dotted-dashes: fixed order NLO; dashes: evolved from the NLO matching conditions 
    with the NLO splitting functions; 
    dots: the same for the NLO matching conditions combined with the NNLO splitting functions).  
    The PDFs obtained in the FFN fit are used throughout.
    Plot taken from~\cite{Alekhin:2020edf}.
  }
  \label{fig:heracvfn}
\end{figure}

In comparison, using a VFN scheme the charm and bottom quarks are considered also in the initial state. They appear as partonic degrees of freedom in the proton, using \autoref{eq:VFNS-hq}, at and above a certain mass scale. This expression is used in the BMSN prescription~\cite{Buza:1996wv} of the VFN scheme 
and determines heavy-quark PDFs at all scales $\mu \geq m_h$ in fixed-order perturbation theory (FOPT), where $m_h$ is the mass of the heavy-quark state under consideration.
Other VFN prescriptions use \autoref{eq:VFNS-hq} only as a boundary condition 
at $\mu=m_h$ and determine the scale dependence with the help 
of the standard QCD evolution equations for massless quarks.
The evolution effectively re-sums logarithms
in the ratio $Q^2/m_c^2$ (or $Q^2/m_b^2$) for the charm (or bottom) PDF.
Note that the corresponding logarithms are not necessarily large.
In addition, the evolution can be either performed at NLO or at NNLO, 
in the latter case utilizing the three-loop splitting functions~\cite{Vogt:2004mw}. 
This variant of the VFN scheme is denoted as NNLO$^{\ast}$, since there is a mismatch in the orders of perturbation theory between the heavy-quark OMEs and the accuracy of the evolution equations. The numerical difference between these approaches is illustrated in \autoref{fig:heracvfn} in a comparison to the combined HERA data 
on DIS charm-quark production~\cite{H1:2018flt}. 

High precision data for DIS heavy-quark production from the EIC will allow the community to probe the applicability and reliability of the different heavy flavor schemes currently in use. This new data should also enable the development or modification of schemes to reflect precision information obtainable from EIC experiments.

\subsubsection{Heavy flavor baryon production}

Heavy flavor baryon production remains a challenging problem that is yet to be understood in QCD. 
Recent data from $p$+$p$, $p$+$A$, and $A$+$A$ collisions at RHIC and LHC showed that 
the $\Lambda_c^+/D^0$ ratio is considerably larger than the fragmentation baseline~\cite{ALICE:2017thy,STAR:2019ank}. The new color reconnection (CR) scheme implemented in the PYTHIA8 generator ~\cite{Christiansen:2015yqa}, together with the baryon-junction scheme, increases the $\Lambda_c^+/D^0$ ratio at low $p_T$ and is comparable to the experimental data. However, the same scheme underpredicts the heavier charm baryon $\Xi_c$ yield~\cite{ALICE:2021bli}.
A detailed investigation of heavy flavor baryon production at high-luminosity EIC collisions will offer an opportunity to enable detailed investigations to understand how the hadronization plays a role from $e^+e^-$ to hadronic collisions.

\begin{figure*}
	\centering
	\includegraphics[width=0.5\textwidth]{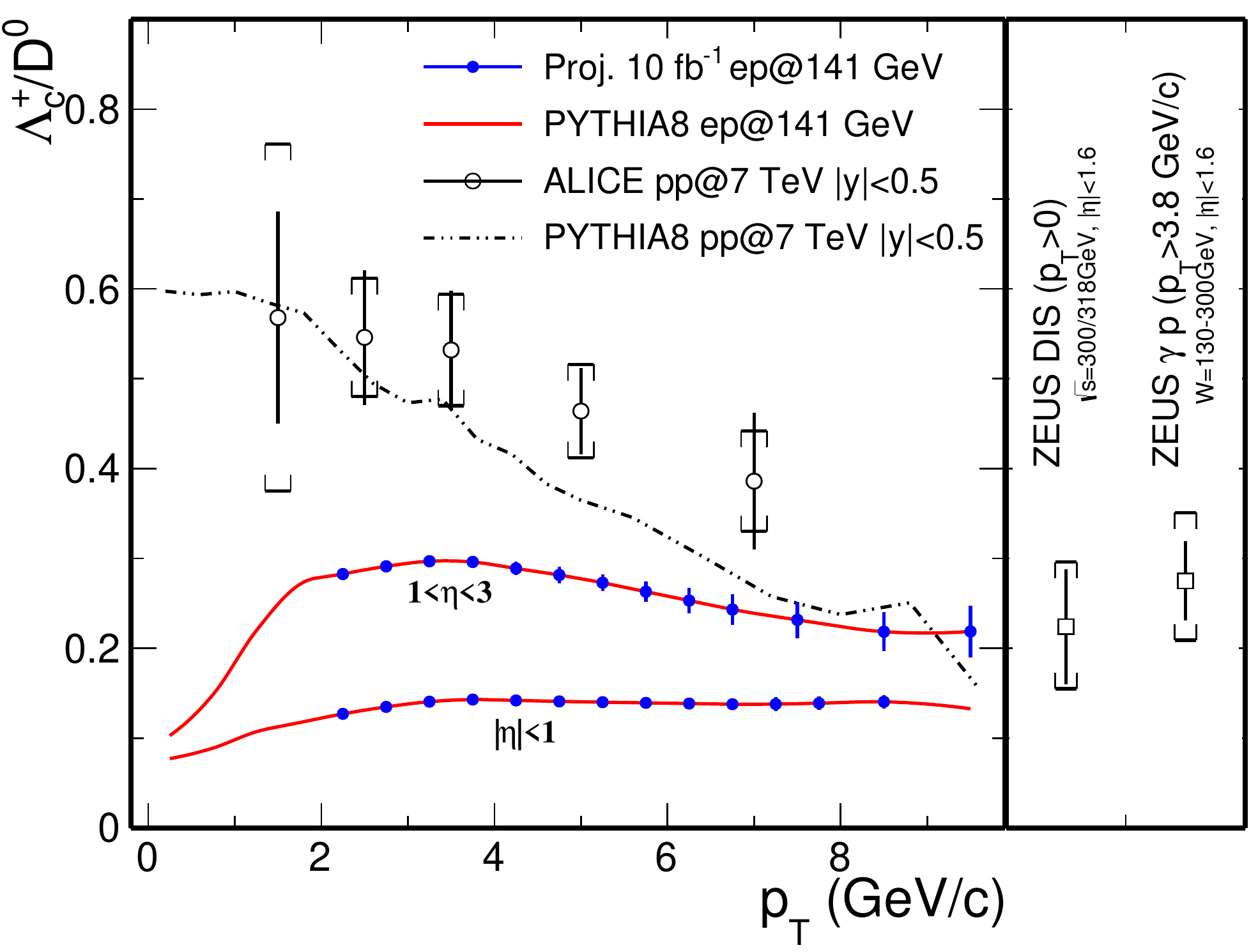}
	\caption{The projected statistical uncertainty of $\Lambda_c^{+}/D^0$ as a function of $p_T$ in $|\eta|<1$ and $1<\eta<3$ in $e$+$p$ 18$\times$275 GeV collisions. The mean values for the projected points are from PYTHIA8.3 calculations. Open circle and open square points are the measurements in $p+p$ $\sqrt{s}=7$\,TeV collisions from ALICE~\cite{ALICE:2017thy} and $e$+$p$ collisions from ZEUS~\cite{ZEUS:2005pvv,ZEUS:2010cic}.}
	\label{fig:Lcproj}
\end{figure*}

Figure \autoref{fig:Lcproj} shows projected statistical uncertainties of $\Lambda_{c}^+/D^{0}$ as a function of $p_T$ for two $\eta$ regions ($|\eta|<1$ and $1<\eta<3$) with 10\,fb$^{-1}$ $e$+$p$ (18$\times$275 GeV) collisions. The $\Lambda_c^+$ cross section used is based on the recent PYTHIA v8.3 calculation which includes the latest development on the color reconnection scheme for baryon production at high energy $p$+$p$ collisions. Also shown in Fig.~\autoref{fig:Lcproj} are the existing measurements in $p$+$p$ collisions from ALICE~\cite{ALICE:2017thy} and $e$+$p$ DIS and $\gamma p$ collisions from ZEUS~\cite{ZEUS:2005pvv,ZEUS:2010cic}. The projection shows that measurements at EIC $e$+$p$ DIS collisions would allow us to systematically investigate the $\Lambda_c$ production over a broad kinematic region, which will shed detail insights on charm hadrochemistry and charm-quark hadronization.

\subsection{Deciphering QCD at the EIC}
\label{sec:ncteq}

The EIC also opens a new era in precision in the investigation of hadronic structure. This endeavor is  enabled by a wealth of data from JLab experiments, RHIC and the LHC, that collectively help us reveal the details of the underlying theory, QCD. The PDF framework has proven remarkably successful in describing processes with hadronic initial states. While the study of proton PDFs has grown exceedingly precise, the need to extend this precision to the \textit{nuclear} sector, involving fits with explicit nuclear degrees of freedom, has become more urgent in recent years. The use of nuclear targets and heavy ion collisions has continued and grown in recent decades, and a new era will open at the EIC. To enhance the accuracy of experimental analyses and theoretical studies of QCD, new information from the EIC can revolutionize our understanding of overall nuclear structure. This can be encapsulated in the determination of nuclear PDFs (nPDF).

\begin{figure}[!htbp]
\centering
\includegraphics[width=0.7\textwidth]{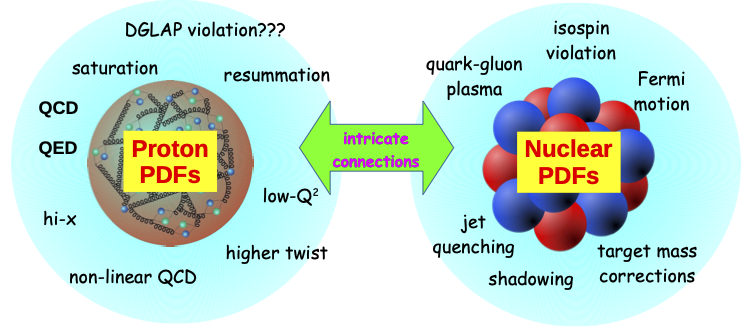}
\caption{Illustration of the intricate connection between proton and nuclear PDFs, and the many phenomena to be considered.}
\label{fig:ncteq}
\end{figure}

Progress in studying QCD dynamics within nuclei has been demonstrated in a number of recent nPDF
analyses~\cite{Kovarik:2015cma,Kusina:2020lyz,Segarra:2020gtj,Hirai:2007sx,deFlorian:2011fp,Eskola:2016oht,Walt:2019slu,AbdulKhalek:2019mzd,AbdulKhalek:2020yuc,Ethier:2020way,Paukkunen:2020rnb,Kusina:2016fxy}. As the community strives to increase its precision approaches and 
extend its  predictions into new frontiers of the kinematic $\{x,Q^2\}$ plane, new phenomena (illustrated in \autoref{fig:ncteq}) will be encountered. These  must be incorporated into theoretical frameworks.  The high statistics and large kinematic reach of the EIC will be instrumental in providing the data necessary to advance global analyses. These, in turn, will yield an improved understanding of nuclear structure and new insights into QCD.
For example of this, a recent study by the nCTEQ Collaboration expanded upon the original nCTEQ15 analysis~\cite{Kovarik:2015cma}
and included  new data from  JLab experiments to explore the high-$x$ and low-$Q$ regime. The resulting nCTEQ15HIX nPDFs investigated the impact of a variety of effects including 
higher-twist and deuteron corrections. As a result, the collaboration was able to provide to the community a reliable description of this new data. 
\begin{figure}[htbp]
\centering
\includegraphics[width=0.5\columnwidth]{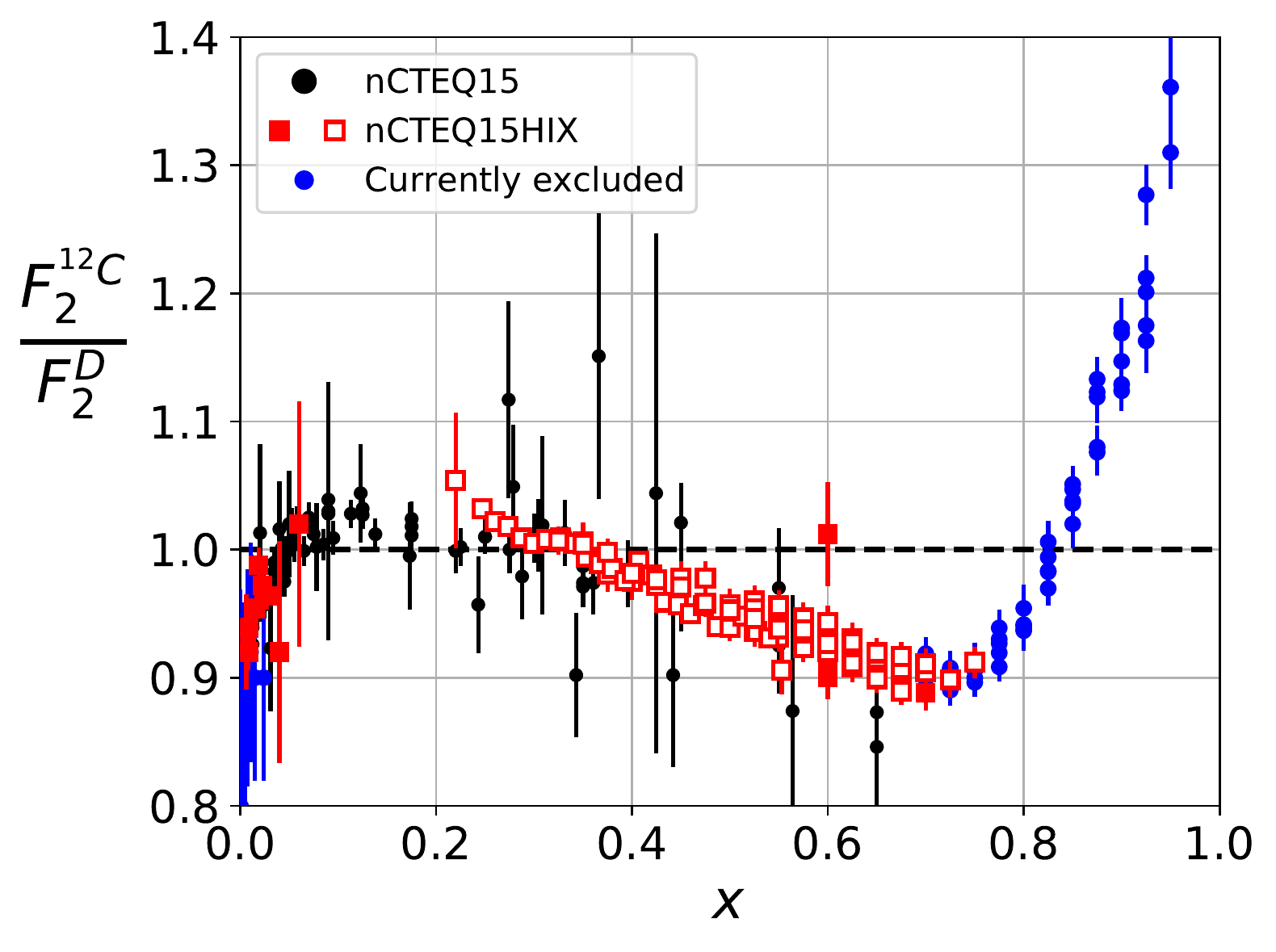} 
\caption{The classic ``nuclear correction factor'' ($F_2^A/F_2^D$) 
for carbon  as a function of the parton momentum fraction $x$.
We show the data used   in the original nCTEQ15 fit (black),
additional data from JLab for the nCTEQ15HIX fit (red),
and data beyond the $\{x,Q^2\}$ cuts. 
}
\label{fig:emc}
\end{figure}

The structure function ``nuclear correction factor'' \mbox{$(F_2^C/F_2^D)$} shown in  \autoref{fig:emc}
provides one measure of how the PDFs inside nuclei are modified relative to those of an individual, isolated proton (or the deuteron, in this particular example). Although the features of \mbox{$(F_2^C/F_2^D)$} display the characteristic EMC shape and have been studied for decades, it is not yet possible to compute this observable from first-principles QCD.  Nevertheless, recent studies suggests that short-range correlations (SRC) could modify the structure of nucleons in neutron–proton pairs and provide a framework for interpreting these  effects~\cite{Schmookler:2019nvf,nCTEQsrc}. Additionally, there are other theoretical paradigms in terms of which one might interpret the nuclear modifications.

PDF global analyses have shown the power to describe hadronic interactions and structure in the context of QCD. They have proven essential computational tools with a range of applications in high-energy physics. Similarly, \textit{global nuclear PDF analyses} will enable a new level of precision in the exploration of nuclear dynamics and particle phenomenology. 
In a broader sense, these tools can validate features of the standard model to the required level of next-generation precision (e.g. to aid in securing the precision era of programs like the LHC or future higher energy collider programs) and enable the search for discrepancies which may signal undiscovered phenomena (by reducing standard model uncertainly to a level sufficient to reveal new processes). Together, these abilities will yield deeper insights into the  QCD theory and the structure of hadronic matter.

Understanding the collective structure and effects of whole nuclei is clearly a frontier of the field. The pursuit of this knowledge presents opportunities and challenges to the community. Studies are ongoing and will continue into and beyond the EIC era. Heavy flavor at the EIC offers a strong avenue into this area.

\subsubsection{Impact of  heavy flavor measurements on gluon nPDFs}

Measurements of heavy flavor hadrons (hadrons containing a charm or bottom quark) in DIS interactions are valuable and provide direct and clean access to the gluonic structure of nucleons/nuclei. (Nuclear) PDFs are an essential ingredient in understanding measurements of nuclear collisions and are of broad interest in the particle and nuclear physics communities. 

\begin{figure*}[!htbp]
    \centering
    \includegraphics[width=0.85\textwidth]{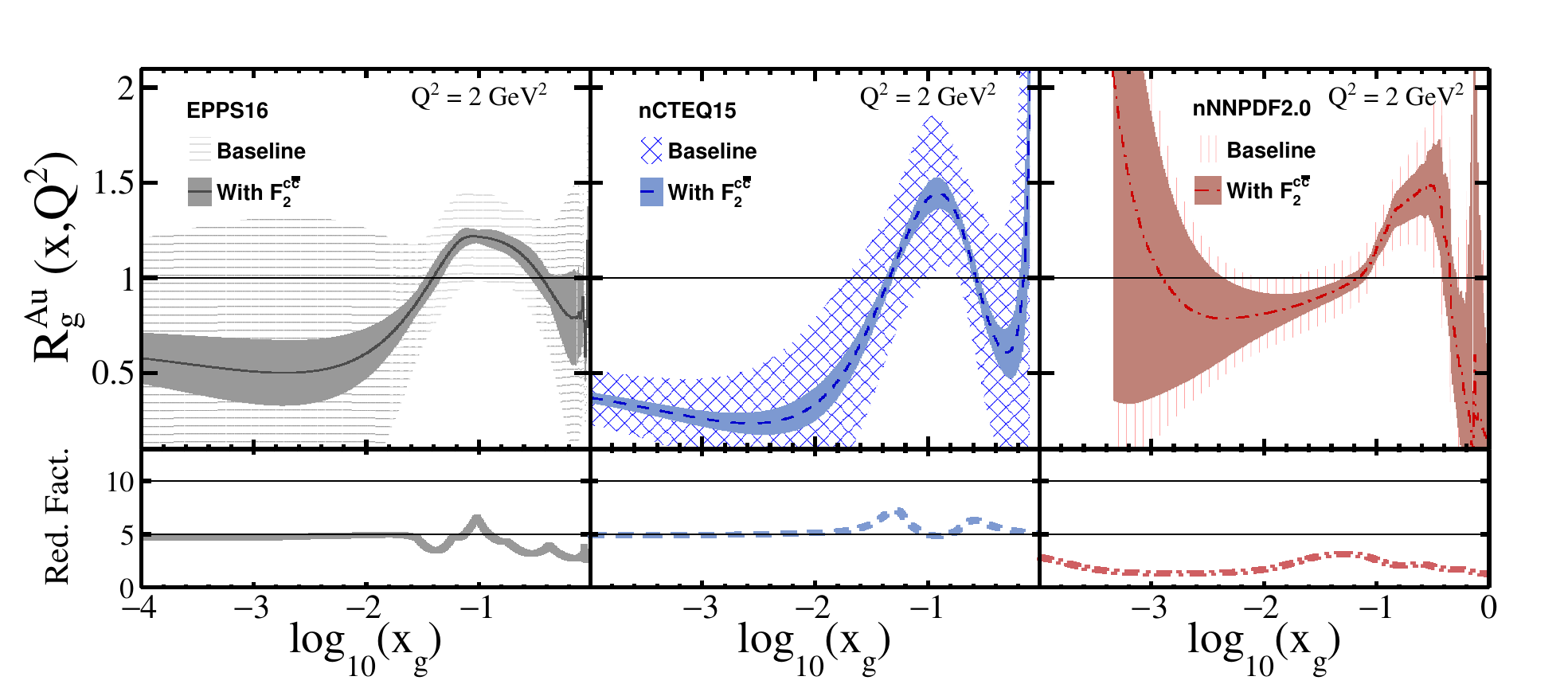}
    \caption{Top: The nuclear gluon ratios as a function of $x_g$ for Au nuclei for the EPPS16~\cite{Eskola:2016oht} (left), nCTEQ15~\cite{Kovarik:2015cma} (middle), and nNNPDF2.0~\cite{Khalek_2020} (right) PDF sets. The hashed curves show the baseline uncertainties and solid curves show the re-weighted uncertainties using the projected charm data with integrated luminosities of 1 fb$^{-1}$/nucleon. Bottom: The reduction factor of the nPDF uncertainties with the inclusion of the EIC charm data.}
    \label{fig:Gluon1}
\end{figure*}

The experimental capabilities of heavy flavor measurements and the constraints to (n)PDFs with charm structure function $F_2^{c\bar{c}}$ at EIC have also been studied~\cite{Aschenauer:2017oxs,Chudakov:2016ytj,Wong:2020xtc,PhysRevD.103.074023}. A recent effort based on the detector design utilizing the silicon tracker outlined in the EIC Yellow Report shows the great potential of measuring charm structure functions over a broad kinematic region~\cite{Kelsey:2021gpk}. The impact on the gluon nPDFs using a Bayesian PDF re-weighting procedure with three different nPDF parametrizations (EPPS16~\cite{Eskola:2016oht}, nCTEQ15~\cite{Kovarik:2015cma} and nNNPDF2.0~\cite{Khalek_2020}) is illustrated in \autoref{fig:Gluon1} at $Q^2$=2\,GeV$^2$ based on 1\,fb$^{-1}$/nucleon data. The bottom panels show the significant expected improvement factor of the nPDF uncertainties. 

\subsection{Charm jets as a probe for strangeness at the EIC}
\label{sec:charmJets}

\def\rslow{\hbox{\textsc{Rs{-}Low}}}
\def\rshigh{\hbox{\textsc{Rs{-}High}}}
\def\rsmid{\hbox{\textsc{Rs{-}Mid}}}

The discussion of heavy flavor has so far focused on the inclusion and description of partons, especially heavy flavor, in nucleons and nuclei. Here, we use a particular example --- the strangeness content of the proton --- as one key highlight of the insight that can be gained from the EIC by combining high-energy physics and nuclear physics approaches to tackle a singular heavy-flavor-related question.

\begin{figure}[hthb]
\centering
\includegraphics[width=0.5\columnwidth]{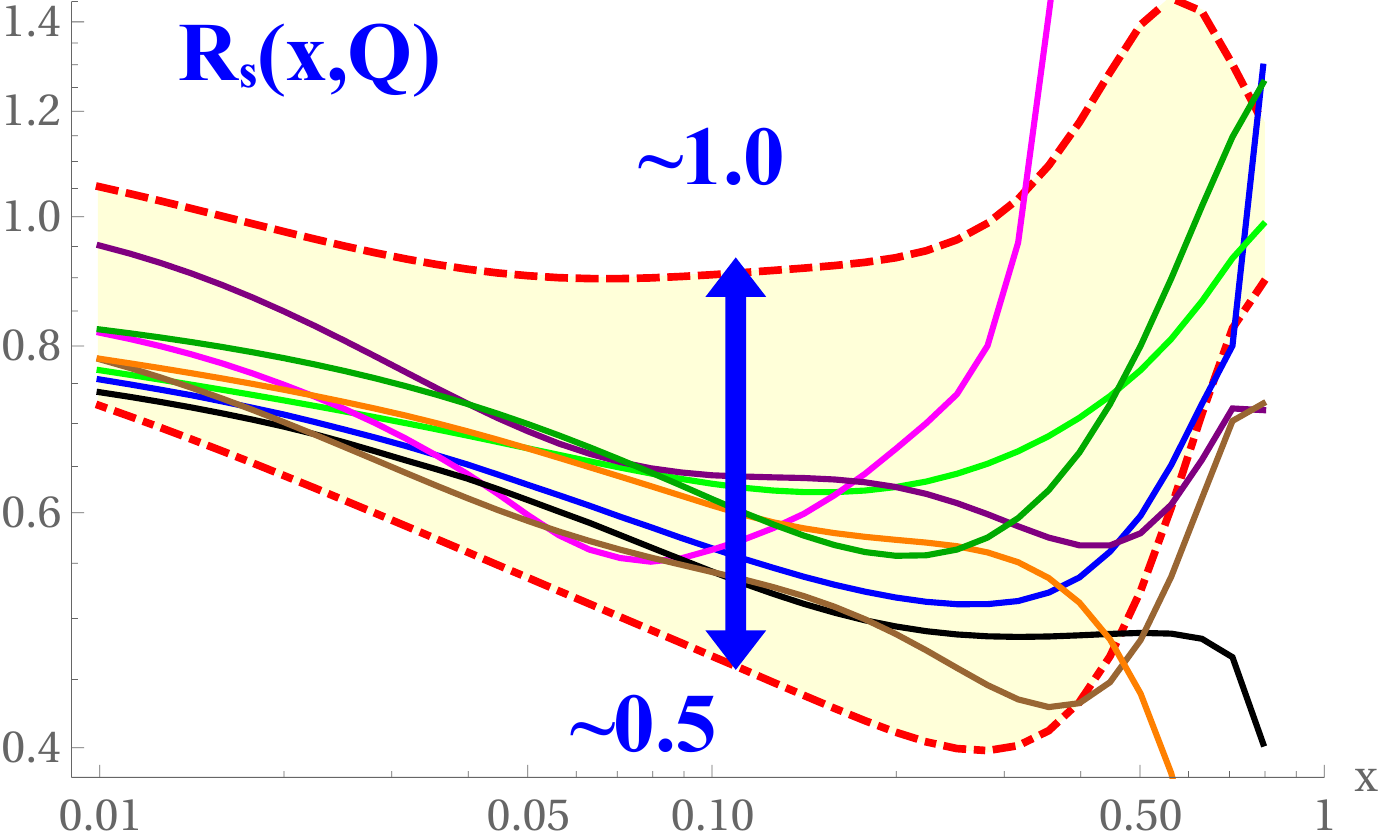}
\caption{\label{fig:rs} 
a)~The strange quark ratio $R_s(x,Q)=(s+\bar{s})/(\bar{u}+\bar{d})$  at $Q{=}10$~GeV 
for a selection of PDFs. 
The region between the   
\rshigh{}{}  (dashed red)
and  \rslow{}   (dot-dashed red) PDF curves 
is consistent with current experiments;
\rsmid{} is shown in blue.
}
\end{figure}

The EIC can  resolve long-standing questions regarding the precise balance of quark flavors contributing to the proton's structure. 
As \autoref{fig:rs} illustrates, 
the strange quark PDF  has large uncertainties 
and can vary by a factor of about 2 in the intermediate $x$ region. 
Thus, a pressing question to address is:
\textit{what is the capability of the EIC to provide improved constraints on the 
nucleon structure in general, and the strange quark in particular?}
As a case study, Ref.~\cite{Arratia:2020azl}  examined the  production of charm-jets in charged-current (CC) DIS  at the EIC . THis was done to determine the sensitivity to the underlying strange PDF. 
To characterize this sensitivity, three PDFs sets were chosen (\textit{cf.} \autoref{fig:rs}): 1)~\rslow{} with suppressed strangeness, 2)~\rsmid{} with intermediate strangeness, and 3)~\rshigh{} with enhanced strangeness.

The process  $e^- p\to \nu_e c$ was simulated in \textsc{Pythia}
for unpolarized electron-proton collisions with beam energies of 10~GeV and 275~GeV.
The \textsc{Delphes} framework was used to obtain a parameterized simulation of detector response  with baseline parameters for the EIC detectors. The performance of a high-impact-parameter track-counting algorithm to tag charm jets was estimated as well as  the potential of particle identification to increment tagging efficiency. Assumptions were generally structured to be conservative compared to what is likely to be far more efficient, etc. performance of real EIC detectors, algorithms, and approaches (based on experimental methods already employed at, for example, LHC experiments).
The detailed detector requirements, including charm tagging, are presented in Ref.~\cite{Arratia:2020azl}.

\begin{figure}[b]
\centering
\includegraphics[width=0.5\columnwidth]{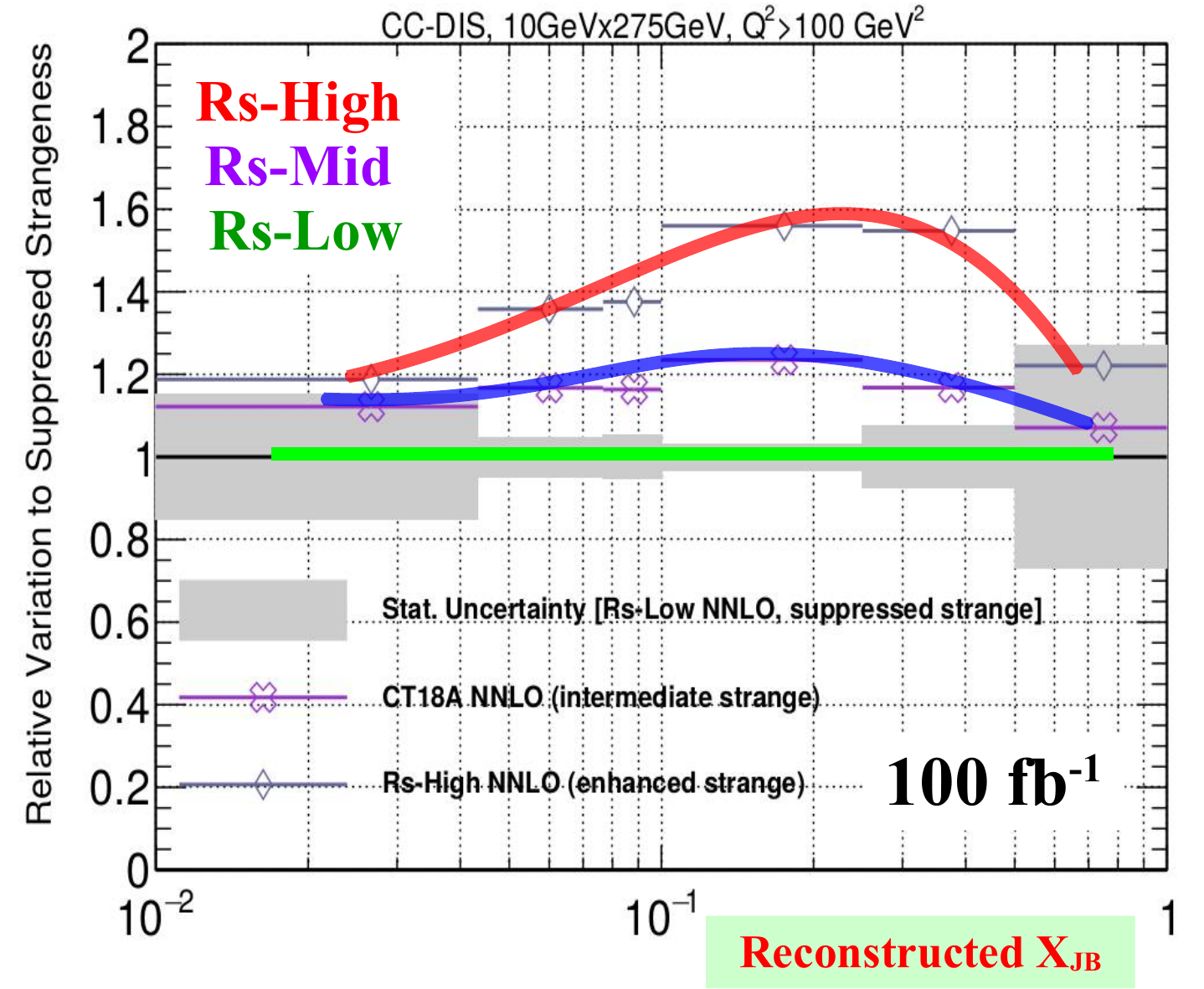}
\caption{\label{fig:sens} 
We compare the impact on the charm-jet yields of three assumed scenarios for $R_s$:
 suppressed strangeness (\rslow{}),
 intermediate strangeness (\rsmid{}),
 and
 enhanced strangeness (\rshigh{})
 {\it vs.}\ the reconstructed $x_{JB}$. 
 Our baseline is $\rslow{}$, and 
the gray band indicates the statistical error.
}\end{figure}

The relative variation (compared to \rslow{}) in the resulting charm-jet yield is displayed inside \autoref{fig:sens} for $100~\mathrm{fb}^{-1}$ of luminosity; the gray band indicates the statistical uncertainty. 
It was observed that these measurements are capable of distinguishing not only the extreme limit of the enhanced strangeness PDF (\rshigh{}), but also perform well for the case of intermediate strangeness (\rsmid{}). 

In summary,  there is strong evidence that the use of charm-tagged jets at EIC will provide new constraints on the strangeness PDF and should be part of a global analysis of strangeness within the EIC program.
Achieving this goal is a challenge that demands both  high luminosity  and a well-designed EIC detector with good capabilities for measuring displaced vertices, dedicated single-particle identification, jets, and missing-transverse energy. As such, it represents a robust platform on which to inform the design of the EIC detectors. 
Additionally, the charm-tagging performance studies advanced in this work  have the potential to extend the rapidly emerging field of jet studies for the future EIC, 
and could be applied to neutral current boson-gluon fusion or photo-production processes.

\subsection{Heavy Flavor and Gluon Spin Structure}

In addition to the kinds of precision approaches that should be possible with direct access to intrinsic heavy flavor, other approaches can shed light on the gluonic structure of nucleons and nuclei. These approaches have implications for knowledge of gluon content and contribution to observables like spin.

\subsubsection{Transverse spin asymmetries and the gluon Sivers}

Exploring and measuring gluon TMD PDFs is one of the primary goals for the future Electron Ion Collider (EIC). Among the gluon TMD PDFs, the so-called \textit{gluon Sivers function} is regarded as one of the “golden measurements” at the future EIC. The gluon Sivers function encapsulates the quantum correlation between the gluon’s transverse momentum inside the proton and the spin of the proton. This provides 3-D imaging of the gluon’s motion.

\begin{figure}[h]
\begin{center}
\includegraphics[width=0.4\textwidth]{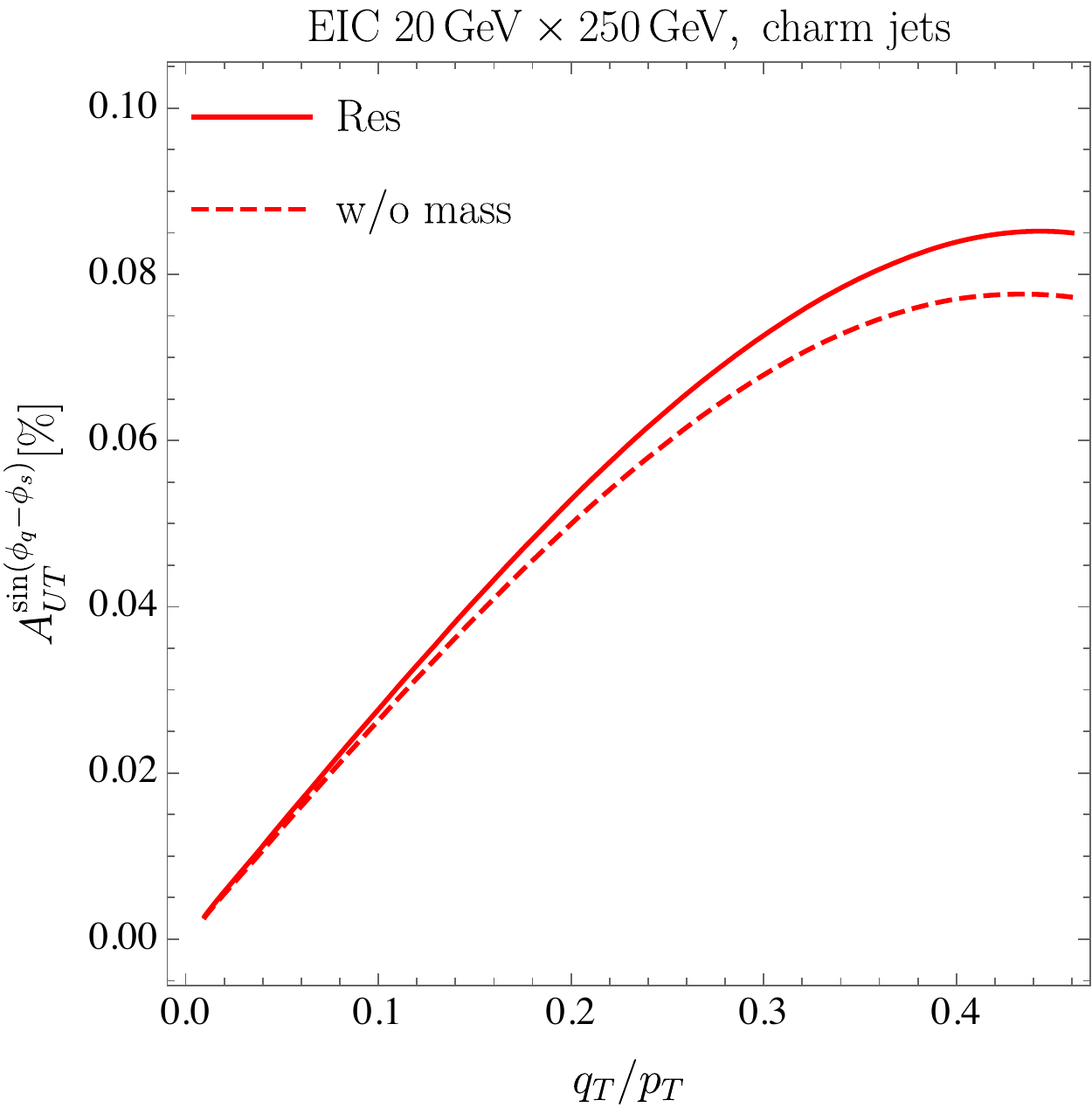}\hspace{0.8cm}
  \includegraphics[width=0.4\textwidth]{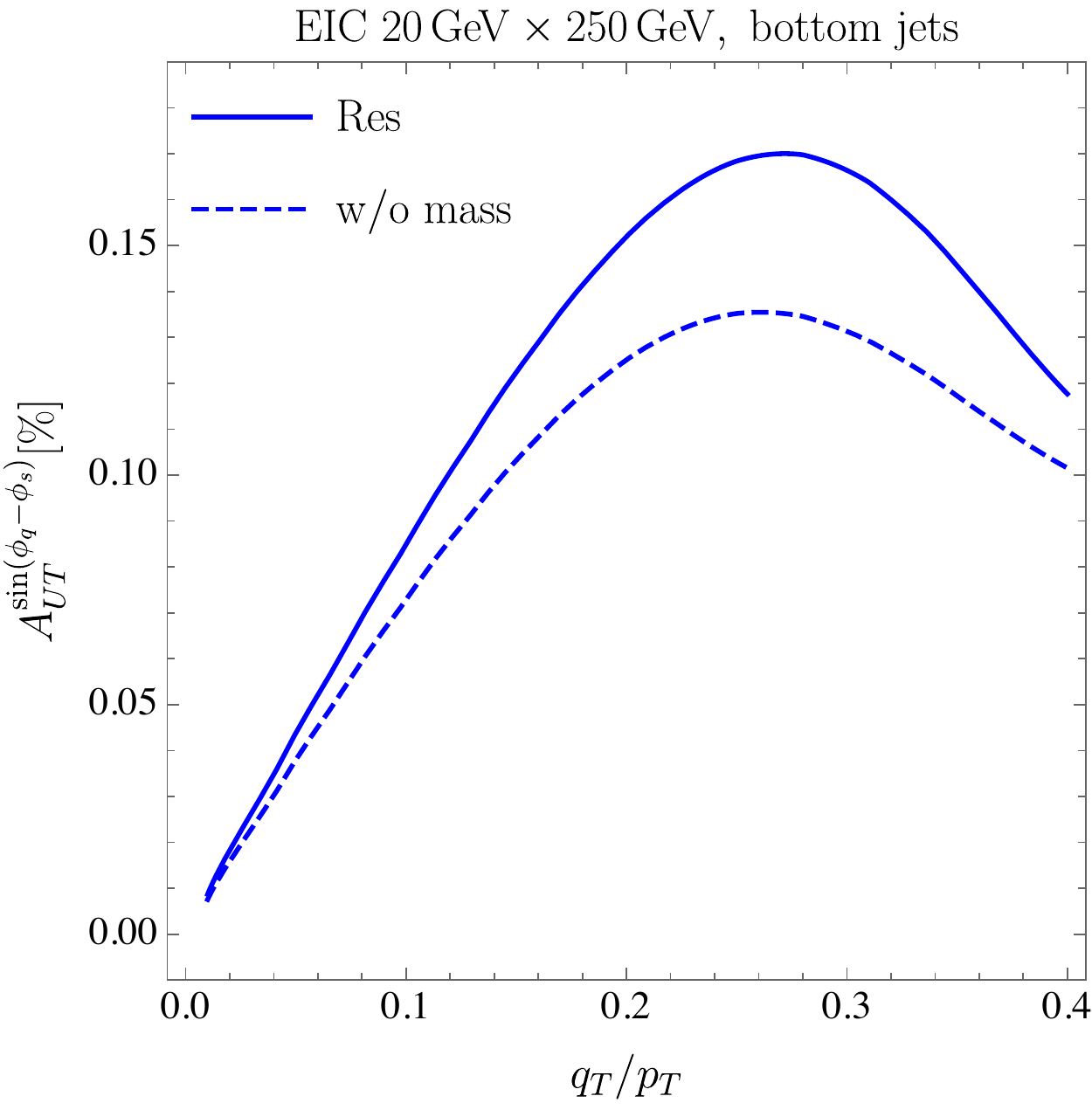}
  \caption{The Sivers spin asymmetry for charm (left plot) and bottom (right plot) dijet production at the EIC is plotted as a function of $q_T/p_T$. The solid curves are the results from using the resummation formula, while the dashed curves represent the resummation prediction using the evolution kernel without finite quark mass corrections.}
\label{fig:siv}
\end{center}
 \end{figure} 

A recent work~\cite{Kang:2020xgk} is used to illustrate this approach. The TMD factorization formalism is developed for heavy flavor (HF) dijet production at the EIC. The authors then investigate the use of back-to-back HF dijet production in transversely-polarized target DIS as a means of probing gluon Sivers function. From this, a  prediction is generated for the Sivers asymmetry for charm and bottom dijets. This preduction can then be used to probe the gluon Sivers function. 

In this example work, the authors carefully study the effects of the HF masses by comparing the mass-dependent predicted asymmetry against the asymmetry in the massless limit.  The Sivers spin asymmetry is presented in \autoref{fig:siv}. The solid curves are the results obtained using the re-summation formula, and the dashed curves represent the re-summation prediction using the evolution kernel without finite quark mass corrections. This work determines that, in the kinematic region considered, the HF masses generate modest corrections to the predicted asymmetry for charm dijet production but \textit{sizable} corrections for the bottom dijet process.

\begin{figure*}[!htbp]
    \centering
    \includegraphics[width=0.45\textwidth]{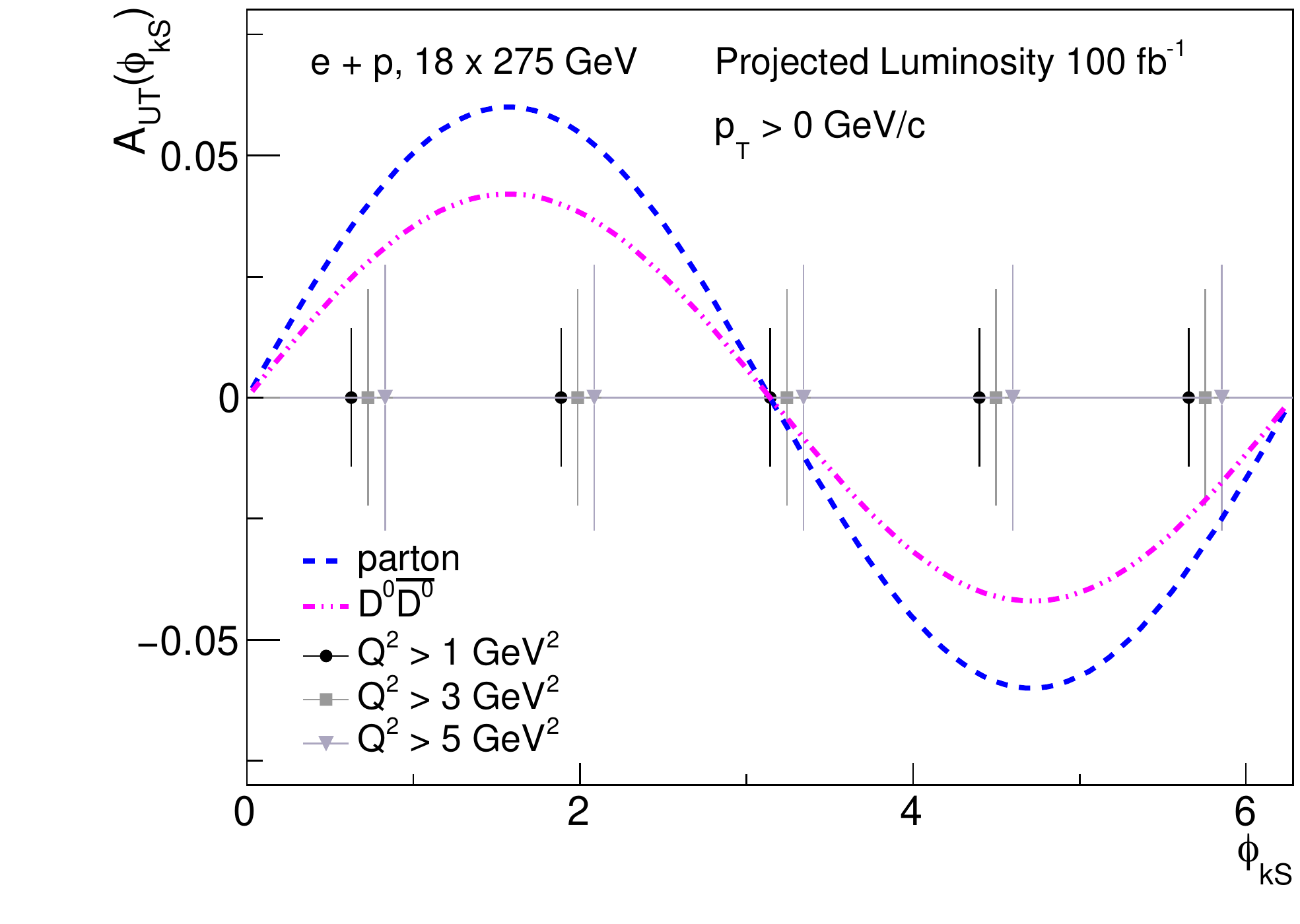}
    \hspace{0.2in}
    \includegraphics[width=0.45\textwidth]{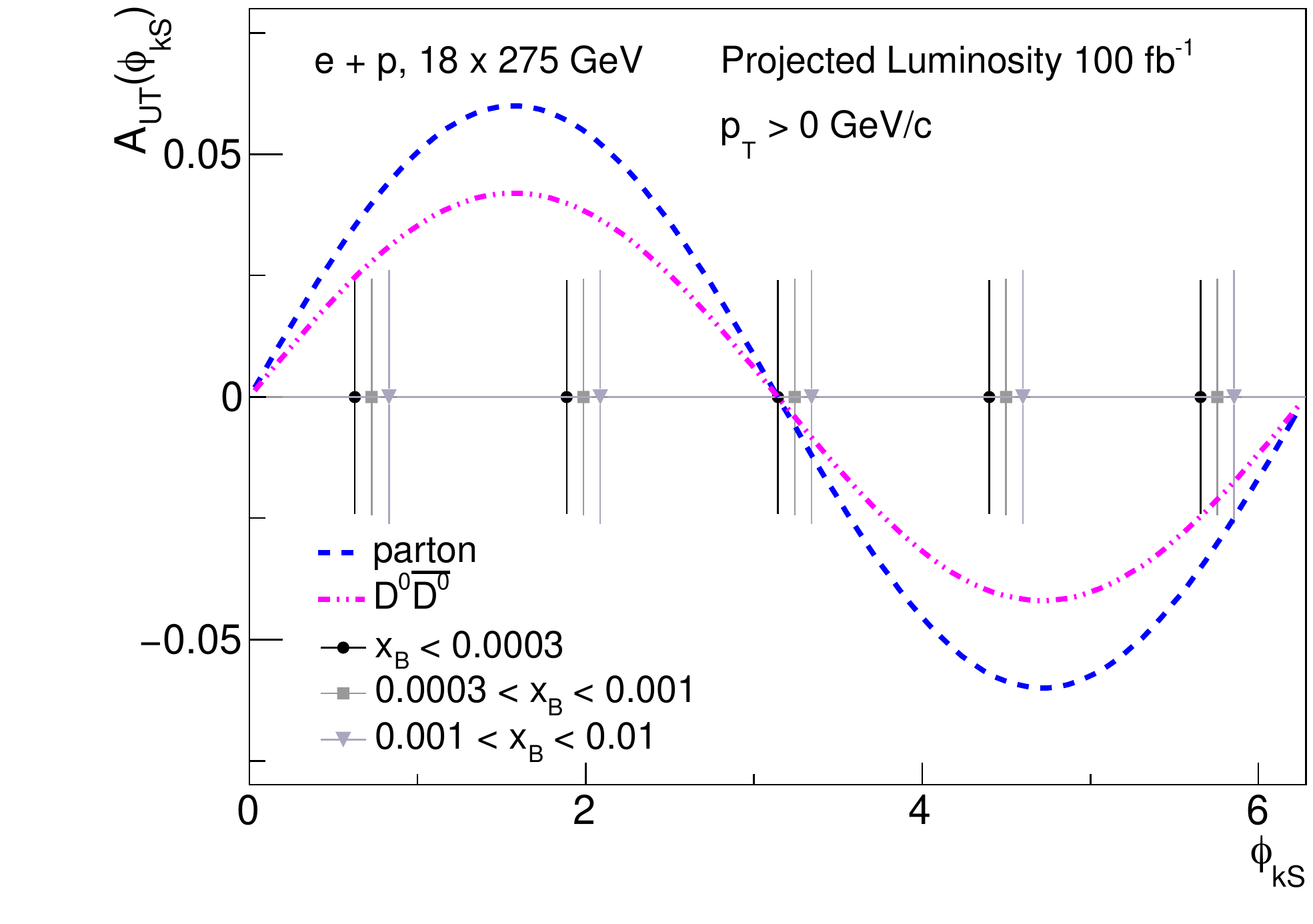}
    \caption{Statistical uncertainty projections for $A_{\rm UT}$ in bins of azimuthal angle of the momentum of the $D^0\overline{D^0}$ pair relative to the spin of the proton ($\phi_{kS}$), for different $Q^2$ (left) and $x_B$ (right) selections. The two curves indicate the signal strength at parton and $D^0\overline{D^0}$ levels. The above results are obtained assuming that the gluon Sivers function takes the magnitude of its 10\% positivity bound.}
    \label{fig:DDbarTMD}
\end{figure*}

The gluon Sivers TMD can be linked to azimuthal anisotropies of the produced charm-anticharm pair. The Sivers asymmetry can be extracted from measurements of the transverse SSA, $A_{\rm UT}$, as a function of the azimuthal angle of the charm-anticharm pair relative to the orientation of the proton spin. The correlation between the azimuthal angle of the charm-anticharm pair momentum and that of the corresponding hadron pair momentum in the case of $D^0\overline{D^0}$ production was studied in PYTHIA 6.4 simulations and was found to be well-preserved during hadronization. The signal strength, $A_{\rm UT}$ at the partonic level, can be reduced by up to 30\% in the heavy-quark production and subsequent hadronization in these simulations. The projections for $A_{\rm UT}$ for different values of $Q^2$ and $x_B$ at future EIC experiment with 100\,fb$^{-1}$ statistics are shown in \autoref{fig:DDbarTMD}. In particular, the uncertainty is shown in comparison with the possible signal size of this thus far poorly constrained quantity.

The work described here utilizes a reconstruction charm hadrons, but a similar analysis can also be conducted by using charm-tagged jets. These approaches are complementary, in that experience from past experiments suggests that the sample of collisions that yield pairs of reconstructable charm hadrons are not 100\% in overlap with the sample of events with two fully tagged charm jets. In addition, the efficiency, acceptance, and backgrounds of these two approaches vary and will each have their own systematic uncertainties. The combination of a range of approaches that can select di-charm-quark production could be used in a global way to further illuminate the picture of the gluon Sivers function. Such portfolios of measurements that can later be combined can be ensured at the EIC by designing and construction multi-purpose experiments with a range of acceptance and technology that allow different approaches to be employed in complementary ways.

\subsubsection{Gluon helicity $\Delta g/g$ through charm-hadron double-spin asymmetry}
The EIC also offers unprecedented opportunities to constrain the $\Delta g$ contribution to the proton spin via the scaling violations of the polarized structure functions. The polarized charm structure function provides direct access to the $\Delta g$ at the leading order. This will complement the inclusive DIS measurements in several important ways, for example, offering a
new ingredient on the $\Delta g$ determination in addition to
the inclusive DIS and providing sensitivity in the moderate $x$ region.

Experimentally, the charm double-spin asymmetry can be measured in the polarized $e+p$ collisions. This has been studied in detail in references~\cite{Adolph:2012ca,Kurek:2011kka}. The asymmetry can be written as
\begin{equation}
A_{LL}^{c} \equiv \dfrac{d\sigma^{++}-d\sigma^{+-}}{d\sigma^{++}+d\sigma^{+-}} = \dfrac{y(2-y)}{y^2+2(1-y)} \dfrac{g_1^c(x,Q^2)}{F_1^c(x,Q^2)} \equiv D(y) \dot A_1^c(x,Q^2),
\end{equation}
where $d\sigma^{++}$ and $d\sigma^{+-}$ are the charm production cross sections for electron and proton beam spin orientation to be parallel and anti-parallel to each other, respectively; and $D(y)$ is the depolarization factor of the virtual photon depending on the inelasticity $y$~\cite{Anderle:2021hpa}. 
The theoretical calculation shows that the $A^c_1$
is sizable at the level of 10-20\% in the moderate $x$ region.

\begin{figure*}[!htbp]
    \centering
    \includegraphics[width=0.46\textwidth]{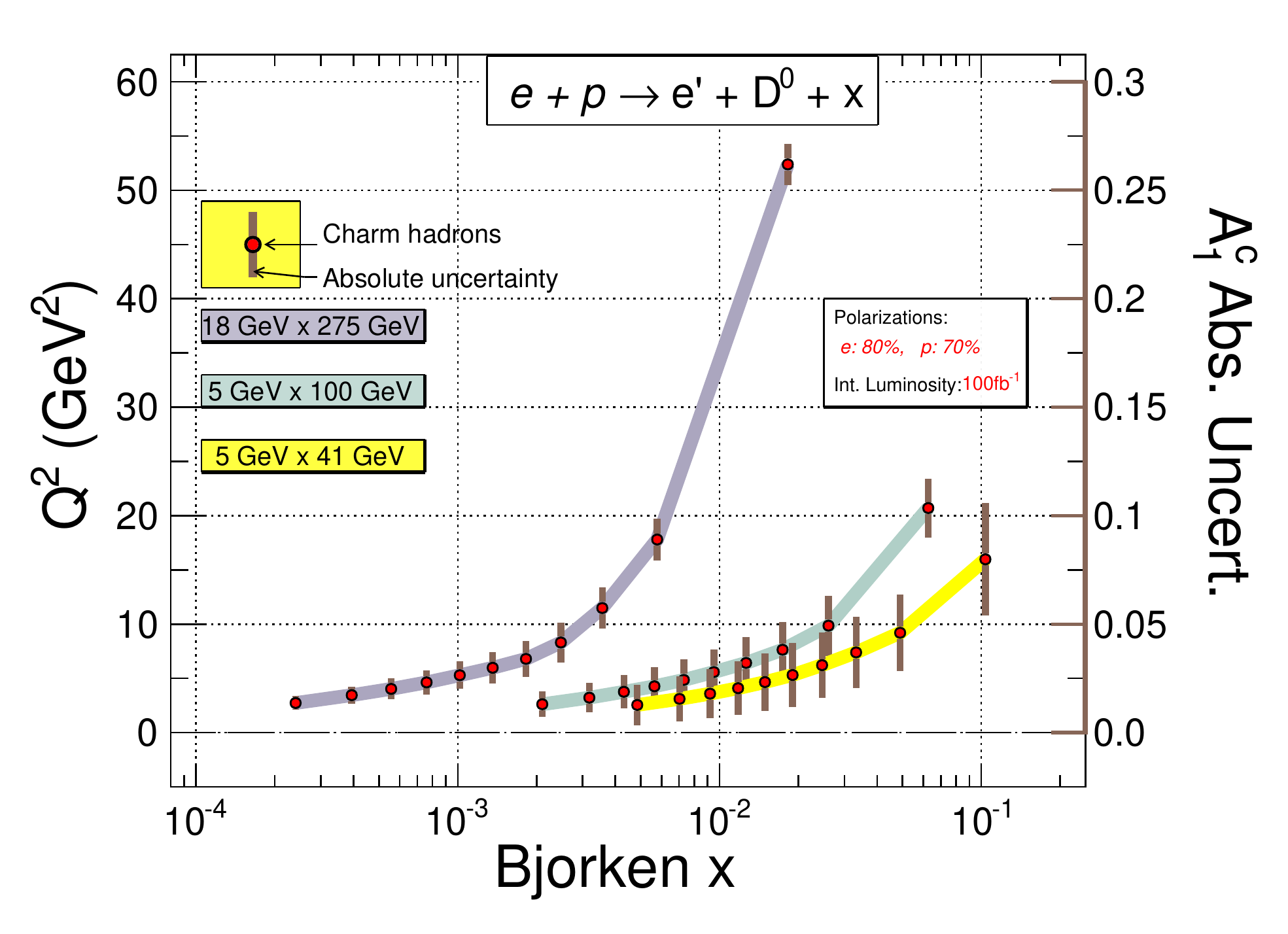}
    \includegraphics[width=0.46\textwidth]{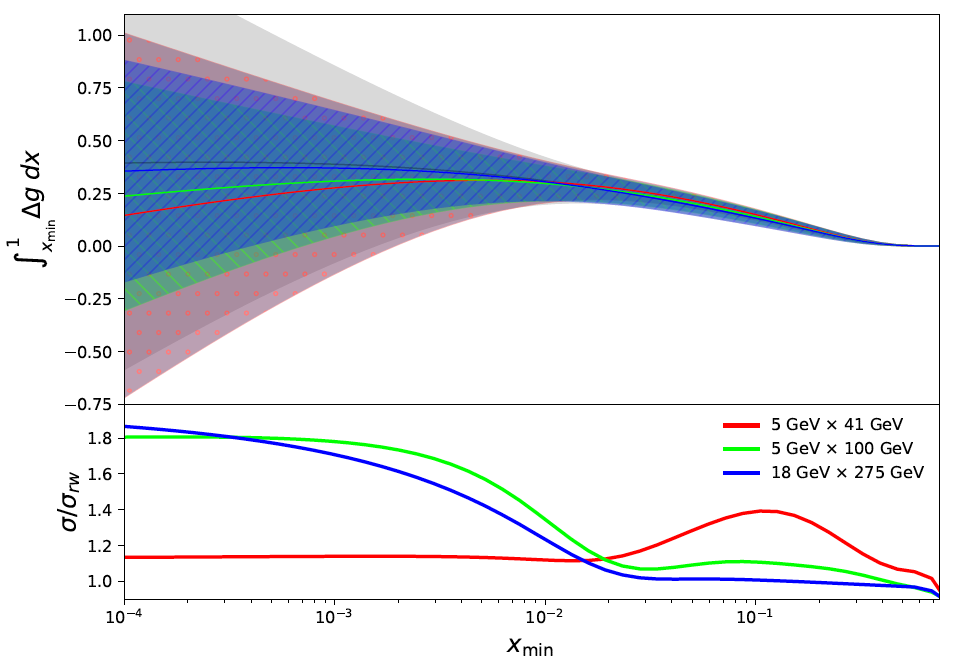}
    \caption{(Left) Projections of the double-spin asymmetry $A^c_1$ in the $\protect\overrightarrow{e} + \protect\overrightarrow{p} \rightarrow e + D^0 + X$ process in bins of $x_B$ for different beam-energy configurations. The integrated luminosity is 100\,fb$^{-1}$ for each configuration in this plot. The electron (proton) beam polarization is assumed to be 80\% (70\%). The position of each data point in the plot is defined by the weighted center of $x_B$ and $Q^2$ for each particular bin. The uncertainty indicated for each data point should be interpreted using the scale shown on the right-side vertical axis of the plot. (Right) Integrals of gluon helicity DSSV distributions as a function of the lower integration limit $x_{\rm min}$ at $Q^2$ = 10\,GeV$^2$. The grey band shows the uncertainty from the original DSSV errorbands, and the red (green, blue) band shows the updated uncertainty by adding EIC pseudo-data. The lower panel shows the ratio between the uncertainties before and after adding the EIC pseudo-data.}
\label{fig:A1c}
\end{figure*}

Future EIC experiments utilizing a precision silicon tracker in parallel with a large dataset will enable this measurement. For example, the results of a recent study on statistical projections of the double-spin asymmetry $A^c_1$ for different beam-energy configurations with 100\,fb$^{-1}$ integrated luminosity~\cite{Anderle:2021hpa} is shown in  \autoref{fig:A1c} left plot. The measurement will provide complementary constraints on the gluon helicity distribution. The right plot in Fig. \autoref{fig:A1c} shows the impact study on the integral of the DSSV gluon helicity distribution as a function of $x_{\rm min}$ using the Bayesian reweighting technique with the current uncertainty band and with the uncertainty bands by adding the pseudo-data from future EIC $A_{\rm LL}$ measurements~\cite{Anderle:2021hpa}.
In the moderate $x$ region, heavy flavor production will offer a unique opportunity. Especially, for a machine that can be operated at the lowest possible center-of-mass energies, it is reasonable to expect significant improvements in the precision of gluon helicity distribution in the $x >$ 0.1 region.

\subsection{Heavy meson tomography of cold nuclear matter at the EIC}

An important part of the physics program at the future electron-ion collider is to understand the nature of hadronization and the transport of energy and matter in large nuclei. Open heavy flavor production in DIS  provides a new tool to address these critical questions. Jets and heavy flavor have been important probes of large nuclei in heavy ion collisions~\cite{Vitev:2006bi,Kang:2014hha} and can also be used in deep inelastic scattering. 

For example, the first calculation of  $D$-mesons and $B$-meson cross sections in electron-nucleus collisions at the EIC, including both next-to-leading order QCD corrections and cold nuclear matter effects, was performed. The theoretical formalism employs generalized DGLAP evolution  to include the contribution of in-medium parton showers. The approach is based on  methods developed in SCET with Glauber gluons that describe inclusive hadron production in reactions with nucleons and  nuclei~\cite{Li:2020zbk}. This comprehensive study  provides the ability to identify optimal observables, center-of-mass energies, and kinematic regions most sensitive to the physics of energy loss and hadronization at the EIC.

To demonstrate the utility of heavy flavor for cold nuclear matter tomography, the authors of this study carried out a comprehensive investigation of the production of various  $D$-mesons and $B$-meson states at different center-of-mass energies and different rapidity ranges at the EIC. The in-medium corrections to the full splitting functions have been calculated to first order in opacity for both massless and massive partons~\cite{Ovanesyan:2011kn,Ovanesyan:2011xy,Kang:2016ofv}. The  fragmentation function evolution in the presence of nuclear matter is then given by
\begin{equation}
 \label{eq:fullevol}
    \frac{d}{d \ln \mu^{2}} \tilde{D}^{h/i}\left(x, \mu\right)= 
    \sum_{j} \int_{x}^{1} \frac{d z}{z}  \tilde{D}^{h/j}\left(\frac{x}{z}, \mu\right)  
   \left( P_{j i}\left(z, \alpha_{s}\left(\mu\right)\right) +  P_{j i}^{\rm med}\left(z, \mu\right)  \right)  \, ,
\end{equation}
where $ P_{j i}^{\rm med}$ are the in-medium contributions to Altarelli-Parisi. This approach has been applied successfully to  heavy ion collision to predict the suppression of hadron production. To investigate the nuclear medium effects, the authors studied the ratio of the cross sections in electron-gold (e+Au) collision to that in e+p collisions.  The cross section of inclusive jet production is used for normalization such that the effect of nuclear PDFs is minimized. This results in
\begin{equation}\label{eq:defRAatEIC}
R_{eA}^{h}(p_T,\eta,z)=\frac{N^{h}(p_T,\eta,z)}{N^{\rm inc}(p_T,\eta)}\Big|_{\rm e+ Au} \Bigg/ \frac{N^{h}(p_T,\eta,z)}{N^{\rm inc}(p_T,\eta)}\Big|_{\rm e+p} \, ,
\end{equation}
where $N^{\rm inc}(p_T,\eta)$ denotes the cross section of large radius jet production with transverse momentum  $p_T$ and rapidity $\eta$.

\begin{figure*}[!t]
    \centering
	\includegraphics[width=0.44\textwidth]{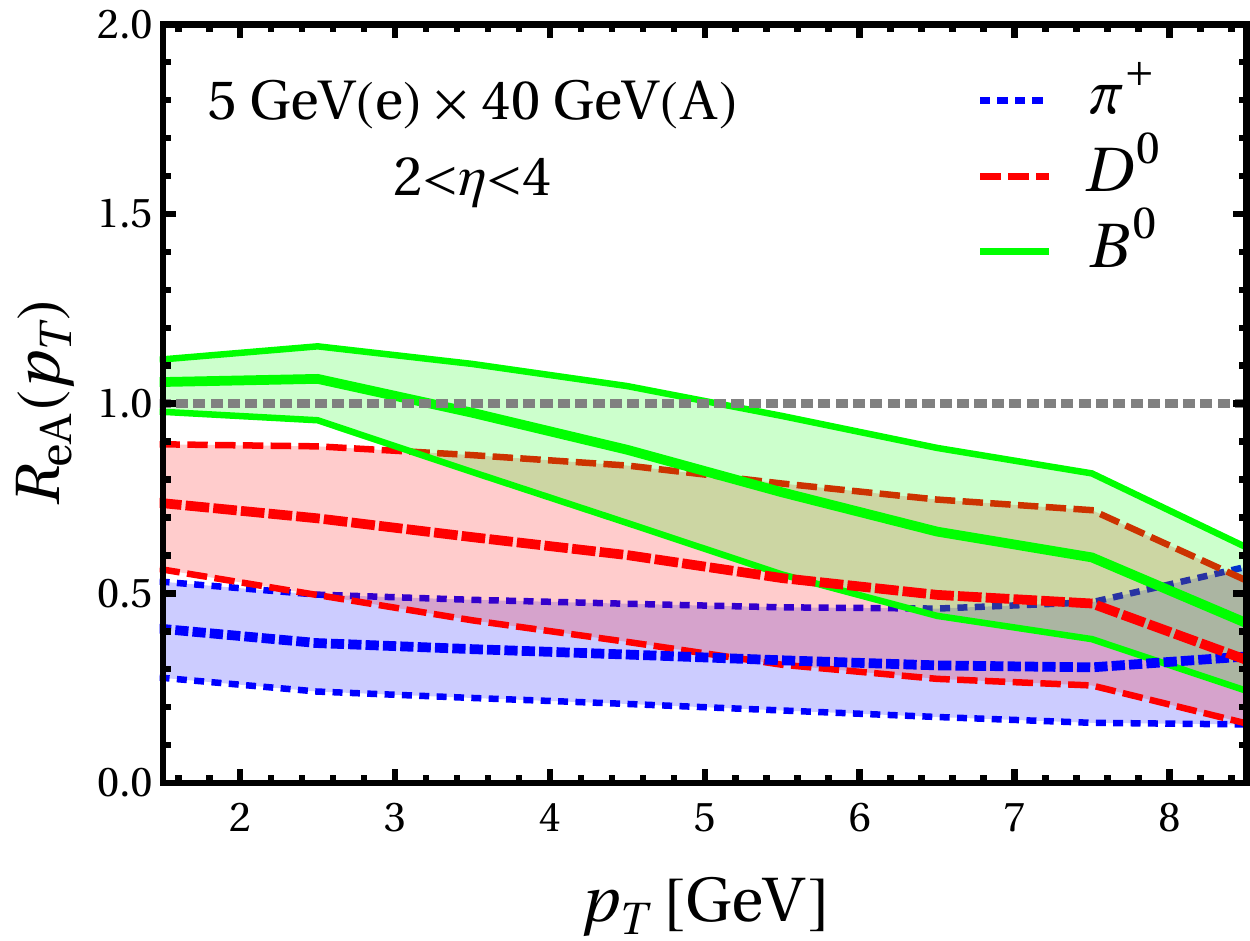}	\,\,\,
	 	\includegraphics[width=0.44\textwidth]{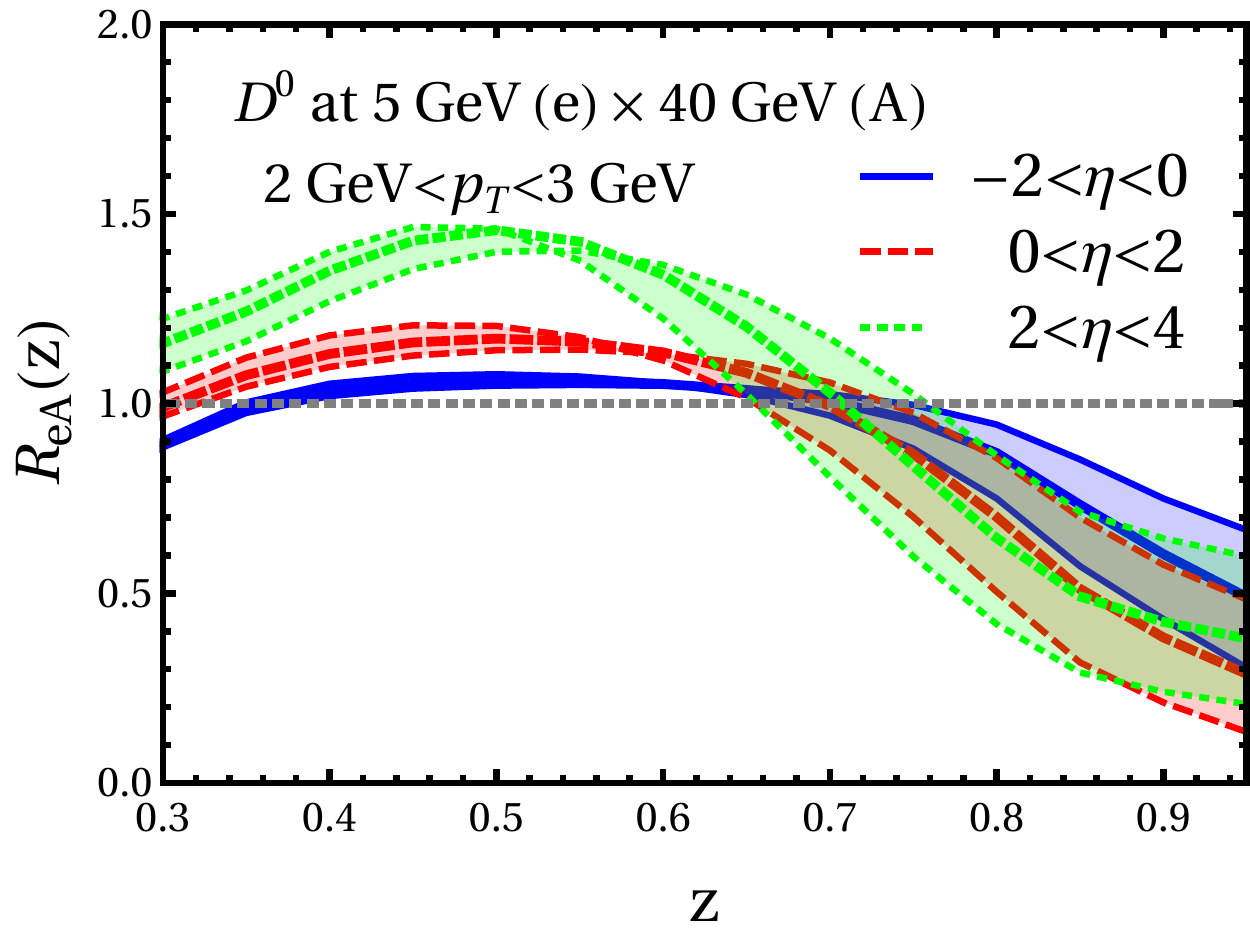}\, 
    \caption{ Left: medium modification of  $\pi^+$, $D^0$ and $B^0$ production on a gold (Au) nucleus at the EIC as a function of transverse momentum for 5 GeV (e)  $\times$ 40 GeV (p) collisions at forward rapidity. Right: in-medium correction for $D^0$  as a function of the momentum fraction $z$ at the EIC in three  rapidity intervals.}
 \label{fig:ptdisEIC}
\end{figure*}

The modification of light and heavy flavor hadron cross sections in reactions with nuclei is sizable and  depends on the electron and proton/nucleus beam energy combinations and the rapidity gap between the produced hadron and the target nucleus. Numerical results show that  the 5~GeV $\times$ 40~GeV  scenario  followed  by  the 10~GeV $\times$ 100~GeV  case and the forward  proton/nucleus going rapidity region   2$<\eta<$4  produce the largest nuclear effects. Results for  $R_{eA}$ for $\pi^+$, $D^0$, and $B^0$ are shown in the left panel of Fig.~\ref{fig:ptdisEIC}. In the future it will be interesting to study the attenuation of di-hadrons and the related modification of their correlations~\cite{Qiu:2003pm,Kang:2011bp} in the large Bjorken-x regime.  
Conversely, semi-inclusive hadron production at large center-of-mass energies, e.g. 18~GeV $\times$ 275~GeV,  and  backward  rapidities, e.g. $-2<\eta<0$, exhibits only small modification in e+A reactions relative to e+p ones. Such kinematics are better suited to explore leading and higher twist shadowing~\cite{Eskola:2016oht,Kovarik:2015cma,Qiu:2003vd,Qiu:2004da,Qiu:2004qk} and the phenomenon of gluon saturation.    
 
This effort has resulted in the identification of experimental observables  that are most sensitive to the details of hadronization. While $p_T$ distributions in the laboratory frame can provide initial information on the quenching of hadrons in cold nuclear matter,  a more differential observable such as the fragmentation fraction, $z$, distribution measured by HERMES is a much better choice, especially for open heavy flavor.  The clear transition from enhancement to suppression at moderate to large values of $z$ will be an unambiguous and quantitative measure of parton shower formation in large nuclei. This is clearly seen in the right panel of \autoref{fig:ptdisEIC}.

\subsection{Heavy flavor tagged jets and jet substructure at the EIC}

 Heavy flavor-tagged jets are complementary probes of the partonic composition and transport coefficients of large nuclei. This approach introduces a new mass scale that modifies the structure of parton showers and must be carefully accounted for in perturbative calculations.  Progress in the framework of SCET with Glauber gluon interactions have allowed  for the first calculation of inclusive charm-jet and bottom-jet cross sections in electron-nucleus collisions at next-to-leading order~\cite{Li:2021gjw}. Furthermore, predictions for the heavy flavor-tagged  jet momentum sharing distributions to further clarify the correlated in-medium modification of jet substructure~\cite{Li:2017wwc}. 

Recently, examples of first calculations of semi-inclusive charm-quark  jet and bottom-quark jet production and substructure  in e+A relative to e+p collisions at the EIC have appeared in the literature~\cite{Li:2021gjw,Li:2021lbj}. The formalism described here allowed for the determinatino of NLO-level results by consistently combining the  parton level cross sections and semi-inclusive jet functions up to  NLO. This also included re-summation for small jet radii in electron-hadron reactions.  

It was found that heavy flavor-tagged jet production is more sensitive to the gluon and sea quark distributions in nucleons and nuclei in comparison to light jets. Thus, in kinematic regions where $R_{eA}$ is dominated by initial-state  nPDF effects the modification was even stronger when compared to inclusive jets. 

\begin{figure}
	\centering
	\includegraphics[width=0.44\textwidth]{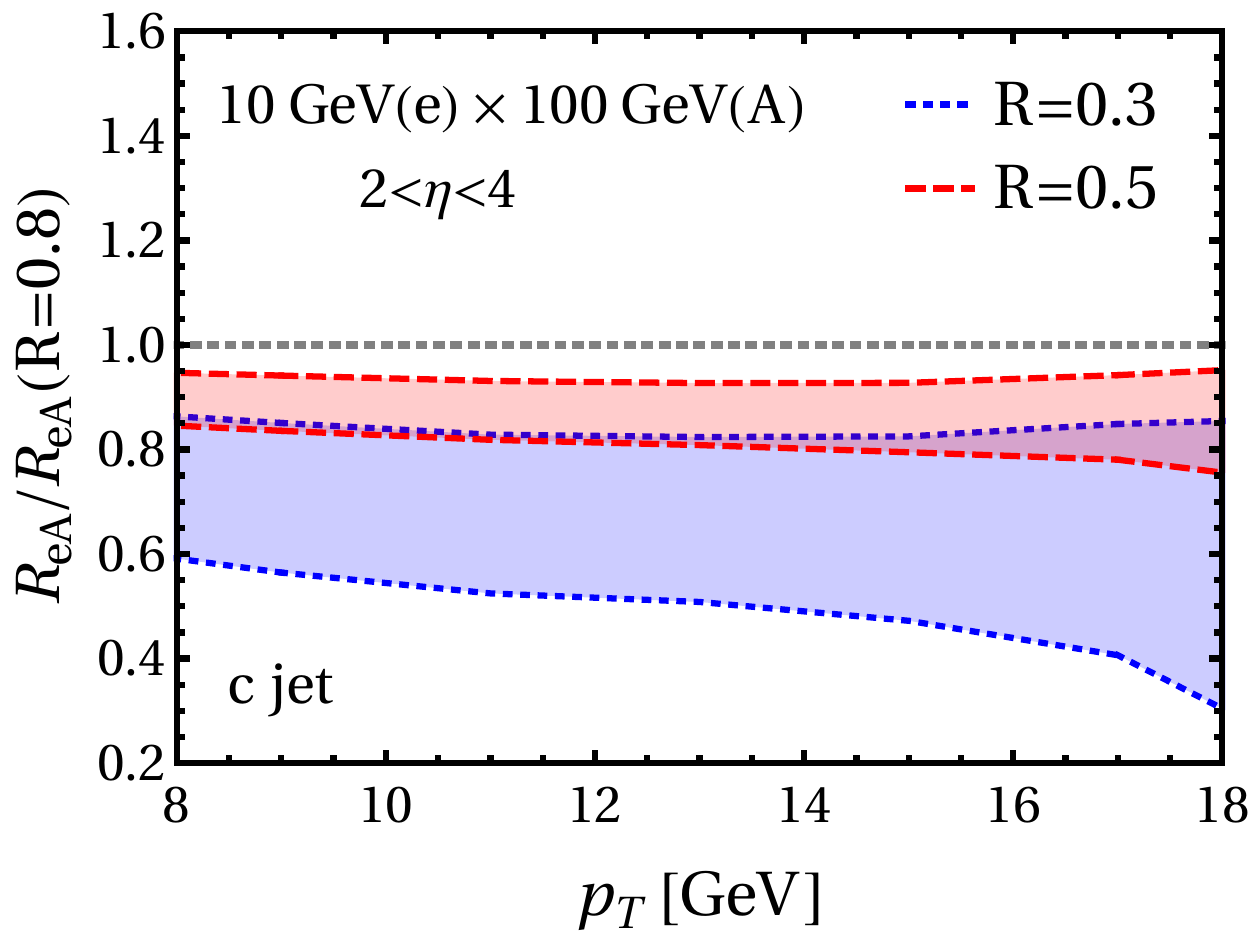}\,\,\,
	\includegraphics[width=0.44\textwidth]{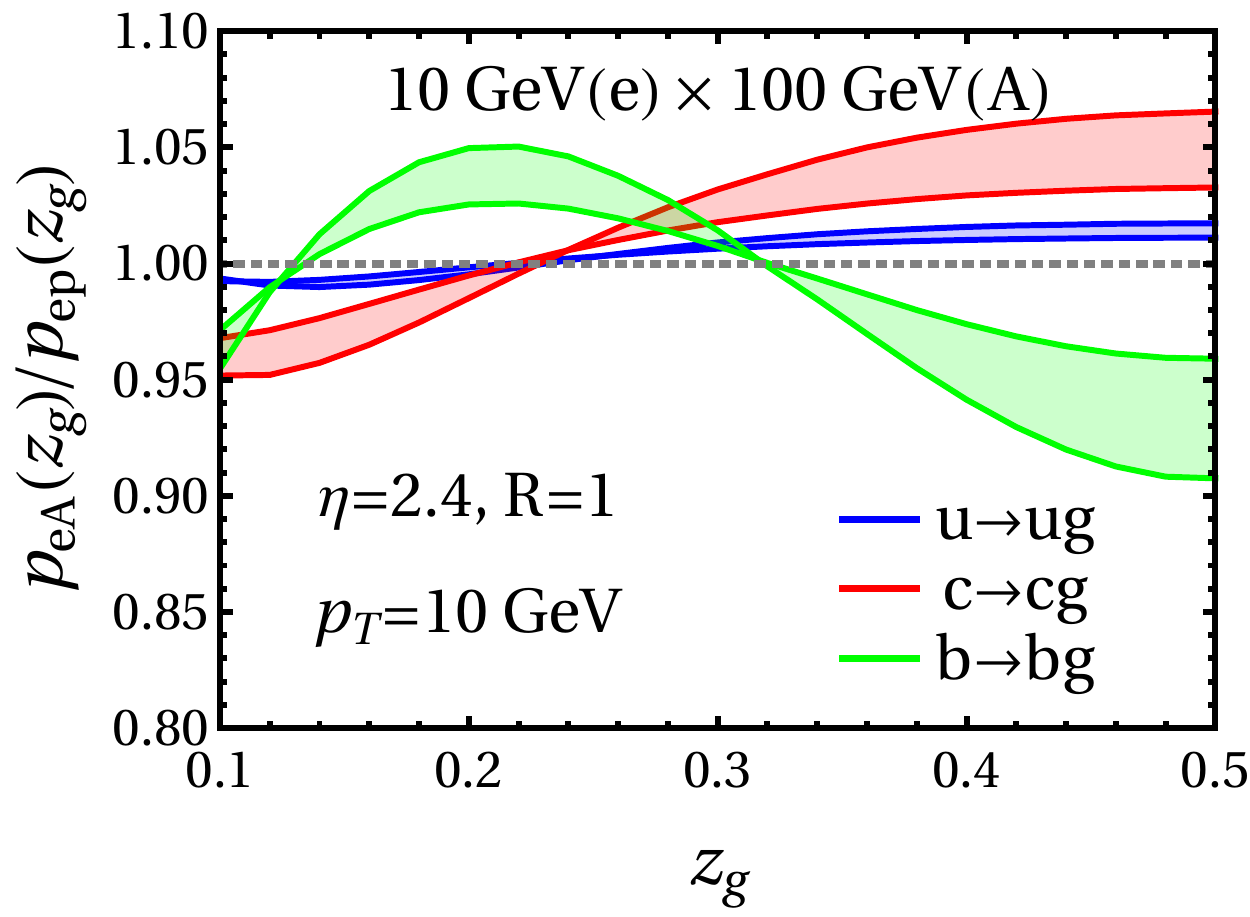} 
	\caption{Left: the ratio of $R_{eA}$ normalized by $R_{eA}(R=0.8)$ for c-jets and b-jets as a function of $p_T$ at the EIC. Blue bands (dotted lines) and red bands (dashed lines) correspond to  $R=0.3$ and $R=0.5$, respectively. Here,  $2< \eta < 4$ for 10 GeV $\times$ 100 GeV collisions. Right: the modification of the jet splitting functions for c-jets and b-jets vs $z_g$ at the EIC.}
	\label{fig:bcj_EIC_nPDF}
\end{figure}

As in the case of light jets, by applying the strategy of studying ratios of the nuclear modification with two different jet radii it is possible to eliminate nPDF effects, primarily the anti-shadowing and the EMC effect in the regions of interest.  The remaining quenching of the jet spectra can be as large as a factor of two for small jet radii, for example $R=0.3$, and can clearly be attributed to final-state interactions and in-medium modification of parton showers containing heavy quarks. This suppression is  comparable to the one predicted for light jets and expected to be observed in the proton/nucleus going direction.  This is illustrated in the left panel of \autoref{fig:bcj_EIC_nPDF}.

In contrast, near the mid-rapidity region and at backward rapidity the  deviation of $R_{eA}(R)/R_{eA}(R=0.8)$  from unity is small since the energy of the parton/jet in the rest frame of the nucleus is very large. This, in turn, strongly reduces the contribution of in-medium parton shower due to the non-Abelian LPM effect. In fact, even at forward rapidity and smaller center-of-mass energies  the parton energies in nuclear rest frame are quite sizeable and, therefore, there isn't much difference in the suppression of c-jets and b-jets.

 Calculation of semi-inclusive jet cross sections are complemented by  a calculation of the groomed, soft-dropped momentum-sharing distribution. The results cited here show that the substructure modification in e+A relative to e+p reactions is relatively small --- on the order of 10\% or smaller. Still, just as in the case of heavy ion collisions at relatively small transverse momenta the differences in the subjet distribution are most pronounced for b-jets, followed by c-jets, as shown in the right panel of \autoref{fig:bcj_EIC_nPDF}. It will be important to extend the studies  of heavy flavor jets to other substructure observables in the future~\cite{Lee:2019lge}. 
 
 In the kinematic regime accessible at the EIC the modification of light jets was found to be the smallest.  In contrast to the heavy ion case, however, there is a significant difference between the energy of the parton in the rest frame of the nucleus  and the jet scale which determines the available phase space for substructure even for large radii $R\sim 1$. Thus, the  jet  momentum sharing distribution at the  EIC probes a different interplay between the heavy quark mass and  suppression of small-angle medium-induced radiation  --- a regime that can only be accessed at the EIC and merits further investigation in the future. Using the theoretical tools that are becoming available it is possible to look at how sub-eikonal corrections to in-medium branching, such as the effects of varying matter density~\cite{Sadofyev:2021ohn}, propagate into the observables that we predicted in this work.

\subsection{Quarkonium production at the EIC}

\newcommand{\Q}[4]{ {}^{#1} #2 ^{[#4]}_{#3} }
\newcommand{\bmat}[1]{{\boldsymbol{#1}}_{T}}

\begin{figure}
\begin{center}
\includegraphics[width= 0.9 \textwidth]{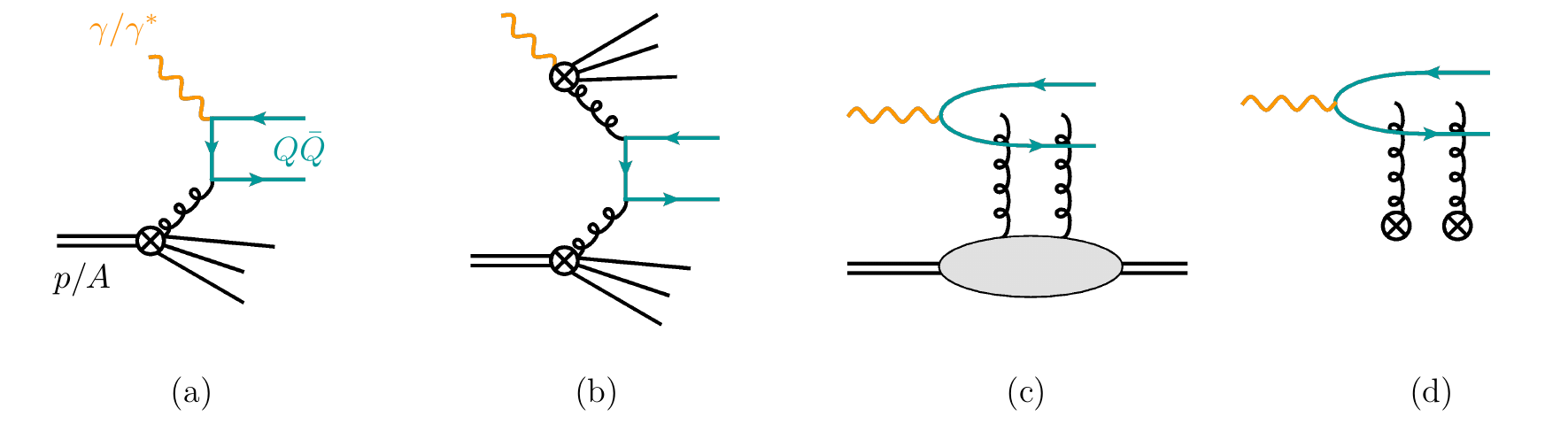}
\caption{\label{fig:quarkonium} Illustrative examples of quarkonium production mechanism in $ep$ and $eA$ colliders: (a) Direct photo/lepto-production, (b) resolved-photon quarkonium production, (c) exclusive quarkonium production, and (d) heavy quark pair production and subsequent Glauber/Coulomb gluon exchanges with nuclear matter. }
\end{center}
\end{figure}

Quarkonia, the bound states of a heavy quark and the corresponding antiquark, have been extensively studied  in collider physics as probes to various aspects of QCD phenomenology.  Nevertheless, the details of their production mechanisms  remain poorly understood. Since the discovery of quarkonia, three main production formalisms have been proposed: 1)~the color evaporation model (CEM)~\cite{Fritzsch:1977ay, Halzen:1977rs}, 2)~color singlet model (CSM)~\cite{Chang:1979nn}, and 3)~the effective theory of non-relativistic-QCD (NRQCD)~\cite{Bodwin:1994jh}.  While all these frameworks enjoyed partial success, NRQCD is the only framework for which systematic corrections can be incorporated order-by-order in a rigorous manner.  In NRQCD, quarkonia are produced when the heavy-quark pair hadronizes into the colorless state with rates that depend on the angular momentum and color configuration of the pair. This is expressed in terms of long distance (non-perturbative) matrix elements (LDMEs) that need to be extracted from experiment and are considered universal.

At the EIC quarkonia can be produced  either through photo-production ($Q^2 \simeq 0$) or lepto-production ($Q^2> 1$ GeV) processes. In these two cases the resolved, diffractive/exclusive, and inclusive productions (see \autoref{fig:quarkonium}(a), (b), and (c) respectively) can be relevant. The various production channels can be disentangled by considering different kinematic regimes, establishing this way DIS as a prime framework for the study quarkonium production channels.  To accomplish this, a crucial role will play the high luminosity expected to be achieved at the EIC.  

One of the most promising applications of quarkonia in the EIC is as probes to hadronic tomography and particularly the gluon content of the nucleons and nuclei.  An example that has recently gathered attention is the gluon TMDs. At small transverse momentum, $q_T \ll Q$ (measured with respect to the beam in the Breit frame) quarkonium  production can be approached from the TMD factorization perspective. In this approach there have been several studies considering both polarized and unpolarized proton beams~\cite{Bacchetta:2018ivt, DAlesio:2019qpk, Kishore:2019fzb, Boer:2020bbd, Echevarria:2020qjk}.  Recent theoretical developments in NRQCD~\cite{Beneke:1997qw,  Fleming:2002rv,  Fleming:2003gt, Fleming:2006cd,  Leibovich:2007vr, Echevarria:2019ynx, Fleming:2019pzj} incorporate the leading perturbative effects from soft radiation in all orders in the strong coupling expansion. This provides a way to safely study the non-perturbative effects that can be accessed in the  semi-inclusive process in the $q_T \to 0$ limit. In this limit the cross section can be expressed in terms of quarkonium TMD-shape functions, $Sh(\bmat{q}, [n])$,
\begin{equation}
 \frac{d\sigma}{d^2 \bmat{q}} = \sigma_0 ([n])\,H(2 m_Q,\mu;[n]) \int d^2 \bmat{k} \,F_{g/P}(x,\bmat{k}) \,Sh(\bmat{q} - \bmat{k};[n])\,,
 \end{equation}
where $[n] \in \{ \Q{1}{S}{0}{8}, \Q{3}{P}{0,2}{8}\}$. Results in this approach are obtained by by fusing NRQCD and SCET.  To date there has been no phenomenological extraction of the TMD shape functions; thus, the EIC will offer a unique opportunity to access those hadronic matrix elements.

Beyond hadronic tomography, in $AA$ and $pA$ collisions quarkonia has been an important tool for accessing the properties of quark-gluon-plasma.  Recent field-theoretic developments, which also rely on the effective theory of NRQCD,  reformulate NRQCD~\cite{Makris:2019ttx,Rothstein:2018dzq} to incorporate the Glauber/Coulomb gluon interactions with heavy quarks in the non-relativistic limit (\autoref{fig:quarkonium} (d)). At the level of the Lagrangian, 
\begin{equation}
\mathcal{L}_{\text{NRQCD-G}} = \mathcal{L}_{\text{NRQCD}}(\chi,\phi)  + \mathcal{L}_{G/C} (A_{G/C}^{\mu},\chi,\phi)
\end{equation}
where $G/C$ stands for Glauber/Coulomb gluons and the coupling of those depends on the properties of the medium. This provides a systematic and formal approach to the inclusion of nuclear effects. Collisional dissociation was shown to lead to a significant and hierarchical suppression of the $J/\Psi$ and $\Upsilon$ states in heavy ion collisions~\cite{Sharma:2012dy,Andronic:2015wma,Aronson:2017ymv}.  The EIC  will provide a complementary environment to those of $AA$ and $pA$ where we will  have the opportunity to observe quarkonium production in $eA$ collisions where one can study the interactions with nuclear matter and the formation of quarkonia in a nuclear medium.

\subsection{Fluctuations and nuclear matter properties}

The nuclear matter produced in heavy-ion collisions (HIC) or probed in DIS is not static but rather evolves in space and time.  Recently,  there were attempts to include the effects of these in-medium processes to the jet energy loss and jet modification calculations in a variety of models,  see e.g.  \cite{Lekaveckas:2013lha, Rajagopal:2015roa, Brewer:2018mpk,Reiten:2019fta}.  A first principle consideration of the effects of the medium motion and in-medium fluctuations  was presented recently, focusing on the jet broadening and soft gluon radiation within the pQCD framework of the Gyulassy-Levai-Vitev (GLV) opacity expansion formalism~\cite{Sadofyev:2021ohn}.  This approach treats the sub-eikonal motion of the medium scattering centers and compute the corresponding modification of the collisional and radiative processes.  It takes into account the change in the scattering center transverse momentum and parameters of the matter between different centers.  This results in a determination of the coupling between the motion of the medium scattering centers and the pattern of medium-induced radiation,  taking a significant step toward full-fledged jet tomography.  

\begin{figure}[!t]
    \centering
	\includegraphics[width=0.45\textwidth]{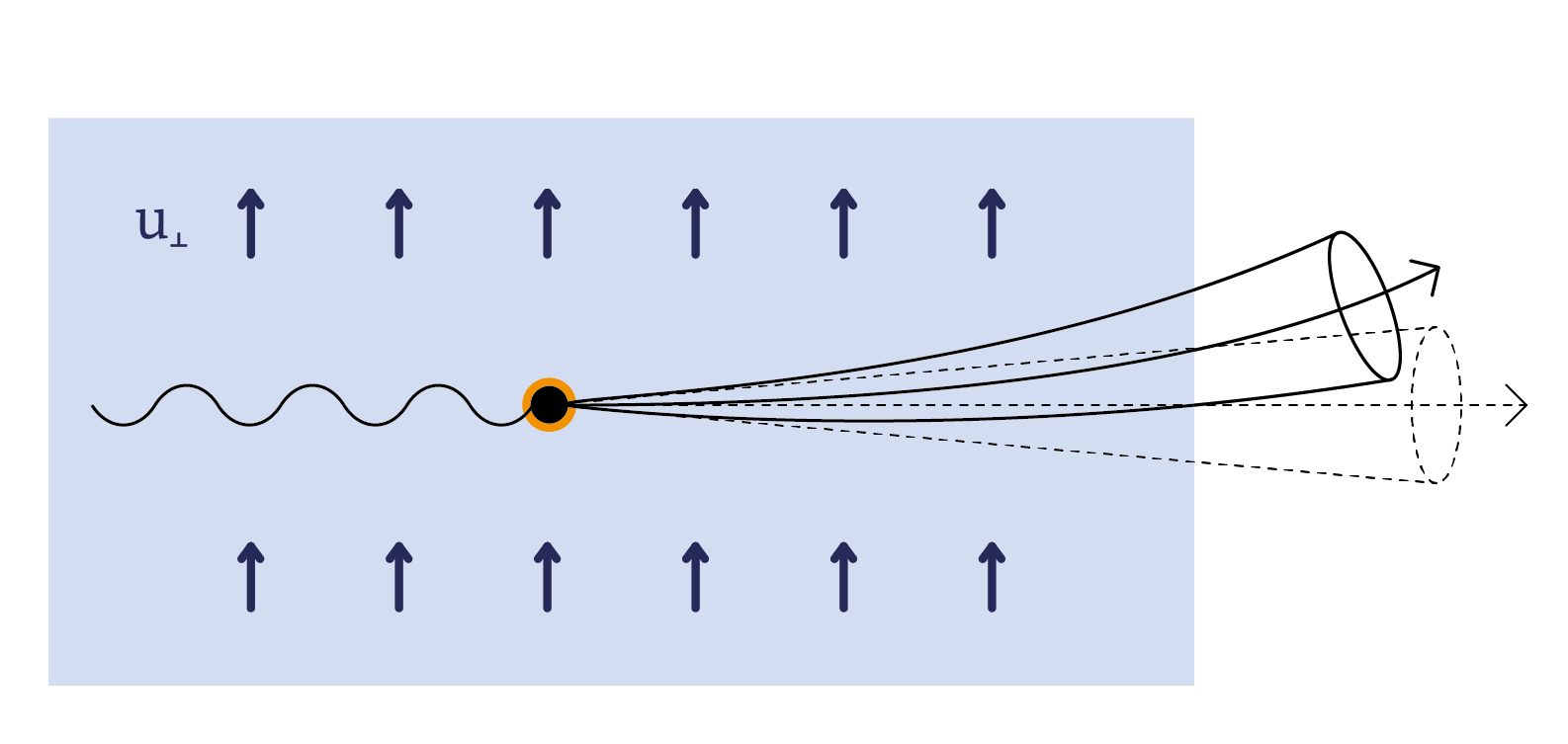}
	\includegraphics[width=0.45\textwidth]{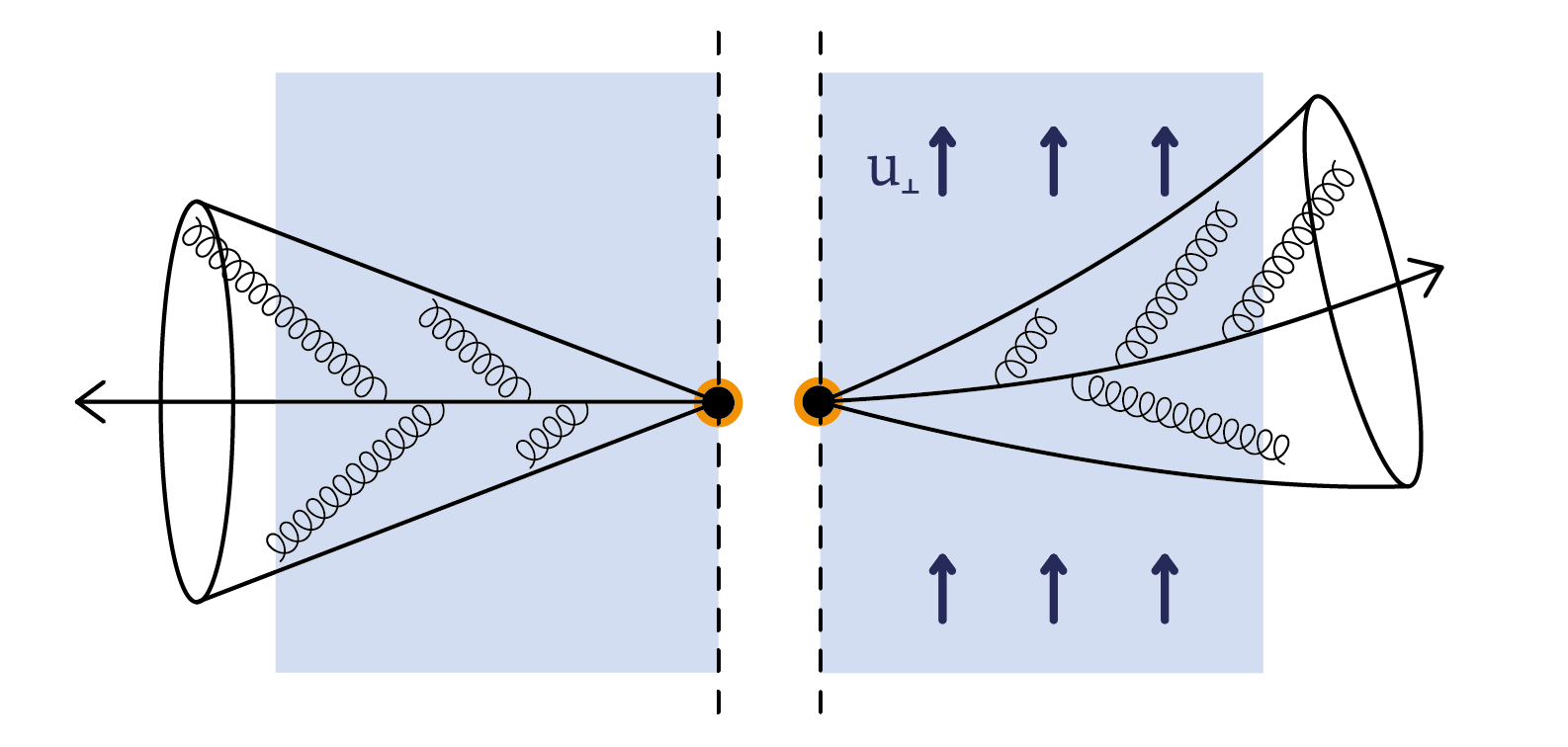}
	\caption{The coupling of jets and their soft radiation to the medium velocity presented in this work.  Left:  The transverse momentum of the jet is deflected in the direction of the medium velocity.  Right: The medium-induced radiation is emitted preferentially in the direction of the medium velocity as schematically shown by the number of gluons along the direction of the flow in comparison to the ``standard" symmetric unbent jet.} 
	\label{f:JETS}
\end{figure}

This example provides a new tool for the community. For example, this can be used  to show that there is no additional contribution to the root-mean-square jet broadening due to a common motion of the scattering centers at the leading sub-eikonal order.  Instead,  the odd moments of the momentum get a contribution proportional to the transverse velocity of the sources.  Thus,  in the context of HIC the most important consequence is that the jet is deflected in the direction of the medium velocity. This deflection depends on how fast the medium flows and grows with the average number of scatterings, but decreases with the jet energy.  This example also provides a modification of the radiation distribution by the medium motion; this aspect is quite detailed and documented in~\cite{Sadofyev:2021ohn}. Work to generalize these effects to DIS is ongoing.

\subsubsection{Heavy flavor cross section modification in e+A}

\begin{figure} [ht]
\centering
\includegraphics[width=0.9\textwidth]{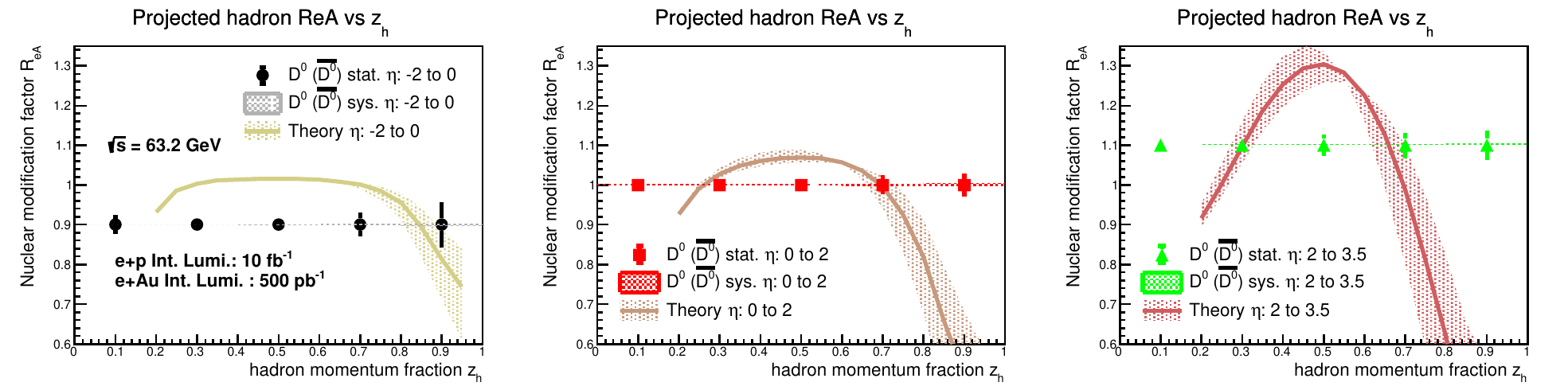}
\caption{\label{fig:recD_ReA}  Projection of reconstructed $D^{0}$ ($\bar{D^{0}}$) nuclear modification factor versus hadron momentum fraction $z_{h}$ in 10 GeV electron and 100 GeV gold collisions with the EIC detector performance including a proposed forward silicon tracker \cite{Li:2020sru}. The projections are shown in three pseudorapidity regions, -2 to 0 (left), 0 to 2 (middle) and 2 to 3.5 (right). Comparison with theoretical calculations \cite{Li:2020zbk} indicates good discriminating power can be obtained by future EIC measurements.}
\end{figure}

The EIC  will provide a clean environment to precisely study the hadronization process within vacuum and the nuclear medium. Heavy flavor hadron and jet measurements at the future EIC can provide enhanced sensitivities to the nuclear transport properties in medium compared to light flavor products \cite{Li:2020sru}. Reconstruction of heavy flavor hadrons (e.g. D-mesons) and jets in e+p collisions have been carried out in simulation studies with different detector conceptual designs for the EIC \cite{Li:2020sru,Wong:2020xtc}. The nuclear modification factor $R_{eA}$ of reconstructed heavy flavor hadrons is not only associated with the accessed nuclear parton distribution functions, but also sensitive to the final fragmentation process within a nucleus. As illustrated in \autoref{fig:recD_ReA}, the reconstructed $D^{0}$ $R_{eAu}$ measurements especially in the most forward pseudorapidity region with around one year EIC operation can provide strong discriminating power on theoretical calculations such as the parton energy loss approach \cite{Gyulassy:1999zd,Gyulassy:2000er,Gyulassy:2000fs,Fickinger:2013xwa}. These studies can constrain the initial state effects for previous and ongoing heavy ion measurements at Relativistic Heavy Ion Collider (RHIC) and the Large Hadron Collider (LHC).

\subsection{Silicon tracking and vertexing for the EIC}

Heavy flavor physics at the EIC requires advances in detector development. This is an area of common interest for HEP and NP.  The ability to reconstruct heavy flavor hadron decays with good signal significance is crucial for most of the proposed physics measurements.  Micro-vertexing and good momentum resolution, combined with low material budget are important requirements for this program.  Collisions at the EIC are asymmetric and, while many of the heavy flavor decay daughters fall in the pseudorapidity region $|\eta| < 3.5$, the pattern of hadron production is skewed in the forward proton/nucleus-going direction.  The community is working on tracker designs including an all silicon detector  and a forward silicon tracker optimized to extend forward rapidity coverage. 

\begin{figure*}[htp]
  \centering
  \includegraphics[width=0.48\textwidth]{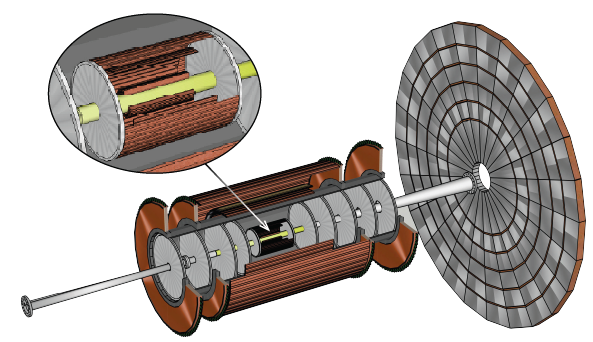}
  \hspace{3ex}
  \includegraphics[width=0.46\textwidth]{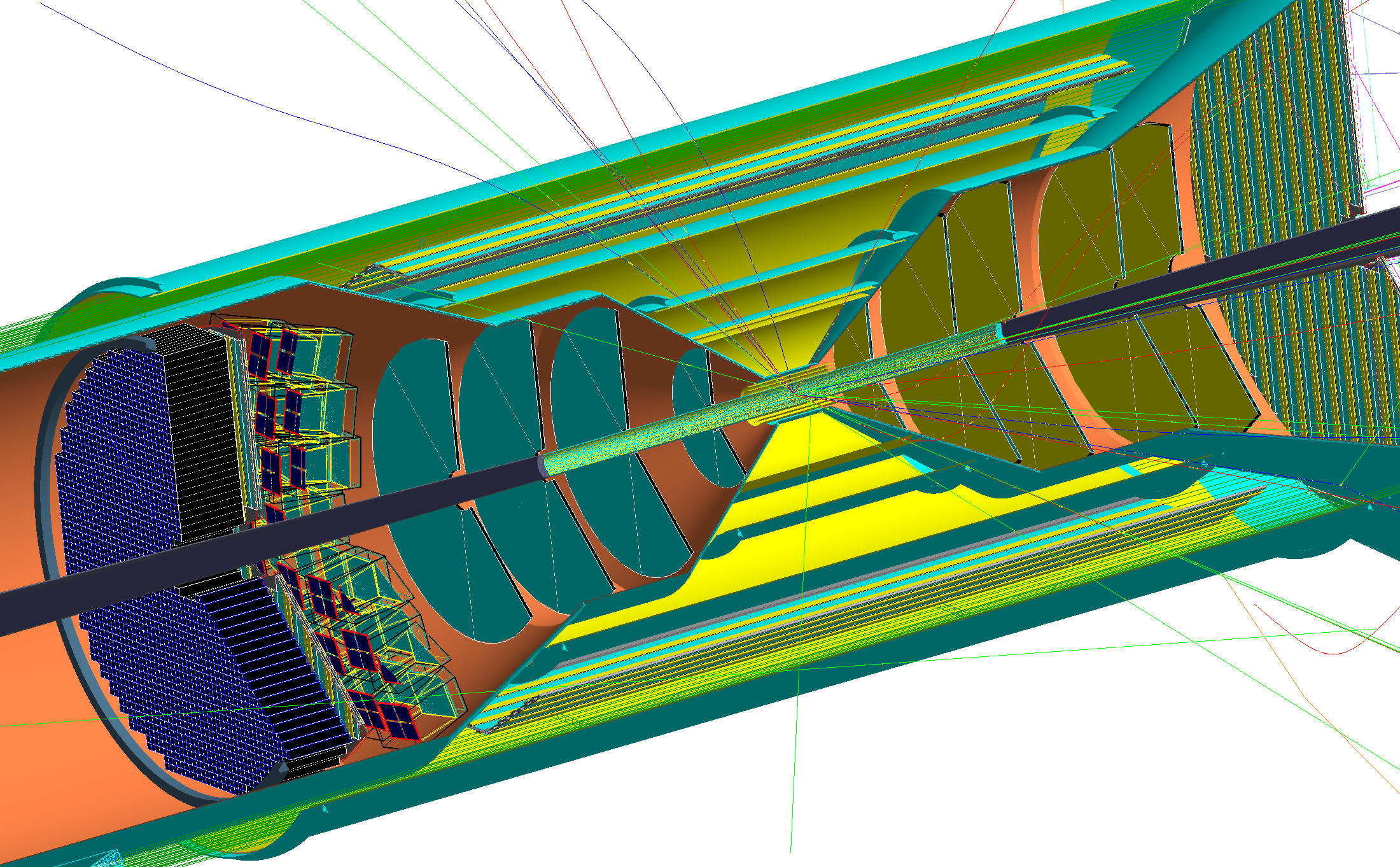}
  \caption{Tracking detector design for the proposed ATHENA (left) and ECCE (right) experiments at EIC.}
  \label{fig:tracker}
\end{figure*}

Physics simulations are essential to back up these design efforts. \autoref{fig:tracker} shows the conceptual designs of the trackers for two proposed EIC detectors: ATHENA (left) and ECCE (right). Both detectors consist a central barrel tracker including silicon vertexing/tracking layers based on MAPS sensors together with gaseous detector at larger radii, and forward/backward silicon disks that extends the tracking coverage up to $|\eta|\sim 3.5$. These trackers are designed to deliver the required tracking and vertexing performance as outlined in the EIC Yellow Report.

\subsubsection{Forward tracking studies}

Heavy flavor and jets in the EIC will help access the parton distribution at the poorly constraint kinematic region and study the nuclear medium effect on the hadronization process~\cite{Li:2020sru}. In order to measure heavy flavor and the tagged jets, a forward silicon tracker (FST) is proposed in the ion-going direction~\cite{Li:2020sru,Wong:2020xtc}. With both good timing and spatial resolutions, the FST can identify heavy flavor mesons via displaced vertex measurement of the products of heavy flavor decay $K^\pm$ and $\pi^\pm$ up to $10$~GeV. The latest FST design consists of five or six disks of silicon detectors with the first, fifth and the sixth disks located at $z=35$~cm, $z=125$~cm and $z=300$~cm, respectively. The outer radius of the FST is up to $45$~cm to cover pseudorapidity from $1$~to~$3$. Three silicon technologies, LGAD~\cite{PELLEGRINI201412,Padilla_2020}, MALTA~\cite{Pernegger_2017,Berdalovic_2018,Hiti:2019ujr} and the ALICE ITS-3 type sensor~\cite{its3det,aliceTDR}, are in consideration. The FST will implement two of these technologies to provide both good spatial and timing resolutions. 

\begin{figure}[!htb]
    \centering
    \includegraphics[height=0.17\textheight]{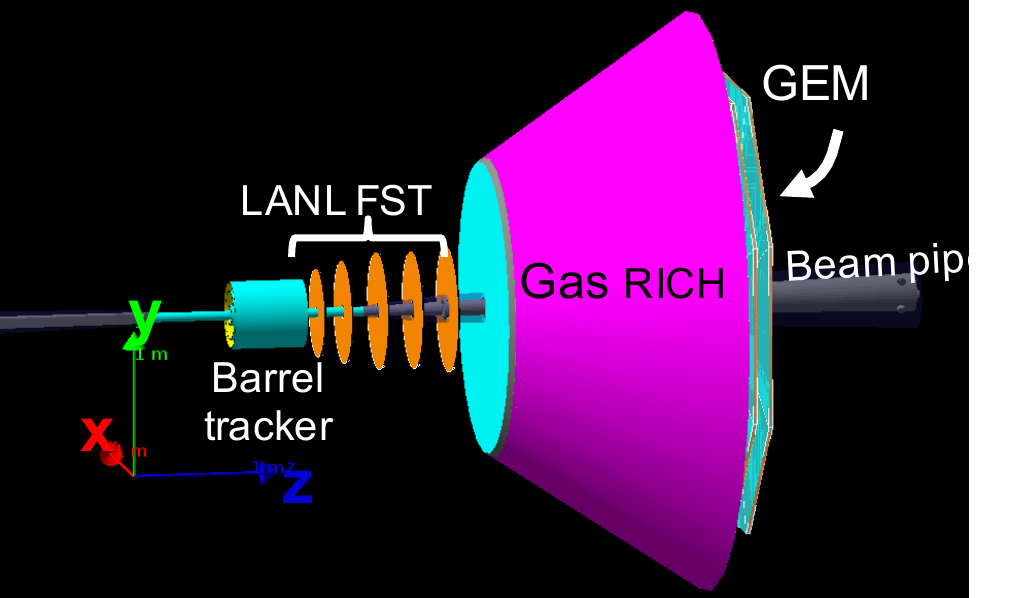}
    \includegraphics[height=0.19\textheight]{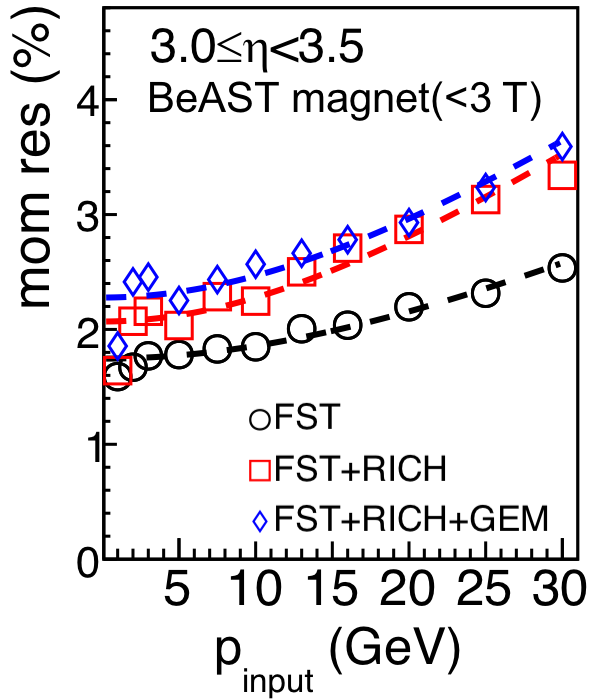}
    \includegraphics[height=0.19\textheight]{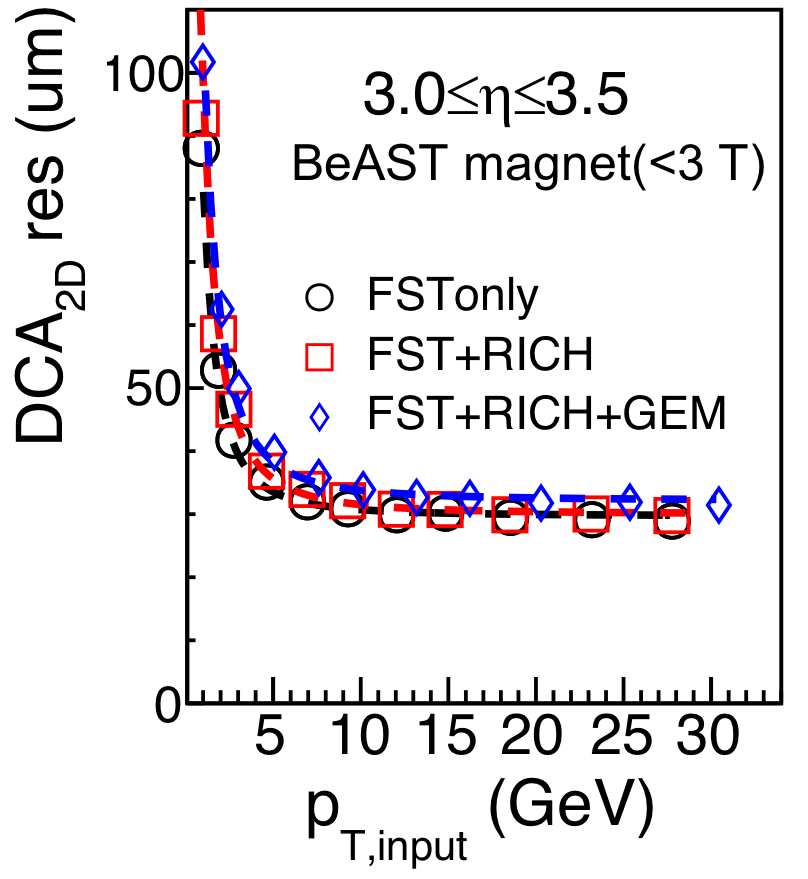}
    \caption{Left: An integrated forward tracking system with a five-disk FST, a barrel tracker, a gas RICH and a three-disk GEM. The center and the right plots are the momentum resolution as a function of true momentum and the $DCA_{2D}$ resolution as a function of transverse momentum of the forward tracking systems, respectively.}
    \label{fig}
\end{figure}

Extensive detector simulations with $\pi^-$ tracks have been carried out. The first three FST disks have a pixel pitch of $20$~$\mu$m and a silicon thickness of $50$~$\mu$m while the last three disks have a pixel pitch and a silicon thickness of $36.4$~$\mu$m and $100$~$\mu$m, respectively. Three integrated detector setups have been studied. The first setup included the six-disk FST and a barrel tracker. The second setup had the addition of a gas RICH. The third setup was similar to the second setup, but replaced the sixth disk of FST with a three-disk GEM, as shown on the left of Figure~\ref{fig}. The simulation showed that the integrated tracking systems with the FST had a momentum resolution between $0.5$\% and $4$\% when the BeAST magnetic field ($<3$~T) is used, as shown at the center of Figure~\ref{fig}. The resolutions of displaced vertex in the transverse direction was measured using distance of closet approach ($DCA_{2D}$) are below $110$~$\mu$m and $50$~$\mu$m at $1<\eta<2$ and $2<\eta<3.5$, respectively. Physics simulations with the consideration of the FST performance shows that the FST will help increase the signal-to-background ratio in heavy flavor reconstruction.

The next stage of simulation studies will include simulated physics events in detector simulation to study the heavy flavor reconstruction performance of the FST design. Moreover, two silicon sensor candidates, LGAD and MALTA, received at LANL will be tested using the LANCE facility. The testing of silicon sensors will help narrow down the selection for the FST sensors.

\subsection{Summary}

Open  heavy-flavor and quarkonium  production in deep-inelastic scattering call for new developments in theory and experiment to maximize the scientific output of the EIC. As heavy quarks introduce a new mass scale, the impact of flavor number schemes --- fixed-flavor number (FFN) scheme and variable-flavor-number (VFN) scheme –-- on charm and bottom distributions has to be better understood.

The EIC will open new windows on our understanding of parton contributions to nucleons and nuclei. For example, via neutral-current (NC) exchange in $e+p/A$ collisions at the EIC, heavy flavor production can be used to probe the initial gluon distributions inside nucleon and nucleus. This can be used to constrain the gluon (nuclear) PDF especially in the large $x_B$ region. In the charged-current (CC) interaction channel with the scattered neutrino, heavy flavor and heavy flavor jet production offer the sensitivity to the strange quark sea.  The interpretation of data from these experiments may be complicated by a subtle interplay of power-suppressed corrections, as well as potential contamination from target fragmentation. This requires significant advances in theory predictions and associated tools in order to extract valuable information (e.g. gluon and sea quark PDFs) from future data. One critical issue to be addressed is how to distinguish nPDF effects from other CNM effects through the analysis of future EIC data together with the data on heavy-flavor production at HERA and the LHC. 

Many ambiguities remain regarding the possible role of heavy quarks --- particularly charm --- in hadronic and nuclear structure.  A prime example of this is the issue of the non-perturbative or intrinsic charm contribution to the proton wave function.  As of the writing of this document there is nearly a complete lack of measurements with direct sensitivity to non-perturbative charm in the nucleon.  The ideal measurement would involve charm structure-function data in the high $x_B > 0.1$ and intermediate $Q \sim 10$~GeV region. The EIC will be ideally poised to extract this with considerable precision.  Similarly, the EIC will be well-positioned to not only constrain/isolate the presence of intrinsic charm but also to potentially determine its detailed origins in QCD.  The EIC could shed light on this subject through a detailed exploration of the scale dependence of the nucleon's charm component.  

An exciting and cross-cutting field that must be further explored is the theory of QCD in dense environments. Comparative studies of light and heavy flavor meson, baryon and jet production on e+p and e+A collisions can shed light in the process of hadronization, the time scales involved and the magnitude of non-perturbative effects.  Modern effective field theories of QCD, such as soft-collinear effective theory, have allowed us to include effects of nuclear matter on the formation of parton showers and to combine those with heavy quark mass effects. Theoretical developments in the past several years have shown that evolution in the nuclear medium can amplify the impact of charm and bottom mass on heavy flavor observables and can constrain the transport properties of large nuclei. This physics can be accessible in the forward proton/nucleus going direction. 

The polarized e+p/A collisions at the future EIC will provide opportunities to further explore the nucleon/nucleus spin structure. Double spin asymmetry measurements of heavy flavor production will offer further constraints to the gluon polarization in addition to the measurements of jets. Compared to the quark transverse momentum distributions (TMDs), our knowledge of the gluon TMDs is much less advanced. Heavy flavor production at the future EIC plays an irreplaceable role in probing gluon TMDs inside unpolarized and polarized nucleons. Processes like open or hidden heavy meson production are induced by photon-gluon fusion at the lowest order, and thus provide opportunities to measure the gluon TMDs. Measurements of charm hadron pairs will be of particular interest. Constraints to gluon Sivers asymmetry can be obtained through measurement of charm hadron pairs in polarized electron-proton collisions. Measurements of azimuthal distribution of the charm hadron pair momentum in unpolarized electron-proton collisions can be used to constrain linearly polarized gluon transverse momentum distributions. In the meantime, theoretical development in studying the QCD effects in these processes as well as their impact on the size of asymmetry will be carried out. 

Quarkonia and exotic states are another important part of the EIC program. The clean DIS environment will help assess the robustness of the NRQCD framework and facilitate better constraints on the long-distance matrix elements and $J/\psi$ and $\Upsilon$ production. It has been suggested that the Color Glass Condensate of QCD has universal properties common to nucleons and all nuclei. Study of heavy flavor production in $e+A$ collisions at the EIC could provide access to the saturation regime. The measurements are possible in two modes, exclusive and inclusive, and the energy range at EIC allows the study both productions of the charmonium and the bottomonium states, which increases the reliability of the searches.  Bound heavy-quark states that are produced in $e+A$ collisions at the EIC are also subject to disruption via interactions with partons inside the nucleus.  Since these interactions are expected to depend on the size and binding energy of the state, measurements of quarkonia suppression at the EIC can provide information on the structure of the heavy quark states.

%

%
%
%
\newpage
\section{Small-$x$ physics at the Electron-Ion Collider} \label{sec:smx}
\vspace{-2ex}
\centerline{\textit{Editors:} 
\href{mailto:renaud.boussarie@polytechnique.edu}{\texttt{Renaud Boussarie}},
\href{mailto:tuomas.v.v.lappi@jyu.fi}{\texttt{Tuomas Lappi}},
\href{mailto:salazar@physics.ucla.edu}{\texttt{Farid Salazar}}, 
\href{mailto:bschenke@bnl.gov}{\texttt{Bjoern Schenke}},
\href{mailto:sschlichting@physik.uni-bielefeld.de}{\texttt{Soeren Schlichting}},
}
\vspace{3ex}

Hadrons and nuclei probed in high-energy scattering experiments feature an increasingly large number of small-$x$ gluons that populate its transverse extent, leading to a dense saturated wave function. This phenomenon of gluon saturation is a consequence of unitarity and can be quantitatively described by an effective theory of Quantum ChromoDynamics (QCD), the Color Glass Condensate~\cite{Iancu:2003xm,Weigert:2005us,Gelis:2010nm}. 
Saturation effects are important for all high-energy collisions of hadrons and nuclei. They need to be taken into account for understanding particle production and multiparticle correlations in proton-proton and proton-nucleus collisions at the LHC and at RHIC, especially at forward rapidities where one probes the smallest values of $x$ in one of the colliding hadrons. Gluon saturation is an extremely important part of  the heavy ion collision program that aims to produce and study deconfined QCD matter in high-energy heavy ion collisions. It is the key ingredient of our current theoretical understanding of the initial stage of the collision process, describing the production and equilibration of the matter that evolves to a quark-gluon plasma~\cite{Gelis:2016upa,Schlichting:2019abc,Berges:2020fwq}.

While experimental data from a large variety of hadronic collision experiments are consistent with the presence of saturation effects, the theoretical interpretation of the experimental signals is always affected by complexities inherent to QCD: hadronization and final state interactions, in particular the whole spacetime evolution of the quark gluon plasma in the case of heavy ion collisions. This situation can only be improved with a program of high-energy deep inelastic scattering on proton and nuclear targets, which give a more direct and precise access to the partonic constituents of protons and nuclei. Such measurements will be performed  at the Electron-Ion Collider (EIC)~\cite{AbdulKhalek:2021gbh}.

\subsection{EIC observables for small-$x$ physics}
Gluon saturation will affect many of the different cross sections measured at the EIC. The CGC effective theory framework provides a way to perform a coherent global analysis of different kinds of scattering processes at small $x$ that provide complementary information on the small-$x$ gluons. Many specific measurements at the EIC  will shed more light on different aspects of the physics of gluon saturation~\cite{Boer:2011fh,Accardi:2012qut,Aschenauer:2017jsk}.

\begin{figure}[tbh]
    \centering
    \includegraphics[scale=0.7]{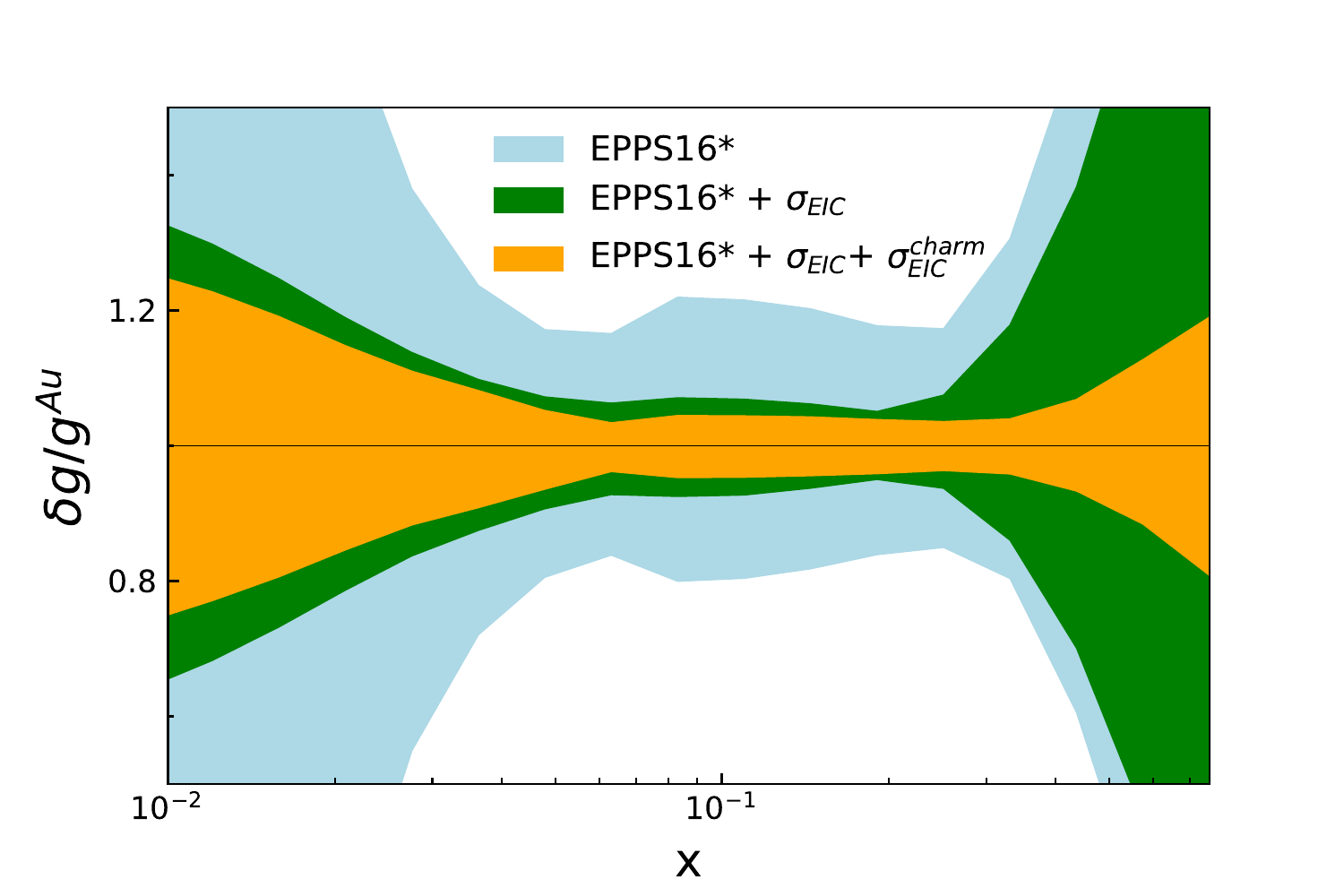}
    \caption{Estimated reduction in the uncertainty for the gluon distribution in Au nuclei at $Q^2=1.69$GeV$^2$ with EIC data.  Figure obtained from \cite{AbdulKhalek:2021gbh}.}
    \label{fig:smallx-EICpdf}
\end{figure}

Inclusive cross sections for electron-nucleus and electron-proton scattering provide the baseline for our understanding of the partonic structure of QCD bound states. The parton densities (parton distribution functions, PDFs, in the collinear factorization picture) extracted from inclusive measurements are required for perturbative calculations of hard QCD probes in all collisions of protons and nuclei. 
The EIC will provide a significant improvement over previous measurements especially for nuclei, where it extends the available kinematical reach towards small~$x$ by more than an order of magnitude. In the perturbative collinear factorization framework this will lead to a significant improvement of extractions of nuclear PDFs (see Fig.~\ref{fig:smallx-EICpdf}). Nuclear targets are particularly important for the physics of gluon saturation, since nonlinear phenomena in high-energy scattering are enhanced by the higher parton density (per unit transverse area) in a heavy nucleus. This enhancement is often parametrized in terms of the nuclear ``oomph'' of the saturation scale $Q_\mathrm{s}^2 \propto A^{1/3}/x^\lambda$ with $\lambda \sim 0.2 \dots 0.3$. Processes with a resolution $Q^2\lesssim Q_\mathrm{s}^2$ are most sensitive to saturation effects. Thus, with the nuclear enhancement of $Q_\mathrm{s}^2$ they are accessible at lower collision energies and higher resolution scales, i.e., are more reliably calculable in the weak coupling regime. A second major advantage of the EIC in comparison to earlier experiments is the access to the transverse and longitudinal structure functions $F_T$ and $F_L$ separately, obtained by combining measurements at different collision energies. In the collinear framework this provides additional constraints that enable a better disentangling of gluon and sea quark PDFs. In the dipole picture that is appropriate for the saturation regime, the longitudinal structure function is a more reliably weak coupling quantity than the transverse one. 

\begin{figure}[tbh]
    \centering
    \includegraphics[width=0.9\textwidth]{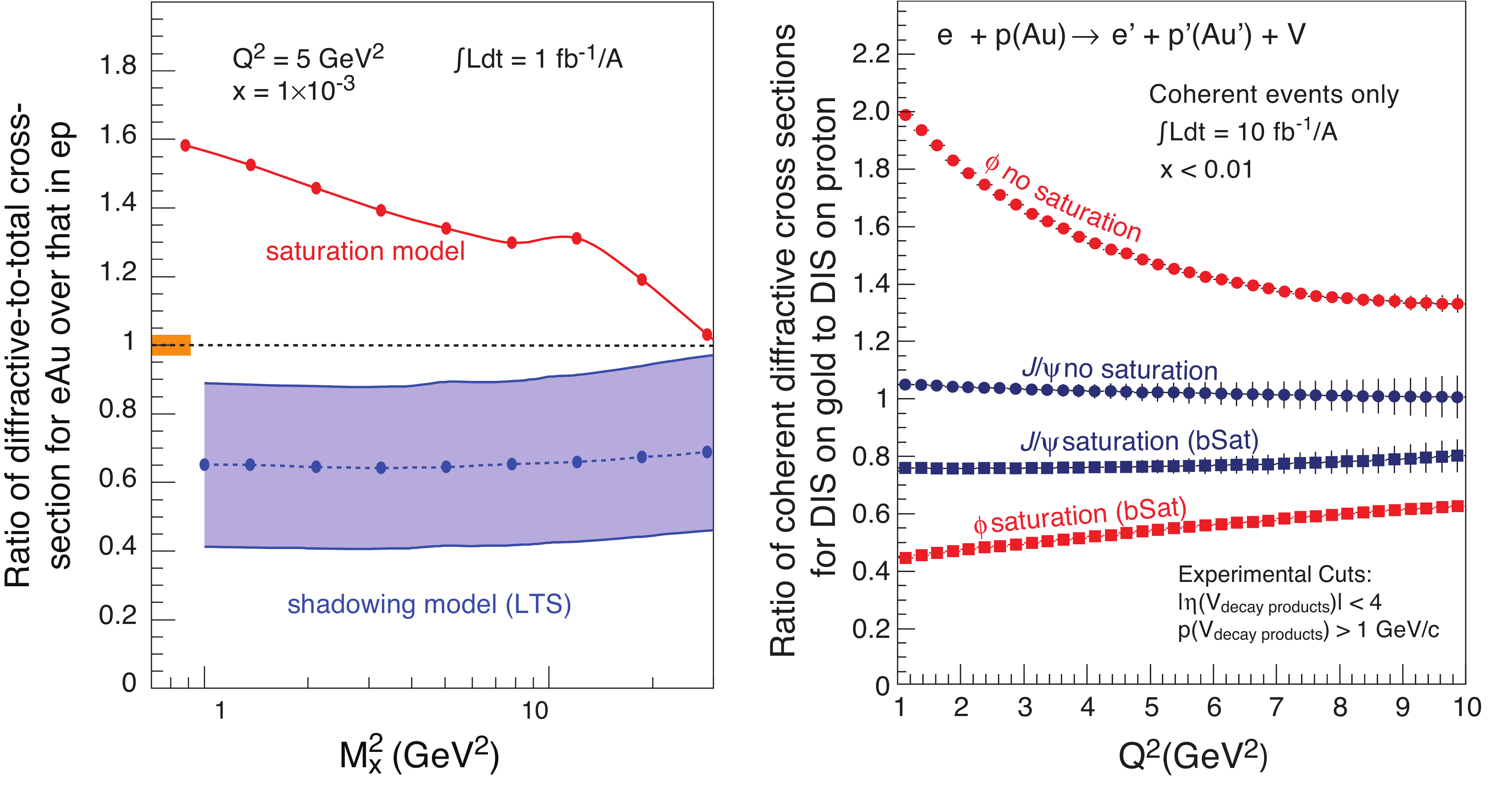}
    \caption{Left: Ratio of diffractive to total cross sections in a specific leading twist (collinear factorization) and saturation model. Right: ratio of exclusive vector meson production cross sections in a nucleus divided by the proton (scaled with $A^{4/3}$) with an without saturation, using the bSat/bNonSat models~\cite{Kowalski:2007rw}. Figure obtained from \cite{Accardi:2012qut}.}
    \label{fig:sigmadoversigmatot}
\end{figure}

Diffractive and exclusive cross sections are, quite generically, more sensitive to the effects of gluon saturation than inclusive cross sections. This is due to the fact that the  total cross section is, by the optical theorem, proportional to the elastic amplitude, whereas exclusive cross sections are quadratic in the same amplitude. 
 The EIC will provide the first ever measurements of nuclear diffractive DIS at small $x$.
The enhancement of the diffractive to total cross section ratio $\sigma_D/\sigma_\mathrm{tot}$ has long been considered as one of the clearest signs of saturation physics playing a role in DIS~\cite{Accardi:2012qut}, see Fig.~\ref{fig:sigmadoversigmatot}.  
The significantly higher luminosity of the EIC compared to HERA will be a major advantage for studies of diffractive scattering, as will the fact that the interaction region design takes the requirements of these processes into account better. The high statistics will enable measurements that are differential in the kinematical variables. Measurements of the total diffractive DIS cross section allow for extraction of the diffractive structure functions $F^{D,(4)}_{T,L}(\beta, x_\mathbb{P},Q^2,t)$ as functions of the size of the rapidity gap between the target and the diffractive system $\ln 1/x_\mathbb{P}$, the mass of the diffractive system $M_X^2\sim Q^2/\beta$ and the momentum transfer $t$.
Of particular interest for several reasons is diffractive dijet production~\cite{Altinoluk:2015dpi,Hatta:2016dxp,Mantysaari:2019csc,Salazar:2019ncp,Boer:2021upt}, which has been argued to be sensitive to saturation physics even at high values of $Q^2$~\cite{Iancu:2021rup}. 

Exclusive production of vector mesons~\cite{Kowalski:2003hm,Aschenauer:2017jsk}, deeply virtual Compton scattering (DVCS)~\cite{Hatta:2017cte,Mantysaari:2020lhf}, and timelike Compton scattering (TCS) are a particularly important subset of diffractive reactions in high-energy DIS. The clear experimental signature of these fully exclusive reactions provides access, in addition to the overall gluon density,  to the distribution of gluon fields in the transverse coordinate plane of the proton or nucleus, and to the fluctuations in this spatial distribution~\cite{Caldwell:2010zza,Lappi:2010dd,Klein:2019qfb}. Here the momentum transfer $t$ is the Fourier conjugate to the impact parameter, i.e., the transverse coordinate inside the proton or nucleus. The effects of gluon saturation have been claimed to  be visible directly in the $t$-dependence of the cross sections (see e.g.~\cite{Armesto:2014sma}). More importantly, the spatial structure of the gluon fields is interesting in itself as a fundamental property of QCD bound states, and useful for understanding and modeling the initial stages of heavy ion collisions~\cite{Mantysaari:2017cni}. In the Good-Walker paradigm~\cite{Good:1960ba,Klein:2019qfb} of high-energy scattering, the elastic scattering (in this case elastic $q\bar{q}$-dipole-target scattering, corresponding to diffractive DIS) the cross section is given by the square of the expectation value of the scattering amplitude. Thus the coherent cross section, where the target proton or nucleus stays in its ground state, measures the average of the target gluon density.

The incoherent cross section, where the target breaks up into color neutral fragments, depends on the variance of the scattering amplitude. Its magnitude and $t$-dependence reveal information about the magnitude and the spatial structure of  the fluctuations of the gluon fields at different length scales. Both coherent and incoherent measurements are necessary to develop a full picture of the structure of the small-$x$ gluon field. Comparing different vector mesons, for example those with different masses, but also using the ratio of exclusive photo-production cross-sections of $\Psi(2s)$ and $J/\Psi$ at the EIC (similar to the analysis done in ultraperipheral collisions \cite{Hentschinski:2020yfm, Bautista:2016xnp, Garcia:2019tne}) should also carry information on saturation effects. While in this case both vector mesons are characterized by a similar hard scale, the wave functions differ \cite{Krelina:2018hmt,Cepila:2019skb} and therefore lead to a different energy dependence for the ratio of photo-production cross-sections for models with and without saturation effects.

The azimuthal angular correlations between two particles in the final state are also a powerful tool to probe the structure and dynamics of gluons at small $x$ \cite{Kharzeev:2004bw}. Potential signatures of gluon saturation have been observed in the measurement of back-to-back dihadrons in deuteron-gold collisions at the Relativistic Heavy Ion Collider (RHIC) \cite{Albacete:2010pg,Lappi:2012nh,STAR:2021fgw}. Similar measurements can be made in nuclear deeply inelastic scattering (DIS) at the EIC, where the ability to use the scattered electron allows to better reconstruct the kinematics of the process \cite{Zheng:2014vka}. These measurements enable one to study linear gluon polarization~\cite{Dumitru:2015gaa,Dumitru:2018kuw}, transverse momentum distributions  \cite{Dominguez:2011wm,Kotko:2015ura,Dumitru:2016jku,Petreska:2018cbf,Kolbe:2020tlq,vanHameren:2021sqc}, multi-gluon correlations \cite{Altinoluk:2019fui,Altinoluk:2019wyu,Mantysaari:2019hkq,Boussarie:2021ybe} and ultimately the  Wigner distribution that encodes the combined spatial and momentum structure of the target~\cite{Hatta:2016dxp,Mantysaari:2019csc}, which at small $x$ are influenced by gluon saturation.

\begin{figure}[h!]
  \centering
  \includegraphics[width=0.6\textwidth]{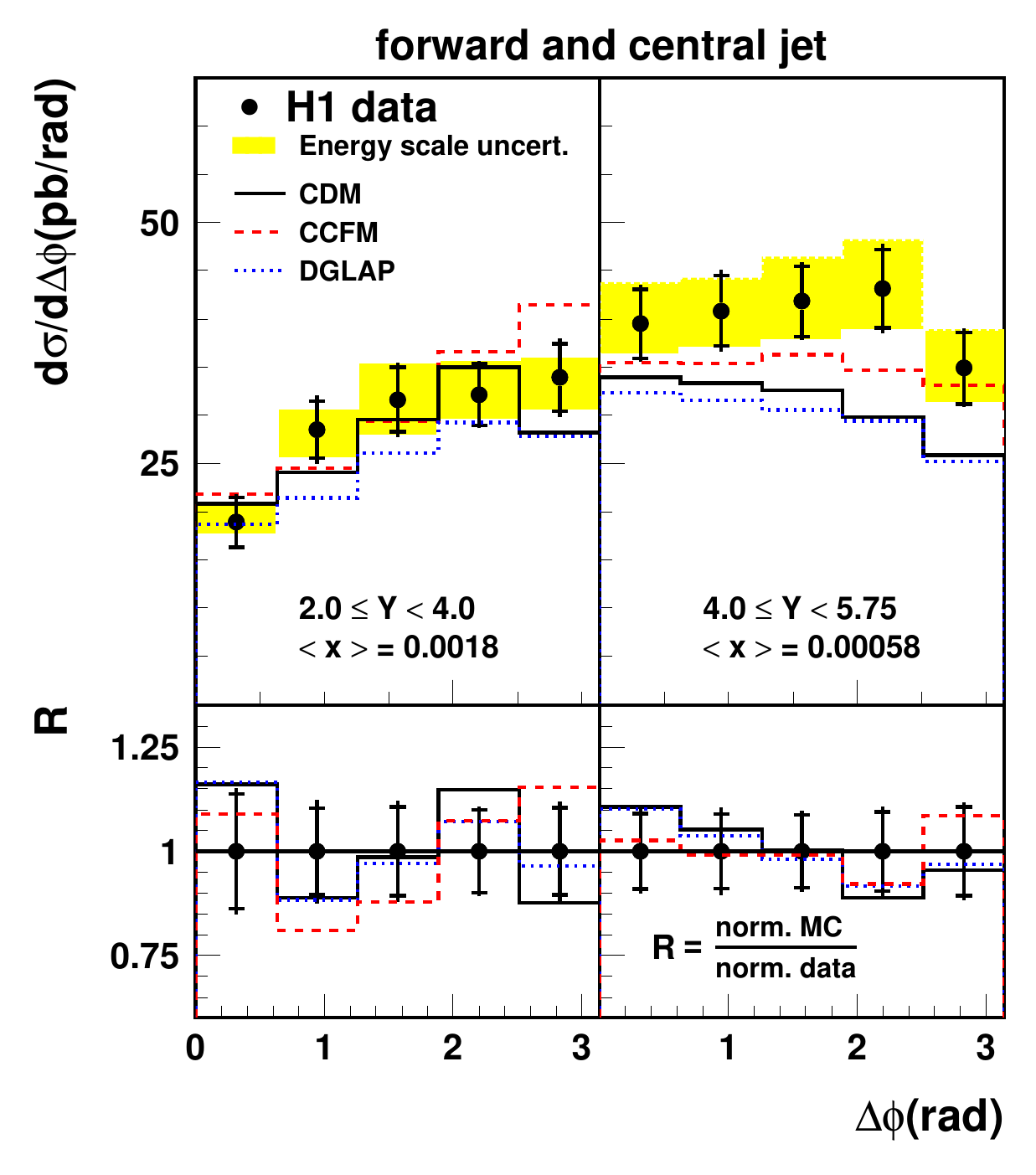}
  \caption{Forward and central jet cross sections as a function of the  azimuthal angle difference $\Delta \phi$ between the most forward jet and the scattered positron
in two intervals of the rapidity distance (reproduced from \cite{2012H1}).
}
\label{forwjH1}
\end{figure}

The effect of the BFKL evolution at low $x$ on various final-state signatures, such as jets and other experimentally-measured characteristics, were discussed in \cite{chekanov2022aspects}. 
Jet measurements at HERA show that the cross sections 
increase steeply towards small $x$, but the
predictions obtained from DGLAP parton shower simulations fall below the measurements.
As an example, some experimental limitations for forward jet measurements \cite{2012H1} at HERA 
are illustrated in Fig.~\ref{forwjH1}.  
The figure shows the cross sections  measured in two intervals of rapidity distance ($Y$),
$2.0 \le Y < 4.0$  and $4.0 \le Y \le 5.75$ compared to Monte Carlo simulations.
The  CASCADE Monte Carlo generator with
the Ciafaloni-Catani-Fiorani-Marchesini (CCFM)
evolution that unifies the DGLAP and BFKL approaches shows  a better agreement 
with the data compared to  DGLAP-based simulations
and the Colour Dipole Model (CDM) (see the corresponding references in \cite{2012H1}).
However,  these forward-jet  
measurements suffer from the lack of statistics
and a significant ($4\%$) hadronic energy scale uncertainty that gives rise to the dominant
uncertainty of $7\%$ to $12 \%$ for the measured cross sections.
The EIC experiment can make a unique contribution to  such measurements 
if jet energy scale uncertainties will be reduced to a percent level.


The gluon-rich environment at low $x$ also has many advantages for various measurements of experimental signatures that can be attributed to  possible instanton-induced processes, short- and long-range correlations, formation of possible glueballs, etc. \cite{chekanov2022aspects}. The long-range correlations \cite{Akushevich:1999dk} in the Breit frame can also be used for the  direct measurements of the spin effects to be discussed in the next section.

\subsection{Spin at small $x$}

Owing to its luminosity and high energy along with its possibility to use polarized beams, the EIC is expected to play a major role in answering one of the greatest questions of QCD: how the spin of a hadron emerges from that of its constituents. It has indeed been known for 35 years that the intuitive picture according to which the proton spin $1/2$ is a basic projection of the spin $1/2$ of its three valence quark is insufficient. The proper decomposition of this $1/2$ spin into contributions from quark and gluon helicity and orbital angular momentum is still an open question. The theoretical approach to the decomposition of hadron spins relies on spin sum rules, which relate the hadron spin to moments of parton distributions inside of it. For example, the Ji~\cite{Ji:1996ek} and Jaffe-Manohar~\cite{Jaffe:1989jz} spin sum rules
for the proton read, respectively
\begin{equation}
\frac{1}{2}=\frac{1}{2}\sum_{q}(\Delta q+L_{q}^{z})+J_{g}^{z},\label{eq:JiSR}
\end{equation}
and
\begin{equation}
\frac{1}{2}=\frac{1}{2}\sum_{q}(\Delta q+L_{q}^{z})+\Delta g+L_{g}^{z}.\label{eq:JMSR}
\end{equation}
Here, $J_{g}^{z}$ is the gluon total angular momentum, $\Delta q,\Delta g$ are helicity contributions from quarks and gluons, and $L_{q},L_{g}$ are their orbital angular momenta. All these quantities are moments of parton distributions. For example 
\begin{equation}
\Delta q=\int_{0}^{1}{\rm d}x\Delta f^{q}(x),\label{eq:DeltaQ}
\end{equation}
where $\Delta f^{q}=f^{q(+)}-f^{q(-)}$ is the quark helicity distribution, given as the difference between the number density of quarks with the same ($f^{q(+)}$) and opposite ($f^{q(-)}$) helicity as the parent hadron.

Experimentally, it is possible to constrain distributions such as $\Delta q$, but for a full understanding of the spin sum rules one has to know the integral of the distributions. In particular, sum
rules require a precise understanding of the $x\rightarrow0$ asymptotics. There are several technical difficulties when considering the small-$x$ asymptotics of helicity distributions. First and foremost, the transfer of longitudinal spin to and from a hadron is null in the
$x\rightarrow0$ limit, which means that one needs to go beyond the CGC approach and incorporate corrections that are subleading in powers of the energy (subeikonal corrections). The second difficulty is actually a consequence of the first one: for some observables, it is necessary to take into account the (energy-suppressed) exchange of quarks in the $t$ channel in addition to the small-$x$ gluons. Finally, we need to quantify the onset of saturation in polarized parton distributions to know how much, if at all, gluon saturation affects spin.

The first historical approach~\cite{Bartels:1995iu,Bartels_1996} to the small-$x$ dependence of helicity distributions relied on the resummation of double logarithms $(\alpha_s \ln^2 1/x)$ through the use of the so-called Infrared Evolution Equations~\cite{Kirschner:1982qf,Kirschner:1983di}. More recently, this approach was generalized in order to extract the small-$x$ dependence of orbital angular momentum distributions as well~\cite{Boussarie:2019icw}.
Modern approaches~\cite{Kovchegov:2015pbl,Kovchegov:2016zex,Kovchegov:2016weo,Kovchegov:2017jxc,Kovchegov:2017lsr,Kovchegov:2018znm,Cougoulic:2019aja,Kovchegov:2020hgb} rely on CGC-like semi-classical descriptions of observables which include saturation effects. Besides evolution equations, these approaches include an extension of the classic McLerran-Venugopalan model \cite{McLerran:1993ni,McLerran:1993ka}, which includes both saturation and longitudinal polarization effects. They led to the first comprehensive analysis taking small-$x$ helicity evolution into account~\cite{Adamiak:2021ppq}. Polarized DIS data from the future EIC will allow to distinguish these CGC-like analyses from more traditional predictions and provide a greater understanding of how much gluon saturation is involved in generating the proton spin.
The presence of the large saturation scale was also argued to lead to ``over the barrier" sphaleron transitions to dominate over instanton--anti-instanton configurations in causing spin diffusion at small $x$. Measurements of the polarized structure function $g_1$ at small $x$ therefore have the potential to provide concrete evidence of sphaleron transitions 
in QCD, with significant ramifications for the chiral magnetic effect in heavy-ion physics and to our understanding of axion  as well as sphaleron dynamics in the early universe \cite{Tarasov:2020cwl,Tarasov:2021yll}.

While longitudinal spin effects have been studied for some time because of their connection with the proton spin puzzle, the interest for transverse spin effects at small-$x$ has recently risen. Indeed, contrary to what could be expected from BFKL resummation schemes, semi-classical descriptions of small-$x$ physics have proven that transverse spin effects are not, in fact, subleading in the energy power expansion. Instead, transverse hadron spin effects relate to the Odderon~\cite{Zhou:2013gsa,Szymanowski:2016mbq,Boer:2015pni} and in fact the Odderon can relate to transverse hadron spin effects~\cite{Boussarie:2019vmk}. Transverse parton spin, as it turns out, also contributes at small $x$. Distributions for linearly polarized gluon pairs have been computed within the MV model~\cite{Metz:2011wb}. In fact, before TMD evolution is added on top of small-$x$ evolution, gluon distributions can be maximally linearly polarized~\cite{Boer:2017xpy}. Such transverse gluon spin distributions can in principle be observed at the EIC in processes with massive quarks by analogy with those described in~\cite{Marquet:2017xwy} for hadron-hadron colliders, or in processes with 3-body final states with appropriately constrained kinematics~\cite{Altinoluk:2020qet}.

\subsection{Towards precision: higher-order loops with gluon saturation} \label{sec:precision}

Hints of gluon saturation have been observed in different collider experiments \cite{Morreale:2021pnn}. While these signatures are consistent with the saturation picture, large theoretical uncertainties inherent to leading order computations, conceal the unambiguous discovery that gluons saturate inside hadronic matter. In recent years, tremendous efforts have been conducted to advance gluon saturation to a precision science, a timely endeavour as we approach the EIC era. These advances can be broadly classified as follows: the determination of evolution equations to next-to-leading logarithmic (NLL) accuracy $(\alpha_s^2 \ln 1/x )$ enhanced with (anti-)collinear resummation, the analytic computation of process-dependent and perturbatively calculable impact factors at next-to-leading order (NLO), and their proper numerical implementation and comparison to experimental data.

The most general set of equations that describe the energy evolution for the fundamental degrees of freedom at small-$x$, the correlators of light-like Wilson lines, are known as the Balitsky-JIMWLK hierarchy of equations. For most phenomenological applications, it is sufficient to consider the energy evolution of the dipole (two-point) correlator, which at large $N_c$ is given by a closed equation known as Balitsky-Kovchegov (BK) equation. At leading logarithmic (LL) accuracy the BK equation has been derived in \cite{Balitsky:1995ub,Kovchegov:1999yj} and the JIMWLK equations in \cite{JalilianMarian:1996xn,JalilianMarian:1997dw,Kovner:2000pt,Iancu:2000hn,Iancu:2001ad,Ferreiro:2001qy}. Their solutions predict too fast evolution with energy \cite{Albacete:2004gw}, suggesting the need to go beyond LL for controlled phenomenological predictions. An important subset of corrections beyond the LL evolution, pertaining to the running of the coupling, have been derived for the BK equation in \cite{Kovchegov:2006vj,Balitsky:2006wa} and for JIMWLK in \cite{Lappi:2012vw}. The former corrections have been extensively applied to phenomenology (see e.g. \cite{Albacete:2009fh,Albacete:2010sy,Lappi:2013zma,Mantysaari:2018zdd}). Presently, the NLL contributions to the BK equation have been derived in \cite{Balitsky:2008zza,Balitsky:2009xg,Balitsky:2009yp}. The first numerical results \cite{Lappi:2015fma}  showed that these corrections are unstable and contain potentially large negative terms. As in the case of BFKL \cite{Kwiecinski:1997ee,Salam:1998tj,Ciafaloni:1999yw,Altarelli:1999vw,Ciafaloni:2003rd,SabioVera:2005tiv}, the BK equation at this order requires resummation of large (anti-)collinear logarithms which have been implemented in its collinearly improved version \cite{Iancu:2015vea,Ducloue:2019ezk,Ducloue:2019jmy}.  The stability of the collinearly improved BK equation has been numerically confirmed in \cite{Lappi:2016fmu}, restoring its predictive power. Likewise, the JIMWLK set of equations at NLL has been obtained in \cite{Balitsky:2013fea,Kovner:2013ona,Kovner:2014lca,Lublinsky:2016meo}, and the collinearly improved version in \cite{Hatta:2016ujq}. The numerical implementation of collinearly improved JIMWLK equation is challenging as it requires the solution to a non-local Langevin equation, unlike the LL JIMWLK which is local in rapidity. Instead of evaluating the energy evolution of higher-point correlators using JIMWLK evolution, one can employ the Gaussian approximation \cite{Dumitru:2011vk,Iancu:2011nj}, which expresses any $n$-point correlator as a non-linear function of the dipole, and evolve the latter with the BK equation.

\begin{figure}[t]
    \centering
    \includegraphics[scale=0.7]{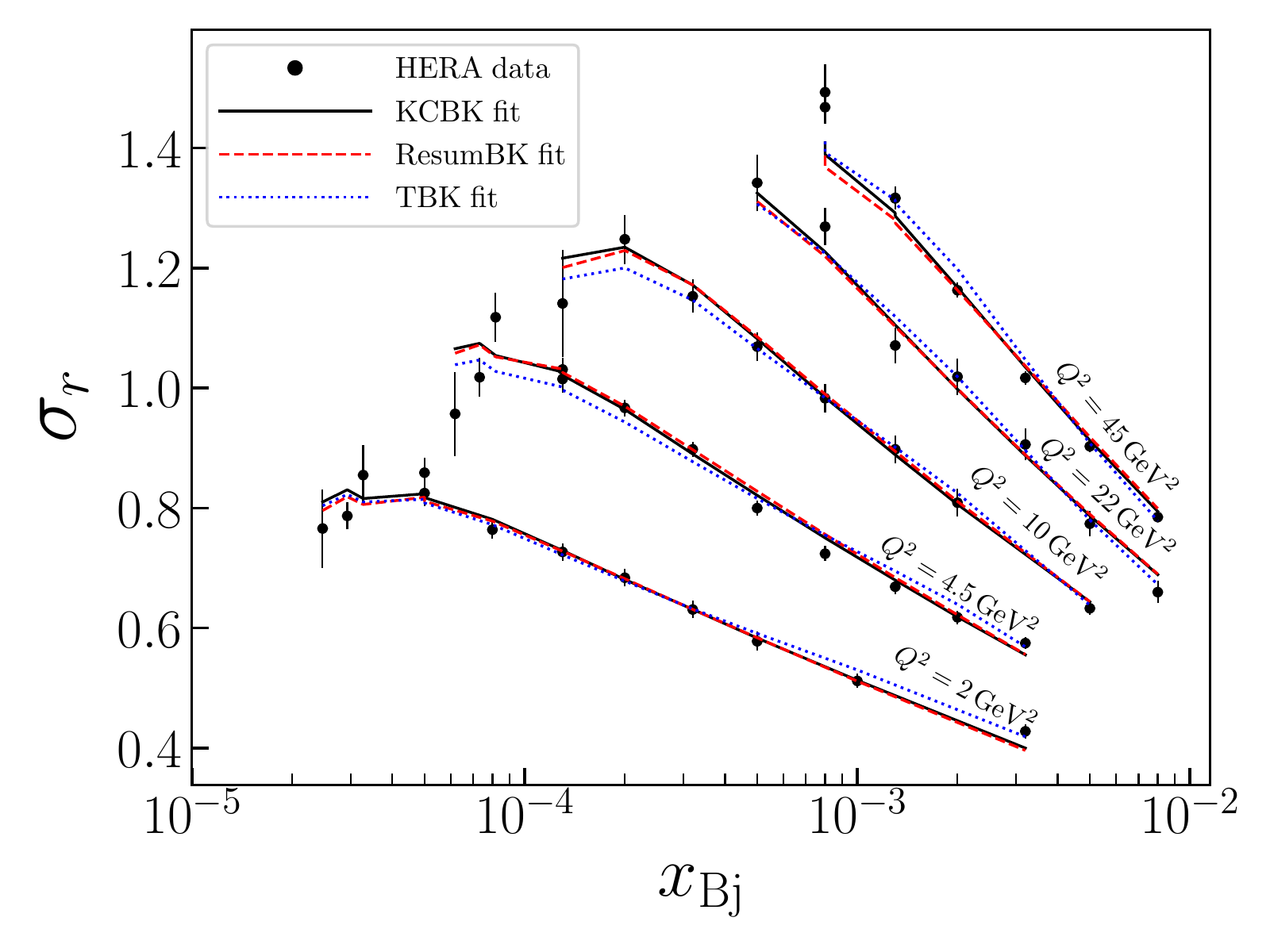}
    \caption{State of the art fit to the HERA reduced cross section within the saturation framework, employing NLO impact factor and NLL BK evolution equations. The curves represent different implementation of the BK evolution. Figure from \cite{Beuf:2020dxl}.}
    \label{fig:smallx-NLO-fit-reducedxsec}
\end{figure}

For precision computations of physical processes, one also requires high order computations of the corresponding impact factors. The impact factors for DIS structure functions, back-bone of small-$x$ observables, have been obtained at NLO for massless quarks in \cite{Balitsky:2010ze,Beuf:2017bpd,Hanninen:2017ddy}, with the first numerical results reported in \cite{Ducloue:2017ftk}. These efforts combined with the most dominant NLL contributions to the BK equation have resulted in the first fit of HERA data at NLO+NLL accuracy within the saturation framework \cite{Beuf:2020dxl}. The fit provides an excellent description of the reduced cross-section across different values of Bjorken-$x$ and $Q^2$ as shown in Fig.\,\ref{fig:smallx-NLO-fit-reducedxsec}. We expect that these results will be promoted to include the impact factor of massive quarks which has been recently derived in \cite{Beuf:2021qqa,Beuf:2021srj}, and could potentially reduce the uncertainties introduced by non-perturbatively large dipoles.

While compelling evidence for the necessity of small-$x$ resummation based on the linear BFKL evolution has been seen in the analysis of HERA structure functions at NLL + NNLO in \cite{Ball:2017otu}, the identification of non-linear effects due to saturation is less clear. This is not surprising since the values of the saturation scale $Q_s$ accessed at HERA remain in the non-perturbative domain \cite{Mantysaari:2018zdd}, thus non-linear effects are small for perturbatively reliable observables. The EIC could change this situation as it provides the opportunity to study nuclear structure functions where saturation effects are expected to be enhanced ($Q_s^2 \propto A^{1/3}$). Tension between non-saturated (BFKL) and saturated (BK) frameworks with a high level of precision (NLL+NLO) could point to the emergence of gluon saturation at the EIC.

The computation of the impact factors, and corresponding numerical implementation, for more differential observables at NLO is underway. Coherent diffractive (exclusive) measurements offer the opportunity to access a three-dimensional picture of saturated gluons inside nuclear matter. The impact factors have been derived for exclusive light vector mesons in \cite{Boussarie:2016bkq}, and exclusive dijets in \cite{Boussarie:2016ogo}. Partial numerical results for the real emissions in dijet production have been reported in \cite{Boussarie:2019ero}. The evaluation of heavy vector meson production has recently become available in \cite{Mantysaari:2021ryb} including numerical results in forward kinematics. A necessary component to extend these studies beyond the strict forward limit is a formalism that accounts for the impact parameter dependence of the evolution equations at NLL. Some progress in this direction can be found in \cite{Cepila:2018faq,Bendova:2019psy}; yet fully satisfactory theoretical treatment is still missing.

Recent developments at small $x$ have also been put forward for semi-inclusive processes, including single hadron production in proton-nucleus collisions \cite{Chirilli:2011km,Chirilli:2012jd}, dijet production \cite{Caucal:2021ent}, and dijet+photon production in DIS \cite{Roy:2019hwr}. The first numerical implementation of single hadron production provided a good agreement with data at low transverse momentum \cite{Stasto:2013cha} but yielded negative results at high $p_T$. The origin of negativity was traced to the subtraction scheme and the lack of kinematic constraints \cite{Beuf:2014uia,Altinoluk:2014eka,Watanabe:2015tja,Stasto:2014sea}. Several studies have attempted to resolve the issue of negativity by exploring different rapidity factorization schemes and the resummation of large collinear logarithms in the kernel of the NLL evolution equations \cite{Ducloue:2016shw,Iancu:2016vyg,Ducloue:2017mpb,Ducloue:2017dit}. Alternatively, it has been suggested that stable results can be obtained via resummation of threshold logarithms \cite{Liu:2020mpy,Shi:2021hwx} while keeping leading logarithmic accuracy in the evolution equations. The recent comprehensive analysis in \cite{Shi:2021hwx} has shown that the saturation framework endowed with NLO impact factor and threshold resummation can describe all the existing data for forward hadron production at RHIC and LHC. We note that this analysis has employed the running coupling BK equation \cite{Kovchegov:2006vj,Balitsky:2006wa} and not the full NLL evolution, whose implementation with simultaneous threshold resummation is unknown. On the other hand, the numerical implementation of inclusive dijet and dihadron production is not available at NLO. We expect that a complete computation will require careful treatment of the rapidity subtraction procedure, and the resummation of potential large logarithms in some kinematic windows, such as Sudakov logarithms in the back-to-back regime \cite{Mueller:2013wwa}. 

We close this section by briefly mentioning a connection between small-$x$ and jet physics that could boost our understanding of higher-order loop computations. Recently, loop computations in high-energy scattering for gauge theories have taken advantage of the remarkable correspondence with the physics of soft wide-angle radiation \cite{Weigert:2003mm,Hofman:2008ar,Hatta:2008st,Mueller:2018llt}. Evidence of this so-called spacelike-timelike correspondence was the fact that the BMS equation \cite{Banfi:2002hw}, which describes non-global
properties of jet decays \cite{Dasgupta:2001sh}, perfectly maps to the BK equation.  This strategy combined with known results from the vast amplitude literature has resulted in the computation of the BFKL equation and BK equation to three loops for $\mathcal{N}$=4 super Yang-Mills \cite{Caron-Huot:2015bja,Caron-Huot:2016tzz}. An exciting possibility is the application of this program to compute non-linear evolution equations in QCD beyond NLL, and to the evaluation of impact factors to higher-order loops.

\subsection{Other theory developments}
\subsubsection{Initial conditions}
All non-perturbative information specific to the quantum numbers of a given nucleus is encoded in the initial conditions of the evolution towards small $x$. While at asymptotically small $x$, at a fixed impact parameter, an external probe would be unable to distinguish between a heavy nuclear target and a proton, for any realistic values of $x$ the initial conditions can affect the gluon distribution significantly. Improving the description of initial conditions consequently has to be an essential component of any program that aims to calculate observables at the EIC with high precision. Currently, common methods employ Gaussian approximations for color charge distributions at the threshold of the gluon dominated regime. These are, however, only valid in the asymptotic limit of infinite nuclei, and corrections to such approximations are expected to be large. 

Recent work has employed light front \cite{Brodsky:1997de} truncation approximations with systematic inclusion of higher Fock states to determine initial conditions for the small-$x$ evolution. This has been done in particular for the proton, which is arguably the least well described by the Gaussian approximation for large nuclei. Its light front wavefunction is dominated by its valence quark state and it can be used to compute color charge correlators, such as the usual two-body term, but also higher-order correlators, such as the odderon term, etc.~\cite{Dumitru:2018vpr}.  The expectation values of quadratic and cubic combinants of the color charge density can be reexpressed in terms of nonperturbative color charge form factors, which are nonperturbative quantities that can e.g.~be determined from exclusive measurements of heavy Quarkonia in DIS at large $x$. First estimates employing a model wave function show that the color charge correlators (as appearing in e.g.~dipole scattering off the target) obtained using this method are far from Gaussian, and depend on impact parameter, distance between the two probes (or their relative momentum), and their relative angular orientation, making it equivalent to a Generalized Parton Distribution (GPD) \cite{Altinoluk:2019wyu}.

The calculation using the three quark Fock state can be systematically improved by inclusion of more partons, i.e., extension to higher Fock states \cite{Dumitru:2020gla}. The inclusion of one gluon was shown for example to have a mild effect on the color charge correlator at $x=0.1$, but a large effect at $x=0.01$ \cite{Dumitru:2021tvw}, the value usually used to initialize the small-$x$ evolution. Three body correlators have also been computed using the three quark plus gluon Fock state. Its knowledge is important for various spin dependent Transverse Momentum Dependent (TMD) distributions such as the (dipole) gluon Sivers function of a transversely polarized proton. The formalism does get increasingly complex when going to larger nuclei and higher Fock states, and will likely only be possible to implement for the smallest nuclei. Alternatively, one may choose to start with the MV model for large nuclei and systematically introduce corrections in the form of non-Gaussian correlations, e.g.~guided by phenomenological constraints as done in \cite{Dumitru:2011ax}.

Lattice QCD provides another venue for determining initial conditions for small-$x$ evolution. Using Large Momentum Effective Theory (LaMET) \cite{Ji:2020ect} parton distribution functions as well as transverse momentum dependent PDFs (TMDs), generalized PDFs (GPDs), and light-front wave functions can in principle be computed on the lattice. This is done via the quasi-PDF \cite{Ji:2013dva}, the momentum distribution of partons in a proton (or nucleus) at large ($z$) momentum, which then has to be matched to the usual PDFs in the limit of infinite momentum. Alternatively, PDFs can be computed via the use of coordinate factorization that leads to pseudo-PDFs \cite{Radyushkin:2017cyf, Balitsky:2019krf}. 

To be useful as input for small-$x$ evolution the partonic structure should be extracted at $x$ values that reside at the threshold of the gluon dominated regime, around $x=0.01$. To push above methods of extracting PDFs to such small $x$ on the lattice is challenging, because small-$x$ partons are generated from long-range correlations of quantum fields along the longitudinal direction. To get $x=0.01$, and using $x\sim \Lambda_{\rm QCD}/P^z$ one needs large $P^z$ of approximately 20 GeV. To resolve the correspondingly large valence quark momentum, the lattice spacing needs to be much less than 0.1 fm, and as the proton is Lorentz contracted to 0.05 fm in the longitudinal direction, proper resolution requires approximately $a=0.01$ fm. At $x=0.01$ to compute the relevant correlation lengths one needs at least a lattice size of 1 fm. Thus, extremely fine lattices, in particular in the longitudinal direction will be needed. Asymmetric lattices may be a first step, also, it might be possible to integrate out modes at small lattice spacing and by that generate an extended momentum dependent source, which, once it has a scale of 0.05 fm, can be put on the lattice to perform simulations with more manageable lattice sizes. It is thus a challenging task to push lattice calculations towards small $x$, but some promising ideas are surfacing.

\subsubsection{Entanglement entropy \& Density matrix of small-$x$ gluons}
One fundamental question associated with the structure of hadrons is the question of how to reconcile the 
nature of the proton as a pure quantum eigenstate of QCD with vanishing entropy, with the phenomenologically successful notion of the proton as an incoherent collection of partons with a non-vanishing entropy. Starting with the work of \cite{Kharzeev:2017qzs} this phenomenon has been attributed to the fact that e.g. in DIS experiments one effectively only probes a part of the hadronic wavefunction, such that a non-vanishing entanglement entropy can be generated by tracing out the unobserved environment. By inspecting the branching cascade associated with the small-$x$ evolution, it was further argued that the entanglement entropy is proportional to the (unpolarized) parton distribution at small-x, implying a picture where DIS experiments in the small-$x$ limit probe the proton as an equiprobable collection of all partonic microstates which maximes the entanglement entropy~\cite{Kharzeev:2017qzs}. By invoking additional assumptions on the hadronization process, phenomenological consequences for hadron multiplicity distributions and derived entropy measures in DIS and hadronic collisions have been explored in ~\cite{Kharzeev:2021yyf,Hentschinski:2021aux,H1:2020zpd,Gotsman:2020bjc,Ramos:2020kaj}. The EIC can be expected to shed further light on this
rapidly developing field.

Entanglement and related concepts from quantum information theory have also started to play a more prominent role for theoretical developments in small-$x$ physics. Generally, the CGC effective theory of high-energy QCD is based on a separation of time scales between the fast non-linear dynamics of small-$x$ gluons, and large-$x$ degrees of freedom, which can be treated as quasi-static eikonal color charges in the high-energy limit. However, since small-$x$ gluons are sourced by the large-$x$ degrees of freedom, the evaluation of any cross section ultimately requires to perform an average over the configurations of large-$x$ color charges. Conventionally, this average has been performed in terms of a classical weight-functional $W[j]$, whose renormalization group evolution is described by the JIMWLK evolution equation. Over the last years, a series of works \cite{Kovner:2015hga,Kovner:2018rbf,Duan:2021clk} has started to formalize the separation between small-$x$ and large-$x$ degrees of freedom by considering the reduced density matrix of small-$x$ gluons upon tracing out the large-$x$ degrees of freedom. Calculations of the entanglement entropy of small-$x$ gluons have been carried out in \cite{Kovner:2015hga,Kovner:2018rbf} and show an entropy growth with decreasing $x$, while explicit diagonalization of the reduced density matrix in \cite{Duan:2021clk} also sheds new insights into the properties of small-$x$ gluons as quasi-particles in the hadronic wave-function. Conceptual developments in \cite{Armesto:2019mna} also include the generalizing concept of a probability density $W[j]$ to a density matrix  $\rho[j,\bar{j}]$, which satisfies a Markovian evolution equation of the Lindblad form for an open quantum system. Small-$x$ evolution of the density matrix in the dilute regime was found to lead to a reduction of purity and increase of entropy with decreasing $x$. Notably, this formulation in terms of a density matrix $\rho$ is also beneficial to formulate the challenging problem of correlated multi-particle production at different rapidities~\cite{Kovner:2006wr,Iancu:2013uva} and may provide a unified framework for future theoretical developments.

\subsection{Event generators}

There are not that many Monte Carlo event generators purely based on small-$x$ phenomenology but there is plenty of common ground with general purpose event generators and generators specializing on collisions with heavy-ions. In addition, generators specialized in nuclear break-up modelling are needed for studying the collision geometry important for saturation studies. Here we discuss some relevant features of (a subset) generators where recent EIC-related improvements have been in the works.

\textsc{Pythia~8} \cite{Sjostrand:2014zea} is a general-purpose event generator which focuses on generating a complete, exclusive, description of final states in hadronic collisions. The generation is divided into different stages which are roughly ordered in terms of energy or transverse momentum scale. The underlying hard scattering process is further evolved with DGLAP-based parton showers (PS), providing an effective leading-logarithmic resummation. The underlying event is formed by multiparton interactions (MPIs) that are based on regularized QCD cross sections and are generated simultaneously with the PS \cite{Sjostrand:2004pf, Sjostrand:2004ef}. The partonic event is then hadronized with the Lund string model. The features relevant for EIC physics include two options for parton showers applicable for high-$Q^2$ DIS processes \cite{Cabouat:2017rzi, Hoche:2015sya} and a framework for photoproduction at $Q^2 = 0~\text{GeV}^2$ \cite{Helenius:2017aqz}. The latter can be applied also for diffractive processes including high- and low-mass single diffraction, hard diffraction and elastic scattering modelled with the vector-meson-dominance model (VMD) \cite{Helenius:2019gbd}. As discussed above, diffractive jet production is one of the key observables also for saturation physics, so having a complementary, factorization-based Monte-Carlo framework that can be used to study hadronization effects for such processes is very useful. Notably the hard diffraction framework in Pythia does provide a natural explanation for the factorization breaking effects in photoproduction of diffractive dijets observed at HERA via the dynamical rapidity gap survival model. The model predicts only a mild reduction in cross sections at the energies accessible with the EIC. The foreseen EIC-related developments include a model for intermediate photon virtualities and electron-ion collisions using the \textsc{Angantyr} heavy-ion model in \textsc{Pythia~8}.

{\textsc{Hijing}} is primarily focused on event generation in heavy-ion collisions \cite{Wang:1991hta, Deng:2010mv}. It borrows many features from \textsc{Pythia} to model the nucleon-nucleon interactions and adds up nuclear shadowing for PDFs and further modelling of jet quenching from interactions of high-energy partons with the QCD medium. As a relevant feature for the EIC, also parton propagation through ``cold'' nuclear matter in DIS events leading to parton energy loss can be modelled with Hijing using a generalized higher-twist approach including the LPM effect \cite{Zhang:2019toi}. This will be included in the upcoming eHIJING version, which can be used to generate DIS-type collisions between electrons and heavy nuclei. The interactions between virtual photons and nucleons are taken from \textsc{Pythia~8} but the parton showers are modified to account for possible interactions with the other nucleons within the target. The cross sections for the multiple collisions are obtained using unintegrated gluon distributions for which a parametrization including a simple model for saturation is applied. The hadronization is done using the Lund string model. This setup has been compared to SIDIS data from CLAS and HERMES with light and heavy nuclear target. Future plans include an implementation of small-$x$ gluon evolution within the eHIJING which is currently under development.

{BeAGLE} is capable of simulating hard process but the main emphasis is to accurately model the nuclear remnants in electron-ion collisions, for further details see e.g. Ref~\cite{Tu:2020ymk, Chang:2021jnu}. The event generation begins by modelling the hard interaction that can be DIS-like or diffractive. BeAGLE applies \textsc{Pythia~6} \cite{Sjostrand:2006za} to generate the partonic scattering processes and relies on DPMJet \cite{Roesler:2000he} to model the target and intra-nucleon cascade. The decays of excited nuclear remants are handled with FLUKA \cite{Ferrari:2005zk}. There are two possible modes which differ on how the small-$x$ nuclear shadowing is accounted for. In the first mode a DIS event is considered as a point-like photon scattering of a single quark and the shadowing is obtained from EPS09 nuclear PDFs \cite{Eskola:2009uj}. The second option is based on a dipole model where the virtual photon fluctuates into a quark-antiquark dipole that can then interact with several nucleons, their number obtained from a Glauber simulation, that results in effective nuclear shadowing. In this case the first interaction is modelled as an inelastic collision with \textsc{Pythia} and the following ones are taken as elastic scatterings between the nucleon and the most forward parton from the first interaction. However, in case of large-$x$ scattering, the generation is reverted back to the point-like interaction mode above. After the hard part is dealt with, the fragmentation is simulated with \textsc{Pythia} using the parton showers and string hadronization and, additionally, it is possible to study fragmentation inside the nucleus by applying the PyQM partonic energy-loss module \cite{Dupre:2011afa}. The stand-out feature of BeAGLE is, however, the accurate modelling of the nuclear remnants. This is achieved with interfacing to FLUKA, which contains the most important evaporation and breakup processes for the excited nuclei. Accurate modelling of the nuclear remnant is crucial for studying the geometry of the collision that allows to classify the events in terms of centrality. Being able to pick out the events with small impact parameter it is possible to increase the sensitivity to saturation phenomena at small-$x$, which is expected to grow with energy and target size. Furthermore, accurate remnant modelling is required also to separate coherent and incoherent diffraction that can be used to study the nuclear structure. Currently the weak point in the simulation chain is the description of nucleonic remnants and the multiple nucleon collisions of the virtual photon, for which further developments are necessary.

\textsc{Sartre} \cite{Toll:2013gda} is a Monte-Carlo event generator for exclusive diffractive vector meson production and deeply-virtual Compton scattering (DVCS) based on the dipole model originally proposed by Golec-Biernat and Wüsthoff \cite{Golec-Biernat:1998zce}. In this picture the incoming (transversely or longitudinally polarized) virtual photon fluctuates into a color dipole (quark-antiquark pair) that interacts with the target hadron where the small-$x$ saturation is built-in to the scattering amplitude. \textsc{Sartre} includes the refined dipole model IPSat \cite{Kowalski:2003hm}, and its linearized version IPNonSat, that can be used to quantify the small-$x$ saturation effects. These models have been extended to handle collisions with a heavy-ion target \cite{Toll:2012mb} by constructing the nuclear scattering amplitude as a combination of proton scattering amplitudes. Also another dipole model, b-CGC \cite{Kowalski:2006hc}, providing an alternative modeling of gluon saturation, has been implemented. The current version of the code (1.33) includes $J/\Psi$, $\phi$, $\rho$ and $\Upsilon$ vector-meson states in addition to real photons for DVCS, and an improved fit to combined HERA data for ep collisions. Later on, the new parameter set will be included also for e+$A$ collisions and a long-term plan is to also include inclusive processes into the same framework.

To conclude, there have been some efforts to improve the existing event generators, and to write new ones, to handle physics relevant for the EIC. However, especially e+$A$ collisions are covered only for a subset of processes and kinematics and more effort is needed to get to the same level of sophistication and selection as what is currently available for LHC physics.

\subsection{Summary}
The study of the small-$x$ regime offers unique discovery potential at the Electron-Ion Collider.
At high energies gluon densities grow quickly, and it is expected that this growth is controlled by non-linear QCD effects at sufficiently small $x$, leading to saturation of the gluon density. Such gluon saturation could be definitively confirmed in collisions of electrons with heavy nuclei, for which we expect to reach the saturated regime at $x$ values accessible at the Electron-Ion Collider. This is because electron-ion collisions offer a cleaner and more controlled environment compared to nuclear collisions, and the nuclear enhancement of the saturation scale offers an important advantage over e+p collisions. Further, a variety of observables with sensitivity to saturation are accessible in e+A collisions. 

This includes structure functions, which are expected to be modified, compared to leading twist results, when saturation effects are included, and semi-inclusive observables, for example forward  dihadron azimuthal correlations, which are expected to show a characteristic nuclear dependence, driven by saturation effects. Furthermore, diffractive observables can be particularly sensitive to saturation effects, as they involve the exchange of at least two gluons, making them sensitive to the gluon distribution squared (at lowest order in perturbation theory). This includes the total diffractive cross section, as well as diffractive vector meson and dijet (dihadron) production. 

To use the Electron-Ion Collider's potential to its fullest, and set the stage for the unambiguous discovery of gluon saturation, important developments are needed on the theory side. These include improved initial conditions for small-$x$ evolution, which can be provided e.g.~by lightcone perturbation theory calculations, or Lattice QCD, the advancement of small-$x$ evolution equations and impact factors to next-to-leading logarithmic accuracy or next-to-leading order (NLO) and beyond, and their numerical implementation, as well as hadronization prescriptions and Monte-Carlo implementations to allow for sophisticated phenomenology. Apart from increasing precision by going to NLO in the coupling constant or further, other improvements, such as sub-eikonal corrections will be highly desirable.

Besides the discovery of saturation, the study of the small-$x$ regime at the EIC has the potential to provide new insights into the spin of protons and nuclei, in particular the contribution from gluons to the spin sum rule of the proton. Also, new insights into the role of entanglement in particle production have the potential to provide a deeper understanding of saturation. This aspect is strongly tied to significant research activity in quantum information science and provides important interdisciplinary connections.

%
%
\newpage
\begin{acknowledgments}
  \begin{itemize}
  \itemsep -1ex
  \item[--] This material is based upon work supported by The U.S. Department of Energy, Office of Science, Office of Nuclear Physics through Contract No. DE-SC0012704.
  \item[--] This material is based upon work supported by The U.S.~Department of Energy, Office of Science, Office of Nuclear Physics through Contract No.~DE-SC0012704 and No.~DE-AC05-06OR23177  under which Jefferson Science Associates, LLC, manages and operates Jefferson Lab.
  \item[--] Argonne National Laboratory's work was supported by the U.S.~Department of Energy, Office of Science, under Contract No.~DE-AC02-06CH11357.
  \item[--] This manuscript has been authored by Fermi Research Alliance, LLC under Contract No.~DE-AC02-07CH11359 with the U.S.~Department of Energy, Office of Science, Office of High Energy Physics.
  \item[--] This project has received funding from the European Union’s Horizon 2020 research and innovation programme under grant agreement STRONG 2020 - No 824093.
  \item[--] In addition to Fermi Research Alliance, the work of T.~J.~Hobbs was supported at earlier stages by a JLab EIC Center Fellowship and the U.S.~Department of Energy under Grant No.~DE-SC0010129.
  \item[--] The work of A.~Prokudin is partially supported by the National Science Foundation under Grant No.~PHY-2012002.
  \item[--] T.~Cridge and R.~S.~Thorne thank the Science and Technology Facilities Council (STFC) for support via grant awards ST/P000274/1 and ST/T000856/1.
  \item[--] L.~Harland-Lang thanks STFC for support via grant awards ST/L000377/1 and ST/T000864/1.
  \item[--] The work of H.-W.~Lin is partially supported by the US National Science Foundation under grant PHY 1653405 and by the  Research  Corporation  for  Science  Advancement through the Cottrell Scholar Award.
  \item[--] The work of D.~Pitonyak has been supported by the National Science Foundation under Grant No.~PHY-2011763.
  \item[--] The work of P.~Schweitzer was partially supported by NSF grants 1812423 and 2111490.
  \item[--] This material is based upon work supported by  is supported by the  Laboratory Directed Research and Development Program at LANL.
  \item[--] This material is based upon work supported by The U.S. Department of Energy, Office of Science, Office of Nuclear Physics through Contract No. 89233218CNA000001
  %
  %
   \item[--] M.~Hentschinski acknowledges support by Consejo Nacional de Ciencia y Tecnología grant number A1 S-43940 (CONACYT-SEP Ciencias Básicas). 
  \item[--] This material is based upon work supported by the U.S. Department of Energy, Office of Science, Office of Nuclear Physics under Award Number DE-SC0004286.
  \item[--] F.~Salazar acknowledges support from the National Science Foundation under grant No. PHY1945471. 
  \item[--] I.~Helenius and T.~Lappi acknowledge support from the Academy of Finland, projects number 331545 and 321840 and have been funded as a part of the CoE in Quark Matter of the Academy of Finland (projects 346324 and 346326).
\end{itemize}
\end{acknowledgments}
%
%
\newpage
\bibliographystyle{apsrev4-2}
\bibliography{bsm/ref,heavyflavor/ref,smallx/ref,tomography/ref,jets/ref}
%
\end{document}